\documentclass[11pt]{article}
\usepackage{hyperref}
\usepackage{graphicx}
\usepackage{graphics}
\usepackage{dsfont}
\usepackage{epsfig}
\usepackage{amsmath,amssymb,amsthm,amscd}
\usepackage[sort&compress,square,numbers]{natbib} 
\usepackage{float}
\restylefloat{table}

\newcommand{\nn}{{\nonumber}}

\def\Mgn[#1]#2{{\overline{\cal M}_{#1,#2}}}
\def\pqs[#1,#2]{{\footnotesize{$\left[\begin{array}{c} #1\\#2  \end{array}\right]$}}} 
\def\pqsu[#1,#2]{\left[\begin{array}{c} #1\\#2  \end{array}\right]} 
\def\pqssu[#1,#2]{{\footnotesize{\left[\begin{array}{c} #1\\#2  \end{array}\right]}}} 
\def\pqh[#1,#2]{{\footnotesize{$\left[\begin{array}{c} #1\\#2  \end{array}\right]$}}} 
\def\pqhu[#1,#2]{\left[\begin{array}{c} #1\\#2  \end{array}\right]} 
\newcommand{\ba}{\begin{eqnarray*}}
\newcommand{\ea}{\end{eqnarray*}}
\newcommand{\ban}{\begin{eqnarray}}
\newcommand{\ean}{\end{eqnarray}}
\newcommand{\be}{\begin{equation}}
\newcommand{\ee}{\end{equation}}
\newcommand{\ben}{\begin{equation}}
\newcommand{\een}{\end{equation}}
\numberwithin{equation}{section}

\newcommand{\del}{\partial}

\newcommand{\Res}{\textrm{Res}}

\numberwithin{equation}{section}

\begin{document}
\begin{titlepage}
{}~ \hfill\vbox{ \hbox{} }\break

\rightline{
USTC-ICTS-13-10}
\rightline{
Bonn-TH-2013-10}

\vskip 3 cm

\centerline{\Large \bf  Refined stable pair invariants for}  \vskip 0.5 cm
\centerline{\Large \bf  E-, M- and $[p,q]$-strings}   \vskip 0.5 cm

\renewcommand{\thefootnote}{\fnsymbol{footnote}}
\vskip 30pt \centerline{ {\large \rm Min-xin Huang\footnote{minxin@ustc.edu.cn}, 
Albrecht Klemm\footnote{aklemm@th.physik.uni-bonn.de}  and Maximilian Poretschkin\footnote{poretschkin@th.physik.uni-bonn.de}} } \vskip .5cm \vskip 20pt

\begin{center}
{$^*$Interdisciplinary Center for Theoretical Study, University of Science and \\ \vskip 0.2 cm Technology of China,\  Hefei, Anhui 230026, China}\\ [3 mm]
{$^{\dagger\ddagger}$Bethe Center for Theoretical Physics and $^\dagger$Hausdorff Center for Mathematics,}\\ 
{Universit\"at Bonn, \  D-53115 Bonn}\\ [3 mm]
\end{center}

\setcounter{footnote}{0}
\renewcommand{\thefootnote}{\arabic{footnote}}
\vskip 60pt
\begin{abstract}
We use mirror symmetry, the refined holomorphic anomaly equation 
and modularity properties of elliptic singularities to calculate the refined 
BPS invariants of stable pairs on non-compact Calabi-Yau manifolds, 
based on del Pezzo surfaces and elliptic surfaces, in particular the 
half K3. The BPS numbers contribute naturally to the five-dimensional 
${N}=$1 supersymmetric index of M-theory, but they can  be 
also interpreted in terms of the superconformal index in six 
dimensions  and upon dimensional reduction the generating 
functions count $N=2$ Seiberg-Witten gauge theory instantons 
in four dimensions.  Using the M/F-theory uplift the additional 
information encoded in the spin content can be used  
in an essential way to obtain information about BPS states 
in physical systems associated to small instantons, tensionless strings, 
gauge symmetry enhancement in F-theory by $[p,q]$-strings as 
well as M-strings. 
\end{abstract}

\end{titlepage}
\vfill \eject


\newpage

\baselineskip=16pt    

\tableofcontents

\section{Introduction}

Decisive information about supersymmetric theories in 
various dimensions is encoded in their BPS spectrum 
and recently much progress has been made in defining  
refined BPS indices. The latter are generating functions  
for the multiplicities of refined BPS states, labelled by 
a set of commuting quantum numbers, which is optimal in 
the sense that they encode as much information as possible, 
while keeping computability and moduli dependence of the BPS spectrum 
under some control.  

In five-dimensional supersymmetric theories with eight supercharges and an 
$U(1)_{\cal R}$ ${\cal R}$-symmetry living on space-time, which have at least an 
$U(1)_L \times U(1)_R$ isometry acting as a subgroup of the 
rotation group $SO(4)$, such an index for refined BPS  states 
has been suggested in~\cite{Nekrasov:2002qd} and mathematical 
rigerously defined in~\cite{CKK}\cite{NekrasovOkounkov} as   
\be
Z_{{\cal BPS}}(\epsilon_L,\epsilon_R,t) =
{\rm Tr}_{{\cal BPS}} (-1)^{2 (J_L+J_R)}
e^{-2 \epsilon_L J_L} e^{-2 \epsilon_R J_R} 
e^{-2\epsilon_{R} J_{\cal R}} e^{\beta H}\  \, .
\label{index}
\ee
In flat space the $U(1)_L \times U(1)_R$ 
acts as the Cartan subgroup of the $SU(2)_L \times SU(2)_R$ 
defining the left and right spin representation. 
One denotes by $J_*$ the Cartan element 
$J^3_*$ of $SU(2)_*$ and by $j_*$ a $SU(2)_*$ spin 
representation or the eigenvalue of the Casimir. 
To make this an index it is essential to twist $J_R$ 
by the $U(1)_{\cal R}$ ${\cal R}$-symmetry.

This index counts the multiplicities $N_{j_L,j_R}^Q$ of five-dimensional  
BPS  states, where $Q$ denotes their charges. These states  
are of particular interest for theories, which come 
from M-theory compactification on local Calabi-Yau 
threefold geometries since this theories are, depending on 
additional  fibration structure on the local geometry, 
related to E-strings~\cite{Klemm:1996hh, Minahan:1997} 
\cite{Minahan:1997a,Minahan:1998,Eguchi:2002nx},
small instantons~\cite{Witten:1995gx,Ganor:1996mu,Klemm:1996hh,Douglas:1996xp}, F-theory gauge 
symmetry enhancement by string junctions~\cite{Johansen:1996am,Gaberdiel:1997ud,Gaberdiel:1998mv,DeWolfe:1998eu}, 
$[p,q]$-brane webs  -- most to the context in~\cite{Benini:2009gi}--, 
M-strings~\cite{Haghighat:2013gba}, flat bundles on elliptic 
curves~\cite{Looijenga1} in heterotic string compactifications~\cite{Friedman:1997yq} 
and ${N}=4$ topological Yang-Mills theory on complex 
surfaces~\cite{VW,Yoshida, Minahan:1998}. Upon further 
compactifications one can reach interesting, especially conformal, 
Seiberg-Witten theories in four dimensions and upon decompactification 
one gets valuable information about $(1,1)$ and $(2,0)$  
theories in six dimensions e.g. about theories living on a stack 
of M5-branes. The topological string on the 
del Pezzo surfaces captures~\cite{Drukker:2010nc} the 
topological sector~\cite{Kapustin:2009kz} of three-dimensional ABJM 
theories proposed for the description of multiple 
M2-branes\cite{Aharony:2008ug} even at strong coupling  
and recently it was found that the $\epsilon_1=0$ 
sector of the refined topological string encodes the 
non-perturbative information of M2-branes in the 
dual $AD_4\times S^7$ geometry~\cite{Hatsuda:2013oxa}.

The aim of this  paper is to calculate these refined BPS 
invariants in the physically  most interesting geometries  
and to interprete them  in the above physical contexts. 
Mostly our local geometries are defined by complex surfaces $S$ 
inside a Calabi-Yau manifold $M$. Due to the Calabi-Yau condition 
the local description is then given by the anti-canonical 
bundle over $S$ and if the canonical class of $S$ 
is ample, $S$ is called a del Pezzo surface, it is 
rigid inside the Calabi-Yau and  may be shrunken to a point. 
Morever in these cases it is easy to construct global elliptic 
Calabi-Yau manifolds fibred over $S$. Ampleness  of the canonical class 
is however not strictly necessary for the set-up, which 
requires only that the $U(1)_{\cal R}$ symmetry is geometrically 
realized as an isometry of the Calabi-Yau space $M$. In particular we consider also the half K3 
surfaces and line bundle geometries ${\cal O}(g -1 +h )
\oplus {\cal O}(g -1 -h )\rightarrow \Sigma_g$ associated 
to a genus $g$ Riemann surface $\Sigma_g$, which generalize 
the conifold singularity.

Mathematically the BPS multiplicities appear as 
cohomological data $H^*({\cal M}_{\underline q},\mathbb{Z})$  
of the moduli space ${\cal M}_{\underline q}$ of supersymmetric 
solutions with the corresponding quantum numbers ${\underline q}$.
These solutions can be solitons, instantons, brane configurations, 
vortices etc. preser-ving some of the supersymmetry. In favorable 
situations one has group actions on ${\cal M}_{\underline q}$, 
which allow to define equivariant cohomologies and 
to calculate the cohomological data by equivariant localization. 

In our case the moduli space is  the one of stable pairs, which 
are defined by a pure sheaf ${\cal F}$ of complex dimension 
one and a section $s$ which generates ${\cal F}$ outside finitely 
many points. Its  topological data are given by ${\rm ch}_2({\cal F})=\beta$ 
and $\chi ({\cal F})=n$, where $\beta\in H_2(M,\mathbb{Z})$ is 
identified with the charge  $Q$ mentionend above, while $n$ 
is related to $J_L$. The $U(1)_{\cal R}$ $\cal{R}$-symmery acts as an 
isometry on $M$ and its induced action on ${\cal M}_{\underline q}$ 
allows to define a motive $[{\cal M}_{\underline q}]$. 
The decomposition w.r.t. this motive yields the refinement 
related to $J_R$.              

As it is typical in geometries with a group action the application 
of the Atiyah-Bott equivariant localization theorem leads to 
formulas of the type 
\begin{equation} 
e({\cal M}_{\underline q})= \int_{ {\cal M}_{\underline{q}}  } \phi =\sum_P \int_P 
\left(\frac{i^*\phi}{e(\nu_P)}\right)\ ,    
\end{equation}
where in the smooth case $\phi$ is an equivariant 
top class of a holomorphic bundle $V$ over ${\cal M}_{\underline q}$.
$P$ is the fixpoint locus under the group action, $i(P)$ the 
embedding map, $\nu_P=N_{P/{\cal M}}$ is the normal 
bundle and $e(\nu_P)$ the Euler class of it. 
In general the integral over ${\cal M}_{\underline q}$ 
has to be replaced by the evaluation of the equivariant 
classes of the deformation complexes over virtual fundamental 
cycles in the equivariant Chow ring of ${\cal M}_{\underline q}$. 
The equivariant classes are then computed by taking alternating 
products of the weights w.r.t. the group action of the sections in 
the automorphism, deformation and obstruction bundles 
of the deformation complex. Physically these alternating 
products are related to the result of integrating out 
Gaussian fluctuations of bosons and fermions around 
BPS solutions in the semi-classical approximation 
to the corresponding supersymmetric path integral.
Remarkably it is often the case that these localization 
calculations can be set up in apparently rather 
different ways of rather different technical complexity  
and nevertheless yield the same result. In particular 
the unrefined BPS states $n^\beta_g=\sum_{J_R} (-1)^{J_R} (2 J_R+1) 
N^\beta_{g,J_R}$ can be either obtained by 
localization in the moduli space of maps 
as originally suggested by the topological 
string approach or technically more favorably by 
localization in the moduli space of stable pairs~\cite{Pandharipande:2007kc,Pandharipande:2007sq,Pandharipande:2007qu}. 
It is also the latter approach that can be refined for $r$-dimensional manifolds 
with $(\mathbb{C}^*)^r$ group actions, i.e. toric varieties,  
by characterizing  $[{\cal M}_{\underline q}]$ 
using the virtual Bialynicki-Birula decomposition of 
${\cal M}_{\underline q}$, w.r.t.  the induced 
$U(1)_{\cal R}$ action~\cite{CKK}.

Unfortunately many of the interesting geometries mentioned 
above are not toric. Moreover the localization approaches 
in toric geometries lead typically to sums over partitions 
whose size increases with $Q$ and are therefore not effective 
in producing global expressions for the BPS generating functions 
on the space-time moduli space ${\cal M}_{phys}$ of the 
effective action. These  generating functions are amplitudes in 
the effective action  and space-time T-duality already predicts 
that they are  modular objects.  The most efficient way to use this modularity is to 
calculate the refined invariants, by mirror symmetry via the 
refined holomorphic anomaly equations.  This formalism is very similar to 
the calculation of the higher genus amplitudes in topological 
string theory on Calabi-Yau manifolds \cite{HK2006}, 
which is based on the holomorphic anomaly equations~\cite{BCOV}. 
One can fix  the holomorphic ambiguity using the gap boundary 
conditions  proposed in  \cite{HK2006, HKQ}. The conventional 
unrefined topological string theory corresponds to the case 
$\epsilon_1+\epsilon_2=0$. In \cite{HK2010,HKK,KW,KW2} the holomorphic 
anomaly equations and the gap boundary conditions are generalized to the 
refined case of arbitrary $\epsilon_1$ and $\epsilon_2$ parameters.

\section{The refined BPS invariants on local Calabi-Yau spaces}   
\label{section2}

Let us start with a description of the refined invariants, their 
physical interpretation and the basic technique of calculation 
we use.   

\subsection{Mathematical definition of the $N^\beta_{J_l,J_R}$}  

A mathematically definition of the $N^\beta_{J_L,J_R}$ 
as well as a new way to determine them on toric Calabi-Yau 
manifolds $M$ was given in~\cite{CKK}, based a mathematical 
description of the the unrefined BPS configurations at 
large radius by stable pairs. Stable pairs are defined by two data
\begin{itemize}
 \item 
A pure sheaf ${\cal F}$  of complex dimension 
one with 
\be 
{\rm ch}_2({\cal F})=\beta, \qquad  \chi ({\cal F})=n \, , 
\ee 
where $n\in\mathbb{Z}$ and $\beta\in H_2(M,\mathbb{Z})$.   
\item A section $s\in H^0(\cal F)$, which generates  
${\cal F}$  outside a finite set of points.      
\end{itemize}
The bound state of even $(D0,D2,D4,D6)$-brane charge $(n,\beta,0,1)$ 
can be written as a complex  
\be 
{\cal I}^\bullet: {\cal O}_M \stackrel{s}{\to} {\cal F} \ 
\ee
and the moduli space of these stable pairs is 
denoted  $P_n(M,\beta)$. 
Stable pairs on Calabi-Yau threefolds 
have a perfect and symmetric obstruction 
theory, which follows from Serre duality 
and the triviality of the canonical class, 
see e.g.~\cite{behrend}.
   
A symmetric obstruction theory implies that 
the first order deformations 
${\rm Ext}^1({\cal I}^\bullet,{\cal I}^\bullet)$ are 
dual to the obstructions  
${\rm Ext}^2({\cal I}^\bullet,{\cal I}^\bullet)$  
and therefore the virtual dimension of 
$P_n(M,\beta)$ is zero. In this situation 
one has a virtual fundamental class $[P_n(M,\beta)]^{\rm vir}$ 
of degree zero, which can be integrated to 
a number
\be
\#^{vir}(P_n(M,\beta))=\int_{[P_n(M,\beta)]^{\rm vir}} 1\  .
\ee
In the smooth cases this number is related to the 
Euler number of the moduli space by 
$\#^{vir}(P_n(M,\beta))=(-1)^{{\rm dim}(P_n(M,\beta))}\chi(P_n(M,\beta))$.

In the case of toric non-compact Calabi-Yau spaces $M$ 
the action of the torus $T=(\mathbb{C}^*)^3$ on $M$ 
descends to an action on  $P_n(M,\beta)$.  For each topological 
class $\beta$ and $n$, which is determined by box configurations,  
equivariant localization expresses 
\be 
{\cal T}_{{\cal I}^\bullet} ={\rm Ext}^1({\cal I}^\bullet,{\cal I}^\bullet)-
{\rm Ext}^2({\cal I}^\bullet,{\cal I}^\bullet)
\label{tangentspace}
\ee 
in terms of a Laurant polynomial of the torus weights $t_i$, 
so that the integration can be performed by taking 
an appropriate coefficient~\cite{ptvertex} of the $t_i$. 

Recently~\cite{CKK} the above symmetric obstruction theory 
has been refined by an extension of the classical 
Bialynicki-Birula decomposition to the virtual 
case. The classical decomposition requires a 
$\mathbb{C}^*$-action on $P_n(M,\beta)$ with finitely 
many isolated fixpoints $p$ on which the tangent 
space can be decomposed into eigenspaces of 
non zero characters $\chi$ of $\mathbb{C}^*$ as 
$T_pM=\oplus_{\chi \in X(\mathbb{C}^*)} T_P^{\chi}$ 
so that the dimensions  $d_p^+={\rm dim}(\oplus_{\chi>0} T^\chi_p)$ 
and  $d_p^-={\rm dim}(\oplus_{\chi<0} T^\chi_p)$ of 
positive and negative subspaces can be defined. 
One can pick a generic enough $\mathbb{C}^*$-action 
so that its fixpoints coincide with one of 
the $T$-action $P_n(M,\beta)^T=P_n(M,\beta)^{\mathbb{C}^*}$ 
and define the virtual motive 
\be 
\label{virtualdecomposition}
[ P_n(M,\beta)]^{vir}=\sum_{p\in P_n(M,\beta)^{\mathbb{C}^*}} 
\left(-\mathbb{L}^{-\frac{1}{2}}\right)^{d_p^+-d_p^-}\ .
\ee
In particular it was shown in~\cite{CKK} that $P_n(M,\beta)]^{vir}$ 
is independent of the choice  of the $\mathbb{C}^*$ subgroup 
of $T$ as long as it preserves the holomorphic three-form.

Given (\ref{virtualdecomposition}) as well as (\ref{tangentspace}), the formalism 
of integrations and the combinatorics of the relevant 
box configurations described in~\cite{ptvertex}, one can 
calculate the refined Pandharipande-Thomas partition 
function $Z_{PT}^{ref}$. Moreover it was shown in~\cite{CKK} that $Z_{PT}^{ref}$ can be expanded 
in terms of the physical multiplicities $N^\beta_{J_L,J_R}$
as 
\be
Z^{ref}_{PT}=\prod_{\beta,J_L,J_R}\prod_{m_{L/R}=-J_{L/R}}^{J_{L/R}}
\prod_{m=1}^\infty\prod_{j=0}^{m-1} \left(1 - \mathbb{L}^{-m/2 + 1/2+j-m_R}
(-q)^{m-2m_L}Q^\beta\right)^{(-1)^{2(J_L+J_R)}N^\beta_{J_L,J_R}} .
\label{motivicproduct}
\ee
With the identifications 
\be 
q_L^{1/2}\to -q^{-1},\ q_R\to \mathbb{L}^{-1},\ e^{-\epsilon_1}\to \mathbb{L}^{-1/2}\left(-q\right),\ e^{-\epsilon_2}\to \mathbb{L}^{1/2}\left(-q\right)
\ee 
this is equivalent to the expression which 
follows from the refined Schwinger-Loop calculation~\cite{IKV}
\begin{equation}
Z=\prod_\beta \prod_{j_{L/R}=0}^\infty \prod_{m_{L/R}=-j_{L/R}}^{j_{L/R}}\prod_{m_1,m_2=1}^\infty \left(1-q_L^{m_L} q_R^{m_r} e^{\epsilon_1(m_1-\frac{1}{2})}
e^{\epsilon_2(m_2-\frac{1}{2})} Q^{\beta}\right)^{(-1)^{2(j_L+j_R)} N^\beta_{j_Lj_R}} .
\label{productrefined}
\end{equation}
in a graviphoton background. Here we parametrized the graviphoton field strength 
by the two-parameters $\epsilon_1$ and $\epsilon_2$, where we also use the notation for the combination 
$\epsilon_{R/L}=\frac{1}{2}(\epsilon_1\pm \epsilon_2)$ and $q_{R/L}=\exp(\epsilon_{R/L})$, so 
that $\epsilon_{R}$ and $\epsilon_{L}$ denote the self-dual and anti-self-dual parts of 
the graviphoton field strength respectively.  $Q^\beta=\exp(-t\cdot\beta)$, where $t$ 
denotes the K\"ahler parameter measuring the volume of a curve in the class $\beta$.  It is 
convenient to expand the topological string amplitude 
as 
\begin{eqnarray} \label{topo2.1}
F(\epsilon_1,\epsilon_2,t)={\rm Log}(Z) = \sum_{n,g=0}^{+\infty} (\epsilon_1+\epsilon_2)^{2n}
(\epsilon_1\epsilon_2)^{g-1} F^{(n,g)}(t).
\end{eqnarray}
Note that the Schwinger-Loop interpretation implies that  only even powers 
of $\epsilon_1$ and $\epsilon_2$ appear, so the summation index $n$ is 
an integer\footnote{For the Nekrasov partition function, there could be naively odd 
power terms in the case of  Seiberg-Witten gauge theory with massive flavors. In any 
case these odd terms can be eliminated by a shift of the mass parameters \cite{HKK, KW, KW2} 
so we will not need to consider them here.}.

\subsection{The direct integration approach}

In~\cite{HK2010,KW} generalized holomorphic anomaly equations were
proposed\footnote{The one in~\cite{KW} contains an additional
term, which is irrelevant for the present  purpose of
counting BPS states.} which take the form
\begin{eqnarray} \label{gen_hol_ano}
\bar{\partial}_{\bar{i}} F^{(n,g)}= \frac{1}{2}\bar{C}_{\bar{i}}^{jk}\big{(}D_jD_kF^{(n,g-1)}
+{\sum_{m,h} }^{\prime}  D_jF^{(m,h)}D_kF^{(n-m,g-h)}\big{)} \,, \quad n+g>1\,,
\end{eqnarray}
where the prime denotes the omission of $(m,h)=(0,0)$ and $(m,h)=(n,g)$
in the sum. The first term on the right hand side is set to
zero if $g=0$. These equations together with the modular
invariance of the $F^{(n,g)}$ and the  gap boundary conditions, 
reviewed in section \ref{gap}, determine $F^{(n,g)}$ recursively 
to any order in $\epsilon_{1,2}$~\cite{HKK}. The equation 
(\ref{gen_hol_ano}) has been given a B-model interpretation in the local
limit~\cite{HKK} in which the deformation direction
corresponds to the puncture operator of topological
gravity coupled to the Calabi-Yau non-linear $\sigma$-model.

\subsection{Elliptic curve mirrors and closed modular expressions} 
\label{ellipticdirectintegration} 

We are mainly concerned with Calabi-Yau manifolds defined as the anti-canonical bundle 
over del Pezzo surfaces $S$, i.e. the total space of ${\cal O}(-K_S)\rightarrow S$. 
The mirror geometry is given by a genus one curve ${\cal C}$ with punctures 
and a meromorphic differential  $\lambda$, with the property that $\partial_u \lambda$ is the 
holomorphic differential of  ${\cal C}$.  For our applications it is sufficient that 
this is true up to exact terms.  The mirror curves are derived for toric del Pezzo 
in section \ref{section:mirror1} and more general in \ref{mirrorsymmetry}.      

These curves can be brought into Weierstrass form
\begin{eqnarray}  
\label{weierstrass}
y^2 = 4x^3 -g_2(u,{\underline m}) x- g_3(u,{\underline m}) \ ,
\end{eqnarray}
i.e. a family of elliptic curves ${\cal C}$ parametrized redundantly 
by $u$ and ${\underline m}$. Our formalism distinguishes 
$u\in {\cal M}=\mathbb{P}^1\setminus \{p_1,\ldots, p_r\}$ 
as the complex modulus of the family of curves, defining 
the monodromy of ${\cal C}$, from the ``mass'' parameters 
${\underline m}=\{m_1,\ldots, m_{n_f}\}$, whose number ranges 
between $0\le  n_f \le 6$ for the toric (almost) del Pezzo surfaces and
between $0\le  n_f \le 8$ for the general del Pezzo surfaces. 

These masses enjoy various interpretations in the  different 
physical context. They are masses of matter in various representations 
in Seiberg-Witten theories with one Coulomb parameter, they  are interpreted 
as non-renormalizable deformations of $[p,q]$ 5-brane webs,
 as Wilson lines in the E-string picture, as bundle moduli of the dual heterotic 
string in the F-theory geometrization or as positions of $[p,q]$ 7-branes in 
the brane probe picture. They are related to K\"ahler parameters 
of the del Pezzo surface, which are obtained for the generic del Pezzo 
surfaces by linear transformations in the homology lattice from 
the volume af the hyperplane class in $\mathbb{P}^2$ and  the 
volumes of the exceptional divisors. Indeed for the Seiberg-Witten limit we have spelled out the connection between mass and K\"ahler parameters in the examples \eqref{sw1}, \eqref{sw2}, \eqref{sw3}, \eqref{sw4} and \eqref{sw5}.

For the almost del Pezzo surfaces, 
see definition after (\ref{centerEn}), the $m_i$ can be related by rational 
transformations to the K\"ahler parameters. Examples for these rational 
transformation occur first for the Hirzebruch surface $\mathbb{F}_2$  in \eqref{zcoordinatesF2} and
(\ref{mirrormapf2}) \footnote{For other geometries they can be found in 
(\ref{zcoordinatesp6}, \ref{mirrormapp6}), (\ref{zcoordinatesp8}, \ref{mirrormapp8}) and (\ref{zcoordinatesp10}, \ref{mirrormapp10}).}. These transformations are 
neccessary,  because the exceptional divisors are not in the K\"ahler cone.  In all 
applications there are additional ``flavor'' symmetries acting on the mass parameters, 
which makes it natural to group them in characters of the Weyl group.         

\subsubsection{The genus zero sector}
With the formalism developed in~\cite{HKK} we can calculate 
the prepotential $F^{(0,0)}(t,{\underline m})$ using its 
relation 
\be 
\label{prepotential} 
\frac{\partial^2}{\partial^2 t} F^{(0,0)}(t,{\underline m}) =-\frac{1}{ 2\pi i}\tau(t,{\underline m})
\ee 
to the $\tau$-function of the elliptic curve. 

Here the relation between the local flat coordinate $t$ at a cusp point 
in ${\cal M}$ and $u,{\underline m}$ is obtained by integrating   
\be
\label{nonlogperiod}
\frac{dt}{du} =\sqrt{\frac{E_4(\tau) g_3(u,{\underline m})}{E_6(\tau) g_2(u,{\underline m})}}\, 
\ee
with vanishing constant of integration. The $g_i$ are not invariants 
of the curve, but can be re-scaled as 
\be 
g_i\rightarrow \lambda^i(u,\underline{m}) g_i,
\label{giscaling}
\ee  
which changes (\ref{nonlogperiod}). One can fix this ambiguity 
so that $\frac{dt}{du}=\frac{1}{2\pi i} \int_\mu \frac{dx}{y}$ 
for the vanishing cycle $\mu$. In praxis this is done by matching 
the leading behaviour of the integral. $E_4$ and $E_6$ are 
the Eisenstein series. We obtain $\tau$ as a function of 
$t,{\underline m}$ by inverting the $j(\tau)$-function 
\be 
j=1728\frac{g_2^3(t,{\underline m})}{\Delta(t,{\underline m})}=
\frac{E_4^2(\tau)}{E_4^3(\tau)-g_3^2(u,{\underline m})}=
\frac{1}{q} + 744 + 192688 q+\ldots\ . 
\ee 
Here $\Delta=27 g_2^3(t,{\underline m})-g_3^2(t,{\underline m})$ denotes 
the discriminant and $q=\exp(2\pi i \tau)$. With this information 
(\ref{prepotential}) determines $F^{(0,0)}(t,{\underline m})$ up to 
classical terms, which can be recovered from properties of constant 
genus zero maps. The prepotential also determines 
the metric on the moduli space as 
\be 
\label{metric}
G_{t,\bar t}= 2\partial_t \bar \partial_{\bar t} {\rm Re}\, \Res\left(\bar t \partial_t  
F^{(0,0)}(t,{\underline m})\right)=  4 \pi \, {\rm Im} \tau \ .  
\ee

\subsubsection{Refining the higher genus sector in  the rigid cases}      

It was shown in~\cite{Klemm:1999gm} in the example of local $\mathbb{P}^2$  how (\ref{gen_hol_ano})  
restricts  to the local mirror geometry and how the  higher genus amplitudes are calculated 
from this local data. A crucial point is that the non-trivial  K\"ahler connection of the global
Calabi-Yau manifold trivializes in the local case iff the period $X_0$, that corresponds to the 
integration over the SYZ torus, i.e. measures  the $D0$-brane charge,   becomes constant in 
the local limit. Physically this corresponds to the limit, which  decouples gravity in Type II compactifications 
 by passing from special geometry of  $N=2$  supergravity, geometrically realized in the complex moduli 
space of Calabi-Yau-manifolds, to  rigid special geometry of  theories with global 
$N=2$ supersymmetry,  geometrically realized in the complex moduli space of Riemann 
surfaces. The latter fact makes it possible to generalize the formalism discussed here 
to higher genus Riemann surfaces in a straightforward way.     

In the case of the half K3 or $n_f=9$ the period $X_0$ does not become 
constant~\cite{Hosono:2002}\cite{HST}, but is instead given by \cite{Klemm:2012}
\be
X_0 (x_e) = \sum_{n \ge 0} a_n x_e^n, \qquad a_n = \frac{(6n)!}{(3n)!(2n)!n!} .
\ee
Here $x_e$ denotes the fibre coordinate. The refinement of  the holomorphic 
anomaly equation is based on a holomorphic anomaly equation, which
uses crucially the rational elliptic fibration structure~\cite{Hosono:2002}\cite{HST}.  
This  form naturally generalizes  to elliptic fibred Calabi-Yau threefolds  with Fano 
bases \cite{Klemm:2012}.  The half K3 is not rigid in the Calabi-Yau, gravity is not completely 
decoupled and in fact the non-trivial K\"ahler connection plays an important role for the derivation of 
the holomorphic anomaly equation.  We discuss the refinement  of this very  interesting 
case seperatly in section~\ref{sec.nb>1} and continue the discussion of the rigid 
cases  below.

$F^{(0,1)}(t,{\underline m})$ fulfills a holomorphic anomaly 
equation, which can be integrated to 
\be \label{genus1a}
F^{(0,1)}=-\frac{1}{2} \log\left( G_{u\bar u} |\Delta u^{a_0} \prod_i m_i^{a_i}|^{\frac{1}{3}}\right),
\ee 
where the integration constants $a_0,\,a_i$ are fixed by constant genus one maps.
The function $F^{(0,1)}(t,{\underline m})$ is holomorphic and its 
form follows from its behaviour at $\Delta=0$ as 
\be \label{genus1b}
F^{(1,0)}=\frac{1}{24} \log(\Delta u^{b_0} \prod_i m_i^{b_i})\ .
\ee 
To determine the constants $b_0,\,b_i$ one requires regularity at 
infinity and needs information about the vanishing of a few 
low degree BPS numbers.
It is convenient for the integration formalism to rewrite the 
holomorphic part of the metric $G_{u\bar u}$ in terms of 
Eisenstein series
\be
F_{hol}^{(0,1)}=-\frac{1}{12} \log\left( \frac{E_6}{g_3}\Delta u^{2 a_0} \prod_i m_i^{2 a_i} \right)\ . 
\ee 
The higher $F^{(n,g)}$  with $n+g>1$ have the general form~\cite{HKK}
\begin{equation} \label{generalformfng}
F^{(n,g)}=\frac{1}{\Delta^{2(g+n)-2}(u,\underline m)} \sum_{k=0}^{3g+2n-3} X^k p^{(n,g)}_k(u,m) .
\end{equation}
Here the non-holomorphic generator $X$ is given by
\begin{eqnarray}
X=\frac{g_3(u, \underline m)}{g_2(u, \underline m)} \frac{\hat{E}_2(\tau)E_4(\tau)}{E_6(\tau)} \, .
\label{Xdef}
\end{eqnarray}
With  $\hat{E}_2$ we denote the non-holomorphic second Eisenstein
series
\be
\hat{E}_2(\tau, \bar{\tau}) = E_2(\tau) - \frac{3}{\pi {\rm Im}(\tau)} \,.
\ee
We note that we choose $\lambda$ in (\ref{giscaling}) so that 
\be
\frac{E_6^2}{E_4^3}=27\frac{g_3^2}{g_2^3} \ .
\label{identity}   
\ee
The proof of (\ref{generalformfng}) proceeds by using 
(\ref{identity}) and (\ref{nonlogperiod}) and 
\begin{equation}
\begin{array}{rl} 
\frac{\rm d}{{\rm d}\tau}E_2 &= \frac{1}{12}(E_2^2-E_4)\ , \\  
\frac{\rm d}{{\rm d}\tau}E_4 &= \frac{1}{3}(E_2 E_4-E_6)\  , \\ 
\frac{\rm d}{{\rm d}\tau}E_6 &= \frac{1}{2}(E_2 E_6-E_4^2)\ , \\ 
\end{array}
\end{equation}
to derive 
\begin{equation}
\begin{array}{rl} 
\frac{\rm d}{{\rm d} t} X &= \frac{1}{\Delta}\frac{{\rm d}u}{{\rm d} t}(A X^2+ B X+ C)\ ,\\ 
\frac{{\rm d}^2u}{{\rm d}^2 t}&= \frac{1}{\Delta}\frac{{\rm d}u}{{\rm d} t}(A X+ \frac{B}{2})\ , \\  
\end{array}
\end{equation}
with 
\begin{equation}
A=\frac{9}{4}( 2 g_2 \partial_u g_3- 3 g_3 \partial_u g_2),\qquad B=\frac{1}{2}( g_2^2 \partial_u g_2- 18 g_3 \partial_u g_3),\qquad C=\frac{g_2 A}{3^3}\ .
\end{equation}
For any family of curves (\ref{weierstrass}) this gives a 
description of the ring of quasi-modular forms in 
which the holomorphic anomaly equation 
(\ref{gen_hol_ano}) can be integrated.     
 
Following \cite{HKK} one can use (\ref{nonlogperiod}) and the fact that the three-point function takes the form
\be
C_{ttt}=\frac{\del^3 F^{(0,0)}}{\del t^3} = - 2 \pi i \frac{d \tau}{dt}
\ee
to rewrite(\ref{gen_hol_ano}) as
\ban
24 \frac{\partial F^{(n,g)} }{\partial X} &=&  \frac{g_2(u)}{g_3(u)} \frac{E_6}{E_4} \Big[ \left(\frac{d u}{d t}\right)^2 
\frac{\del^2 F^{(n,g-1)}}{\del u^2} +\frac{d^2 u}{d t^2} \frac{\del F^{(n,g-1)}}{\del u} \nn \\
&& +\left(\frac{d u}{d t}\right)^2 {\sum_{m,h} }^{\prime} \frac{\del F^{(m,h)}}{\del u} \frac{\del F^{(n-m,g-h)}}{\del u} \Big] \,.   \label{hol_an_X}
\ean
As discussed in \cite{HKK} one can deduce inductively 
that the r.h.s.\ of (\ref{hol_an_X}) is a polynomial of $X$ of 
maximal degree  $2(g+n)-3$ and a rational function in
$(u,{\underline m})$ with denominator $\Delta^{2(g+n)-2}(u,{\underline m})$. 
Equation (\ref{hol_an_X}) can also be used to integrate  the holomophic 
anomaly efficiently up to the polynomial $p^{(n,g)}_k(u,{\underline m} )$, 
which is undetermined after the integration.

\subsection{The gap condition}  
\label{gap} 
For completeness we note that the boundary conditions 
for the higher genus invariants are given by the 
leading behavior of $F(\epsilon_1,\epsilon_2,t)$ at 
the nodes of the curve (\ref{weierstrass}). Let us denote by $t$ 
the vanishing coordinate at the node under investigation, then 
the leading behavior reads
\begin{eqnarray}  \label{expansionschwinger}
F(s,g_s,t)&=&\int_0^{\infty} \frac{ds}{s}\frac{\exp(-s t)}{4\sinh(s\epsilon_1/2)\sinh(s\epsilon_2/2)} +\mathcal{O} (t^0) \\
&=& \big{[}-\frac{1}{12}+\frac{1}{24} (\epsilon_1+\epsilon_2)^2 (\epsilon_1\epsilon_2)^{-1}\big{]}\log(t) \nonumber \nonumber \\ &&
+ \frac{1}{\epsilon_1\epsilon_2} \sum_{g=0}^\infty \frac{(2g-3)!}{t^{2g-2}}\sum_{m=0}^g \hat B_{2g} \hat B_{2g-2m} \epsilon_1^{2g-2m} \epsilon_2^{2m} +\ldots \nonumber \\ &&
=\big{[}-\frac{1}{12}+\frac{1}{24} s g_s^{-2}\big{]}\log(t)
+ \big{[} -\frac{1}{240}g_s^2+\frac{7}{1440} s-
\frac{7}{5760} s^2g_s^{-2} \big{]} \frac{1}{t^2} \nonumber \\ &&
+ \big{[} \frac{1}{1008}g_s^4-\frac{41}{20160} s g_s^2 +\frac{31}{26880} s^2 -\frac{31}{161280} s^3 g_s^{-2}\big{]} \frac{1}{t^4}   +\mathcal{O} (t^0)  \nonumber\\[ 7 mm]
&& +  \,\,\mbox{contributions to $2(g+n)-2>4$}\, ,  \nonumber
\end{eqnarray}
where\footnote{Here
$\hat B_{m}=\left(\frac{1}{2^{m-1}}-1\right)\frac{B_m}{m!}$ and the Bernoulli numbers $B_m$
are defined by $t/(e^t-1)=\sum_{m=0}^\infty B_m \frac{t^m}{m!}$.} $g_s^2= (\epsilon_1 \epsilon_2)$  and $s= (\epsilon_1 + \epsilon_2)^{2}$. 
The expansion (\ref{expansionschwinger}) is simply obtained by 
evaluating the Schwinger-Loop integral under the asumption 
that a single hypermultiplet with mass $m=t$ becomes 
massless at the node.

\section{Mirror symmetry for non-compact Calabi-Yau spaces 
with  del Pezzo basis}

In this section we present an analysis of the 
non-compact Calabi-Yau spaces $X_{nc}$, 
which are given as the total space of the fibration of the 
anti-canonical line bundle 
\be
{\cal O}(- K_B)\rightarrow  B,
\label{ncy}
\ee 
over a Fano variety $B$ and their mirror manifolds. 
A Fano manifold is a smooth rational variety that has an ample anti-canonical class. 
By the adjunction formula (\ref{ncy}) defines a non-compact Calabi-Yau $d$-fold 
for  $(d-1)$-dimensional Fano varities $B_{d-1}$. It is also the normal geometry of 
$B$  inside a compact Calabi-Yau space. Since $-K_B=-c_1(TB)$ is negative $B$ can be shrunken  
to a point inside the Calabi-Yau space.  

\subsection{The Del Pezzo surfaces and the half K3 surface} \label{delpezzohalfk3} 
Del Pezzo surfaces $S=B_2$ are two-dimensional smooth Fano manifolds and 
enjoy a finite classification\footnote{A classic review is~\cite{demazure}.
For a modern review see~\cite{dolgachev}. A physically motivated one is presented in ~\cite{Friedman:1997yq}.}.
The list is  $\mathbb{P}^2$ and 
blow-ups of $\mathbb{P}^2$ in up to $n=8$ points, called ${\cal B}_n$, 
as well as $\mathbb{P}^1\times \mathbb{P}^1$. 

\subsubsection{The general structure} 

By definition of a 
rational surface  $h_{2,0}=h_{1,0}(S)=0$ hence the arithmethic genus 
$\chi_0=\sum_i h_{i,0}=1$. The Hirzebruch-Riemann-Roch theorem gives for the 
arithmetic genus and the signature $\sigma= b_{2}^+-b_{2}^-$
\be 
\chi_0=1=\frac{1}{12}\int_S (c_1^2 +c_2), \quad \sigma=\frac{1}{3}  \int _S(c_1^2 -2 c_2)\  
\label{twoindex}
\ee
respectively. Blowing up increases the Euler number $\chi(S)$ and the second Betti number $b_2(S)$ by $1$. 
From $\chi(\mathbb{P}^2)=3$ and the first equation of \eqref{twoindex} follows 
$k=\int_{{\cal B}_n } c_1^2 =9-n$, a quantity often called the degree of the del 
Pezzo surface. Further from $b_2(\mathbb{P}^2)=1$ and $\sigma({\cal B}_n) =1-n$ it follows 
that the middle cohomology lattice 
\be
\Lambda=H_2({\cal B}_n,\mathbb{Z}) 
\ee 
has signature $(1,n)$. Let  $h$ denote the hyperplane class in $\mathbb{P}^2$ and  
$e_i$ the exceptional divisors associated to the $i$'th blow-up, then the intersection pairing 
``$\cdot$'' is defined by the non-vanishing intersections $h^2=-e_i^2=1$. 
The anti-canonical class is 
\be
K=c_1({\cal B}_n)= 3 h -\sum_{i=1}^n e_i, 
\ee
so that again $k=K^2_{{\cal B}_n}=9-n$, i.e. the positivity of $K_{{\cal B}_n}$ restricts the number of blow-ups 
to $n<9$.  Let us denote by  $\Lambda'\subset \Lambda$ the sublattice orthogonal to 
$c_1({\cal B}_n)$ 
\be 
\Lambda'=\{x \in \Lambda | x\cdot K=0\} \ .   
\label{Lambdaprime} 
\ee  
The intersection form ``$\cdot$'' is negative on $\Lambda'$ and since all coefficients 
in $K$ are odd it has even intersections. The determinant is equal to the degree $9-n$, so for $n=8$, 
$ \Lambda'$ is the unique even self-dual lattice of rank 8, the $E_8$ lattice and 
for $n=9$ it becomes the $\hat E_8$ lattice.  Similar one can see that for $n=2,\ldots, 8$ the lattice 
$\Lambda'$ corresponds to the root (or co-root lattice) of the exceptional Lie algebras as follows 
\begin{equation} 
  \begin{array}{|c|ccccccccc|cc|} 
   {\rm degree}=9-n&9&8&7&6&5&4&3&2&1&0&\\ 
       G & -& -&A_1& A_1\times A_2& A_4& D_5& E_6& E_7& E_8&  \hat E_8&  \\
  \end{array} \, .
\label{groupscosandweights}
\end{equation} 
In particular the simple roots are $\{e_i-e_j\}$ for $n=2$ and $\{e_i-e_j,h-e_i-e_j-e_k\}$ for 
$3\le n\le 8$. 

It is  also convenient to introduce the weight lattice 
\be 
\Lambda'':=\Lambda/(K \mathbb{Z}) , 
\label{Lambdaprimeprime} 
\ee 
so that the pairing on $\Lambda$ yields a perfect pairing 
\begin{equation} 
\Lambda''\otimes \Lambda'\rightarrow \mathbb{Z}\ .
\label{lattices}
\end{equation}
Further the center of $E_n$ is given by 
\be 
\mathbb{Z}_{9-n}\sim \Lambda''/\Lambda' \ .
\label{centerEn}
\ee 

In addition $\mathbb{P}^1\times \mathbb{P}^1$ is a del Pezzo surface  
with $\Lambda_2=\Gamma^{1,1}$ the hyperbolic lattice.  For us it is 
natural to include examples in which $c_1(B_2)$ is only semi-positive, 
which we call almost Fano varities. These are numerically effective, 
but not Fano, as it is dicussed e.g. in section 15.4 of~\cite{mirrorbook} 
for the Hirzebruch surface $\mathbb{F}_2$.  This is notion also used 
in~\cite{Friedman:1997yq}. Another important generalization 
is to the half K3, also denoted by  $\frac{1}{2} K3:={\cal B}_9$ that has the lattice 
\be 
\Lambda\left(\frac{1}{2}K3\right)=
\Gamma^{1,1}\times E_8 \ .
\ee  

To each del Pezzo surface ${\cal B}_n$ one can associate an elliptic 
pencil 
\be 
\{a C_1 + b C_2\}\in \mathbb{P}^2\times \mathbb{P}^1
\label{ellipticpencil}
\ee  
of sections $C_1$, $C_2$  of the canonical sheaf of $\mathbb{P}^2$ with $9-n$ base 
points. The base point free pencil for ${\cal B}_9$ defines a rational  
elliptic surface fibred over $\mathbb{P}^1$, which is isomorphic to ${\cal B}_9$ as can be 
seen by projection to the first factor. Hence the $\frac{1}{2}K3$ is a 
rational elliptic surface. If all base points of the elliptic pencil are blown 
up these $(-1)$-curves $e_i$ become sections of the elliptic surface and the 
corresponding Mordell-Weyl group is a free abelian group of rank 
8~\cite{Neron}\footnote{Rank eleven Mordell-Weyl 
groups have also been constructed by N\'eron~\cite{Neron2}.}
\be 
MW=\mathbb{Z}_8 \times {\rm Weyl}(E_8)
\ee
while the torsion part is the Weyl group of $E_8$~\cite{Manin}. 
Indeed the Weyl group of $E_n$ acts already on the cohomology 
of ${\cal B}_n$ \cite{Manin} and beside becoming the Mordell-Weyl 
group of the rational elliptic surface in the last blow-up, it 
is also extended to the affine Weyl group of $\hat E_8$ on the full 
cohomology of the $\frac{1}{2}K3$~\cite{demazure}. 

In families of del Pezzo surfaces the action of the Weyl group can be 
generated explicitly by deforming $S$ to a singular surface so that 
the vanishing cycle corresponds to a simple root $\alpha$. By the 
Picard-Lefshetz monodromy theorem the monodromy in the moduli space 
around the point where the cycle $\alpha\in H_2(S,\mathbb{Z})$ vanishes, 
generates a Weyl reflection\footnote{For ADE singularities these 
inner automorphisms generate the Weyl group. Singularities corresponding to 
non-simply laced Lie algebras  are obtained by a suitable outer automorphism action acts 
on the classes in the Hirzebruch-Jung sphere configuration, e.g. by monodromy 
in a family, see~\cite{slowody} for review. In this article we only consider 
simply laced singularities.} on the hyperplane defined by $\alpha=0$, 
i.e.  on any cycle $\beta\in H_2(S,\mathbb{Z})$ with non-trivial 
intersection with $\alpha$ the monodromy  action  is 
$S_\alpha (\beta) =\beta - (\beta \cdot \alpha) \alpha$.  
For the $\frac{1}{2} K3$ the intersection of the irreducible components 
of the singular fibres are given by Kodairas classification with affine 
intersection form and the corresponding monodromies generate $\hat E_8$.

In order to explicitly specify the action 
of the Weyl group $\hat {E}_8$ on the moduli parameters, 
denote the ``volumes\footnote{The $e_i$ do not lie in the K\"ahler cone 
and the ``volumes'' can formally be negative for flopped $\mathbb{P}^1$'s.}'' 
of exceptional $\mathbb{P}^1$'s by $m_i$ and the modulus of the elliptic 
fibre emerging in the ninth blow-up by $\tau$. Then the Weyl group is 
generated by the reflexions
\begin{equation}
\begin{array}{rl} 
\label{weyle8onparameters}
m_i & \longleftrightarrow  m_j, \hspace{1.05cm} \text{for any pair $(i,j)$}, \\ [1mm]
m_i & \longleftrightarrow  -m_j, \qquad \text{for any pair $(i,j)$}, \\ [ 1mm] 
m_i & \longrightarrow  \frac{1}{4}\sum_{j=1}^8 m_j, 
\end{array}
\end{equation} 
which defines the Weyl group of $E_8$. For the affine $\hat {E}_8$ there is an 
additional infinite shift symmetry
\begin{equation}
\begin{array}{rl} 
\label{weylaffinee8onparameters}
m_i & \longrightarrow  m_i + 2\pi \alpha_i  \\ [1 mm]
m_i & \longrightarrow  m_i + 2\pi \alpha_i \tau \, .
\end{array}
\end{equation} 
Here $\vec \alpha$ $=$ $(\alpha_1,\dots, \alpha_8)$ is an element of the root 
lattice of $E_8$. Recall that the latter is defined as the sublattice of 
$\mathbb{R}^8$ whose elements have either all integer or half-integer entries, 
such that the sum of all entries adds up to an even integer. In addition there 
is a ${SL}(2,\mathbb{Z})$ symmetry acting on the fibre modulus
\begin{equation}
\begin{array}{rl} 
\label{sl2zfibres}
\tau & \longrightarrow  \tau + 1   \\ [1 mm]
\tau & \longrightarrow  -\frac{1}{\tau}, \qquad  \vec m  \longrightarrow  \frac{\vec m}{\tau}
\end{array}
\end{equation} 
making the affine characters Jacobi forms with $\tau$ their modular parameter
and $\vec m$ a tuple of elliptic parameters. The ring of these 
forms relevant for the direct integration approach of our 
refined holomorphic anomaly equation~(\ref{refinedmodular})
is summarized in appendix~\ref{appendixJacobi}.

\subsubsection{Algebraic realizations} 

The $D_5$, $E_6$, $E_7$ and $E_8$ del  Pezzo can be represented 
by the zero locus of two quadrics in $\mathbb{P}^4$, the cubic in $\mathbb{P}^3$, the 
quartic in $\mathbb{P}^3(1,1,1,4)$ and the sixtic in $\mathbb{P}^3(1,1,2,3)$. 
By use of  the adjunction formula the Euler number can be calculated to be $8,9,10$ 
and $11$ while $c_1(S)=(\sum w_i-\sum d_k) h=h$. Here $w_i$ denote the weights, $d_i$ are degree(s) 
of the defining polynomial constraints  and $h$ is the hyperplane class of the ambient space. 
Generic anti-canonical models for higher degree del Pezzo surfaces cannot be realized as 
hypersurfaces or complete intersections. E.g. the 
degree six $A_1\times A_2$ del Pezzo is a determinantal variety in $\mathbb{P}^6$ 
and the degree five  $A_4$ del Pezzo is given by five linear quadrics in $\mathbb{P}^5$~\cite{dolgachev}. 

Finding these algebraic realizations is closely related to the problem of 
constructing ample families of elliptic curves ${\cal E}$ with $d$ rational 
points $Q_i$, $i=1,\ldots,d$, i.e. such that ${\cal E}$ is embeddable in some  
(weighted) projective space $\mathbb{P}^n({\underline w})$~\footnote{In fact, as recalled 
in section \ref{verticaldelpezzo}, these families of elliptic curves 
can be promoted to elliptic fibred Calabi-Yau spaces with several 
sections and in an extremal transitions the del Pezzo in the corresponding 
algebraic realization shrinks to zero size.}.The construction of the ample 
families of elliptic curves proceeds as follows. 
Assume the embedding exists, consider the bundle ${\cal L}={\cal O}(\sum_{i}^d Q_i)$ 
over ${\cal E}$ and match $m {\cal L}= K_{\mathbb{P}^n({\underline w})}$ so that  
a trivial canonical class is obtained. The ideal of the relations of the sections in 
$m {\cal L}$, which has according to the Riemann-Roch theorem $\delta=h^0(m {\cal L})= 
{\rm deg}(m {\cal L})$ sections, defines the desired embedding of ${\cal E}$ into 
$\mathbb{P}^n({\underline w})$. To be explicit call $x_k$, $k=1,\ldots,d$ 
the sections of the degree $w_k$ line bundles $L_k$ of $\mathbb{P}^n({\underline w})$. 
We can assume that $x_{k_i}$ vanishes at $Q_i$ (in general $d\le n+1$ 
with the strict inequality for weighted projective space), so that
${\rm deg}(m {\cal L})=m\sum_{i} w_{k_i}$. It is easy to see that the 
case $d=1$ requires weights $w=(1,2,3)$ and one gets $m=6$, $\delta=6$ and 
the seven sections of $6 {\cal L}$ are represented by the monomials 
of polyhedron 10 of figure 1 made explicit in figure 4.  The case $d=2$ 
requires $w=(1,1,2)$ with $m=4$, $\delta=8$ and the sections are the monomials 
of polyhedron 13. $d=3$ requires $m=3$, $3{\cal L}= K_{\mathbb{P}^2}$ has degree $\delta=9$, 
leading to one relation among the ten sections, the monomials of degree 
three in $x_1,x_2,x_3$, i.e. the monomials of polyhedron 15 and the embedding 
is the cubic in $\mathbb{P}^2$. For $d=4$ one gets $m=2$,  $\delta=8$ and ten 
quadratic monomials, leading to two quadrics in $\mathbb{P}^3$. For $d=5$, $m=2$ and 
$\delta=10$ one gets five linear independent quadrics in $\mathbb{P}^4$ and $d=6$ has 
$m=2$ $\delta=12$ i.e. nine conditions in $\mathbb{P}^5$.                              

The anti-canonical divisor defines for all Fano varieties of dimension $d$ a Calabi-Yau 
manifold  $X_c$ of dimension $d-1$, while as discussed ${\cal O}(- K_B)\rightarrow  B$ 
defines a non-compact Calabi-Yau  of dimension $d+1$. In the del Pezzo case 
the anti-canonical bundle in $S$ is of course an elliptic curve ${\cal E}$. 
If we use the above model for the algebraic realization it allows $d$ rational 
points. The anti-canonical model determines the local mirror geometry of ${\cal O}(- K_S)\rightarrow  S$. 
Moreover the geometry of this anti-canonical class captures also the moduli of 
semi-stable $G$-bundles for $E_{9-d}$ groups on ${\cal E}$ as we review 
in the next section.

\subsection{Del Pezzo surfaces and semi-stable $G$-bundles on elliptic curves ${\cal E}$}
\label{stablebundles}
As mentioned in the last section for the del Pezzo surface $S$, the compact Calabi-Yau 
manifold $X_c$ is just an elliptic curve ${\cal E}$. We will explain in \ref{mirrorsymmetry} 
first in the toric context and then more generally, that the complex geometry of the pair  
$(B,X_c)$ describes the K\"ahler geometry of ${\cal O}(- K_B)\rightarrow  B$, therefore 
these geometries are mirror dual to each other.   

On the other hand for $B$ a del Pezzo surface $S$ there is a beautiful construction 
which relates the moduli of the pair $(S,{\cal E})$, $S$ being a del Pezzo $S$ with fixed canonical class,  
 to semi-stable $G$-bundles, for $G$ as in \ref{groupscosandweights}, on 
${\cal E}$~\cite{Looijenga1,Looijenga2,Friedman:1997yq}. 
The heterotic string on ${\cal E}$ requires a choice of such gauge bundles, 
and the construction explains how the moduli of the bundle are geometrized 
under the heterotic F-theory duality into moduli of the pair 
$(S,{\cal E})$ first in eight dimensions. Of course the  
program of~\cite{Friedman:1997yq} is to extend this construction 
fibre-wise and to relate the heterotic string on elliptically fibred Calabi-Yau  $d$-folds 
$Z_d$ over $B_{d-1}$ and F-theory on elliptically fibred 
Calabi-Yau spaces $Y_{n+1}$ over $B_{d}$, which is in turn rationally 
fibred over $B_{d-1}$. For a short overview see also ~\cite{donagitl}. 

The point explained in~\cite{Looijenga1,Friedman:1997yq} is 
that the choice of a semi-stable $G$-bundle over ${\cal E}$ is equivalent to a choice   
\be 
{\rm Hom}(\Lambda'',{\cal E}) \cong (\Lambda'')^* \otimes  {\cal E}=\Lambda' 
\otimes  {\cal E}\ ,
\label{homlambda}  
\ee 
up to an action of the Weyl group on the lattices. It comes from the fact that 
semi-stable $G$-bundles are equivalent to flat connections $A$ 
with values in the maximal torus $T$  of $G$. Every  weight $w\in \Lambda''$ 
defines a representation $\rho_w$ of $T$ and the flat connection $A$ in 
this representation defines uniquely a line bundle  ${\cal L}_w$, 
i.e. a point on  $Jac({\cal E})\cong {\cal E}$. Vice versa 
giving (\ref{homlambda}) up to ${\rm Weyl}(G)$  determines the $G$-bundle.        

If ${\cal E}$ is embedded as anti-canonical class in a del Pezzo $S$ 
there is a natural homomorphism from $\Lambda$ to line bundles over 
$S$, i.e. each $v\in \Lambda$ defines a line bundle ${\cal L}_v$ on $S$ 
and ${\cal L}_{v_1+v_2}={\cal L}_{v_1}\otimes {\cal L}_{v_2}$. Because 
of the definition (\ref{Lambdaprimeprime}) and the definition of 
${\cal E} $  as anti-canonical bundle, any $v\in \Lambda'$  defines a 
line bundle ${\cal L}_v$ on ${\cal E}$ and thereby a 
${\rm Hom}(\Lambda',{\cal E})$. By the Torelli theorem the moduli 
of the latter mod ${\rm Weyl}(G)$ are equivalent to the geometric 
moduli of $(S,{\cal E})$. For $G=E_8$ the lattice $\Lambda'$ 
is self-dual $\Lambda''\cong \Lambda'$ and the moduli of $(S,{\cal E})$ 
determine the moduli of the  $G$-bundle. If $\mathbb{Z}_{9-n}$ is 
non-trivial one has to pick a bundle ${\cal Q}$ as the root of the 
restriction  ${\cal L}_K$ so that ${\cal Q}^{-(n-9)} {\cal L}_K$ is trivial.  
Then $v\rightarrow {\cal Q}^{-K\cdot v} {\cal L}_v $  is the 
desired homomorphism ${\rm Hom}(\Lambda'',{\cal E})$.

As a consequence one can study the moduli space of stable $G$-bundles 
simply as the moduli space of del Pezzos surfaces with fixed canonical 
class, by counting the corresponding deformations. This is done by 
counting the coefficients of monomials modulo reparametrizations. 
E.g. for $G=E_8$ the del Pezzo is realized as mentioned above 
as degree six hypersurface $P^{E_8}=0$ in $\mathbb{P}^4(1,1,2,3)$, say 
with coordinates $(v,w,x,y)$. The hypersurface 
has $m_S=23$ momomials and in total $r_S=15$ weighted 
reparametrizations. One obtains ${\cal E}$ by taking the anti-canonial 
divisor $v=0$. ${\cal E}$ is realized as degree six hypersurface 
in $\mathbb{P}^3(1,2,3)$ having $m_{\cal E}=7$ monomials and 
$r_{\cal E}=6$ reparametrizations and therefore as expected for an 
elliptic curve one complex structure parameter say $u$. 
Now from $15$ weighted reparametrizations only $r_S^{\cal E}=8$ 
vanish at $v=0$, hence leave ${\cal E}$ fix. Among these is the action  $v\mapsto  
\mu v$ with $\mu \in \mathbb{C}^*$ which clearly respects  
the locus ${\cal E}$, the rest may be used to set seven of $23-7=16$ monomials 
of $S$ not belonging to ${\cal E}$ to zero. That leaves us nine 
monomials describing the perturbation of ${\cal E}$ all 
vanishing at $v=0$ and therefore multiplied by various  
powers of $v$. As a consequence the objects describing 
the deformation of $S$ with fixed ${\cal E}$  are the 
independent coefficients of the nine remaining monomials, 
which are identified with weights under the $\mathbb{C}^*$-action 
and hence form a weighted projective space $\mathbb{P}^8[w]$. 

The precise monomials can be constructed as follows. 
Use the $r_{\cal E}=6$ reparametrizations to write ${\cal E}$ 
in Weierstrass form 
\be 
P^{E_8}_{\cal E}=y^2-(4 x^3- g_2(u)x w^4 -g_3(u)w^6)=0
\ee 
and consider the ring of perturbations of the hypersurface 
singularity $P({\underline{x}})=0$ in $\mathbb{C}^3$.  
This  given by  
\be 
{\cal R} =\frac{\mathbb{C}[ {\underline{x}}]}{\{\partial_i P({\underline{x}})\}}\ ,
\ee 
where $\{\partial_i P({\underline{x}})\}$ is the ideal of 
the partial derivatives of $P({\underline{x}})$. Dividing 
out the ideal, essentially means that the perturbations 
can be described by monomials in $y,x,v$ that have no higher 
powers the coordinates as $y^0$, $x$ or $w^4$. The $w^4$ 
constraint comes  from  the $g_3(u)w^6$ term and holds only 
if $g_3(u)\neq 0$. The geometry of the deformations 
fixing $P_{\cal E}$ in $S$ is hence encoded in 
\be
\begin{array}{rl} 
P^{E_8}=&P_{\cal E}^{E_8} + (m^{(4)}_6 v^4+ m^{(3)}_4 v^3 w+ m^{(2)}_2 v^2 w^2+m^{(1)}_1 w v^3)x +\\
&(m^{(6)}_p v^6+ m^{(5)}_8 v^5 w+ m^{(4)}_7 v^4 w^2+  m^{(3)}_5 w^3 v^3+m^{(2)}_3 w^2 v^4), 
\label{SEpairE8}
\end{array} 
\ee
such that the $m^{(i)}_k$ have scaling weight $i$ under the 
$v\mapsto  \mu^{-1} v$ action and fit into the weighted 
projective space $\mathbb{P}^8[1,2,2,3,3,4,4,5,6]$. The 
weights are the Coxeter labels of the corresponding 
(untwisted) affine Lie algebra~\cite{Fuchs}, see section \ref{e8delpezzo} 
for the numbering of the roots.          

Originally Looijenga~\cite{Looijenga1} constructed $A=\Lambda' \otimes  
{\cal E}$ modulo the Weyl group precisely to study the deformations 
of elliptic surface singularities such as $P_{\cal E}=0$ at the
origin with the behaviour at infinity kept fixed. 
Again it is natural in this context to parametrize the $m^{(i)}_k$ 
in a Weyl invariant way by the characters of the affine Weyl 
group~\cite{Looijenga2,Saito,BS,Looijenga3,Wirthmueller}. The corresponding ring of Jacobi forms 
is summarized in appendix~\ref{appendixJacobi}.

If the behaviour at infinity is not 
fixed or equivalently the volume of the elliptic curve 
goes to zero, the weight restriction on the $m^{(k)}_i$ 
disappears and one gets rather $k$ sections of a degree 
$k$ line bundle representing  deformations. 
In the physical context the volume is the Coulomb 
parameter and the  $\tilde m_i$ in the characters are the  Wilson
line parameters. So if the volume vanishes the sum of the $k$'s add 
to $30$, the dual coxeter numbers yielding the dimension of the 
Higgs branch of the theory.  

Note that $P^{E7}$ has $m_S=22$, $r_S=16$, $r_S^{\cal E}=6$, 
$m_{\cal E}=9$ and $r_{\cal E}=8$ leading to  
$\mathbb{P}^{22-6-9}[{\underline w}]$. With $P^{E_7}_{\cal E}=
y^2-(4 w x^3- g_2(u)x w^3 -g_3(u)w^4)=0$ one sees that the 
weights are Coxeter labels of $E_7$, i.e. one gets 
$\mathbb{P}^7[1,1,2,2,2,3,3,4]$. Similar $P^{E_6}$ has $m_S=20$, $r_S=16$, 
$r_S^{\cal E}=4$, $m_{\cal E}=10$ and $r_{\cal E}=9$ leading to  
$\mathbb{P}^{20-4-10}[{\underline w}]$. With $P^{E_6}_{\cal E}=
w y^2-(4 x^3- g_2(u)x w^2 -g_3(u)w^3)=0$ one gets 
$\mathbb{P}^6[1,1,1,2,2,2,3]$. In all cases one has in  
${\cal Q}={\cal O}(p)$ with $p$ given by $(w,x,y)=(0,0,1)$ 
the correct root restriction of the anti-canonical bundle 
of $S$, see \cite{Saito,Friedman:1997yq}.     

\subsection{Reduction of the structure group and almost del Pezzo surfaces}
\label{reductionofG} 
As explained in the last section the moduli space $(S,{\cal E})$ of the del 
Pezzo surface $S={\cal B}_n$ with fixed anti-canonical divisor is the moduli 
space of $E_n$ vector bundles on an elliptic curve~\cite{Looijenga1,Friedman:1997yq}. 
It was further argued in~\cite{Friedman:1997yq} that the semi-simple commutant group 
$G\in E_n$ of the structure group of the $E_n$ vector bundle is always 
realized as an ADE singularity of $S$, obtained by specializing 
the complex structure of $S$. Resolving the latter by the 
corresponding Hirzebruch sphere tree, with an intersection matrix, 
which is the negative of the Cartan matrix of $G$, leads to an almost 
del Pezzo surface. The cohomology classes of the spheres lie in 
$\Lambda'$.   

The non-generic toric del Pezzo surfaces are exactly examples of this 
type. The toric description does not allow for complex structure 
parameters, hence for $n>3$ we deal with such specialized and 
resolved  almost del Pezzo surfaces. E.g. the polyhedron 10 in 
fig. \ref{poly} corresponds to a reduction of the $E_8$ structure 
group to the commutant of $G=SU(2)\times SU(3)$.                           
           
This construction provides a dual interpretation of 
the gauge bundle moduli and the complex structure moduli 
of a singularity on the same moduli spaces, which 
can be promoted to F-theory, by resolving the 
elliptic pencil to the rational elliptic fibration. 
In particular it gives the exact map of the heterotic 
bundle moduli on $T^2$ to the geometric moduli of  
F-theory on $Y_2=K3$ in the stable degeneration limit 
$Y_2={\cal B}^1_9\cup_{T^2} {\cal B}^2_9$. This can 
be fibred provided that the heterotic Calabi-Yau space 
$Z_d$ is an elliptic fibration ${\cal E}\rightarrow B_{d-1}$,
then the F-theory manifold $Y_{d+1}$ has the structure 
$Y_{d+1}=({\cal B}^1_9 \rightarrow B_{d-1}) \cup_{Z_d} 
({\cal B}^2_9 \rightarrow B_{d-1})$.

Describing this geometry using mirror symmetry as discussed 
in the next chapter, adds a new aspect to this  picture, 
because in mirror symmetry the K\"ahler structure 
deformations are generally described by complex structure 
deformations and secondly for mirror symmetry in two complex 
dimensions these are again described by the same complex 
moduli, the difference that the mirror decription makes is 
merely a different choice of the polarization.

Apart from the physical implications that the geometric 
invariants of stable pairs that we calculate are the BPS states 
associated to the $[p,q]$ 7-branes configurations specified by the 
affine singularity $\hat G$, they should also find a 
natural interpretation as geometric invariants associated
to gauge bundles on elliptically (fibred) manifolds.

\subsection{Toric Fano varieties and non-compact Calabi-Yau spaces}
The $d$-dimensional toric\footnote{We refer to~\cite{Fulton,Oda} for a general  background in 
toric geometry.}  Fano varieties are most easily classified by 
$d$-dimensional reflexive polyhedra. Toric almost del Pezzo surfaces are given by 
reflexive polyhedra in two dimensions, which are depicted 
in figure 1, where also the reflexive pairs $(\Delta_2,\Delta_2^*)$ 
are indicated. The anti-canonical class is only semi-positive 
if there is a point on one edge of the toric diagram, 
otherwise positive and ample. In particular the polyhedra 1,2,3,5,6 
correspond to toric del Pezzo surfaces, by the construction 
explained below.

\begin{figure}[h!] 
\begin{center} 
\includegraphics[angle=0,width=1.1\textwidth]{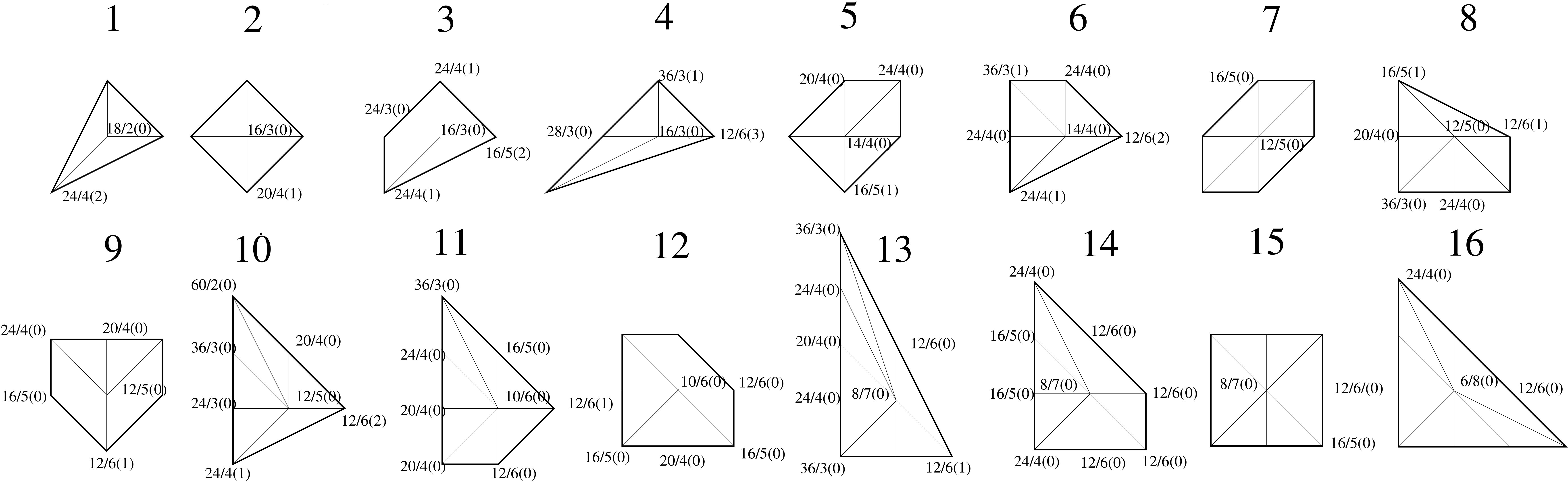} 
\begin{quote} 
\caption{These are the 16 reflexive polyhedra $\Delta$ in two dimensions, which  
build $11$ dual pairs $(\Delta,\Delta^*)$. Polyhedron $k$ is dual to  
polyhedron $17-k$ for $k=1,\ldots,6$. The polyhedra $7,\ldots,10$  
are self-dual. We denote on inequivalent corners the 
numbers $a_F/b_F(t_F)$, which determine the structure of fibrations 
by the elliptic curve defined by the anti-canonical class in 
$\mathbb{P}_\Delta$ and its sections over toric Fano bases 
of any dimension. E.g. Euler and Hodge number $h_{11}$ for the total 
space of these fibrations over two-dimensional bases are given by 
(\ref{hodgefibre}) in terms of $a_F$ and $b_F$. 
The points are labelled counter-clockwise with the one right 
to the origin with label $1$. The origin has label $0$.
\vspace{-1.2cm}} \label{poly} 
\end{quote} 
\end{center} 
\end{figure} 

We fix the following conventions in arbitrary dimensions. 
If the dimension $d$ of $\Delta$ is important we indicate it 
as a subscript. $\Delta$ is  a lattice polyhedron 
in the lattice $\Gamma$ (whose real completion is denoted by
$\Gamma_{\mathbb{R}}$), if it is the convex hull of points 
$\nu^{(i)}\in \Gamma$ containing the origin $\nu^{(0)}$ 
and spanning $\Gamma_{\mathbb{R}}$. Analogeous conventions 
are made for the dual polyhedron $\Delta^*$, where the above 
data are all marked with a star. We denote by 
$\langle \nu,\nu^* \rangle \in \mathbb{Z}$ 
the pairing between $\Gamma$ and the dual lattice $\Gamma^*$.
The dual polyhedron $\Delta^*$ is defined by~\cite{Batyrev}
\be 
\Delta^*=\{y \in \Gamma^*_{\mathbb{R}} | \langle y,x\rangle \ge -1, \  \forall x \in \Delta \} \ .
\label{reflexivity} 
\ee           
$(\Delta^*)^*=\Delta$ and $\Delta$ contains only $\nu^{(0)}$ as 
inner point. A pair is called reflexive if both $\Delta$ 
and $\Delta^*$ are lattice polyhedra.

$\Delta$ together with a triangulation defines  a complete toric 
fan $\Sigma_{\Delta_d}$  spanned from the origin $\nu^{(0)}$. 
The latter  describes for a reflexive polyhedron in real dimension $d$ an (almost)  
Fano variety $\mathbb{P}_{\Sigma_{\Delta_d}}$ of complex dimension $d$. For 
simplicity we denote $\mathbb{P}_{\Sigma_{\Delta}}=\mathbb{P}_{\Delta}$, 
explicitly given in (\ref{Pdelta}).
E.g. in the two-dimensional case $\mathbb{P}_{\Delta_2}$ is a toric (almost) del Pezzo surface $S$. 
In this construction a point in $\Delta$ different from the origin specifies a
ray in the fan $\Sigma_{\Delta}$. Generally the rays $\Sigma(1)$ of a 
fan $\Sigma$ correspond to the toric divisors $D_i$ in the Chow group $A_{d-1}(P_\Sigma)$  
of the $d$-dimensional toric variety $\mathbb{P}_\Sigma$ and we can 
assign a coordinate $Y_i$, whose vanishing $Y_i=0$ specifies 
the divisor $D_i$.

The non-compact toric Calabi-Yau space $X_{nc}=X_{\bar \Delta}$ is 
canonically obtained from $\Delta_d$ by a similar construction: In a
($d$+1)-dimensional lattice $\bar \Gamma$ spanned by  
$\bar\nu^{(i)}=(1;\nu^{(i)})$,  $\Delta_d$ is canonically embedded  
in the hyperplane at distance one from the origin $O=(0; 0,\ldots,0) 
\in\bar \Gamma$ as the convex hull $\bar \Delta$ of the points 
$\bar \nu^{(i)}$. From $O$ one can span an incomplete fan  
through $\bar \Delta$, which defines $X_{\bar \Delta}$ as a 
non-compact toric variety with trivial canonical bundle, i.e. 
$\Delta_d$ defines a ($d$+1)-dimensional non-compact toric 
Calabi-Yau variety ${\cal O}(-K_{\mathbb{P}_{\Delta_d}}) 
\rightarrow \mathbb{P}_{\Delta_d}$. The toric fans for the 
compact twofold and the non-compact threefold are  shown in figure 2.   

Note that since we add the origin  $O$  the construction  for the 
non-complete fan and hence the non-compact Calabi-Yau threefold 
does not require that $\Delta_d$ is reflexive.  Any maximally 
triangulated convex polyhedron, whose points span the 
lattice $\bar \Gamma$, will lead to a smooth non-compact CY 
($d$+1)-fold, otherwise to a singular one, which can be crepantly 
resolved to  a  smooth non-compact CY ($d$+1)-fold.   In 
particular $\Delta_s$ can have an arbitrary number of  
inner points.

\begin{figure}[h!] 
\begin{center} 
\includegraphics[angle=0,width=.9\textwidth]{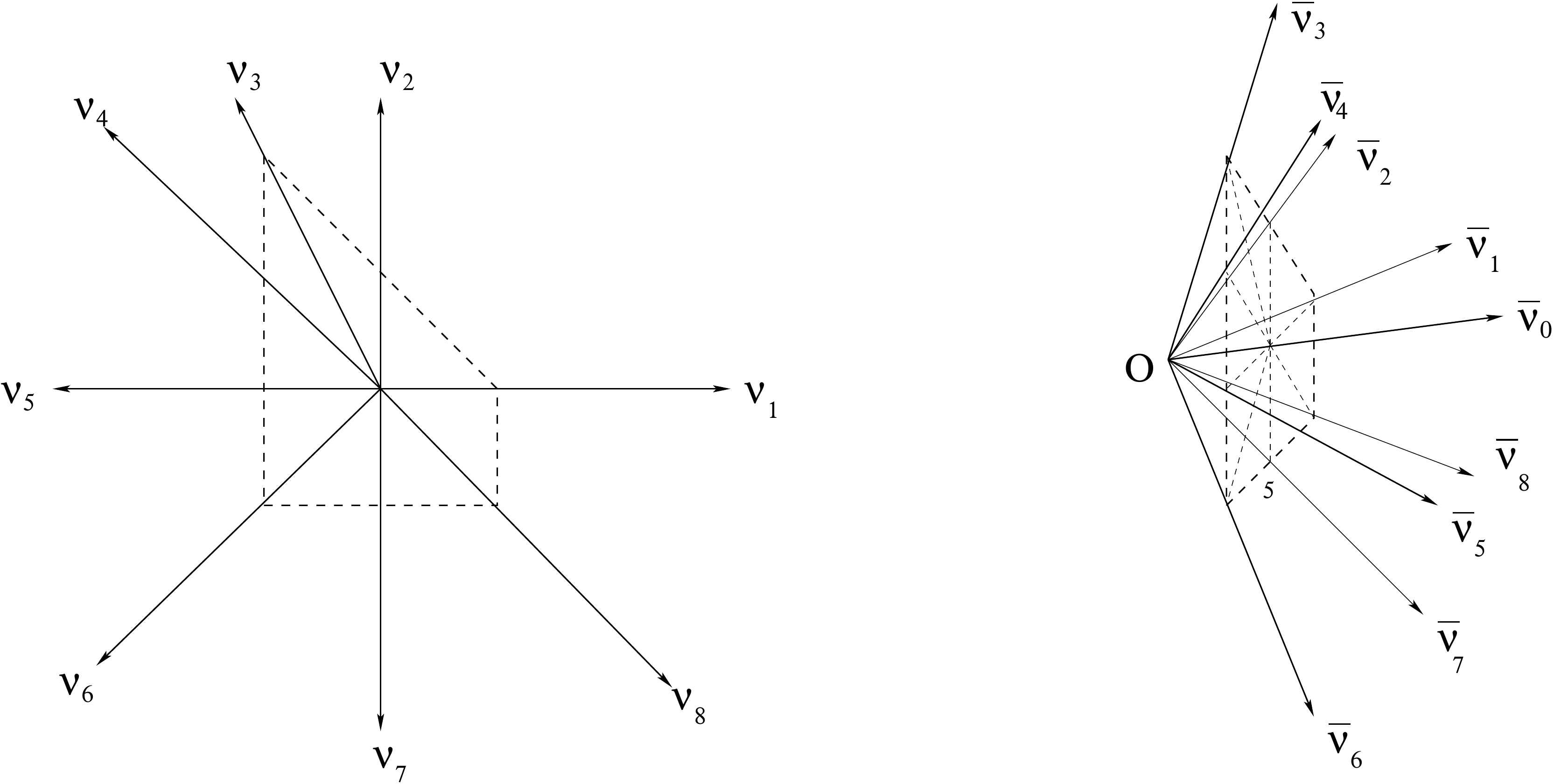} 
\begin{quote} 
\caption{Here we show the fan that yields the compact almost del Pezzo surface 
associated to polyhedron $\Delta^{(14)}_2$ on the right and ${\overline {\Delta^{(14)}_2}}$ 
one that yields ${\cal O}(-K_{B_2})\rightarrow B_2$ on the left. 
\vspace{-1.2cm}} \label{poly}  
\end{quote} 
\end{center} 
\end{figure}

\begin{figure}[h!] 
\begin{center} 
\includegraphics[angle=0,width=.8\textwidth]{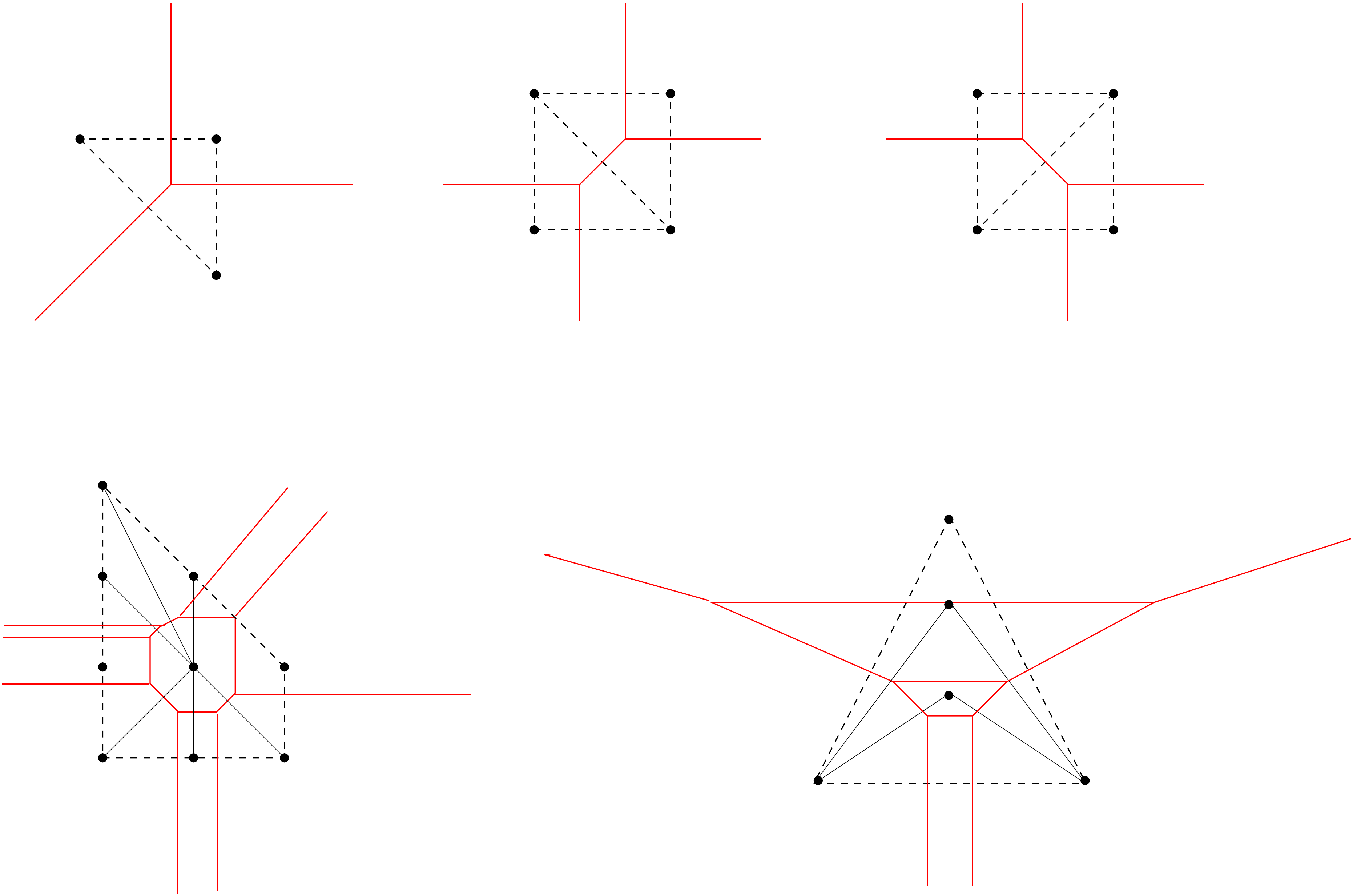} 
\begin{quote} 
\caption{Here we show the graph of the toric diagram in light black  and dashes black and the 
dual graph in solid red. The latter is interpreted  as web of $[p,q]$ 5-branes of the type IIB 
string.  The geometries are $\mathbb{C}^3$, ${\cal O}(-1)\times {\cal O}(-1)\rightarrow \mathbb{P}^1$, 
with the $\mathbb{P}^1$ flopped,    ${\cal O}(-K_{P_{\Delta^{(14)}}})\rightarrow {P_{\Delta^{(14)}}}$ , 
and an $A_2$ fibration over $\mathbb{P}^1$.   
\vspace{-1.2cm}} \label{pqwebs} 
\end{quote} 
\end{center} 
\end{figure}

The coordinate ring of $\mathbb{P}_{\Delta_d}$ is defined
most explicitly in~\cite{Cox} as follows.  Let $l(\Delta)$ denote all 
integer points in $\Delta$ and the vectors $l^{(k)}$, $k=1,\ldots,l(\Delta_d)-d-1$  specify a basis
of linear relations among the points of $\Delta_d$, i.e. 
\be 
\sum_i l^{(k)}_i \nu^{(i)}=0\ .
\label{chargevectors}
\ee
The toric variety can be defined in the coordinate ring $Y_i$ as 
\be 
\mathbb{P}_{\Delta}={\mathbb{C}^{|\Sigma(1)|}\setminus Z_{\Delta}\over {\rm Hom}(A_{d-1}
(\mathbb{P}_\Delta),\mathbb{C}^*)}\ .
\label{Pdelta}
\ee
Here ${\rm Hom}(A_{d-1}(\mathbb{P}_\Delta),\mathbb{C}^*)=(\mathbb{C}^*)^{{\rm rank} 
A_{d-1}(\mathbb{P}_\Delta)}\times A_{d-1}(\mathbb{P}_\Delta)_{\rm tors}$ 
and the $\mathbb{C}^*$-action is specified by the $l^{(k)}$ 
as 
\be 
Y_i\mapsto Y_i (\mu^{(k)})^{l^{(k)}_i} \quad  \forall i ,
\label{scalings}
\ee
with $\mu^{(k)}\in \mathbb{C}^*$. $Z_{\Delta}$ is the  
Stanley-Reisner ideal. Its substraction guarantees 
well-defined orbits under the torus action. It is 
determined from a triangulation of $\Delta$ and consists of all 
loci in the intersection of divisors $D_{i_1}\cap \ldots 
\cap D_{i_r}$ for which the set of corresponding points 
$\{ \nu_{i_1},\ldots,\nu_{i_k}\}$ are not on a common 
triangle. The triangulation  determines also the  generators 
$l^{(k)}$ of the Mori cone, which  is dual to the 
K\"ahler cone, i.e. to each $l^{(k)}$ there is a dual 
curve whose volume vanishes at the boundary of the K\"ahler 
cone. Since all Mori cones and triangulations have 
been calculated in~\cite{Klemm:2012} we just give an 
example: Let us label the points of the ${\cal O}(-1)\times {\cal O}(-1)\rightarrow \mathbb{P}^1$ 
polyhedron in figure 3 counter-clockwise starting from the right lower corner, 
the first triangulation corresponds to a Mori vector  $l^{(1)}=(-1,1,-1,1)$. The coordinates 
$(y_2:y_4)$ are homogeneous coordinates of the compact  $\mathbb{P}^1$, 
whose positive volume is the K\"ahler cone and  $y_2=y_4=0$ is the Stanley-Reisner ideal. 
$y_1$ and $y_3$ are the line bundle coordinates. The coordinates of the flopped $\mathbb{P}^1$ with  
$l^{(1)}=(1,-1,1,-1)$ are correspondingly given by $(y_1:y_3)$ etc. 

The coordinate ring of $X_{\bar \Delta_d}$ is defined similarly by (\ref{Pdelta}), 
with $\Delta$ replaced with $\bar \Delta$ in the definition of  (\ref{chargevectors}).
Note that the inner point  $\bar \nu^{(0)}=(1,0,\ldots,0)$ spans now a 
ray $\Sigma(1)$ and corresponds to a new coordinate belonging to 
the new non-compact direction of $X_{\bar \Delta}$.  
Note that in this case $\sum_i l^{(k)}_i=0$, as the points
 lie in a plane\footnote{In the equivalent description by an abelian  2d  
gauged linear $\sigma$-model it ensures the cancellation of the 
axial anomaly.}. It is easy to see from (\ref{scalings}) that this 
condition ensures the existence of a globally defined $(d+1,0)$-form, 
hence $X_{nc}=X_{\bar \Delta_d}$ is a non-compact CY $(d+1)$-fold.

\subsection{Global and local mirror symmetry}
\label{section:mirror1}  
As it is familiar from Batyrev's mirror construction~\cite{Batyrev} we can 
view each polyhedron $\Delta$ in two ways. Firstly as defining $\mathbb{P}_\Delta$ as 
explained above  and  secondly  as defining the Newton polyhedron for a polynomial 
$W_\Delta(Y)$, where the coordinates of the points determine the exponents of the 
$Y_i$.   

It is useful for the following to recall the difference between compact and non-compact toric 
mirror symmetry. In the compact case the Calabi-Yau $X$, or $X_{\Delta^*}$ -- 
the anti-canonical divisor in $\mathbb{P}_{\Delta^*}$ -- 
is defined as a section of the anti-canonical bundle
\be 
W_\Delta=\sum_{i=0}^{l(\Delta)-1} a_i X_i=\sum_{\nu^{(i)}\in \Delta}a_i \prod_{\nu^{*(k)}\in\Delta^*} Y_k^{\langle \nu^{(i)}, \nu^{*(k)}\rangle+1}=0
\label{WDelta}
\ee   
in the coordinate ring $Y_k$ of $\mathbb{P}_{\Delta^*}$. 
Here the coefficients $a_i$ parametrize (redundantly) 
the complex structure of $X$. In compact mirror symmetry 
points $\nu^{(i)}$ ($\nu^{*(k)}$) inside codimension one
faces of $\Delta$ ($\Delta^*$)  can be excluded from the sum (products) above, 
because the corresponding monomials $X_i$ can be removed by the automorphism 
group acting on $\mathbb{P}_{\Delta^*}$ while the corresponding variables $Y_k$ describe 
exceptional divisors in the resolution  of singularities that vanish outside of  $W_\Delta=0$.

Similarly the mirror to $X$ called $X^*$ (or more specifically $X_{\Delta}$)  
is defined as a hypersurface 
\be 
W_{\Delta^*} =\sum_{i=0}^{l(\Delta^*)-1} a^*_i Y_i=
\sum_{\nu^{*(i)}\in \Delta^*} a_i^* \prod_{\nu^{(k)}\in\Delta} X_k^{\langle \nu^{*(i)}, 
\nu^{(k)}\rangle+1}
\label{WDelta*}
\ee   
in the coordinate ring of $\mathbb{P}_{\Delta}$. 
 
Note that $\mathbb{P}_{\Delta^*}$ can be embedded as a singular 
variety in $\mathbb{P}^{l(\Delta^*)}$ with the constraints
\be 
\prod_{i} Y_i^{l^{(k)}_i}= 1\qquad \forall k \ .
\label{mirrorrelations}
\ee         
By construction the $Y_i$ in (\ref{WDelta*}) viewed as a function of the 
$X_k$ fulfill this constraint. The quotient construction 
of mirror symmetry can be realized, if there is an embedding map 
$\Phi:(\Delta^*,\Gamma^*)\rightarrow  (\Delta,\Gamma)$. 
This defines the {\^e}tale map from the $X_i$ to the $Y_i$.
For the example of the quintic the relevant charge vector (\ref{chargevectors}) is $l^{(1)}=(-5,1,1,1,1,1)$ 
and the    {\^e}tale map is 
\begin{equation}
\phi:(X_0:X_1:\ldots : X_5)\mapsto  (\prod_{i=1}^5X_i:X_5^5:\ldots : X_5^5)= (Y_0:Y_1:\ldots : Y_5)\ , 
\label{etale}
\end{equation}  
which is many to one and is made unique by identifying 
the $X_k$ with the mirror quotient group $G$. I.e. ${\rm kern} (\Phi)=G$  and the 
order of $G$ is the degree of $\Phi$. 

For example the pairs $(\Delta_2,\Delta_2^*)$ define 
one-dimensional compact Calabi-Yau hypersurfaces $W_{\Delta_2}(Y)=0$ 
in (almost) del Pezzo surfaces $\mathbb{P}_{\Delta^*_2}$, 
 i.e. elliptic curves and all $a_i$ up to one can be set to $0$ 
or $1$ by the automorphism group of 
$\mathbb{P}_{\Delta^*_2}$ and rescalings of $Y_i$ 
leaving (\ref{mirrorrelations}) invariant.

The construction of non-compact mirror symmetry described first in \cite{KKV} restricts this construction
to the coordinate ring defining $X_{nc}$. One starts therefore with\footnote{In the following sections 
we drop the $^*$  for notational convenience.}            
\be 
W_{\Delta_d^*}=\sum_{i=0}^{l(\Delta_d)-1} a^*_i Y_i=0\ ,
\label{mirrorcurve}
\ee
where $\Delta_d^*$ is a $d$-dimensional polyhedron, 
not necessarily reflexive.  In the case of a global embedding 
$X_{nc}\subset X$, $\Delta^*_d\subset \Delta^*_{d'\ge d+2}$  is at least a two-codimensional 
face of a $d'$-dimensional reflexive polyhedron $\Delta^*_{d'}$. 
In this case $\Delta^*_d$ defines a ($d$+1)-dimensional  
lattice $\bar \Gamma_{d+1}$ as described at the 
end of the last section and by the reflexivity of $\Delta^*_{d'}$ it lies as 
$\bar \Delta^*_{d}$ in a hyperplane at distance one from the origin. 
The corresponding in-complete fan describes the non-compact CY ($d$+1)-fold 
$X_{\bar \Delta_d}$ inside the compact CY  ($d'$-1)-fold.  
In contrast to the compact mirror symmetry discussed above 
there are no automorphisms in $X_{\bar \Delta_d}$ to remove 
monomials in (\ref{mirrorcurve}), hence the sum runs over all points in $\Delta_d$.

Let $l^{(i)}$ 
generate a basis of linear relations $\sum_i l^{(k)}_k 
\bar \nu^{(i)}=0$ among the points of $\bar \Delta_n$, 
which define (\ref{mirrorrelations}). These relations restrict 
the possibility to undo deformations of the $a_i^*$  by rescalings 
of $Y_i$, leaving   $l(\Delta^*_d)-d-1$ independent deformations 
of the B-model. A convenient way to introduce these in the 
curve is  to set all $a^*_i=1$  and modify (\ref{mirrorrelations}) to
\be 
\prod_{i} Y_i^{l^{(k)}_i}= z_k \qquad \forall k \  .
\label{modmirrorrelations}
\ee 
Here we use Batyrev's coordinates    
\be 
 z_k=\prod_i {a^*}_i^{l^{(k)}_i}
 \label{batyrevcoordinates}
\ee
so that $z_k=0$ is the large complex structure point. 

In this description  (\ref{mirrorcurve}) with $a^*_i=1$,  (\ref{modmirrorrelations})
and a $\mathbb{C}^*$-identification $Y_i\sim  \mu Y_i$ with $\mu \in \mathbb{C}^*$ 
define the mirror geometry.  It can be written as a $(d-1)$-dimensional affine variety by 
adding to the singularity $W_{\Delta^*}=0$  trivial non-compact normal 
directions as quadratic coordinates. E.g. for $\Delta^*_2$  it is    
\begin{equation}
H(X,Y,{\underline{z}}):=W_{\Delta^*_2}(X,Y,{\underline{z}})=uv\ .
\label{mirrorequation} 
\end{equation} 
Note that in order to solve (\ref{modmirrorrelations}) in 
favor of two variables say $X,Y$ we have to view $Y_i$  
as $\mathbb{C}^*$-variables. $W_{\Delta^*}(X,Y)=0$ becomes
in general a Laurant polynomial in $\mathbb{C}^*$-variables
 defining a genus $g$ Riemann surface $\Sigma$ with 
$h$ punctures. Here $g$ is the number of inner points 
in $\Delta^*_2$ and $h=l(\Delta^*_2)-g$. The nowhere 
vanishing holomorphic $(3,0)$-form can be defined 
in a coordinate patch of the $(d+1)$-dimensional ambient 
space by a contour integral 
\be 
\Omega=\frac{a_0}{(2 \pi i)} \oint_{W=0} \left(\prod_{p} Y_p\right)\frac{\wedge_{j=1}^{d+1} \frac{d Y_j}{Y_j}}{W}
\ee 
and restricts to the Riemann-surface $H(X,Y,{\underline{z}})$ as~\cite{Katz:1996fh}
\be 
\lambda=\log(X) \frac{d Y}{Y}\ . 
\ee
In local mirror symmetry we study the variation of mixed Hodge structures 
of the non-compact local Calabi-Yau spaces using this logarithmic 
form on $H(X,Y,{\underline{z}})$ in particular by analysing its 
Picard-Fuchs equation. 

The formalism was certainly well known in the study of mixed Hodge 
structure assocuated to singularties, see e.g.~\cite{Saitodiff} or~\cite{Hertling} 
for a review. The variation of the mixed Hodge structure for log Calabi-Yau 
spaces $(X,D)$ with $D$ a divisor in particular the isomorphism
\be 
\phi:H^3(X\setminus  D) \rightarrow \bigoplus_{p+q=3} H^q(X,\Omega^p_{\bar X}(\log(D))        
\ee
to the log cohomology has been used to calculate 
superpotentials in~\cite{Grimm:2008dq} and a recent 
application to stable degenerations~\cite{Donagi:2012ts} is 
similar to the local mirror symmetry for vertical divisors with a 
transition to del Pezzo surfaces discussed in section 
\ref{verticaldelpezzo}.

The inner points deform the complex structure of $\Sigma$, 
while the punctures define by the counting of independent
deformations $l(\Delta)-g-3$ independent residue values
of $\lambda$ refered to as masses $m_i$ and $i=1,\ldots,l(\Delta)-g-3$.

In the case of a del Pezzo basis there is only one 
inner point whose coefficient $a_0$ is 
identified  with the complex structure of the elliptic 
curve  $W_\Delta^*(X,Y)=0$, physically related to 
the gauge coupling of the $U(1)$ theory on the 
Coulomb branch while $l(\Delta^*_2)-4$ of the $a^*_i$ 
are identified with mass parameters. 

There  is a  physical interpretation for the dual graph associated 
to a general triangulated polyhedron. It can be viewed as a web of $[p,q]$5-branes 
for the type IIB string. These 5-branes fill the $0,\ldots,5$ directions 
of the five-dimensional space-time. The figure corresponds to the $(5,6)$-plane, where the 
5-branes   extend as lines, whose slope is given by the $SL(2,\mathbb{Z})$ 
charge  $[p,q]$.

\subsection{Global embeddings of the local geometries}
Let us now discuss two kinds of global embeddings of 
local geometries in compact Calabi-Yau spaces $X$. 
Both are related to elliptic fibrations. In the first the 
del Pezzo appears as the base and all $(1,1)$-classes 
of the del Pezzo are are $(1,1)$-classes in $X$ in the 
second a rational elliptic fibration typically a half 
K3 appears as a so-called vertical divisor over blow-ups 
in the base. After flopping out a number of 
$\mathbb{P}^1$ the rational elliptic fibration 
becomes a del Pezzo, which can be blown down.                    

Both global embeddings can be studied in very concrete global 
embeddings of the reflexive polyhedra $(\Delta^B_n,\Delta^{B*}_n)$ 
into a pair  of reflexive polyhedra  $(\Delta_{n+2},\Delta^*_{n+2})$,
so that the anti-canonical hypersurface in $\mathbb{P}_{\Delta^*_{n+1}}$ gives
rise to an elliptically fibred Calabi-Yau $(n+1)$-fold 
over the toric base $\mathbb{P}_{\Delta_n^{B*}}$ with an 
interesting structure of global sections. 

For notational simplicity and because the virtual dimension of 
the moduli space of stable pairs is only zero for threefolds,     
we outline the embedding of two-dimensional polyhedra in a 
four-dimensional polyhedron, which gives rise to an 
elliptically fibred threefold over a toric  
del Pezzo base $\mathbb{P}_{\Delta_2^{B*}}$, 
specified by $\Delta_2^{B*}\in \Delta_4^*$. However 
everything in this section, except for (\ref{hodgefibre})\footnote{For 
which aspects of the  generalization have been discussed in~\cite{Klemm:2012}.}, 
generalized trivially to arbitrary dimension.      

The reflexive pair $(\Delta^*_4,\Delta_4)$ is the convex hull of 
the following points   
\begin{equation} 
  \footnotesize 
  \begin{array}{|ccc|ccc|} 
   \multicolumn{3}{c}{ \nu^*_i\in \Delta_4^*} &\multicolumn{3}{c}{ \nu_j\in \Delta_4}  \\ 
    &            &\nu_i^{F*} &             &\nu_j^{F} & \\
    & \Delta^{B*}_2 & \vdots                &s_{ij}\Delta^{B}_2&\vdots & \\
    &            &\nu_i^{F*} &                   & \nu_j^{F}& \\
    & 0 \ldots 0      &                       & 0\dots 0              &                          & \\
    & \vdots     &\Delta^{*F}_2             &   \vdots          & \Delta^{F}_2            & \\
    & 0 \ldots 0      &                       & 0\ldots 0              &                          & \\
  \end{array} \, .
\label{polyhedrafrombaseandfibre}
\end{equation} 
Here we consider all points $\nu_j^{F*}\in \Delta_F^*$ 
and define 
\begin{equation} 
s_{ij}= \langle \nu_i^F,\nu_j^{F*} \rangle+1\in \mathbb{N}\ .
\label{scalingpoly}
\end{equation}
Note that  we scaled $\Delta^B_2 \rightarrow s_{ij}\Delta_2^{B}$. This 
means to scale the coordinates of the points of $\Delta^{B}_2$ 
by $s_{ij}$ while keeping the original lattice basis, i.e. 
$s_{ij}\Delta_2^{B}$, contains in general more lattice points.  
Note that the vertices of $\Delta^*$ ($\Delta$)  are given by the vertices 
of the polyhedra $\Delta^{*F}_2$ ($\Delta_2^{*F})$  and $\Delta^{*B}_2$ ($s_{ij} \Delta_2^B$) 
respectively.

Both polyhedra $\Delta $ and $\Delta^*$ have 
the following features in common, which we spell out only 
for $\Delta$ in this paragraph, where we also 
call $s_{ij}\Delta_2^B$ simply $\Delta_2^B$. 
They contain a  polyhedron $\Delta_2^{F}\subset 
\Delta_4$ that is a sub-polyhedron of $\Delta_4$ and  
shares the unique inner point.  It also
implies an exact sequence of the lattices 
$0\rightarrow \Gamma_F\rightarrow \Gamma \rightarrow 
\Gamma_B\rightarrow 0$, where  $\Gamma_F$ is the sublattice 
associated to $\Delta_2^F$. Further the base 
polyhedron $\Delta_2^B$ is a two-face of $\Delta_4$ and the image 
of a projection along the fibre polyhedron, i.e. obtained by identifying 
all other points modulo $\Delta_2^F$. These  are necessary conditions
for $P_{{\Delta_4}}$ to have a fibration 
\begin{equation} 
\mathbb{P}_{{\Delta_4}}\rightarrow \mathbb{P}_{\Delta_2^B}
\label{fibration}
\end{equation}
over $P_{\Delta_2^B}$ with $P_{\Delta_2^F}$ as the generic fibre.
As stated in \cite{Fulton} a sufficient condition (F1)  
for the existence of the above fibration as a (smooth)
and flat one, is the existence of a  fan  $\Sigma_{\Delta}$ 
(whose cones have lattice volume one) and which  
is defined by a triangulation of $\Delta$  that 
lifts from a fan $\Sigma_{\Delta^B_2}$. The 
hypersurface $W_{\Delta_4*}=0$ in $P_{{\Delta_4}}$ 
becomes then an elliptic fibration whose generic fibre 
is defined as the section of the anti-canonical bundle in  
$P_{\Delta_2^F}$.        

It easy to see that (F1) is fulfilled  for $P_{{\Delta^*_4}}$ 
so that $X$ given by $W_{\Delta_4}=0$ is a smooth and flat 
elliptic fibration.   The problem in establishing  (F1)  
for $X^*$ is the scaling of $\Delta_2^B$. In this case 
$W_{\Delta^*_4}=0$ is in general only a non-flat elliptic fibration.      

The Euler number and the Hodge number of $X$ depend in a simple way 
on the base and the type of the fibration as  
\be 
\chi(X)= - a_F \int_B c_1^2, \qquad h_{11}(X)=l(\Delta^*_B)-4+b_F+t_F\ .
\label{hodgefibre}  
\ee   
where $a_F,b_F,T_F$ depend only on the fibre type, which is in turn specifed by  
$\Delta^{*F}$ and $\nu^{*F}$. We therefore give $a_F/b_F(t_F)$  at the 
inequivalent corners  of the polyhedra  in figure 1. The contribution of the 
base classes to $h_{11}(X)$ is $l(\Delta^*_B)-3$. They correspond to vertical 
divisors. These are rational surfaces. One of the $b_F+t_F+1$ classes 
comes from the zero-section of the base in the fibre or if $t_f>0$ 
this can be a $t_f+1$ multi-section.  The rest can come either from additional 
rational sections or from gauge symmetry enhancements. Since the Euler 
number is proportional to $c_1^2$ the gauge symmetry enhancement  
occurs along a multiple of the canonical divisor. Which of these possibilities 
is realized  can be distinguished by analyzing which Kodaira fibre 
occur in $W_{\Delta_4}=0$,  we will discuss this further for 
some specific fibrations below.        

\subsubsection{The del Pezzo surface as base}  

Let us take the tenth polyhedron $\Delta(10)$ as  
$\Delta_2^{*F}$ i.e. $\Delta_2^{*F}=\text{conv}\{(1,0)$, $(-1,2)$, $(-1,-1)\}$ and  
for $\nu_3^{F*}=(-1,-2)$. Since $\Delta_2^{*F}$ is self-dual,
$\Delta_2^{F}=\text{conv}\{(1,0)$, $(-1,2)$, $(-1,-1)\}$ and  $\nu_j^{F}=(-1,-2)$, 
so $s_{ij}=6$. Start for the base with the first polyhedron, i.e.  
$\Delta_2^{*B}=\Delta(1)$. In a hopefully obvious notation 
referring to figure 1 we denote this manifold $X_{(\Delta^F(10)\times_F^3\Delta^B(1))}$.    
Since $\mathbb{P}_{\Delta(1)}=\mathbb{P}^2$,
this yields an elliptic fibration over $\mathbb{P}^2$. On the 
corresponding corner we find $a_F=60$ and  $b_F=2$, i.e. we get $\chi(X)=-60\times 9=-540 $, 
$h_{11}(X)=2$ and this is an elliptic fibration with a single 
section, since there are no additional `twisted` states, if we 
chose $\Delta_2^{*B}=\Delta(3)$, we obtain an elliptic fibration with 
a single section over the blow-up of $\mathbb{P}^2$ the Hirzebruch surface  
$\mathbb{F}_1$ with $\chi(X)=-60 \times 8 =-480$ 
and $h_{11}(X)=4$ etc. This creates a first series of 15 fibrations with one section. 
Denoting the coordinates associated to the points $CO=\{(0,0,0,0)$, $(0,0,1,0)$, $(0,0,-1,-1)$, $(0,0,-1,2)\}$ 
by $x_0$, $y$, $x$ and $z$ we see from (\ref{WDelta}) that 
$W_\Delta$ is in the Tate form 
\begin{equation}
x_0 W_\Delta= x_0 (y^2+h_1({\underline Y}_B) x y z+ 
h_3({\underline Y}_B) y z^3-(x^3+ h_2({\underline Y}_B) x^2 z^2+ h_6({\underline Y}_B)z^6))     
\end{equation}
and the pure monomials in the variables ${\underline Y}_B$ corresponding 
to the coordinate ring of the base are multiplied with $z$. 
Hence  at $z=0$ we get a section, which is the 
un-constrained Fano variety $\mathbb{P}_{\Sigma_{\Delta^{B*}}}$.  The 
constraint $W_\Delta=0$ is solved by $x^3=y^2$, which has a 
unique solution, up to automorphisms in chosen $P_\Delta^{*F}$, 
so that this fibration has a single section. I.e. we get a 
fibration map
\be
X_\Delta \rightarrow \mathbb{P}_{{\Delta^{*B}_2}} \, ,
\ee
whose generic fibre is the elliptic curve ${\cal E}= X_{\Delta_2^{*F}}$.
The non-compact Calabi-Yau manifold ${\cal O}(-K_{P_{{\Delta^{B*}_2}}})
\rightarrow {P_{{\Delta^{B*}_2}}}$ is obtained by 
scaling the volume of the ${\cal E}$ fibre to infinity. 
This limit can be made very precise, because the Mori vector, that 
corresponds to the fibre is given by $l^{(E)}=(-6,3,2,1,0,\ldots,0)$, 
w.r.t. the points in $CO$ and has no entry at other points. By 
(\ref{batyrevcoordinates}) $z_E \sim e^{-t_E}$ where $t_E$ is the 
volume of the fibre, so the limit is $z_E=0$. We discuss the 
local mirror further in section~\ref{mirrorsymmetry}.

\subsubsection{The del Pezzo as a transition 
of a vertical divisor and rational sections} 
\label{verticaldelpezzo} 
If one blows up a $\mathbb{P}^1$ in the base one 
gets as  additional divisor a vertical 
divisor~\cite{Morrison:1996pp,Klemm:1996hh}, 
which is a half $K3$ realized as a rational elliptic fibration 
${\cal E}\rightarrow \mathbb{P}^1$ over the  
exceptional $\mathbb{P}^1$. It inherits the fibration 
structure of the generic fibre, e.g. the number of 
rational or multi-sections, which makes general 
fibrations ${\cal E}= X_{\Delta_2^{*F}}$ interesting 
to study. 

We give an example below, but discuss first the most generic 
case ${\cal E}= X_{\Delta(10)}$. The simplest case is when 
we blow up the $\mathbb{P}^2$ and obtain a Hirzebruch surface $\mathbb{F}_1$. 
With the choice of the Mori cone 
\be
l^{({\cal E})}=(-6;3,2,1,0,0,0,0),\quad l^{(\mathbb{F}_1^F)}=(0;0,0,-1,-1,1,1,0),\quad 
l^{(\mathbb{F}_1^B)}=(0;0,0,-2,1,0,0,1)\ ,
\label{coneembede8} 
\ee
$X$ has an elliptic as well as a K3 fibration, where $l^{({\cal E})}$ corresponds to the elliptic fibre, 
$l^{(\mathbb{F}_1^B)}$ represents the base of the K3 fibration and the base of $\mathbb{F}_1$, 
while $l^{(\mathbb{F}_1^F)}$ corresponds to the exceptional $\mathbb{P}^1$, 
the fibre of the Hirzebruch surface, and the base of the elliptic 
surface. As explained in section 3.1, if we flop this $\mathbb{P}^1$ 
then the elliptic surface ${\cal E}\rightarrow \mathbb{P}^1$ becomes the elliptic 
pencil (\ref{ellipticpencil}) with exactly one base base point 
i.e. the $E_8$ del Pezzo. As explained in~\cite{Morrison:1996pp,Klemm:1996hh} 
the new phase is characterized by $l^{(1)}=
l^{({\cal E})}+ l^{(\mathbb{F}_1^F)}=(-6;3,2,0,-1,1,1,0)$ and  
representing the canonical class in the del Pezzo, 
$l^{(2)}= -l^{(\mathbb{F}_1^F)}$ representing the flopped curve 
and $l^{(2)}= l^{(\mathbb{F}_1^F)}+ l^{(\mathbb{F}_1^B)}$ 
representing the hyperplane class in $\mathbb{P}^2$.     
Now if a del Pezzo shrinks the Higgs branch of the 
corresponding $N=2$ field theory opens up and by deforming 
the singularity one gets, according to (\ref{hodgefibre}) 
with $a_F=60$, a transition to a Calabi-Yau manifold  
$X'$ with $h_{21}(X')=h_{21}(X)+29$, $h_{11}(X')=h_{11}(X)-1$, 
i.e. since one vector multiplet is going to be massive by the Higgs effect,   
the Higgs branch is of dimension $a_F/2$. The fibres  
\be 
{\cal E}=X_{\Delta(10)},\ \ X_{\Delta(4)},\ \ X_{\Delta(1)}                   
\ee
correspond to the $E_8$, $E_7$ and $E_6$ fibres. In 
general the dimension of the Higgs branch is $a_F/2=30,\ 18$ 
and $12$ appearing at one corner of these polyhedra is the dual 
Coxeter number of the group. These models have one, two and three 
multi-sections, i.e. if we blow down the base of the elliptic surface  
as before, we get elliptic pencils with one, two and three 
base points, corresponding to the $E_8$, $E_7$ and 
$E_6$ del Pezzo surfaces. 

In the case of vertical divisors not all classes of rational surfaces are classes in the 
Calabi-Yau space $X$, i.e. the image of the inclusion map $i^*:H^{1,1}(X) 
\rightarrow H^{1,1}(S)$ has rank $1\le k\le h^{1,1}(S)$. According to the theorem  
of N\'eron, sections of the elliptic surface are in the 
image and as already observed in~\cite{Morrison:1996pp} they 
can be extended as sections over the entire base. Once they are flopped the 
elliptic pencil develops base points and the 
corresponding del Pezzo can be shrunken. This was 
studied in~\cite{Klemm:1996hh} for $E_7$ and $E_6$ with  
toric divisors and as a an explicetly mentioned by-product 
fibrations with a holomorphic zero-section and additional 
rational toric sections were constructed and their K\"ahler
classes identified as Wilson line parameters that break the 
$E_8$ of the tensionless string sucessively to $E_7$ and $E_6$. 
These K\"ahler parameters correspond to the new rational global 
sections. Moreover the precise 
breaking of the $E_8$ representations into $U(1)\times E_7$ and 
further  into $U(1)_1\times U(1)_2 \times  E_6$ representations 
was studied on the level of genus 
zero BPS states~\cite{Klemm:1996hh}.   
The $U(1)$'s are supported at the new global sections.
The $U(1)_1\times U(1)_2 \times  E_6$  model is based
$\Delta^{*F}_2=\Delta(5)$ the double blow-up of 
$\mathbb{P}^2$ with  $\nu_2^{F*}=\nu_2(\Delta(5))$. 
Then $a_F/b_f(t_f)=24/4(0)$ and if we fibre over 
$\mathbb{P}^2_{\Delta(3)}=\mathbb{F}_1$ the $\chi=-192$, 
$h_{11}=5$ manifold $X_{(\Delta^F(5)\times_F^2\Delta^B(3))}$  
allows for an $E_6$ del Pezzo transition to  $X_{(\Delta^F(5)\times_F^2\Delta^B(1))}$  
the elliptic fibration over $\mathbb{P}^2$ with $\chi=-216$, $h_{11}=4$. 
Since the discussion of the toric realization of the 
transition with more sections in~\cite{Klemm:1996hh} was 
too sketchy we give the Mori cone data in the appendix.

These examples have been re-discovered and interesting variants 
that have three rational sections have been 
discussed in detail in ~\cite{Cvetic:2013nia,Cvetic:2013uta}. 
The dual $\Delta^{F}_2$ has the Newton polynom (\ref{generalcubic}) 
with $s_4=s_{10}=0$. This curve has therefore three sections 
at $[x:y:z]=\{[0:0:1],[0:1:0],[0:s_9:-s_7]\}$, where in $X$ $s_7$ and 
$s_9$ are line bundles over the base. Calling the associated 
divisors $S_7$ and $S_9$ a straightforward application of the adjunction 
formula \cite{Cvetic:2013uta} yields $\chi(X)=\int_B 
(-24 c_1^2+ 8 c_1 S_7 -4 S_7^2 +8 c_1 S_9 +2 S_7 S_9- 4 S_9^2)$, i.e. 
the four cases $S_7=S_9=0$,  $S_7=c_1, S_9=0$, $S_7=0, S_9=c_1$ 
and $S_7=S_9=c_1$ in which $\chi(X)$ becomes proportional to $c_1^2$ 
reproduce the $X$ with the Euler numbers that are realized toric 
hypersurfaces given by the combination of the data in Fig 1 
with (\ref{polyhedrafrombaseandfibre}).

A more extrem case is to choose $\Delta_2^{*B}=\Delta_2^{*F}=
\Delta(15)$ and $\nu_i^{F*}=(0,0)$. Then $\nu_i^{F}=(0,0)$ and 
we get $\chi(X)=-6 \times 3=-18$ and $h_{11}(X)=10-4+8=14$. 
It is also easy to see that in these cases, where we fibre using 
the anti-canonical class  $\nu_i^{F*}=(0,0)$, we have  
$s_{ij}=1$ and $X^*$ is also a flat elliptic fibration 
with the dual fibre polyhedron. In fact both $X$ and $X^*$ exhibit two 
elliptic fibrations with base and fibre polyhedron exchanged. 
Using the formula for intersections for the second elliptic fibration one 
discovers in this case an $SU(3)^3$ gauge enhancement 
over the anti-canonical divisor over the base.         

\subsection{A short cut to the mirror geometry}
\label{mirrorsymmetry}
The mirror of the del Pezzo in the base must occur as a specialization of  
the constraint (\ref{WDelta*}). Due to the form of $\Delta$  
we can always find a triangulation that leads 
to an elliptic fibration, not necessarily a smooth 
and flat one. However for the discussion of the complex 
deformations of the mirror geometry, this is good enough. 
Denoting  the coordinates associated to  $(0,0,-1,-1),\,
(0,0,2,-1)\, \text{and}\, (0,0,-1,1)$ again by $y$, $x$ and $z$, $W_{\Delta*}$ is realized 
in the  Tate form. In the mirror geometry the 
restriction is given by $y=x=0$ implying 
$z\neq 0$ because of the form of the Stanley-Reisner ideal and hence the constraint  
\begin{equation}
W_{\Delta^*}=z^6 h^6_{\Delta^*_B}({\underline X}'_B)=0\  
\end{equation}
implies $h^6_{\Delta^*_B}({\underline X}'_B)=0$.
Further note  the possibility of rescaling $z$ 
which leads to the afore mentioned $\mathbb{C}^*$-identification 
$X_{iB}'\sim \mu X_{Bi}'$ with $\mu \in \mathbb{C}^*$. 
Secondly the scaling (\ref{scalingpoly}) is only due to  
the global embedding and the corresponding refinement of 
the  lattice of $\Delta'$ w.r.t to $\Delta$ can be
undone in the local case  by an {\^e}tale map  
\begin{equation}
\phi_{s_{ij}}:(X_0',\ldots:X_{l(\Delta_B)}')\mapsto  
(X_0^{s_{ij}}:\ldots:X_{l(\Delta_B)}^{s_{ij}})\ .
\label{scale}
\end{equation}
Hence the mirror geometry to ${\cal O}(- K_{\mathbb{P}_{\Delta^*}})\rightarrow  
K_{\mathbb{P}_{\Delta}^*})$ is simply given as the Newton polynomial  
\begin{equation} 
H({\underline X})=h^6_{\Delta^*_B} ({\underline X}_B)=0
\end{equation} 
of $\Delta^{*B}_2$ itself.

We define therefore the coordinates of Newton polynomials of $\Delta^{*B}$ 
for the biggest three polyhedra in which all other polyhedra are 
embeddable. These numbres of polyhedra are $16$ yielding the most general cubic in $\mathbb{P}^2$, 
$13$ for the most general quartic in $(\mathbb{P}(1,1,2)$ and  $15$ for the most 
general bi-quadratic in $\mathbb{P}^1 \times \mathbb{P}^1$. 
The Newton polynom is defined by (\ref{WDelta*}) letting  
$\nu^{*(i)}$ run over $\Delta_2^{B*}$ and $\nu^{*(i)}$ over 
the corners of the dual polyhedron $\Delta_2^{B}$ and the 
coordinate ring is subject to (\ref{scale}). This yields the 
coordinates as indicated in figure \ref{poly}.   
\begin{figure}[h] 
\begin{center} 
\includegraphics[angle=0,width=.7\textwidth]{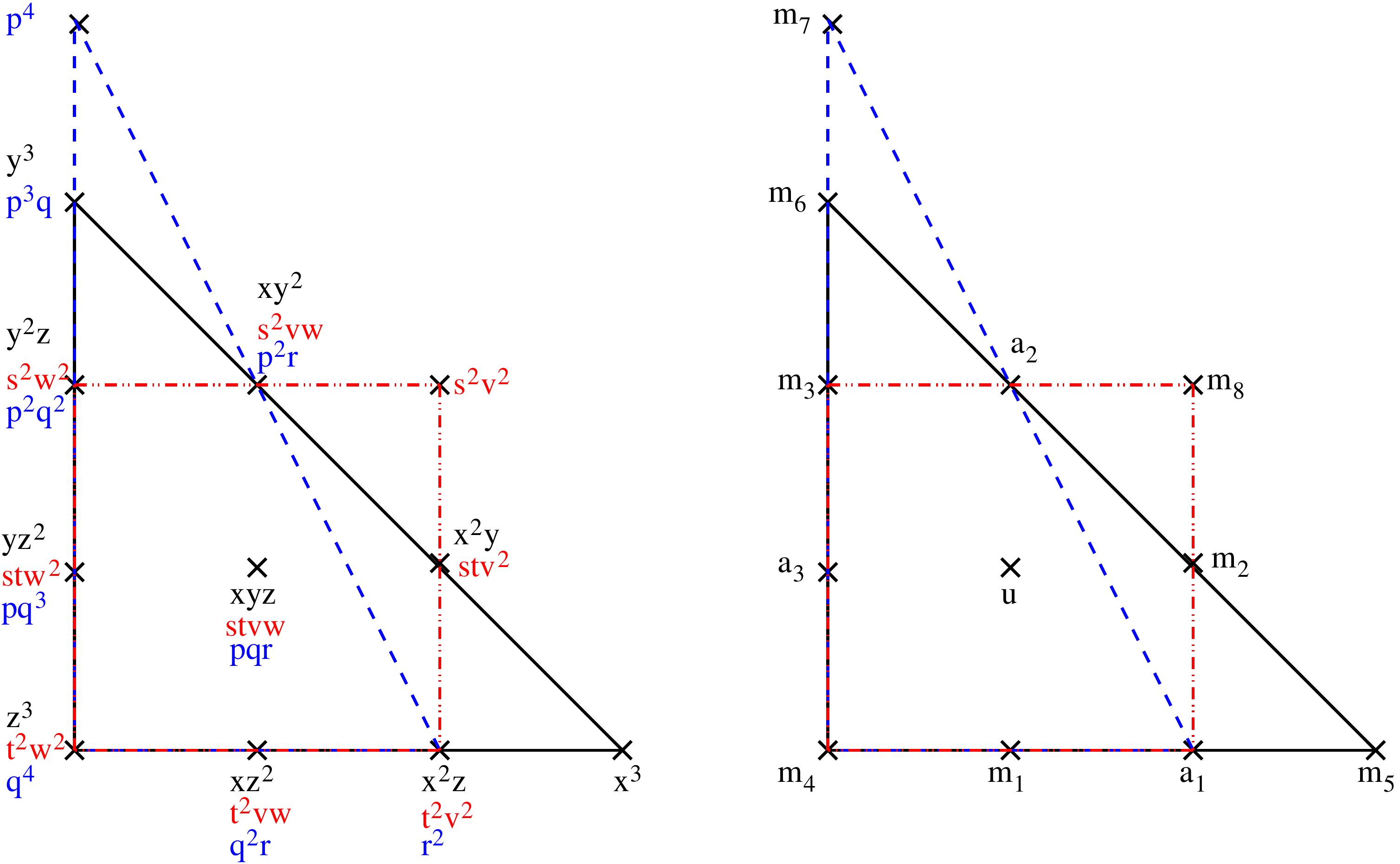} 
\begin{quote} 
\caption{All 16 reflexive polyhedra can be embedded into this diagram.\vspace{-1.2cm}} \label{poly} 
\end{quote} 
\end{center} 
\end{figure} 

Using the remaining scaling of the above projective spaces we can write 
\begin{equation}
H(X,Y,{\underline{a^*}})= h^6_{\Delta^*_B}(X,Y,{\underline{a^*}})
\end{equation}
as an inhomogeous equation. Note that there as many independent 
$a^*_i$ as there are relations between the points on $\Delta^{*B}$. 
So in two dimensions we can gauge away three $a_i^*$. The formalism 
does not depend on the existence of a global embedding and in 
particular $\Delta^{*B}$ must not be reflexive.  However for 
reflexive polyhedra the corresponding  elliptic curves can be readily 
brought into Weierstrass form using simple transformation 
algorithms such as Nagell's algorithm, which is very useful for further 
calculations and will be summarized in Appendix \ref{section:weierstrass}.
Moreover the Mori cones and triangulations have been 
calculated. These data will be used to relate the parameters  
$a_i^*$ in the Newton polynomials to the K\"ahler parameters. 
The upshot is that the compact part, i.e. the elliptic curve, of the mirror 
to the local del Pezzo geometry is the anti-canonical class in the del 
Pezzo surface defined by the Newton polynomial of $\Delta_B^*$ which
fixes a choice of the automorphism group.

It follows from the above and the general discussion at the end of section 
\ref{section:mirror1} that the mirror curves to toric del Pezzo 
surfaces have one complex structure parameter called $u$ and 
$l(\Delta)-4$ mass parameters called  $m_i$,  corresponding to the 
canonical  class of del Pezzo and the $e_i$-curves respectively. 
If more then three  points are blown up, the del Pezzo surfaces 
have in addition to the K\"ahler structure moduli, complex structure 
moduli and the toric description by the reflexive polyhedra with 
$l(\Delta)-4>3$ holds only at a special fixed value of the 
complex structure.  This is not a problem for the goal to describe 
the full K\"ahler structure moduli space of the del Pezzo surface by 
the elliptic curve as long as $h_{1,1}(S)\le 7$ (the bound 
comes simply from polyhedron 16 which has the maximal $l(\Delta)=10$), 
because K\"ahler and complex structure moduli decouple in $N=2$ theories. Above 
$h_{1,1}(S)>7$ i.e. for the $E_8$ and $E_7$ del Pezzo we find torically 
no mirror in which all masses can be turned on. 

However we can construct the full anti-canonical model $(S,{\cal E})$ 
for the $E_8$ and $E_7$ del Pezzo by completing the Weyl orbits for 
the mass parameters in polyhedron 10 and 14.  We note that by the 
construction in section \ref{stablebundles} and \ref{reductionofG} 
the description of the mirror of the del Pezzo and the description 
of the gauge bundle over ${\cal E}$  are on the same moduli space. 
In particular within the half $K3$ the moduli of the Kodaira singularities, 
i.e. $[p,q]$ 7-branes positions and the heretoric bundle moduli  
are unified in one moduli space.

\section{Physical interpretations} 

The physical aspects of the refined BPS states on local del Pezzo surfaces were 
reviewed in~\cite{CKK} from the point of view of five-dimensional $N=1$ gauge theory compatified 
on $\mathbb{S}^1_5\times \mathbb{R}^4$, but our ability to explore here the full 
moduli space of an elliptically fibred $\frac{1}{2}K3$ locally in an elliptically fibred 
Calabi-Yau manifold makes them in fact central in the following string/M-/F-theory 
dualities.

\subsection{Small instanton and E-string perspective}

In this section we will argue that the refined stable pair 
invariants count the higher spin partition function of the 
tensionless string that is the F-theory dual to a small 
$E_8$ instanton on the heterotic side.     

To cancel the anomaly in the heterotic string on $K3$ one has to have 
$\sum_{a} c_2(V_a)=c_2(T_{K3})$ and in particular the total instanton 
number of the vector bundle(s) $V_a$ has to be $24$. Due to the absence of 
vector multiplet moduli, the dynamics of the six-dimensional $N=1$ field theory with {\sl eight} 
supercharges in two left spinors $(2,0)$  is described by the Higgs effect. 
It was first argued~\cite{Witten:1995gx} that when one of these $SO(32)$ 
instanton shrinks to zero size, i.e. its curvature is concentrated 
in a point on the K3, space-time develops an infinite tube in which an 
unbounded increasing  dilaton profile develops, a $SP(1)$ gauge group 
is enhanced and hypermultiplets in the $({\bf 32}, {\bf 2})$ become massless.  
A single shrinking instanton corresponds to the nucleation of a 
solitonic heterotic five-brane that scales with ${1\over g_s}$ 
exactly as the Dirichlet-brane in the Type I theory, with which it can be 
identified under heterotic-Type I duality. The phenomenon is 
independent of the heterotic string coupling outside the tube.  
In accordance with the unoriented type I open string sector the 
maximal non-perturbative gauge symmetry enhancement in the heterotic $SO(32)$ string, 
when all instantons shrink at one point is $Sp(24)$. 

However the shrinking of instantons in $E_8\times E_8$ heterotic string 
cannot be described by the Higgs dynamics, because the dimensions of 
the $E_8$ representation are too big relative to the 
dimension of the moduli space of a single $E_8$ instanton which  
is $29$. It has therefore been suggested that the dynamical effect is 
due to a tensionless string. This string is similar in nature 
as the self-dual string of type IIB on K3 from 
a D3-brane wrapping a holomorphic curve, that becomes 
tensionsless when the latter shrinks.                

Upon compactification on a circle $\mathbb{S}^1_5$ one can use 
T-duality on this circle between heterotic $SO(32)$ and 
$E_8\times E_8$ to relate the small instanton dynamics in 
five dimensions~\cite{Ganor:1996mu}. Massless new states have to appear 
in five dimensions, which confirms the picture of a tensionless string. 
The strong coupling of the $E_8\times E_8$ string theory is conjectured 
to be M-theory on an interval $S_{11}^1/\mathbb{Z}_2$ and the 
solitonic heterotic five-brane is identified with 
the M5-brane. Purely based on the ten-dimensional anomaly 
cancellation mechanism in was argued in \cite{HW} purely 
that on the two fix points there are a novel kind of 9-branes 
with an $E_8$ super Yang-Mills theory on each of them.  Since 
the dilaton profile grows near the gauge bundle singularity, the 
dynamical effect related to the shrinking in one $E_8$ 
instanton occurs for any value of the 
asymptotic heterotic string coupling and can been 
interpreted in the strong coupling description as the 
nucleation of a M5-brane close to one $E_8$-branes. 
M2-branes can end on the M5-branes \cite{Strominger:1995ac} 
and it has been argued that they can end on the 
$E_8$-branes~\cite{Ganor:1996mu}, yielding the 
tensionless string.  

The analysis of the spectrum of this six-dimensional tensionsless string 
has been iniated in~\cite{Ganor:1996gu,Klemm:1996hh}. 
In~\cite{Klemm:1996hh} it starts with analyzing the BPS 
states encoded in (\ref{genuszero3.1}), which describes 
winding one $n_b=1$ in the base of the $\frac{1}{2} K3$ 
discussed at the beginning of section (\ref{verticaldelpezzo}). 
Here the image of $i^*$ has rank two: The class of the section of 
the base $\mathbb{P}^1$ and the class of the fibre and 
the modularity of this expression is due to (\ref{sl2zfibres}).    

Using the hypothesis that the right-movers of the tensionsless 
string are as the ones for the M-string or Green-Schwarz string in six 
dimensions that couples to the six-dimensional tensor $N=2$ tensor 
multiplet that arises from two parallel M5-branes, i.e. using an 
${\cal O}(4)$ lightcone quantization one gets from the $252$ unrefined 
BPS genus zero states at winding $d_f=1$ the space-time 
spectrum~\cite{Klemm:1996hh} 
\be
\left[{\bf 248}; 4(0,0)\oplus 
\left(\frac{1}{2},0\right)\oplus \left(0,\frac{1}{2}\right)\right]+ \left[{\bf 1};  
4\left(\frac{1}{2},\frac{1}{2}\right)\oplus \left(1,\frac{1}{2}\right) \oplus 
\left(\frac{1}{2},1\right)\oplus \left(0,\frac{1}{2}\right) \oplus  
\left(\frac{1}{2},0\right) \right]\ .
\label{6dstates}   
\ee
Here the left representations refer to the $E_8$ representations 
and  the right ones to the space-time representations in six dimensions. 
As has been argued using the M/F-theory duality in~\cite{Klemm:1996hh}, 
all diagonal states $(d_b,d_f)=(n,n)$ become part of the massless 
spectrum of the tensionless string at the transition point, where 
the volume of the curve $t_E+t_{\mathbb{F}^F_1}$ specified by 
(\ref{coneembede8}) becomes zero. As explained below (\ref{coneembede8}) 
this requires that the $\mathbb{F}^F_1$ is flopped so that its volume 
formally becomes negative. The first $n=1,2,3,4,\ldots$ diagonal 
unrefined BPS states are at genus zero $n^{(0)}_n= 252, -9252, 848628, 
-114265008,\ldots$ at genus one $n^{(1)}_n= -2,760,-246790,76413833,\ldots$ 
at genus two  $n^{(1)}_n= 0,-4, 30464, -26631112,\ldots$  etc. 
Clearly they can hardly be interpreted in themselves as individual 
BPS states of the tensionless string. First of all, since they 
are not positive, they can be at most an index, secondly they 
do not fall in any obvious way into representations of $E_8$ or 
the space-time spin.         

These problems all evaporate if we consider the refined stable pair  
invariants, as decribed in section~\ref{e8delpezzo}. First of 
all they are all positive, secondly they do fall in a simple 
way in $E_8$  representations and finally they do reproduce 
the only individual BPS states that could be inferred in~\cite{Klemm:1996hh}, 
namely the one above (\ref{6dstates}) from the splitting of the $E_8$ 
representation into the five-dimensional spins  $(j_L,j_R)$ 
\be
\left[{\bf 248}; (0,0)\right]+ 
\left[{\bf 1};\left(\frac{1}{2},\frac{1}{2}\right) \right]\ .
\ee
We conclude that the refined stable pair invariants do count the 
full tower of massless BPS states including all spins!  With the 
hypothesis above the six-dimensional space-time representations can be 
reconstructed at all levels $n$. Of course in this application we 
directly count stable pair invariants in the positive K\"ahler cone 
of the local del Pezzo and argue that they become all massless 
if the latter shrinks to zero size. It is very suggestive but not 
entirely clear that these states are stable bound states in 
this limit. Of course our formalism allows to calculate the $F^{(n,g)}$ 
at any place in the K\"ahler moduli space of the geometry. 
In particular as we argue in section \ref{e8delpezzo} that the spectrum 
at the conifold, where the volume vanishes is identical to the large volume point 
due to the self-duality  of the $E_8$ lattice. This underlines the claim 
that we found the stable spectrum of the tensionless string. So it is a quite 
concrete proposal for the spectrum for a conformal higher spin theory of the 
type recently analyzed in \cite{Vasiliev,MZ}.

As proposed in \cite{Klemm:1996hh} we can turn on Wilson 
lines parametrized by vectors ${\bf W}_\alpha$  in the Cartan 
algebra of $E_8$. In our local description of the $E_n$ curves 
mirror to $E_n$ del Pezzo surfaces this literally means to 
shift the masses $m_i$. To be concrete we have to choose a basis 
of the wheight lattice to parametrize the characters by $m_i$ in the 
same basis and call $\Lambda$ the charge of a BPS state. Then 
the shift of the masses by the Wilson lines is simply given by
\be
{\bf  m} \rightarrow {\bf  m} + \sum_{\alpha} {\bf \Lambda}\cdot W_{\alpha}\ . 
\label{Wilsonline} 
\ee
This will in general break $E_8$ in $U(1)^{k}\times E_{8-k}$, where 
the $U(1)^k$ can be globalized in F-theory as discussed in section 
(\ref{verticaldelpezzo}).

\subsection{The $[p,q]$-string perspective}   

F-theory describes the varying axion-dilaton background 
$\tau=C_0+ i e^{-\phi}$ of type IIB compactifications on manifolds 
$B_d$ with positive canonical class by the complex structure 
$\tau$ of an elliptic fibration ${\cal E}\rightarrow B_d$, 
so that the total space is a Calabi-Yau manifold $Y_{d+1}$. 
The latter condition requires the fibre to degenerate over 
divisors $d_i\in B_d$ so that the canonical class fulfills  
\be 
K_{B_d}=\sum_i a_i d_i,
\ee
where $a_i$ are rational numbers associated to the possible 
Kodaira types of the singular fibres, which are elliptic singularities 
whose Hirzebruch-Jung sphere configurations intersect in an affine 
$ADE$ Dynkin diagram, the simplest one, called $I_0$, 
beeing just a nodal curve for which $a=\frac{1}{12}$ making 
$Y_2=K3$ over $B_1=\mathbb{P}^1$ with $24$ $I_0$ fibres at 
points $u_i\in \mathbb{P}^1$, $i=1,\ldots,24$ the simplest 
example. The single vanishing cycle $\gamma=$\pqh[p,q]$\in H_1({\cal E},\mathbb{Z})$ 
at say $u_1\in \mathbb{P}^1$ ($d_1 \in B_d$) determines the $[p,q]$-charge 
of the 7-branes that extend over $d_1$ and the 
non-compact directions. The Picard-Lefshetz monodromy 
action on $H_1({\cal E},\mathbb{Z})$ along a counter-clockwise 
loop encircling $u_1$ is like in the rank one Seiberg-Witten 
families over $z$ given by~\cite{Klemm:1995wp}  
\be 
M_{p,q}=\left(\begin{array}{cc} 1- pq& p^2\\ - q^2 & 1+ 
p q\end{array}\right)\ .
\label{pqmonodromy}
\ee
It acts as subgroup of the ${\rm SL}(2,\mathbb{Z})$-symmetry of 
type II on the doublet $(H_3,F_3)$ of the NS and RR three-forms 
and their sources: the fundamental \pqs[p,q]=\pqs[1,0]- and $D$  
\pqs[p,q]=\pqs[0,1]- string and as ${\rm PSL}(2,\mathbb{Z})$-transformation 
on $\tau$.  This happens at a cut emanating 
from the $[p,q]$ 7-brane position, whose precise 
position must be irrelevant for physical questions, 
in particular regarding BPS states from string junctions.  

The global monodromy is encoded in the  Weierstrass form of the family as 
in (\ref{weierstrass}), where we view $u$ as parameter on the base 
while $\vec m$ parametrizes the position of the 7-branes, and 
is not trivial. As a consequence there are mutually non-local 
$[p,q]$ 7-branes and no global perturbative description 
of F-theory, not even an understanding of the full spectrum 
of its BPS states in space-time, like the one we inferred for 
the tensionless string in the last section. What comes 
closest to it is to consider groups of in general non-local $[p,q]$ 7-branes 
and construct BPS states such that the gauge bosons that 
correspond to the roots are given by string junctions. The simplest 
group of such branes is the configuration of Sen in which he sets 
$g_2=c f^2$ and $g_3=f^3$ with $f=\prod_{i=1}^4 (u-u^0_i)$, so 
that $j$ and hence $\tau$ become constant. The monodromy 
at each $u=u^0_i$ $i=1,\ldots ,4$ is $M=-{\bf 1}$, so the configuration has no 
net charge and since $M$ is the involution on ${\cal E}$ the 
configuration at this point must  be four 7-branes of charge $[1,0]$ and 
an O7-brane, with charge $[-4,0]$, the latter splits into a 
$[-2,-2]$ and $[-2,2]$. Of course the orientifold brane 
configuration gives a $SO(8)$ gauge symmetry, the breaking of 
which is described by moving the 7-branes away from $u^0_i$.
Moreover since $g_2\sim u^2$, $g_3\sim u^3$ and the 
gauge symmetry acts as flavour symmetry on the $m_i$ the problem 
of  constructing the local deformation of the brane configuration is the same as constructing 
the $SU(2)$, $N_f=4$ Seiberg-Witten curve. Physically 
this can be argued more intuitively in the D3-probe-brane
picture in which always the gauge and flavour 
symmetry are exchanged. 

The general approach to construct the non-Cartan gauge bosons, 
which in particular extends to the $E_n$ groups, is by string 
junctions. For them one has the follwing key properties 
\begin{itemize}
 \item J.1 String junctions are configurations of $\pqssu[p,q]_i$-strings $i=1,\ldots,N$, 
which meet in a single point, subject to a non-force condition\footnote{This is analogous to no-force condition 
in the $[p,q]$ 5-brane webs shown in figure \ref{pqwebs}, which are dual to the local del Pezzo triangulation.} 
\be 
\sum_{i=1}^N \left[\begin{array}{c}p \\ q\end{array} \right]_i=0\, .
\label{chargeconservation}
\ee
\item J.2 Two $\pqssu[p,q]$-strings can end on each other iff 
they are compatible  
\be 
\pqsu[p,q]_i \wedge \pqsu[p,q]_k= p_i q_k- q_i p_k=\pm 1\ , 
\label{compatibility}
\ee 
the sign depending on the orientation of the corresponding cycle 
in $H_1({\cal E},\mathbb{Z})$. 3-junctions are BPS configurations iff 
(\ref{compatibility}) holds for each pair~\cite{Dasgupta:1997pu}. 
\item J.3  Each $\pqssu[p,q]$-string line emerging  from the junction can end 
on a $[p,q]$ 7-brane, where $\gamma=\pqssu[p,q]\in H_1({\cal E},\mathbb{Z})$ shrinks.    
\end{itemize}
The general idea is that the string junctions extend the fundamental 
\pqs[1,0] open strings that lead to the $A_n$ gauge symmetry on 
stacks of $n+1$ fundamental branes to stacks of 
non-mutual $[p,q]$ 7-branes connected by string junctions.  
$N=3$ turns out to be enough to get all roots of the exceptional 
groups~\cite{Gaberdiel:1997ud}. 

The most relevant questions are what are the low energy states of these 
configurations, what is their moduli space,  how to quantize them, 
i.e. what are the associated BPS states. Unfortunately the anwers  
are to a large extent unknown. It is known of course 
that in the presence of a 7-brane a configuration, with a given 
asymptotic charge at the ends, is independent of the position of 
the $[p,q]$ 7-brane cut, i.e. 
(\ref{pqmonodromy}, \ref{chargeconservation}, \ref{compatibility}) 
are compatible, as explained e.g in~\cite{Gaberdiel:1997ud} and shown in 
figure \cite{HW}.  

\begin{figure}[h!] 
\begin{center} 
\includegraphics[angle=0,width=.4\textwidth]{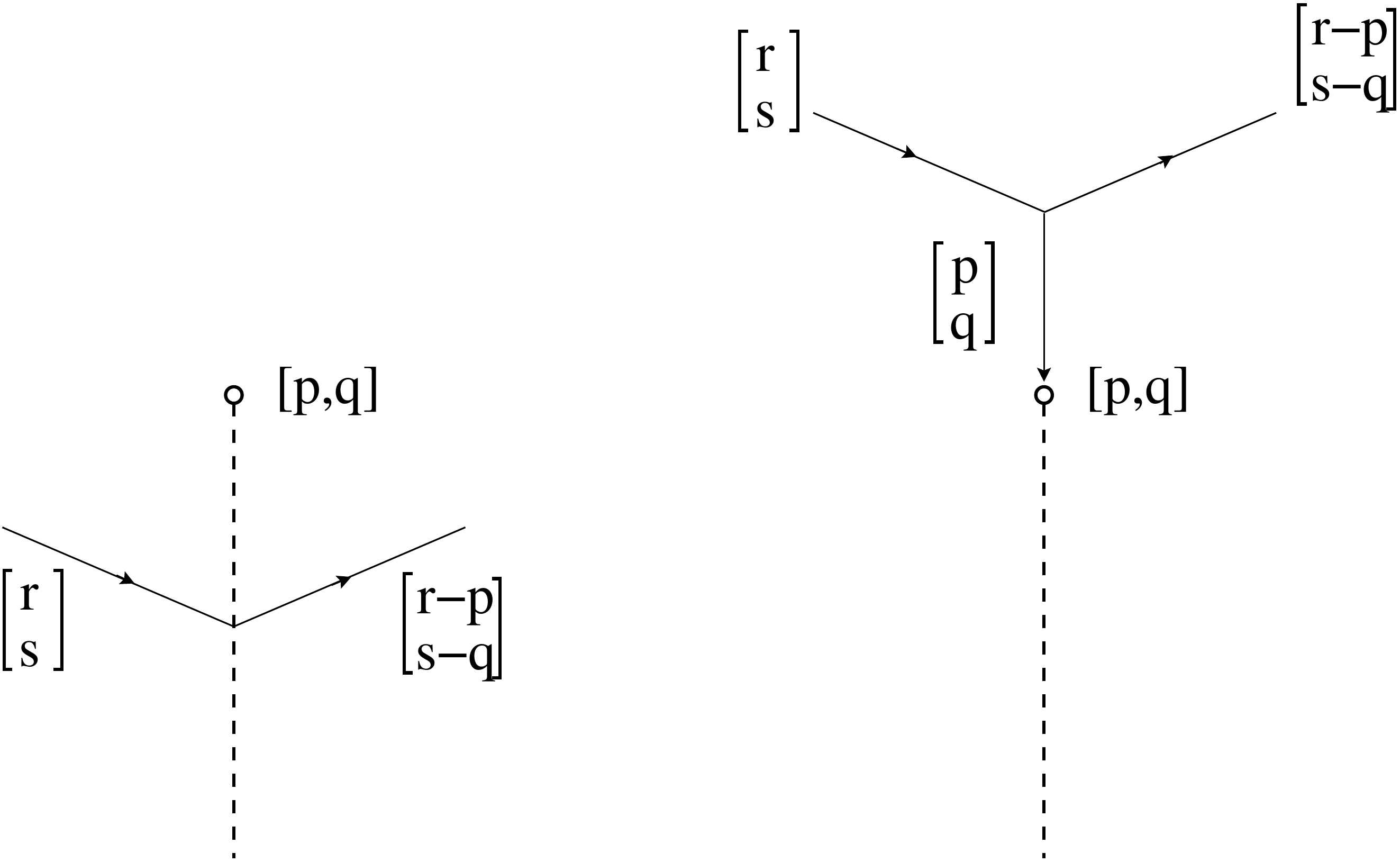} 
\begin{quote} 
\caption{
If a $[r,s]^t$  string is moved over the $[p,q]$ 7-brane 
leaving its cut and if  $[p,q]^t$ is compatible with $[r,s]^t$ then 
a $[p,q]^t$-string, connecting $[p,q]$ with $[r,s]^t$ and forming there 
a junction, is created.   
\vspace{-1.2cm}}  
\label{HW}
\end{quote} 
\end{center} 
\end{figure} 

This fact as  well as J.1-J.3  makes it possible to lift topological 
configurations of string junctions ending on $[p,q]$ 7-branes 
to closed curves in the total space of a two complex dimensional 
${\cal E}$ fibration and search for their minimal energy 
configuration~\cite{Mikhailov:1998bx}. It is claimed~\cite{Mikhailov:1998bx} that the 
polarisation can be can be chosen, so that the geodesic minimality in 
the $u$ plane is equivalent to the lifted curve being holomorphic 
in the complex structure. Moreover the natural intersection on 
junctions defined on junctions~\cite{Mikhailov:1998bx, DeWolfe:1998eu}
should yield the intersection on 
holomorphic curves, which gives simple constraints on 
certain  BPS junctions~\cite{Mikhailov:1998bx}. The metric 
for the geodesic minimality is the same then the one used 
in~\cite{Klemm:1996bj} only that there the lift is supposed 
to yield special Lagrangian 3-cycles. 

So in order to search for the BPS states in our geometries we 
might attempt to identify stable pairs in the rational elliptic 
surface,  whose pure sheaf of complex dimension one  is supported 
on a holomorphic curve\footnote{In two dimensions the question 
can be addressed in the symplectical or the holomorphic approach, 
so whether we work with resolutions or deformations in local problems is 
a matter of taste. For obvious reason we make the second choice, even 
so it is interesting to study the meaning of geometric invariants 
related to the refined stable pair invariants in the symplectic category.}  
${\rm ch}_2({\cal F})=\beta$ in a mixed class of fibre and base of 
the massless $\frac{1}{2}K3$. 

These classes precisely decompose into Weyl orbits by formula 
(\ref{Weylorbit}) and the list of results for the diagonal 
classes in the section below is quite encouraging. 
The numbers with the lowest degree and spin yields for the 
$E_8$ case the $8+240$ in the trivial and the first non-trivial 
Weyl orbit, i.e. the $E_8$ gauge bosons for which 
the \pqs[p,q]-string junctions were designed for. Moreover 
it is well-known in the heterotic/type II duality, most noticable 
in the YZ formula~\cite{Yau:1995mv}, that the BPS states of those 
string oscillation encoded the elliptic genus are mapped to 
$\chi({\cal F})=n$ i.e. the ``$D0$ brane'' content of the stable pair, 
which yields the spin content of the refined stable pair invariant. Indeed in
the simplest case, as in~\cite{Yau:1995mv},  they are counted by the 
G\"ottsche formula for Hilbert schemes of points on the symmetric 
product, a structure that is refined for the $n_b=1$ classes 
in the $\frac{1}{2}K3$ in section \ref{sec.nb=1}. This makes it 
reasonable to assume that the exitation of  \pqs[p,q]-strings are 
encoded in the $(j_L,j_R)$ spin content. Beside the issue of stability, 
which we do not address here, it should be clear that due the 
flavor symmetries the $\mathbb{C}^*$-action used in the Bialynicki-Birula 
decomposition there can be shifts in the association of mass and 
spin, e.g. the diagonal K\"ahler class could be shifted as 
$t\rightarrow  t-a + c (\epsilon_1+\epsilon_2)$.     
The mirror construction relates the position of the 
$[p,q]$ 7-branes in the curve (\ref{weierstrass}) and turning them 
on splits the representrations of the  $E_8$ into massive ones 
and massless ones associated  the unbroken subgroups. All this seems 
sufficient evidence to conclude that refined BPS-states do 
capture properties of the infinite towers of BPS states 
associated to the \pqs[p,q]-strings suspended between 
mutual non-local $[p,q]$ 7-branes.        

\section{The $D_5,E_6,E_7$ and $E_8$ del Pezzo surfaces} 
\label{sec.delpezzo}

The formalism described in section \ref{ellipticdirectintegration}, 
the description of mirror symmetry of local del Pezzo surfaces in 
section \ref{mirrorsymmetry} together with the general Weierstrass form 
given in section \ref{section:weierstrass} allows to recursively calculate 
the amplitudes $F^{(n,g)}$. Then the formulae \eqref{productrefined} and 
\eqref{topo2.1} can be used to extract the invariants $N_{j_L,j_R}^\beta$. 
As a warm-up we consider special one parameter del Pezzo's  of the type indicated above. In this one 
parameter family one sums over all classes $\Lambda'$ of the del Pezzo surface, 
by setting  the corresponding K\"ahler classes to $t_i\rightarrow 0$, 
i.e. $q_i=e^{t_i}=1$. Since the Weyl group of the corresponding 
Lie algebra acts on $\Lambda'$ we expect to find the states organized in the 
dimensions of the Weyl orbits.  Physically the specialization corresponds 
to setting the mass  parameters in the five-dimensional theory to zero. 
We will denote  $\beta \in H_2(M,\mathbb{Z})$  simply by the positive integer 
$d$, the degree of the holomorphic maps.  

\subsection{The $E_8$ del Pezzo surface}  
\label{e8delpezzo} 
According to section \ref{mirrorsymmetry} the massless $E_8$ can be obtained 
from the polyhedron 10 with all mass parameters on the edges set to zero. 
This is simply done by setting in \eqref{generalcubic} (see table) all 
parameters to zero except $m_2=m_4=m_6=1$ while keeping $\tilde u$. The right 
large complex structure variable $u=\frac{1}{\tilde u^\frac{1}{6}}$ is found based on 
the analysis of the Mori cone below (\ref{ellipticMori}). Then we get after a 
rescaling $g_i\rightarrow \lambda^i g_i$ with $\lambda= 18 u^{7/3}$   
\begin{equation} 
g_2= 27 u^4, \qquad g_3=-27 u^6 (1-864 u)\ .
\end{equation} 
so that near $u=0$, we get $\frac{d t}{du}=\frac{1}{u}+60+13860 u+4084080 u^2+ {\cal O}(u^3)$. 
The $j$-function 
\begin{equation} 
j=\frac{1}{1728 u ( 1- 432 u)} \ . 
\end{equation}
identifies this as the special family whose monodromy group is classic and has already 
been discussed in~\cite{Klein}. As a consistency check we can also take the curve 
(\ref{sakaie8}) and turn off all the Wilson lines by setting the $\chi_i(0)$ 
to the values of the dimensions of the weight moduls. Let us define the Dynkin 
diagram of the affine $E_8$ as 
\begin{figure}[H] 
\begin{center} 
\includegraphics[angle=0,width=.6\textwidth]{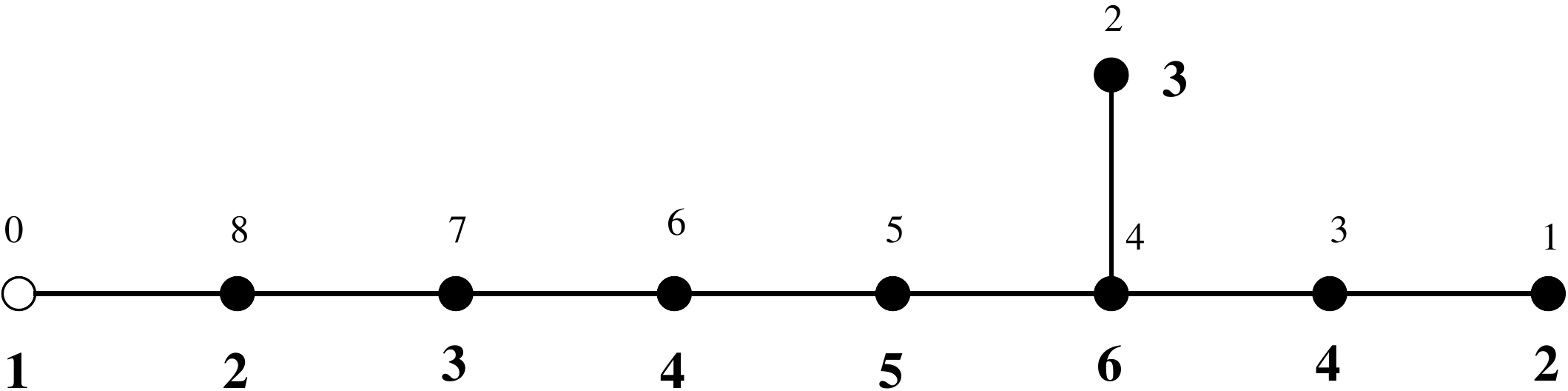} 
\vspace{.2cm} 
\end{center} 
\label{e8} 
\end{figure} 
where we denote by the bold numbers the Coxeter labels. The smaller 
numbers  give simply an ordering of the basis of Cartan generators and 
the basis for the weights. Let us denote by $w_i$ the weight of the classical Lie algebra 
with a $1$ at the $i$th entry and $w_0$ the trivial weight. We record the dimensions of the 
corresponding weight modules 
\be
\begin{array}{rl} 
\chi_1(0) &= 3875,\quad \chi_2(0)= 147250, \quad \chi_3(0) = 6696000, \quad \chi_4(0) = 6899079264,\\ 
\chi_5(0)& =146325270, \quad \chi_6(0) = 2450240, \quad \chi_7(0) = 30380, \quad \chi_8(0) = 248\ .
\end{array}
\label{e8dimensions}
\ee 
Specializing (\ref{sakaie8}) gives indeed the same family as can be seen by 
comparing the $j$-functions.

For the BPS states $N^d_{j_L,J_R}$  at $d=1$ one gets:   
\begin{table}[H]
\centering
{\footnotesize 
\begin{tabular} 
{|c|c|c|} \hline $2j_L \backslash 2j_R$  & 0 & 1 \\  \hline  0 & 248 &  \\  \hline1 &  & 1 \\  
\hline \end{tabular}} \vskip 3pt  $d=1$ 
\label{tableE8d1}
\end{table} 
\vspace{-0.5cm}
It is obvious that the adjoint represention $248$ of $E_8$ appears as the spin $N_{0,0}^1$, 
which decomposes into two Weyl orbits with the weights $w_1+ 8 w_0$.   I. e. we are 
counting exactly the BPS numbers  of the $[p,q]$-string configurations, which are relevant for the 
gauge theory enhancement in F-theory. Note that the contributions of different Weyl orbits come 
in general from curves with  different genus. In this way also the higher spin invariants fall 
systematically into Weyl orbits of weights of $E_8$. E.g. $3876=1+3875$, where the latter 
decomposes in the Weyl orbits  of $w_1 + 7w_8 +35 w_0$.
The multiplicities of the Weyl Orbits are encoded in the solution of the $\frac{1}{2} K3$ 
model by the formula (\ref{masslessEn}), where we report the dimension 
of some lower $E_8$ Weyl orbits in equation (\ref{Weylorbit}).      

\begin{table}[H]
\centering
{\footnotesize \begin{tabular} {|c|c|c|c|c|} \hline $2j_L 
\backslash 2j_R$  & 0 & 1 & 2 & 3 \\  \hline  0 &  & 3876 &  &  \\  \hline1 &  &  & 248 &  \\  \hline2 &  &  &  & 1 \\  \hline 
\end{tabular}} \vskip 3pt  $d=2$ 
\label{tableE8d2}
\end{table}
\vspace{-0.5cm}
At $d=3$ we see the decompositions into representaions $4124=1+248+3875$, $34504=1+248 +30380$, 
$34504=1+ 248+ 30380$, $151374=1+248+3875+147250$ and $30628=248+30380$ 
\begin{table}[h!]
\centering
{\footnotesize\begin{tabular} {|c|c|c|c|c|c|c|c|} \hline $2j_L \backslash 2j_R$  & 0 & 1 & 2 & 3 & 4 & 5 & 6 \\  \hline  0 & 30628 &  & 151374 &  & 248 &  &  
\\  \hline1 &  & 4124 &  & 34504 &  & 1 &  \\  \hline2 & 1 &  & 248 &  & 4124 &  &  \\  \hline3 &  &  &  & 1 &  & 248 &  
\\  \hline4 &  &  &  &  &  &  & 1 \\  \hline  
\end{tabular}} \vskip 3pt  $d=3$ 
\label{tableE8d3}
\end{table}
\vspace{-0.0cm}
while for higher degree the geometric multiplicities of the Weyl orbits become bigger with the lower spins  farer away from the maximal spin, 
still it obvious how the states  decompose into Weyl orbits, e.g.   
\begin{table}[h!]
\centering
{\footnotesize 
\begin{tabular} {|c|c|c|c|c|c|c|c|c|c|c|c|} \hline $2j_L \backslash 2j_R$  & 0 & 1 & 2 & 3 & 4 & 5 & 6 & 7 & 8 & 9 & 10 \\  
\hline  0 &  & 3480992 &  & 7726504 &  & 212879 &  & 248 &  &  &  \\  \hline1 & 185878 &  & 1209127 &  & 3632614 &  & 38876 &  & 1 &  &  \\  
\hline2 &  & 38876 &  & 251755 &  & 1030753 &  & 4373 &  &  &  \\  \hline3 & 
248 &  & 4373 &  & 39125 &  & 217003 &  & 249 &  &  \\  \hline4 &  & 1 &  & 249 &  & 4373 &  & 35000 &  & 1 &  \\  
\hline5 &  &  &  &  & 1 &  & 249 &  & 4125 &  &  \\  \hline6 &  &  &  &  &  &  &  & 1 &  & 248 & \\  
\hline7 &  &  &  &  &  &  &  &  &  &  & 1 \\  \hline \end{tabular}} \vskip 3pt  
$d=4$  
\label{tableE8d4}
\end{table}
\vspace{-0.5cm}
$7726504=2+9\times 248+6\times 3875+6\times 147250+669600$.  

\subsection{The $E_7$ del Pezzo surface }
\label{e7delpezzo} 
The massless $E_7$ del Pezzo corresponds to the polyhedron 13 with all 
parameters on the edges set to zero. Again this is simply done by 
specializing the Weierstrass form (\ref{thequartic}) to 
\be 
a_1=1,\quad m_4 = 1,\quad  m_5 =1,\quad  u=\frac{1}{ (-\tilde u)^\frac{1}{4}}\  
\label{e7restr}
\ee
while setting all other parameters to zero. 
Again the inverse quartic root identification of $u=\frac{1}{(-\tilde u)^\frac{1}{4}}$ can be 
predicted from the Mori cone vector $l=(-4,1,1,2)$. It could be also obtained  
by firstly requiring at large radius $t(u)\sim \log(u)$ and at the conifold $t_D(u)\sim \Delta$. 
This also fixes the $-1$ in  (\ref{e7restr}), in fact that $t(u)= \log(u) - 12 u + 210 u^2+{\cal O}(u^3)$ and 
secondly knowing that genus zero curves exist at $d=1$. 

Relative to (\ref{e7restr}) we have to scale the  
$g_2$ and $g_3$ by $\lambda= 18 i  u^\frac{2}{5}$ yielding\footnote{The labels $b$ 
and $s$ refer as big and small  to the size of the polyhedra used to define the geometries.}   
\be 
g^b_2=27 u^4 (1 - 192 u),\quad  g_3^b=27 u^6 (1 + 576 u)\  
\ee 
and the $j$-function as
\be 
j_b=\frac{(192 u-1)^3}{1728 u (64 u+1)^2}\ . 
\ee 
It is well-known that massless theories can be 
formulated on isogenous curves~\cite{Klemm:1995wp}. These curves are not 
distinguished by their Picard-Fuchs equation, 
neither for the holomorphic nor the meromophic 
differential, but they are distinguished by a choice of 
a relative factor $\kappa\in \mathbb{N}_+$  in the  
normalization of the $a$- and the $b$-cycles. 
As pointed out in~\cite{Klemm:1995wp} this exchanges 
the two cusp points -- corresponding to the large radius and conifold points --
of the curves, but is not a symmetry of the $N=2$ 
theory neither of the topological string. In the context
of the del Pezzo surfaces the existence of isogeneous
curves has been discussed in~\cite{Haghighat:2008gw}. 
It finds a natural interpretation in terms of the center 
of $E_n$ given in (\ref{centerEn}) as follows. Since the 
Picard-Fuchs equations depend only 
on the linear relations among the points in the polyhedra, the 
polyhedron 4 with one mass at the edge of the corner 
set to zero will lead to the same Picard-Fuchs operator. 
Now with the Weierstrass form obtained by embedding polyhedron 4 into 
polyhedron\footnote{Of course this family can be also realized as 
special cubic by embedding polyhedron 4 into polyhedron 15. That does not 
change the analysis.}  13 by setting all coefficients to zero exept  
\be 
a_1=1,\quad a_2 = 1,\quad  m_5 =1,\quad  u=\frac{1}{ (\tilde u)^\frac{1}{4}}\, ,
\label{e7restr2}
\ee
we can precisely understand the relation between the two geometries. With $\lambda= 18  u^\frac{2}{5}$ we get now  
\be 
g^s_2=27 u^4 (48 u+1) ,\quad  g_3^s=27 u^6 (72 u+1) \  
\ee 
and the $j$-function as
\be 
j_s=\frac{(48 u+1)^3}{1728 u^2 (64 u+1)}\  
\ee 
so that the $\mathbb{Z}_2$ transformations 
\be 
\mathbb{Z}_2: u\mapsto -\frac{1}{64} - u,
\quad \mathbb{Z}_2:j_b \leftrightarrow j_s, \quad \mathbb{Z}_2:\tau_s  \leftrightarrow 2 \tau_b 
\label{z2}
\ee 
exchanges as in~\cite{Klemm:1995wp} the conifold with the 
large radius point and identifies $j_b \leftrightarrow j_s$ and 
rescales the $U(1)$-coupling. However to get integral charges 
for the matter representations, or equivalently integral 
K\"ahler classes, one has to choose the curve corresponding 
to the big polyhedron\footnote{This is exactly the same reason 
that in first paper of~\cite{SWIandII} they use the $\Gamma(2)$, 
while once they want to include masses to study chiral symmetry 
breaking they use the isogenous $\Gamma_0(4)$ curve.}.
Note that the last relation in (\ref{z2}) can be already seen 
from the fact that $\Delta=1+64 u$ appears quadratically in 
the denominator of $j_b$. The story is analogous for the $E_6$ 
group with the big polyhedron being polyhedron 15 and the 
small polyhedron being polyhedron 1. So we conclude that the 
volumes of polyhedra $P_b$ and $P_s$ the are related to the center 
of the groups, or the volumes of the fundamental cell in the  lattices $\Lambda'$ and 
$\Lambda''$ as  
\be 
\frac{{\rm Vol}(P_b)}{{\rm Vol}(P_s)}=\frac{{\rm Vol}(\Lambda'')}{{\rm Vol}(\Lambda')}
\ee 
and the existence of the self-dual polyhedron 10 is a consequence 
of the  self-duality of the $E_8$ lattice.  We further notice that 
the $j$-function of the massless $E_7$  curve of \ref{sakaie8}
\be 
j_{E7}^{es} =\frac{(u_{es}-36)^3}{(1728 (u_{es}-52 )} 
\ee 
is not very naturally related to $j_b(u_b)$  or $j_s(u_s)$ 
\be 
u_{es} = -12 - \frac{1}{u_s},\quad u_{es} = \frac{52 - 768 u_b}{1 + 64 u_b}\ . 
\ee
Let us agree on the Dynkin diagram of $E_7$ in the following conventions
\begin{figure}[H] 
\begin{center} 
\includegraphics[angle=0,width=.5\textwidth]{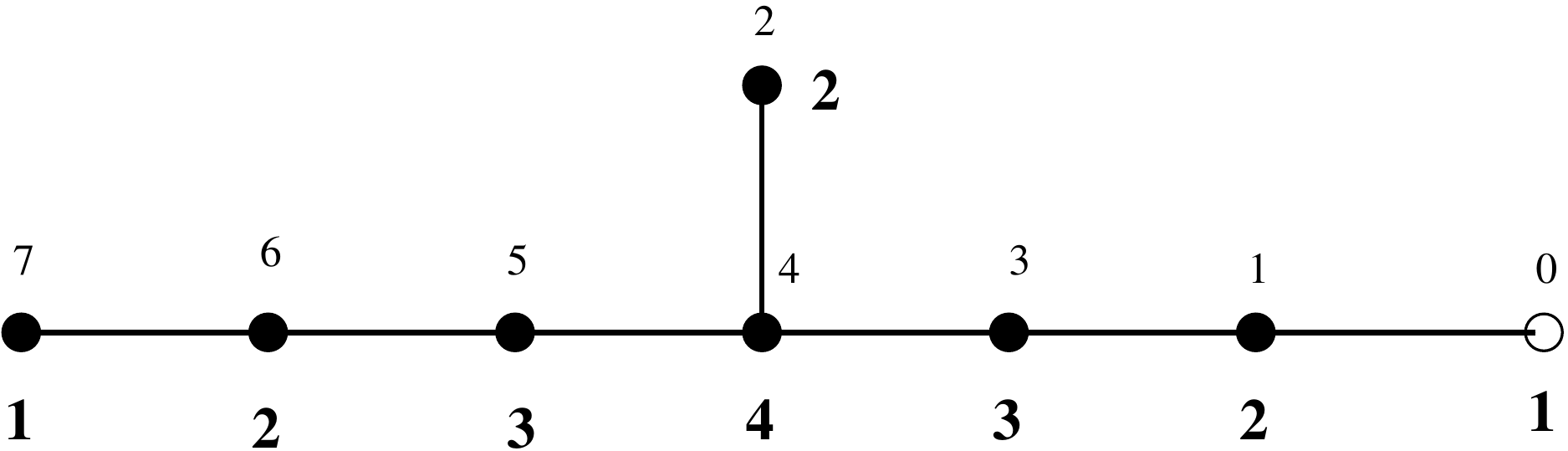} 
\vspace{.2cm} 
\end{center} 
\label{e7} 
\end{figure} 
\be
\begin{array}{rl} 
&\chi_1(0) = 133,\quad \chi_2(0)= 912, \quad \chi_3(0) = 8645, \quad \chi_4(0) = 365750, \quad \chi_5(0) =27664, \quad \\ 
& \chi_6(0) =1539 , \quad \chi_7(0) = 56\ .
\end{array}
\label{e7dimensions}
\ee 
From either the big or the small polyhedron we get the following 
refined BPS invariants. Again there are the Weyl orbits for curves 
in different genera which to combine in simple representations of $E_7$.        
\begin{table}[H]
\centering
{\footnotesize\begin{tabular} {|c|c|} \hline $2j_L \backslash 2j_R$  & 0 \\  \hline  0 & 56 \\  \hline 
\noalign{\vskip 2mm} 
\multispan{2} $d$=1 \\  \end{tabular}} \hspace{0.5cm}
{\footnotesize 
\begin{tabular} {|c|c|c|c|} \hline $2j_L \backslash 2j_R$  & 0 & 1 & 2 \\  \hline  0 &  & 133 &  \\  \hline1 &  &  & 1 \\  \hline 
\noalign{\vskip 2mm} 
\multispan{4} $d$=2 \\ \end{tabular}} \hspace{0.5cm} 
{\footnotesize
\begin{tabular} {|c|c|c|c|c|} \hline $2j_L \backslash 2j_R$  & 0 & 1 & 2 & 3 \\  
\hline  0 & 56 &  & 912 &  \\  \hline1 &  &  &  & 56 \\  \hline  \noalign{\vskip 2mm} 
\multispan{4} $d$=3 \\ \end{tabular}}
\end{table}
\vspace{-0.5cm}
We note that there is a periodicity with the degree mod 2 
in the  contributions of the BPS  states  with highest spins.  
In even degree we always find for the highest spin the trivial and 
the adjoint representation $133$ in the Weyl orbits $w_1+7w_0$  of $E_7$,  
while in odd degrees we find the $56$ representation in a single Weyl orbit. 
This is a consequence of the nontrivial center of $E_7$  (\ref{centerEn}), 
which is reflected on the square root of the line bundle ${\cal Q}$ for 
the $E_7$ case.

At $d=3$ the $912$, $w_2+6 w_7$ representation  appears and  
again we find the behaviour that the higher degree stable pair invariants 
decompose in a simple fashion into representations and hence Weyl orbits.
The systematic can again be understood form the solution of the $\frac{1}{2}K3$ and 
formula (\ref{masslessEn}). The relevant dimensions of the Weyl orbit 
for the $E_5=D_5,\ldots,E_7$ goups are summarized in table \ref{Enorbit}. 

E.g. at $d=4$:  $8778=8645+ 133$, with  $8645$ decomposes as 
$w_3 + 5w_6 +22 w_1 +77 w_0$ and $1673= 1539 + 133 + 1$ with  
$1539=w_6  + 6 w_1 +27 w_0$. 
\begin{table}[H]
\centering
{\footnotesize
\begin{tabular} {|c|c|c|c|c|c|c|c|} \hline $2j_L \backslash 2j_R$  & 0 & 1 & 2 & 3 & 4 & 5 & 6 \\  \hline  0 &  & 1673 &  & 8778 &  & 1 &  \\  
\hline1 &  &  & 134 &  & 1673 &  &  \\  \hline2 &  &  &  & 1 &  & 133 &  \\  \hline3 &  &  &  &  &  &  & 1 \\  \hline \end{tabular}} 
\vskip 3pt  $d=4$ 
\end{table}
\vspace{-0.5cm}
At degree $d=4$ we have the following decomposition $1024=912+ 2\times 56$, 
$7504=4\times 1539 + 912+ 2\times 133 + 3 \times 56 + 2$, $8472=5\times 1539+ 5\times 133+ 2\times 56$, 
$36080=27664+5\times 1539+ 5\times 133+56$ and $93688 =3 \times 27664 + 
8645+ 1539 + 3\times 133  + 2\times 56  + 1$.
\begin{table}[h!]
\centering
{\footnotesize
\begin{tabular} {|c|c|c|c|c|c|c|c|c|c|} \hline $2j_L \backslash 2j_R$  & 0 & 1 & 2 & 3 & 4 & 5 & 6 & 7 & 8 \\  \hline  0 & 6592 &  & 36080 &  
& 93688 &  & 968 &  &  \\  \hline1 &  & 968 &  & 8472 &  & 36080 &  & 56 
&  \\  \hline2 &  &  & 56 &  & 1024 &  & 7504 &  &  \\  \hline3 &  &  &  &  &  & 56 &  & 968 &  \\  \hline4 &  &  &  &  &  &  &  &  & 56 \\  \hline 
\end{tabular}} \vskip 3pt  $d=5$ 
\end{table}
\vspace{-0.5cm}

\begin{table}[h!]
\centering
{\footnotesize
\begin{tabular} {|c|c|c|c|c|c|c|c|c|c|c|c|c|c|} \hline $2j_L \backslash 2j_R$  & 0 & 1 & 2 & 3 & 4 & 5 & 6 & 7 & 8 & 9 & 10 & 11 & 12 \\  
\hline  0 &  & 225912 &  & 650050 &  & 1062065 &  & 54419 &  & 133 &  &  &  \\  
\hline1 & 10451 &  & 73839 &  & 289109 &  & 650184 &  & 13588 &  & 1 &  &  \\  \hline2 &  & 1807 &  & 13855 &  & 75512 &  & 234691 &  & 1807 &  &  &  
\\  \hline3 & 1 &  & 134 &  & 1808 &  & 13855 &  & 61924 &  & 134 &  &  \\  \hline4 &  &  &  & 1 &  & 134 &  & 1808 &  & 12048 &  & 1 &  \\  
\hline5 &  &  &  &  &  &  & 1 &  & 134 &  & 1674 &  &  \\  \hline6 &  &  &  &  &  &  &  &  &  & 1 &  & 133 &  \\  \hline7 &  &  &  &  &  &  &  &  &  &  &  &  & 1 \\  \hline \end{tabular}}   $d=6$ 
\end{table}
\vspace{0.5cm}

\subsection{The $E_6$ del Pezzo surface }
\label{e6delpezzo} 
As we mentioned before, we specialize the polyhedron 15 to the 
massless case by setting  all coefficients in (\ref{thecubic}) to zero 
except of 
\be 
m_4=1,\quad m_5=1,\quad m_6=1,\quad  u= \frac{1}{\tilde u^{1/3}}\ . 
\ee 
With $\lambda=18 u^\frac{8}{3}$  we get 
\be 
g_2=27 u^4 (1-216 u),\quad    g_3=27 u^6 (1 + 540 u - 5832 u^2)\  
\ee 
hence the $j$-function of the $\Gamma_0(3)$ curve. 
\be
j_b=-\frac{(1-216 u)^3}{1728 u (1+27 u)^3}\ .
\ee 
Similar the isogeneous $\Gamma(3)$ curve is obtained by considering the small polyhedron 1 by  
setting $a_2=1$, $m_2=1$, $m_4=1$ and $u= \frac{1}{\tilde u^{1/3}}$, which yields with 
the same scaling $\lambda$, $g_2=27 u^4 (24 u+1)$  
and  $g_3=27 u^6 \left(216 u^2+36 u+1\right)$ so 
\be
j_s=-\frac{(1+ 24 u)^3}{1728 u^3 (1+27 u)}\ 
\ee 
 and the  relating data between the isogeneous curves are 
\be 
\mathbb{Z}_2: u\mapsto -\frac{1}{27} - u,
\quad \mathbb{Z}_2:j_b \leftrightarrow j_s, \quad \mathbb{Z}_2:\tau_s  \leftrightarrow 3 \tau_b \, .
\label{z2e6}
\ee 
Note that the massless $E_6$  curve of \ref{sakaie8}
\be 
j_{E6}^{es} =\frac{(u_{es}-18)^3 (u_{es}+6)}{1728 (u_{es}-21)}
\ee
is again not completely naturally related to $j_{b/s}(u_{b/s})$ 
\be 
u_{es} = -6 - \frac{1}{u_s},\quad u_{es} = \frac{21 - 162 u_b}{1 + 2b u_b}\ . 
\ee
\begin{figure}[H] 
\begin{center} 
\includegraphics[angle=0,width=.3\textwidth]{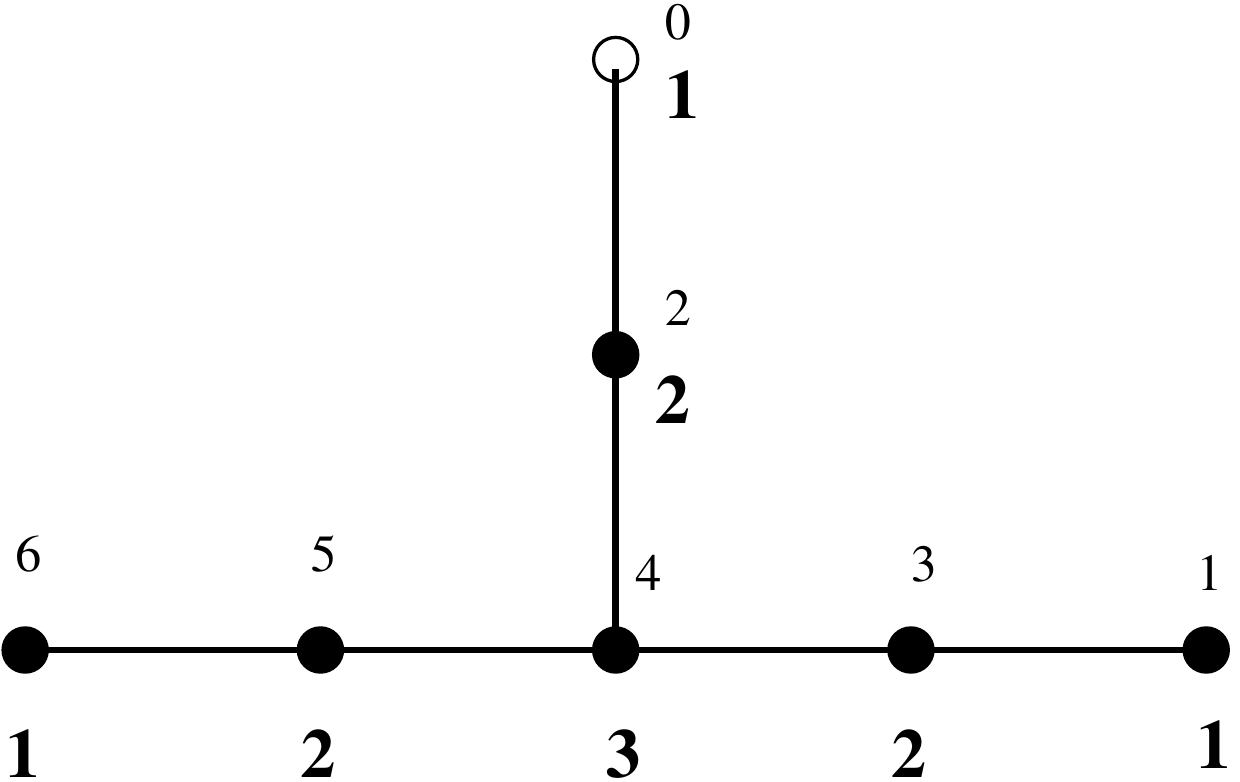} 
\vspace{.2cm} 
\end{center} 
\label{e6} 
\end{figure} 
We record the characters according to the above basis of weights 
\be
\chi_1(0) = 27,\quad \chi_2(0)= 78, \quad \chi_3(0) =351, \quad \chi_4(0) = 2925, \quad \chi_5(0) =351,\quad  \chi_6(0) = 27\ .
\label{e6dimensions}
\ee 
The low degree spin invariants fall in these representations.
\begin{table}[H] 
\centering
\footnotesize{\begin{tabular} {|c|c|} \hline $2j_L \backslash 2j_R$  & 0 \\  \hline  0 & 27 \\  \hline \noalign{\vskip 2mm} 
\multispan{2} $d$=1 \\
\end{tabular}} \  
\begin{tabular} {|c|c|c|} \hline $2j_L \backslash 2j_R$  & 0 & 1 \\ 
 \hline  0 &  & 27 \\  \hline \noalign{\vskip 2mm}   \multispan{3} $d$=2 \\ 
\end{tabular}  \   
\begin{tabular} {|c|c|c|c|c|} \hline $2j_L \backslash 2j_R$  & 0 & 1 & 2 & 3 \\  \hline  0 & 1 &  & 78 &  
\\  \hline
1 &  &  &  & 1 
\\  \hline  \noalign{\vskip 2mm} 
\multispan{5} $d$=3 \end{tabular}  \   
\begin{tabular} {|c|c|c|c|c|c|} \hline $2j_L \backslash 2j_R$  & 0 & 
1 & 2 & 3 & 4 \\  \hline  0 &  & 27 &  & 351 &  \\  \hline1 &  &  &  
&  & 27 \\  \hline   \noalign{\vskip 2mm} 
\multispan{6} $d$=4 \end{tabular} 
\end{table}
\vspace{-0.5cm}
Note that the periodicity in which the adjoint representation appears is 
now the  degree $d$ modulo 3 as expected from the center of $E_6$. 

The first splitting representation that appears is the $378=351+27$ where  
$351$ splits in the Weyl orbits $w_3+5 w_6$ and  further $1755=5\times 351$, see (\ref{masslessEn}) 
and table \ref{Enorbit}. 
\begin{table}[H]
\centering
\footnotesize{\begin{tabular} {|c|c|c|c|c|c|c|c|} \hline $2j_L \backslash 2j_R$  & 
0 & 1 & 2 & 3 & 4 & 5 & 6 \\  \hline  0 & 27 &  & 378 &  & 1755 &  &
  \\  \hline1 &  &  &  & 27 &  & 378 &  \\  \hline2 &  &  &  &  &  &  
& 27 \\  \hline \end{tabular}} \vskip 3pt  $d=5$ 
\end{table}
\vspace{-0.5cm}

\begin{table}[H]
\centering
\footnotesize{
\begin{tabular} {|c|c|c|c|c|c|c|c|c|c|c|} \hline $2j_L \backslash 
2j_R$  & 0 & 1 & 2 & 3 & 4 & 5 & 6 & 7 & 8 & 9 \\  \hline  0 &  & 730 
&  & 3732 &  & 8984 &  & 78 &  &  \\  \hline1 &  &  & 79 &  & 808 &  
& 3732 &  & 1 &  \\  \hline2 &  &  &  & 1 &  & 79 &  & 730 &  &  \\  
\hline3 &  &  &  &  &  &  & 1 &  & 78 &  \\  \hline4 &  &  &  &  &  & 
 &  &  &  & 1 \\  \hline \end{tabular}} \vskip 3pt  $d=6$ 
\end{table} 
\vspace{-0.5cm}
\begin{table}[H]
\centering
\footnotesize{\begin{tabular} {|c|c|c|c|c|c|c|c|c|c|c|c|c|} \hline $2j_L 
\backslash 2j_R$  & 0 & 1 & 2 & 3 & 4 & 5 & 6 & 7 & 8 & 9 & 10 & 11 
\\  \hline  0 & 2133 &  & 10584 &  & 30240 &  & 47439 &  & 2133 &  &  
&  \\  \hline1 &  & 378 &  & 2889 &  & 12717 &  & 30240 &  & 405 &  &
  \\  \hline2 &  &  & 27 &  & 405 &  & 2889 &  & 10584 &  & 27 &  \\  
\hline3 &  &  &  &  &  & 27 &  & 405 &  & 2484 &  &  \\  \hline4 &  & 
 &  &  &  &  &  &  & 27 &  & 378 &  \\  \hline5 &  &  &  &  &  &  &  
&  &  &  &  & 27 \\  \hline \end{tabular}} \vskip 3pt  $d=7$ 
\label{tableE6}
\end{table}
\vspace{-0.5cm}

\subsection{The $D_5$ del Pezzo surface}
\label{d5delpezzo} 
Finally we discuss the case of the $D_5$ surface which can be obtained from the 
polyhedra 2 (small) and 15 (big). We consider again the massless limit by the following choice of coefficients and redefinition of $u$ 
\begin{eqnarray}
    a_1 = i,  \quad    m_3 = i, \quad
 m_4 = i, \quad  m_8 = i, \quad u = \tilde{u}^{-\frac{1}{2}} \quad  (\text{big}) \nn \\
 a_2 = i,   \quad a_3 = i, \quad m_1 = i,   \quad m_2 = i, \quad u = \tilde{u}^{-\frac{1}{2}}   \quad (\text{small}).
\end{eqnarray}
All the other mass parameters vanish.
Accordingly, one obtains the respective Weierstrass normal forms
\begin{eqnarray}
g_2 &=& 27 u^4 \left(256 u^2+16 u+1\right), \nn \\
g_3 &=& -27 u^6 \left(4096 u^3+384 u^2-24 u-1\right), \quad  \text{big polyhedron}, \\
g_2 &=& 27 \left(\frac{1}{u^2}+\frac{16}{u}+16\right) u^6, \nn \\
g_3 &=& 27 \left(\frac{1}{u^3}+\frac{24}{u^2}+\frac{120}{u}-64\right) u^9, \quad  \text{small polyhedron}.
\end{eqnarray}
In both cases we have performed a recaling with
\be
\lambda = 18 u^3
\ee
in order to arrive at the respective expressions for $g_2$ and $g_3$.
Finally the $j$-functions are given as
\be
j_b = \frac{\left(256 u^2+16 u+1\right)^3}{1728 u^2 (16 u+1)^2}, \qquad j_s = \frac{\left(16 u^2+16 u+1\right)^3}{1728 u^4 (16 u+1)}.
\ee
In contrast to the previous cases ($E_6,E_7,E_8$) we observe a different behaviour concerning the exchange of conifold locus and large radius point
\be
\mathbb{Z}_2: u \mapsto -u-\frac{1}{16} , \quad \mathbb{Z}_2: j_b \leftrightarrow j_b, \quad j_s \leftrightarrow j'_s =  -\frac{\left(256 u^2-224 u+1\right)^3}{1728 u (16 u+1)^4}, 
\quad \mathbb{Z}_2: \tau_s \leftrightarrow 4\tau_s'.
\ee
Instead the $j$-functions of the two polyhedra are related by the map
\be
u_b \mapsto -\frac{16 u-(8 u+1) \sqrt{16 u+1}+1}{32 (16 u+1)}.
\ee
We end the discussion by comparing the curves to the massless $D5$ curve given by Sakai and Eguchi. 
This is given by setting again the characters to the dimensions of the fundamental representations in \eqref{SakaiD5curve}. The Weierstrass data of this curve are given by
\be \label{SakaiD5massless}
g_2 = \frac{1}{12} (u+4)^2 \left(u^2-8
   u-32\right), \qquad g_3 = \frac{1}{216} (u+4)^3 \left(u^3-12 u^2-24
   u+224\right),
\ee
and the $j$-function reads
\be
j^{es}= \frac{\left(32 u_{es}^2+8
   u_{es}-1\right)^3}{1728 u_{es}^4
  \left(48 u_{es}^2+8
  u_{es}-1\right)}
\ee
In contrast to the previous cases, this curve is not connected to any of the 
two previous curves by a birational coordinate transformation.

The Dynkin diagram of $\hat D_5$ 
\begin{figure}[H] 
\begin{center} 
\includegraphics[angle=0,width=.3\textwidth]{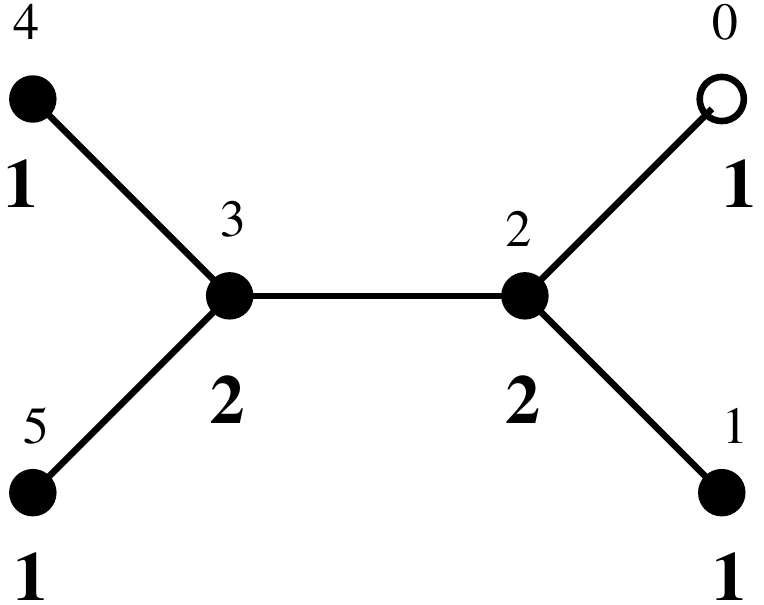} 
\vspace{.2cm} 
\end{center} 
\label{d5} 
\end{figure} 
leads to the following dimensions of the weight moduls 
\be
\chi_1(0) = 10,\quad \chi_2(0)= 45, \quad \chi_3(0) =120, \quad \chi_4(0) = 16, \quad \chi_5(0) =16\ .
\label{dfive-dimensionalimensions}
\ee

We see the periodicity with respect to the degree is now modulo four and the adjoint 
representation of $D5$ appears for the first time at $d=4$. The representation 
$45$ falls in the Weyl orbits $w_2+5 w_0$. The Weyl orbit of $w_2$ is $40$-dimensional and gets contributions only from genus zero curves, while the 
$w_i$ get contributions from a genus two curve, whose leading contribution 
is at spin $[1/2,2]$.     
\begin{table}[H]
\centering
\footnotesize{\begin{tabular} {|c|c|} \hline $2j_L \backslash 2j_R$  & 0 \\  \hline 
 0 & 16 \\  \hline \noalign{\vskip 2mm} 
\multispan{2} $d$=1 \\ \end{tabular}} \hspace{0.3cm}
\footnotesize{\begin{tabular} {|c|c|c|} \hline $2j_L \backslash 2j_R$  & 0 & 1 \\  
\hline  0 &  & 10 \\  \hline \noalign{\vskip 2mm}   \multispan{3} $d$=2 \\ \end{tabular}}  \hspace{0.3cm}
\footnotesize{\begin{tabular} {|c|c|c|c|} \hline $2j_L \backslash 2j_R$  & 0 
& 1 & 2 \\  \hline  0 &  &  & 16 \\  \hline \noalign{\vskip 2mm} 
\multispan{4} $d$=3 \end{tabular}} \hspace{0.3cm}
\footnotesize{
\begin{tabular} {|c|c|c|c|c|c|} \hline $2j_L 
\backslash 2j_R$  & 0 & 1 & 2 & 3 & 4 \\  \hline  0 &  & 1 &  & 45 &  
\\  \hline1 &  &  &  &  & 1 \\  \hline \noalign{\vskip 2mm} 
\multispan{6} $d$=4 \end{tabular}  } 
\end{table}
\vspace{-0.5cm}
\begin{table}[H]
\centering
\footnotesize{\begin{tabular} {|c|c|c|c|c|c|c|} \hline $2j_L 
\backslash 2j_R$  & 0 & 1 & 2 & 3 & 4 & 5 \\  \hline  0 &  &  & 16 &  
& 144 &  \\  \hline1 &  &  &  &  &  & 16 \\  \hline \end{tabular}} 
\vskip 3pt  $d=5$ 
\end{table}
\vspace{-0.5cm}
\begin{table}[H]
\centering
\footnotesize{
\begin{tabular} {|c|c|c|c|c|c|c|c|c|} 
\hline $2j_L \backslash 2j_R$  & 0 & 1 & 2 & 3 & 4 & 5 & 6 & 7 \\  
\hline  0 &  & 10 &  & 130 &  & 456 &  &  \\  \hline1 &  &  &  &  & 
10 &  & 130 &  \\  \hline2 &  &  &  &  &  &  &  & 10 \\  \hline 
\end{tabular}} \vskip 3pt  $d=6$ 
\end{table}
\vspace{-0.5cm}

\begin{table}[H]
\centering
\footnotesize{\begin{tabular} {|c|c|c|c|c|c|c|c|c|c|c|} \hline $2j_L \backslash 
2j_R$  & 0 & 1 & 2 & 3 & 4 & 5 & 6 & 7 & 8 & 9 \\  \hline  0 & 16 &  
& 160 &  & 736 &  & 1440 &  & 16 &  \\  \hline1 &  &  &  & 16 &  & 
176 &  & 736 &  &  \\  \hline2 &  &  &  &  &  &  & 16 &  & 160 &  \\  
\hline3 &  &  &  &  &  &  &  &  &  & 16 \\  \hline \end{tabular}} 
\vskip 3pt  $d=7$ 
\end{table}
\vspace{-0.5cm}

\begin{table}[H]
\centering
\footnotesize{\begin{tabular} 
{|c|c|c|c|c|c|c|c|c|c|c|c|c|c|} \hline $2j_L \backslash 2j_R$  & 0 & 
1 & 2 & 3 & 4 & 5 & 6 & 7 & 8 & 9 & 10 & 11 & 12 \\  \hline  0 &  & 
311 &  & 1345 &  & 3431 &  & 4726 &  & 257 &  &  &  \\  \hline1 &  &  
& 46 &  & 357 &  & 1602 &  & 3431 &  & 46 &  &  \\  \hline2 &  &  &  
& 1 &  & 46 &  & 357 &  & 1345 &  & 1 &  \\  \hline3 &  &  &  &  &  & 
 & 1 &  & 46 &  & 311 &  &  \\  \hline4 &  &  &  &  &  &  &  &  &  & 
1 &  & 45 &  \\  \hline5 &  &  &  &  &  &  &  &  &  &  &  &  & 1 \\ 
\hline \end{tabular}} \vskip 3pt  $d=8$ \vskip 10pt    
\label{tableE5}
\end{table}
\vspace{-0.5cm}

\subsection{An alternative approach to the massless cases}

Alternatively we use the Picard-Fuchs equations, the Yukawa couplings, i.e. the usual B-model 
methods, that also apply in the compact cases.  The complex geometry of the mirror manifolds 
are described by the Picard-Fuchs differential equations 
\begin{eqnarray} \label{PF2.4}
(\theta_z ^2 +c_0 z \prod_{i=1}^2 (\theta_z+1-a_i))\theta_z   \int_{\gamma_i}   \Omega =0, 
\end{eqnarray}
where $z$ is the complex structure modulus in the mirror manifold and $\theta_z=z\partial_z$. $a_1, a_2$ and $c_0$ are classical constants of the the Calabi-Yau manifolds. 
$c_0$ is a normalization constant  for the complex structure parameter $z$ such that $t=\log(z)+\mathcal{O}(z) $ around $z\sim 0$ corresponds to the K\"ahler modulus 
in the large volume limit.  The vectors $\vec{a}=(a_1,a_2)$ satisfy $a_1+a_2=1$ and are given as follows for various one-parameter families of Calabi-Yau manifolds we consider 
\begin{eqnarray} \label{cons}
&&  \mathbb{P}^2: ~ \vec{a}= (\frac{1}{3},\frac{2}{3}),~~~~ 
\mathbb{P}^1\times \mathbb{P}^1: ~ \vec{a}= (\frac{1}{2},\frac{1}{2}),~~~~
D_5: ~ \vec{a}= (\frac{1}{2},\frac{1}{2}),~~~ ~\nonumber \\
&& E_6: ~ \vec{a}= (\frac{1}{3},\frac{2}{3}),~~~~
E_7: ~ \vec{a}= (\frac{1}{4},\frac{3}{4}),~~~~~~~~~
E_8: ~ \vec{a}= (\frac{1}{6},\frac{5}{6}). 
\end{eqnarray}

The $E_n$ ($n=5,6,7,8$) del Pezzo surfaces can be represented as complete intersections of degree $(2,2)$ 
in $\mathbb{P}^4$, a degree 3 hypersurface in $\mathbb{P}^3$, a degree 4 hypersurface in weighted projective space $\mathbb{P}^3(1,1,1,2)$ 
and a degree six hypersurface in $P^3(1,1,2,3)$. In these cases the normalization constant $c_0$ can be computed as $c_0=(\prod_i d_i^{d_i} )/(\prod_j w_j^{w_j}) $ 
where $d_i$ are the degree(s) of hypersurfaces or complete intersections, 
and $w_j$ are weights of the  ambient projective space. The constant is $c_0=27$ for the $\mathbb{P}^2$ model and $c_0=-16$ for the $\mathbb{P}^1\times \mathbb{P}^1$ model. 

The prepotential $F^{(0,0)}(t)$ is determined by the Picard-Fuchs (PF) equation (\ref{PF2.4}) from the fact that the mirror map $t(z)$ and derivative $\partial_t F^{(0,0)}(t)$ 
are solutions to the PF equation besides the constant solution. The normalization of the prepotential is fixed by the classical intersection 
number $\kappa$ as $F^{(0,0)}(t)= -\frac{\kappa}{6}t^3 +\cdots$. The intersection number can be calculated by the formula 
$\kappa=(\prod_i d_i )/(\prod_j w_j) $ in the $E_n$ models.  The numbers are $\kappa=\frac{1}{3}$ for the $\mathbb{P}^2$ model and $\kappa=1$ for the $\mathbb{P}^1\times \mathbb{P}^1$ model. 
We list the constants $c_0$ and $\kappa$ for the Calabi-Yau models in Table \ref{table1}.

\begin{table}
\begin{center}
\begin{tabular} {|c |c |c |c|c |c|c| }
\hline
  CY & $\mathbb{P}^2$ & $\mathbb{P}^1\times \mathbb{P}^1$ & $D_5$ & $E_6$ & $E_7$ & $E_8$ \\ \hline
 $c_0$ & 27 & -16 & 16  & 27 & 64 & 432  \\ \hline
  $\kappa$ & $\frac{1}{3}$ & 1 & 4 & 3 & 2 & 1 \\ \hline
\end{tabular}
\caption{The constants $c_0$ and $\kappa$ for the Calabi-Yau models.}
\label{table1}
\end{center}
\end{table}

We discuss next the genus one amplitudes $F^{(1,0)}$ and $F^{(0,1)}$. The $F^{(1,0)}$ amplitude is holomorphic while the amplitude $F^{(0,1)}$ has a holomorphic anomaly 
which is determined by the genus one holomorphic anomaly equation \cite{BCOV}. Both amplitudes have logarithmic cuts for the discriminant $\Delta (z)= 1+ c_0z$ 
whose coefficients are determined by the genus one gap boundary conditions at the conifold point $\Delta(z)=0$. Furthermore, it turns out that the amplitudes 
also contain a logarithmic piece $\log(z)$. We can write the amplitudes as 
\begin{eqnarray}
F^{(1,0)} &=& \frac{\log(\Delta(z)) - c^{(1,0)} \log( z) }{24},  \nonumber \\
F^{(0,1)} &=& -\frac{1}{2}\log(\partial_z t(z)) -\frac{1}{12}  (\log(\Delta(z)) +c^{(0,1)} \log( z) ),
\end{eqnarray} 
where we use the constants $c^{(1,0)}$ and $c^{(0,1)}$ to denote the coefficients for $\log(z)$ terms in the refined amplitudes. 
We determine the constants for the Calabi-Yau models and list them in Table \ref{table2}.

\begin{table}
\begin{center}
\begin{tabular} {|c |c |c |c|c |c|c| }
\hline
  CY & $\mathbb{P}^2$ & $\mathbb{P}^1\times \mathbb{P}^1$ & $D_5$ & $E_6$ & $E_7$ & $E_8$ \\ \hline
 $c^{(1,0)}$ & 1 & 2 & 8  & 9 & 10 & 11  \\ \hline
  $c^{(0,1)}$  & 7 & 7 & 4 & 3 & 2 & 1 \\ \hline
\end{tabular}
\caption{The constants $c^{(1,0)}$ and $c^{(0,1)}$  for the Calabi-Yau models.}
\label{table2}
\end{center}
\end{table}

The three-point Yukawa coupling and the K\"ahler metric in the moduli space are given up to an anti-holomorphic factor by
\begin{eqnarray}
C_{zzz} = -\frac{\kappa} {z^3 (1+c_0z)}, ~~~ G_{z\bar{z}} \sim \partial_z t\, .
\end{eqnarray}
The Christoffel connection in the holomorphic limit $\Gamma_{zz}^z =\partial_t z( \partial^2_z t  )$ is not a rational function of $z$. There is a relation with the propagator which satisfies 
${\partial} _{\bar{z}} S^{zz} =\bar{C}_{\bar{z}}^{zz}$, 
\begin{eqnarray}
\Gamma _{zz}^z= -C_{zzz} S^{zz} +f_z
\end{eqnarray}
where $f_z$ is a rational function of $z$ since the anti-holomorphic derivatives $\bar{\partial}_{\bar{z}}$ of both sides are the same. For the one-parameter models we simply denote the propagator as $S\equiv S^{zz}$. The rational function $f_z$ is a holomorphic ambiguity that we can choose such that the propagator $S$ has a nice behavior near the special singular points in the moduli space 
\begin{eqnarray}
f_z= -\frac{6a_1+5}{6z} -\frac{c_0}{6(1+c_0z)}
\end{eqnarray}
where $a_1$ is the constant in (\ref{cons}) and $c_0$ is the normalization constant in table \ref{table1}. With this choice of ambiguity $f_z$, the propagator $S$ is regular at the conifold point $z=-\frac{1}{c_0}$. Near the orbifold point $z^{-1}\sim 0$, the propagator generically scales as $S\sim z^3$ and we have chosen the constant $(6a_1+5)$ in $f_z$  to cancel the leading $z^3$ term so that the scaling behavior is less singular near the orbifold point as $S\sim z^2$. The cancellation can be seen by noting that the flat coordinate scales as $t\sim z^{-a_1}$ near the orbifold point $z^{-1}\sim 0$, and accordingly the Christoffel connection scales as $\Gamma_{zz}^z \sim -(a_1+1) z^{-1}$ and cancels the leading term in $f_z$.  
 
The derivative of the propagator can be derived from the special geometry relation, 
\begin{eqnarray} \label{deri2.10}
D_z S= -C_{zzz} S^2 +\tilde{f}(z), 
\end{eqnarray}
where the covariant derivative reads $D_z S =(\partial_z +2 \Gamma_{zz}^z ) S$.  The holomorphic ambiguity $\tilde{f}(z)$ is a rational function with a simple pole at $\Delta(z)$, 
and it can be  fixed by computing $S$ and $\Gamma_{zz}^z$ in the holomorphic limit. 

The propagator $S$ is the only an-holomorphic component in the higher genus amplitudes, and the generalized holomorphic anomaly for the refined theory is 
\begin{eqnarray} \label{holo2.11}
\partial_S F^{(n,g)} (S, z) =\frac{1}{2} [D_z^2 F^{(n,g-1)} +\sum_{n_1=0}^n \sum_{g_1=0}^g D_z F^{(n_1,g_1)} D_z F^{(n-n_1,g-g_1)}],   
\end{eqnarray} 
where the first term on the RHS is defined to be zero if $g=0$, and the sum in the second term does not include the two cases $n_1=g_1=0$ and $n_1=n, g_1=g$.  
Since the derivative of the propagator forms a closed algebra as seen in equation (\ref{deri2.10}), the higher genus amplitudes $F^{(n,g)}$ with 
$n+g\geq 2$ are polynomials of the propagator $S$ and the coefficients of the polynomials are rational function of $z$.

The holomorphic anomaly equation determines the $S$-dependent part in the higher genus 
amplitudes $F^{(n,g)}$, but not the $S$-independent holomorphic ambiguity which is a rational function of $z$ and we can denote as $f_0^{(n,g)}(z)$.
 To further fix this function we consider the boundary conditions at the special points in the moduli space, 
the large volume point $z\sim 0$, the conifold point $z\sim -\frac{1}{c_0}$ and the orbifold point $z\sim \infty$.  

The behaviors near the large volume point  and the conifold point are universal for all models.
 The amplitude  $F^{(n,g)}$ and the ambiguity $f_0^{(n,g)}(z)$ go to a constant $\mathcal{O}(z^0)$ near the large volume point.
 The leading constant term in the conventional unrefined theory is the constant map contribution in Gromov-Witten theory.
 This constant does not affect the calculations of the refined GV invariants which only contribute to the world-sheet instantons of positive degrees, and here we will not determine the constant for the refined theory.

Near the conifold point, the amplitude $F^{(n,g)}$ satisfies the gap condition $F^{(n,g)} \sim \frac{1}{t_D^{2(n+g)-2}} \\ + \mathcal{O}(t_D^0)$, where the $t_D$ is the flat coordinate near the conifold point and scales like $t_D\sim z+\frac{1}{c_0}$.
 Accordingly the ambiguity scales as $f_0^{(n,g)}(z)\sim \frac{1}{(1+c_0z)^{2(n+g)-2}}$ and the gap conditions fix $2(n+g)-2$ constants in the holomorphic ambiguity $f_0^{(n,g)}(z)$.

The boundary conditions near the orbifold point $z\sim \infty$ are more tricky,  and needed to be classified into several cases, similar to the situation studied in \cite{HKQ}. 

For the $\mathbb{P}^2$ model, the higher genus amplitude $F^{(n,g)}$ is regular at the orbifold point.
 Since we have chosen the propagator $S$ to have a nice scaling behavior $S\sim z^2$ at the orbifold point, there is no singularity at the orbifold point from the $S$-dependent part in $F^{(n,g)}$.
 Therefore the holomorphic ambiguity  $f_0^{(n,g)}(z)$ is also regular at the orbifold point, and we can write an ansatz 
\begin{eqnarray}
 f_0^{(n,g)}(z) =\sum_{k=0}^{2(n+g)-2}  \frac{x_k}{(1+c_0z)^k}. 
\end{eqnarray}
The gap condition fixes the $2(n+g)-2$ constants $x_k$ for $k=1,2,\cdots, 2(n+g)-2$, and we do not need to fix the 
constant $x_0$.
 So in this model we can in principle compute the refined topological string amplitudes to any genus and extract the corresponding refined GV invariants.

For the other five models, the amplitude $F^{(n,g)}$ is singular at the orbifold point but is less singular than $\frac{1}{t_o^{2(n+g)-2}}$, where $t_o$ is the flat coordinate near the orbifold point and scales as $t_o\sim z^{-a_1}$, where $a_1$ is the fractional number in (\ref{cons}).
 So the ansatz for the ambiguity is 
\begin{eqnarray} \label{ambi2.13}
 f_0^{(n,g)}(z) =\sum_{k=0}^{2(n+g)-2}  \frac{x_k}{(1+c_0z)^k} +\sum_{k=1}^{[2a_1(n+g-1)]} y_kz^k \, .
\end{eqnarray}
In these cases that are similar to the $\mathbb{P}^2$ model, the conifold gap condition fixes the $2(n+g)-2$ constants $x_k$ for $k=1,2,\cdots, 2(n+g)-2$.
 However we still need to fix the $[2a_1(n+g-1)]$ constants $y_k$ in order to solve the refined topological string amplitudes (up to a constant $x_0$). 

For the $\mathbb{P}^1\times \mathbb{P}^1$ model, there is also a further gap condition at the orbifold point similar to the conifold point which implies that $F^{(n,g)}\sim \frac{1}{t_o^{2(n+g)-2}} + \mathcal{O}(t_o^0)$.
 Since only even powers of $t_o$ appear due to the leading scaling behavior $t_o\sim z^{-\frac{1}{2}}$, this provides $n+g-1$ boundary conditions which exactly fix the constants $y_k$ with $k=1,2,\cdots, (n+g-1)$.
 So in this model we can also in principle compute the refined topological string amplitude to any genus.

For the remaining $E_n$ ($n=5,6,7,8$) models, there is no nice boundary condition at the orbifold point to fix the constants $y_k$ in (\ref{ambi2.13}).
 Here we can use the nice behavior of refined GV invariants at the large volume point to provide boundary conditions to fix these constants.
 It is often the case that some low degree refined GV invariants $\tilde{n}^d_{g_L,g_R}$ vanish at a given genus $g_L, g_R$.
 If we have computed the refined GV invariants $\tilde{n}^d_{g_L,g_R}$ for all the genera $g_L+g_R\leq  n+g$, $g_R\leq n$ and up to degree $d\leq [2a_1(n+g-1)]$ either by the B-model method or by their vanishing property,  then we would have enough boundary conditions to fix the constants $y_k$ with $k=1,2,\cdots, [2a_1(n+g-1)]$ in  $f_0^{(n,g)}(z)$ in (\ref{ambi2.13}) and would have solved the refined amplitude  $F^{(n,g)}$ as well.
 Using this technique we can solve the refined topological string amplitudes to some finite but not arbitrary high genus.

Using the B-model techniques we compute the refined topological string amplitudes to some higher genus and we fix the complete refined GV invariants up to some finite degrees for the various models.
 We list the results in the tables \ref{tableE5} - \ref{tableP1P1}.
 The refined GV invariants for the local $\mathbb{P}^2$ and $\mathbb{P}^1\times \mathbb{P}^1$ models have been computed before in the previous papers \cite{HKK, IKV}.
 Here we also include them for completeness. The blank elements in the tables represent vanishing GV invariants.

We discuss some salient features of the refined GV invariants. For degree $d$ which is a positive integer as an element in $H_2(M,\mathbb{Z})$, there is a non-vanishing positive integer $n_{j_L,j_R}^d = \tilde{n}_{2j_L,2j_R}^d$ at the top genus $(2j_L,2j_R)=(g_L^{top},g_R^{top})$.
 All higher genus invariants vanish so the non-vanishing GV invariants form a rectangular matrix, and we find that the left top genus is always less than the right top genus $g_L^{top} \leq g_R^{top}$.
 For a Calabi-Yau model, the top genus of higher degree is always larger than that of the lower degree, i.e. we always find  $g_L^{top}(d)\geq g_L^{top}(d-1)$ and  $g_R^{top}(d)\geq g_R^{top}(d-1)$. 

In the basis of integers $\tilde{n}^d_{g_L,g_R}$, the GV invariants do not generically vanish if the genus pair lies in the rectangular matrix, i.
e. $g_L\leq g_L^{top}$ and $g_R\leq g_R^{top}$.
 So we can determine the top genus $(g_L^{top},g_R^{top})$ as the smallest integer pair such that $\tilde{n}^d_{g_L^{top}+1,0}=\tilde{n}^d_{0,g_R^{top}+1}=0$.
 The vanishing of a GV invariant  $\tilde{n}^d_{g_L,g_R}=0$ implies that its higher genus neighbors also vanish $\tilde{n}^d_{g_L+1,g_R}=\tilde{n}^d_{g_L,g_R+1}=0$. 

However in the $j$-spin basis $n^d_{j_L,j_R}$, there is furthermore a large number of vanishing GV invariants  $n^d_{j_L,j_R}$ inside  the rectangular matrix $2j_L\leq g_L^{top}$ and $2j_R\leq g_R^{top}$.
 The genus pairs of these non-vanishing integers follow certain patterns as we go up in higher degrees.
 More precisely, suppose at degree $d-1$ we find $n^{d-1}_{g_L/2,g_R/2}\neq 0$, then for the corresponding genus pair $(g_L^{\prime}, g_R^{\prime})=(g_L+g_L^{top}(d)-g_L^{top}(d-1), g_R+g_R^{top}(d)-g_R^{top}(d-1))$ at degree $d$, we always find that the GV integer is also non-vanishing  $n^d_{g_L^{\prime}/2, g_R^{\prime}/2}\neq 0$.
 On the the hand, if the integer   $n^{d-1}_{g_L/2,g_R/2}$ vanishes, it is also usually but not always the case  that the vanishing $n^d_{g_L^{\prime}/2, g_R^{\prime}/2}= 0$ also happens at the higher degree $d$.

The non-vanishing GV invariants seem to cluster together, but no two non-vanishing GV invariants are next neighbors to each others.
 More precisely, we define the distance as  $|g_L-g_L^{\prime}|+ |g_R-g_R^{\prime}|$ between two  GV invariants $n^d_{g_L/2, g_R/2}$ and $n^d_{g_L^{\prime}/2, g_R^{\prime}/2}$.
 We find that the distance of a non-vanishing GV invariant $n^d_{g_L/2, g_R/2}$ to its nearest non-vanishing neighbor is almost always $2$.
 Only two exceptions occur in the $\mathbb{P}^2$ model where the distance with the nearest non-vanishing neighbor is $4$.

With the B-model method we can extract the GV invariants in the basis $\tilde{n}^d_{g_L,g_R}$ from the refined topological string amplitudes.
 We find that by utilizing the pattern in the $j$-spin basis, we do not need to solve all non-vanishing $\tilde{n}^d_{g_L,g_R}$ inside the top genus rectangular matrix in order to fix the complete GV invariants.
 It is still necessary to compute a number of non-vanishing $\tilde{n}^d_{g_L,g_R}$ from the B-model which is larger than the number of non-vanishing $n^d_{j_L,j_R}$ in the $j$-spin basis.
 However, since we do not know a priori the number and the positions of the non-vanishing $n^d_{j_L,j_R}$ before we find that the solution, we usually need to compute the B-model to a few more genus higher.
 In practice we find that we can usually fix the complete GV invariants when we compute the refined amplitudes $F^{(n,g)}$ up to the total genus $n+g$ a little bigger than the left top genus $g_L^{top}$.
 We consider a solution for the GV  integers $n^d_{j_L,j_R}$ that passes non-trivial consistency 
checks, if the number of non-vanishing integers $\tilde{n}^d_{g_L,g_R}$ obtained from the B-model is larger than the number of non-vanishing integers $n^d_{j_L,j_R}$ in the solution.

\section{Toric del Pezzos and mass deformations} 
\label{toricandmass}

In this section we discuss the calculation of refined BPS numbers of geometries that have a toric realization. In particular we consider the toric del Pezzo surfaces $F_0$, $\mathbb{P}^2 \cong \mathcal{B}_0$,  $\mathcal{B}_1$, $\mathcal{B}_2$ and $\mathcal{B}_3$ as well as almost del Pezzo surfaces and mass deformations of the local $E_8$-geometry.

An important observation is that the GKZ-system \eqref{generalPFequations} that can be easily determined from the toric diagram can be reduced to a single ordinary differential operator depending on only one variable $u$ and some mass parameters $m_i$. These mass parameters correspond to trivial solutions of the Picard-Fuchs equations\footnote{I.e. these solutions take the form of a linear combination of logarithms of the variables.}. The geometrical interpretation is that the local mirror geometry has just one complex modulus corresponding to the base whereas the other moduli correspond to isomonodromic deformations. We have determined this differential operator for the cases $F_0$ \eqref{PFEp1p1}, $\mathbb{P}^2$ \eqref{PFEp2},  $\mathcal{B}_1$ \eqref{PFEf1} and $\mathcal{B}_2$ \eqref{PFEf2}. 

As discussed in section \ref{section2}, a crucial ingredient for the computation is the Weierstrass normal form of the mirror geometry that can be obtained by embedding the toric diagram into either one of the polyhedra 13, 15 or 16, see also the discussion in \ref{mirrorsymmetry}. This procedure is explicitly demonstrated for the embedding of the toric del Pezzo surfaces in the picture below and is the starting point for the subsequent discussion. Following the procedure described in section \ref{section2}, we have determined the free energies for the first genera. However in the following discussion we just discuss the important steps to set up the calculation and mostly restrict ourselves to just pointing out new phenomena when passing from one geometry to another\footnote{The calculated free energies are available on request.}.

\begin{figure}[h] 
\begin{center} 
\includegraphics[angle=0,width=.9\textwidth]{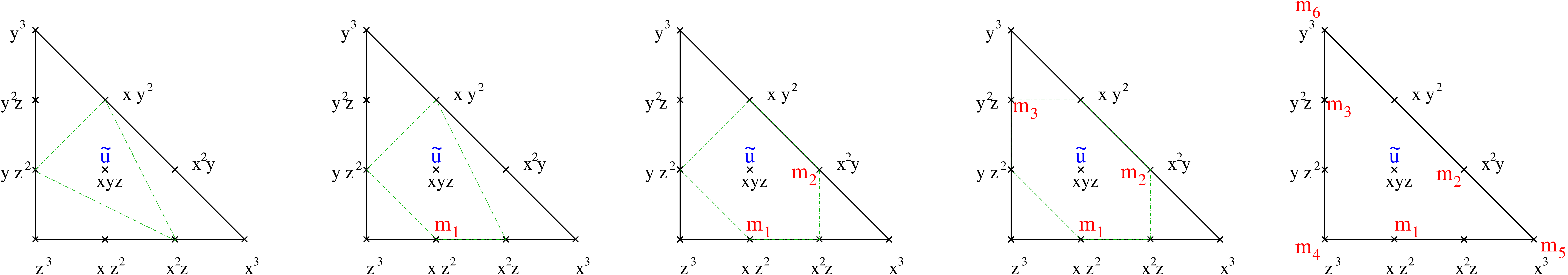} 
\begin{quote} 
\caption{Here we depict the polyhedral embedding of $\mathcal{B}_0,\ldots, \mathcal{B}_3$ 
into polyhedron 16. The Weierstrass form of the general Newton 
polynom to polyhedron 16 is calculated in Appendix A1.
\vspace{-1.2cm}} \label{polyemdedding} 
\end{quote} 
\end{center} 
\end{figure}

\subsection{${\cal O}(-K_{\mathbb{P}^2}) \rightarrow \mathbb{P}^2$}
\begin{figure}[h!] 
\begin{center} 
\includegraphics[angle=0,width=.1\textwidth]{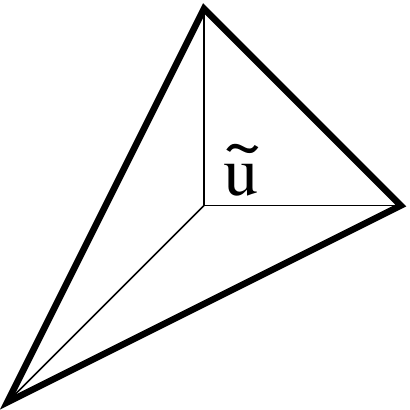}
\begin{quote} 
\caption{The polyhedron 1. 
with the modulus $\tilde u$  
\vspace{-1.2cm}} \label{poly1} \end{quote} 
\end{center} 
\end{figure} 

We start by providing the data of the Mori cone
\begin{equation} 
 \label{datap2} 
 \begin{array}{c|crrr|rl|} 
    \multicolumn{5}{c}{\nu_i }    &l^{(1)}&    \\ 
    D_u    &&     1&     0&   0&         -3&        \\ 
    D_1    &&     1&     1&   0&         1&         \\ 
    D_2    &&     1&     0&   1&         1&          \\ 
    D_3    &&     1&    -1&   -1&         1&           \\ 
  \end{array} \, . 
\end{equation}
This gives us the invariant coordinate
\be
z = \frac{a_1 a_2 a_3}{u^3} = \frac{1}{\tilde{u}^3}.
\ee
We set $a_1=a_2=a_3=1$ and denote the complex modulus by
$\tilde u$. Note that the coordinate $z$ is small at the large radius point, whereas the coordinate $\tilde u$ is small at the orbifold point.
 However, from the point of view of embedding the toric diagram of interest into either one of the polygons 13, 15 or 16, it more natural to use the coordinate $\tilde{u}$.
 As we proceed with blowing up $\mathbb{P}^2$ we will always use this coordinate and finally pass to the small coordinate $1/\tilde{u}^\alpha$ where $\alpha$ has to be suitably determined.

As explained above instead of starting like in 
\cite{KKV} with (\ref{mirrorcurve}) and 
eliminating $X_i$ by  (\ref{modmirrorrelations}) 
we solve this equations more geometrically by 
embedding $\Delta^*$ into polyhedra, so that
the Newton polyhedron solves immediately the above
constraints 
\be
X Y^2 + Y Z^2 + X^2 Z + \tilde{u} X Y Z = 0
\ee
and yields $H(X,Y)$ by setting $Z=1$.
By Nagell's algorithm  its Weierstrass normal 
form is given by
\ban
y^2 &=& x^3+  \frac{1}{12}\Big(-24 \tilde{u} - \tilde{u}^4   
  \Big) x + \frac{1}{216} \Big(-216 - 36 \tilde{u}^3 - \tilde{u}^6  \Big)\, .
   \ean
It is easy to show that the period integrals $\int_\gamma \lambda$  over the meromorphic 
differential
\begin{equation} 
\lambda = \log(x) \frac{d y}{y},   
\end{equation}
which describe the closed string moduli fulfill the differential 
equations $i=1,\ldots, \# \ {\rm moduli}$ 
\begin{equation} 
\left(\prod_{l^{(i)}_k > 0} \partial_{a_k}^{l^{(i)}_k}-\prod_{l^{(k)}_k < 0} \partial_{a_k}^{-l^{(i)}_k}\right) \int_\gamma \lambda=0\ .
\label{generalPFequations} 
\end{equation}
The $a_i$, $i=1,\ldots, \# \ {\rm points}$ are subject to 
symmetries of the geometry 
and can be `gauge'-fixed to the variables $z_i$ using $a_i\partial_{a_i}= l_i^{(k)} z_k \partial_{z_l}$. 
In the case at hand there is just one Picard Fuchs equation \eqref{generalPFequations} which has third order
\be \label{thetaformp2}
\mathcal{L}_{l.r.} = \theta^3 + 3 z \theta (3 \theta+1)(3\theta +2) ,
\ee
where $\theta$ denotes the logarithmic derivative $z \frac{d}{dz}$, s.t. \eqref{thetaformp2} reads in terms of $z$
\be \label{zformp2}
\mathcal{L}_{l.r.}=(1+60 z)\, \partial_z + (3 z + 108 z^2)\, \partial_z^2 + (z^2 + 27 z^3)\, \partial_z^3 .
\ee
Recall that the solutions to this differential operator give the periods at the large radius point. As already discussed above, it will often 
more natural to use the coordinate $\tilde{u}$, in which \eqref{zformp2} takes the form
\be
\mathcal{L}_{orb} = \tilde{u}\, \partial_{\tilde{u}} + 3 \tilde{u}^2\, \partial_{\tilde{u}}^2 + \left( 27 + \tilde{u}^3 \right) \partial_{\tilde{u}}^3 .
\label{PFEp2}
\ee
The corresponding solutions give the periods at the orbifold point.

We end the discussion of $\mathbb{P}^2$ by writing down the prepotential up to degree 7 in $Q_1$, denoting ${\rm L}_\beta = \text{Li}(Q_1^\beta)$
\be
F = \text{class} + 3 {\rm L}_1 - 6 {\rm L}_2 + 27 {\rm L}_3  - 192 {\rm L}_4  + 
 1695 {\rm L}_5 - 17064 {\rm L}_6 + 188454 {\rm L}_7 .
\ee
The refined invariants have been calculated in~\cite{HKK}. We 
list a few for reference with the blow-up cases. The connection 
to our solution of the $\frac{1}{2} K3$ is given by 
(\ref{masslessEn}) and table \ref{Enorbit}.  

\begin{table}[h!]
\centering{{
\begin{tabular}[h]{|c|c|cccccccccccccccccccccccccccc|}
\hline
d\!&\!\!$j_L\backslash j_R$\!&\!\!0\!&\!\!\!$\frac{1}{2}$\!\!\!&\!\!\!1\!\!\!&\!\!\!$\frac{3}{2}$\!\!\!&\!\!\!2\!\!\!&\!\!\!$\frac{5}{2}$\!\!\!&\!\!\!3\!\!\!&\!\!\!$\frac{7}{2}$\!\!\!&\!\!\!4\!\!\!&
\!\!\!$\frac{9}{2}$\!\!\!&\!\!\!5\!\!\!&\!\!\!$\frac{11}{2}$\!\!\!&\!\!\!6\!\!\!&\!\!\!$\frac{13}{2}$\!\!\!&\!\!\!7\!\!\!&\!\!\!$\frac{15}{2}$\!\!\!&\!\!\!8\!\!\!&\!\!\!$\frac{17}{2}$\!\!\!
&\!\!\!9\!\!\!&\!\!\!$\frac{19}{2}$\!\!\!&\!\!\!10\!\!\!&\!\!\!$\frac{21}{2}$\!\!\!&\!\!\!11\!\!\!&\!\!\!$\frac{23}{2}$\!\!\!&\!\!\!12\!\!\!&\!\!\!$\frac{25}{2}$\!\!\!&\!\!\!13\!\!\!
&\!\!\!$\frac{27}{2}$\!\!\\
\hline
1\!\!\!&\!\!\!0\!\!\!&\!\!\!\!\!\!&\!\!\!\!\!\!&\!\!\!1\!\!\!&\!\!\!\!\!\!&\!\!\!\!\!\!&\!\!\!\!\!\!&\!\!\!\!\!\!&\!\!\!\!\!\!&\!\!\!\!\!\!&\!\!\!\!\!\!&\!\!\!\!\!&\!\!\!\!\!\!&\!\!\!\!\!\!&\!\!\!\!\!\!&\!\!\!\!\!\!&\!\!\!\!\!\!&\!\!\!\!\!&\!\!\!\!\!&\!\!\!\!\!\!&\!\!\!\!\!&\!\!\!\!\!\!&\!\!\!\!\!\!&\!\!\!\!\!\!&\!\!\!\!\!\!&\!\!\!\!\!\!&\!\!\!\!\!&\!\!\!\!\!\!&\!\!\\
\hline
2\!\!\!&\!\!\!0\!\!\!&\!\!\!\!\!\!&\!\!\!\!\!\!&\!\!\!\!\!\!&\!\!\!\!\!\!&\!\!\!\!\!\!&\!\!\!1\!\!\!&\!\!\!\!\!\!&\!\!\!\!\!\!&\!\!\!\!\!\!&\!\!\!\!\!\!&\!\!\!\!\!\!&\!\!\!\!\!\!&\!\!\!\!\!\!&\!\!\!\!\!\!&\!\!\!\!\!\!&\!\!\!\!\!\!&\!\!\!\!\!\!&\!\!\!\!\!\!&\!\!\!\!\!\!&\!\!\!\!\!\!&\!\!\!\!\!\!&\!\!\!\!\!\!&\!\!\!\!\!\!&\!\!\!\!\!\!&\!\!\!\!\!\!&\!\!\!\!\!\!&\!\!\!\!\!\!&\!\!\\
\hline
3\!\!\!&\!\!\!0\!\!\!&\!\!\!\!\!\!&\!\!\!\!\!\!&\!\!\!\!\!\!&\!\!\!\!\!\!&\!\!\!\!\!\!&\!\!\!\!\!\!&\!\!\!1\!\!\!&\!\!\!\!\!\!&\!\!\!\!\!\!&\!\!\!\!\!\!&\!\!\!\!\!\!&\!\!\!\!\!\!&\!\!\!\!\!\!&\!\!\!\!\!\!&\!\!\!\!\!\!&\!\!\!\!\!\!&\!\!\!\!\!\!&\!\!\!\!\!\!&\!\!\!\!\!\!&\!\!\!\!\!\!&\!\!\!\!\!\!&\!\!\!\!\!\!&\!\!\!\!\!\!&\!\!\!\!\!\!&\!\!\!\!\!\!&\!\!\!\!\!\!&\!\!\!\!\!\!&\!\!\!\!\\
\!\!\!&\!\!\!$\frac{1}{2}$\!\!\!&\!\!\!\!\!\!&\!\!\!\!\!\!&\!\!\!\!\!\!&\!\!\!\!\!\!&\!\!\!\!\!\!&\!\!\!\!\!\!&\!\!\!\!\!\!&\!\!\!\!\!\!&\!\!\!\!\!\!&\!\!\!1\!\!\!&\!\!\!\!\!\!&\!\!\!\!\!\!&\!\!\!\!\!\!&\!\!\!\!\!\!&\!\!\!\!\!\!&\!\!\!\!\!\!&\!\!\!\!\!\!&\!\!\!\!\!\!&\!\!\!\!\!\!&\!\!\!\!\!\!&\!\!\!\!\!\!&\!\!\!\!\!\!&\!\!\!\!\!\!&\!\!\!\!\!\!&\!\!\!\!\!\!&\!\!\!\!\!\!&\!\!\!\!&\!\!\\
\hline
4\!\!\!&\!\!\!0\!\!\!&\!\!\!\!\!\!&\!\!\!\!\!\!&\!\!\!\!\!\!&\!\!\!\!\!\!&\!\!\!\!\!\!&\!\!\!1\!\!\!&\!\!\!\!\!\!&\!\!\!\!\!\!&\!\!\!\!\!\!&\!\!\!1\!\!\!&\!\!\!\!\!\!&\!\!\!\!\!\!&\!\!\!\!\!\!&\!\!\!1\!\!\!&\!\!\!\!\!\!&\!\!\!\!\!\!&\!\!\!\!\!\!&\!\!\!\!\!\!&\!\!\!\!\!\!&\!\!\!\!\!\!&\!\!\!\!\!\!&\!\!\!\!\!\!&\!\!\!\!\!\!&\!\!\!\!\!\!&\!\!\!\!\!\!&\!\!\!\!\!\!&\!\!\!\!\!\!&\!\!\!\!\\
\!\!\!&\!\!\!$\frac{1}{2}$\!\!\!&\!\!\!\!\!\!&\!\!\!\!\!\!&\!\!\!\!\!\!&\!\!\!\!\!\!&\!\!\!\!\!\!&\!\!\!\!\!\!&\!\!\!\!\!\!&\!\!\!\!\!\!&\!\!\!1\!\!\!&\!\!\!\!\!\!&\!\!\!1\!\!\!&\!\!\!\!\!\!&\!\!\!1\!\!\!&\!\!\!\!\!\!&\!\!\!\!\!\!&\!\!\!\!\!\!&\!\!\!\!\!\!&\!\!\!\!\!\!&\!\!\!\!\!\!&\!\!\!\!\!\!&\!\!\!\!\!\!&\!\!\!\!\!\!&\!\!\!\!\!\!&\!\!\!\!\!\!&\!\!\!\!\!\!&\!\!\!\!\!\!&\!\!\!\!&\!\!\!\!\\
\!\!\!&\!\!\!2\!\!\!&\!\!\!\!\!\!&\!\!\!\!\!\!&\!\!\!\!\!\!&\!\!\!\!\!\!&\!\!\!\!\!\!&\!\!\!\!\!\!&\!\!\!\!\!\!&\!\!\!\!\!\!&\!\!\!\!\!\!&\!\!\!\!\!\!&\!\!\!\!\!\!&\!\!\!1\!\!\!&\!\!\!\!\!\!&\!\!\!\!\!\!&\!\!\!\!\!\!&\!\!\!\!\!\!&\!\!\!\!\!\!&\!\!\!\!\!\!&\!\!\!\!\!\!&\!\!\!\!\!\!&\!\!\!\!\!\!&\!\!\!\!\!\!&\!\!\!\!\!\!&\!\!\!\!\!\!&\!\!\!\!\!\!&\!\!\!\!\!\!&\!\!\!\!\!\!&\!\!\!\!\\
\!\!\!&\!\!\!$\frac{3}{2}$\!\!\!&\!\!\!\!\!\!&\!\!\!\!\!\!&\!\!\!\!\!\!&\!\!\!\!\!\!&\!\!\!\!\!\!&\!\!\!\!\!\!&\!\!\!\!\!\!&\!\!\!\!\!\!&\!\!\!\!\!\!&\!\!\!\!\!\!&\!\!\!\!\!\!&\!\!\!\!\!\!&\!\!\!\!\!\!&\!\!\!\!\!\!&\!\!\!1\!\!\!&\!\!\!\!\!\!&\!\!\!\!\!\!&\!\!\!\!\!\!&\!\!\!\!\!\!&\!\!\!\!\!\!&\!\!\!\!\!\!&\!\!\!\!\!\!&\!\!\!\!\!\!&\!\!\!\!\!\!&\!\!\!\!\!\!&\!\!\!\!\!\!&\!\!\!\!&\!\!\!\!\\
\hline
5\!\!\!&\!\!\!0\!\!\!&\!\!\!\!\!\!&\!\!\!\!\!\!&\!\!\!1\!\!\!&\!\!\!\!\!\!&\!\!\!\!\!\!&\!\!\!\!\!\!&\!\!\!1\!\!\!&\!\!\!\!\!\!&\!\!\!1\!\!\!&\!\!\!\!\!\!&\!\!\!2\!\!\!&\!\!\!\!\!\!&\!\!\!2\!\!\!&\!\!\!\!\!\!&\!\!\!2\!\!\!&\!\!\!\!\!\!&\!\!\!1\!\!\!&\!\!\!\!\!\!&\!\!\!\!\!\!&\!\!\!\!\!\!&\!\!\!\!\!\!&\!\!\!\!\!\!&\!\!\!\!\!\!&\!\!\!\!\!\!&\!\!\!\!\!\!&\!\!\!\!\!\!&\!\!\!\!\!\!&\!\!\!\\
\!\!\!&\!\!\!$\frac{1}{2}$\!\!\!&\!\!\!\!\!\!&\!\!\!\!\!\!&\!\!\!\!\!\!&\!\!\!\!\!\!&\!\!\!\!\!\!&\!\!\!1\!\!\!&\!\!\!\!\!\!&\!\!\!1\!\!\!&\!\!\!\!\!\!&\!\!\!2\!\!\!&\!\!\!\!\!\!&\!\!\!2\!\!\!&\!\!\!\!\!\!&\!\!\!3\!\!\!&\!\!\!\!\!\!&\!\!\!2\!\!\!&\!\!\!\!\!\!&\!\!\!1\!\!\!&\!\!\!\!\!\!&\!\!\!\!\!\!&\!\!\!\!\!\!&\!\!\!\!\!\!&\!\!\!\!\!\!&\!\!\!\!\!\!&\!\!\!\!\!\!&\!\!\!\!\!\!&\!\!\!\!&\!\!\!\\
\!\!\!&\!\!\!1\!\!\!&\!\!\!\!\!\!&\!\!\!\!\!\!&\!\!\!\!\!\!&\!\!\!\!\!\!&\!\!\!\!\!\!&\!\!\!\!\!\!&\!\!\!\!\!\!&\!\!\!\!\!\!&\!\!\!1\!\!\!&\!\!\!\!\!\!&\!\!\!1\!\!\!&\!\!\!\!\!\!&\!\!\!2\!\!\!&\!\!\!\!\!\!&\!\!\!2\!\!\!&\!\!\!\!\!\!&\!\!\!2\!\!\!&\!\!\!\!\!\!&\!\!\!1\!\!\!&\!\!\!\!\!\!&\!\!\!\!\!\!&\!\!\!\!\!\!&\!\!\!\!\!\!&\!\!\!\!\!\!&\!\!\!\!\!\!&\!\!\!\!\!\!&\!\!\!\!\!\!&\!\!\!\\
\!\!\!&\!\!\!$\frac{3}{2}$\!\!\!&\!\!\!\!\!\!&\!\!\!\!\!\!&\!\!\!\!\!\!&\!\!\!\!\!\!&\!\!\!\!\!\!&\!\!\!\!\!\!&\!\!\!\!\!\!&\!\!\!\!\!\!&\!\!\!\!\!\!&\!\!\!\!\!\!&\!\!\!\!\!\!&\!\!\!1\!\!\!&\!\!\!\!\!\!&\!\!\!1\!\!\!&\!\!\!\!\!\!&\!\!\!2\!\!\!&\!\!\!\!\!\!&\!\!\!1\!\!\!&\!\!\!\!\!\!&\!\!\!1\!\!\!&\!\!\!\!\!\!&\!\!\!\!\!\!&\!\!\!\!\!\!&\!\!\!\!\!\!&\!\!\!\!\!\!&\!\!\!\!\!\!&\!\!\!\!&\!\!\!\!\\
\!\!\!&\!\!\!2\!\!\!&\!\!\!\!\!\!&\!\!\!\!\!\!&\!\!\!\!\!\!&\!\!\!\!\!\!&\!\!\!\!\!\!&\!\!\!\!\!\!&\!\!\!\!\!\!&\!\!\!\!\!\!&\!\!\!\!\!\!&\!\!\!\!\!\!&\!\!\!\!\!\!&\!\!\!\!\!\!&\!\!\!\!\!\!&\!\!\!\!\!\!&\!\!\!1\!\!\!&\!\!\!\!\!\!&\!\!\!1\!\!\!&\!\!\!\!\!\!&\!\!\!1\!\!\!&\!\!\!\!\!\!&\!\!\!\!\!\!&\!\!\!\!\!\!&\!\!\!\!\!\!&\!\!\!\!\!\!&\!\!\!\!\!\!&\!\!\!\!\!\!&\!\!\!\!\!\!&\!\!\!\!\\
\!\!\!&\!\!\!$\frac{5}{2}$\!\!\!&\!\!\!\!\!\!&\!\!\!\!\!\!&\!\!\!\!\!\!&\!\!\!\!\!\!&\!\!\!\!\!\!&\!\!\!\!\!\!&\!\!\!\!\!\!&\!\!\!\!\!\!&\!\!\!\!\!\!&\!\!\!\!\!\!&\!\!\!\!\!\!&\!\!\!\!\!\!&\!\!\!\!\!\!&\!\!\!\!\!\!&\!\!\!\!\!\!&\!\!\!\!\!\!&\!\!\!\!\!\!&\!\!\!1\!\!\!&\!\!\!\!\!\!&\!\!\!\!\!\!&\!\!\!\!\!\!&\!\!\!\!\!\!&\!\!\!\!\!\!&\!\!\!\!\!\!&\!\!\!\!\!\!&\!\!\!\!\!\!&\!\!\!\!\!&\!\!\!\!\\
\!\!\!&\!\!\!3\!\!\!&\!\!\!\!\!\!&\!\!\!\!\!\!&\!\!\!\!\!\!&\!\!\!\!\!\!&\!\!\!\!\!\!&\!\!\!\!\!\!&\!\!\!\!\!\!&\!\!\!\!\!\!&\!\!\!\!\!\!&\!\!\!\!\!\!&\!\!\!\!\!\!&\!\!\!\!\!\!&\!\!\!\!\!\!&\!\!\!\!\!\!&\!\!\!\!\!\!&\!\!\!\!\!\!&\!\!\!\!\!\!&\!\!\!\!\!\!&\!\!\!\!\!\!&\!\!\!\!\!\!&\!\!\!1\!\!\!&\!\!\!\!\!\!&\!\!\!\!\!\!&\!\!\!\!\!\!&\!\!\!\!\!\!&\!\!\!\!\!\!&\!\!\!\!\!\!&\!\!\!\!\\
\hline
6\!\!\!&\!\!\!0\!\!\!&\!\!\!\!\!\!&\!\!\!1\!\!\!&\!\!\!\!\!\!&\!\!\!1\!\!\!&\!\!\!\!\!\!&\!\!\!3\!\!\!&\!\!\!\!\!\!&\!\!\!2\!\!\!&\!\!\!\!\!\!&\!\!\!6\!\!\!&\!\!\!\!\!\!&\!\!\!4\!\!\!&\!\!\!\!\!\!&\!\!\!8\!\!\!&\!\!\!\!\!\!&\!\!\!5\!\!\!&\!\!\!\!\!\!&\!\!\!7\!\!\!&\!\!\!\!\!\!&\!\!\!2\!\!\!&\!\!\!\!\!\!&\!\!\!2\!\!\!&\!\!\!\!\!\!&\!\!\!\!\!\!&\!\!\!\!\!\!&\!\!\!\!\!\!&\!\!\!\!&\!\!\!\!\\
\!\!\!&\!\!\!$\frac{1}{2}$\!\!\!&\!\!\!\!\!\!&\!\!\!\!\!\!&\!\!\!1\!\!\!&\!\!\!\!\!\!&\!\!\!2\!\!\!&\!\!\!\!\!\!&\!\!\!3\!\!\!&\!\!\!\!\!\!&\!\!\!5\!\!\!&\!\!\!\!\!\!&\!\!\!6\!\!\!&\!\!\!\!\!\!&\!\!\!9\!\!\!&\!\!\!\!\!\!&\!\!\!9\!\!\!&\!\!\!\!\!\!&\!\!\!10\!\!\!&\!\!\!\!\!\!&\!\!\!7\!\!\!&\!\!\!\!\!\!&\!\!\!5\!\!\!&\!\!\!\!\!\!&\!\!\!1\!\!\!&\!\!\!\!\!\!&\!\!\!1\!\!\!&\!\!\!\!\!\!&\!\!\!\!&\!\!\!\!\!\\
\!\!\!&\!\!\!1\!\!\!&\!\!\!\!\!\!&\!\!\!\!\!\!&\!\!\!\!\!\!&\!\!\!1\!\!\!&\!\!\!\!\!\!&\!\!\!1\!\!\!&\!\!\!\!\!\!&\!\!\!3\!\!\!&\!\!\!\!\!\!&\!\!\!3\!\!\!&\!\!\!\!\!\!&\!\!\!7\!\!\!&\!\!\!\!\!\!&\!\!\!7\!\!\!&\!\!\!\!\!\!&\!\!\!11\!\!\!&\!\!\!\!\!\!&\!\!\!9\!\!\!&\!\!\!\!\!\!&\!\!\!9\!\!\!&\!\!\!\!\!\!&\!\!\!4\!\!\!&\!\!\!\!\!\!&\!\!\!2\!\!\!&\!\!\!\!\!\!&\!\!\!\!\!\!&\!\!\!\!\!\!&\!\!\!\!\!\\
\!\!\!&\!\!\!$\frac{3}{2}$\!\!\!&\!\!\!\!\!\!&\!\!\!\!\!\!&\!\!\!\!\!\!&\!\!\!\!\!\!&\!\!\!\!\!\!&\!\!\!\!\!\!&\!\!\!1\!\!\!&\!\!\!\!\!\!&\!\!\!1\!\!\!&\!\!\!\!\!\!&\!\!\!3\!\!\!&\!\!\!\!\!\!&\!\!\!4\!\!\!&\!\!\!\!\!\!&\!\!\!7\!\!\!&\!\!\!\!\!\!&\!\!\!7\!\!\!&\!\!\!\!\!\!&\!\!\!10\!\!\!&\!\!\!\!\!\!&\!\!\!6\!\!\!&\!\!\!\!\!\!&\!\!\!4\!\!\!&\!\!\!\!\!\!&\!\!\!\!\!\!&\!\!\!\!\!\!&\!\!\!\!\!&\!\!\!\!\!\!\\
\!\!\!&\!\!\!2\!\!\!&\!\!\!\!\!\!&\!\!\!\!\!\!&\!\!\!\!\!\!&\!\!\!\!\!\!&\!\!\!\!\!\!&\!\!\!\!\!\!&\!\!\!\!\!\!&\!\!\!\!\!\!&\!\!\!\!\!\!&\!\!\!1\!\!\!&\!\!\!\!\!\!&\!\!\!1\!\!\!&\!\!\!\!\!\!&\!\!\!3\!\!\!&\!\!\!\!\!\!&\!\!\!4\!\!\!&\!\!\!\!\!\!&\!\!\!7\!\!\!&\!\!\!\!\!\!&\!\!\!6\!\!\!&\!\!\!\!\!\!&\!\!\!6\!\!\!&\!\!\!\!\!\!&\!\!\!2\!\!\!&\!\!\!\!\!\!&\!\!\!1\!\!\!&\!\!\!\!\!\!&\!\!\!\!\\
\!\!\!&\!\!\!$\frac{5}{2}$\!\!\!&\!\!\!\!\!\!&\!\!\!\!\!\!&\!\!\!\!\!\!&\!\!\!\!\!\!&\!\!\!\!\!\!&\!\!\!\!\!\!&\!\!\!\!\!\!&\!\!\!\!\!\!&\!\!\!\!\!\!&\!\!\!\!\!\!&\!\!\!\!\!\!&\!\!\!\!\!\!&\!\!\!1\!\!\!&\!\!\!\!\!\!&\!\!\!1\!\!\!&\!\!\!\!\!\!&\!\!\!3\!\!\!&\!\!\!\!\!\!&\!\!\!3\!\!\!&\!\!\!\!\!\!&\!\!\!5\!\!\!&\!\!\!\!\!\!&\!\!\!3\!\!\!&\!\!\!\!\!\!&\!\!\!2\!\!\!&\!\!\!\!\!\!&\!\!\!\!&\!\!\!\!\!\\
\!\!\!&\!\!\!3\!\!\!&\!\!\!\!\!\!&\!\!\!\!\!\!&\!\!\!\!\!\!&\!\!\!\!\!\!&\!\!\!\!\!\!&\!\!\!\!\!\!&\!\!\!\!\!\!&\!\!\!\!\!\!&\!\!\!\!\!\!&\!\!\!\!\!\!&\!\!\!\!\!\!&\!\!\!\!\!\!&\!\!\!\!\!\!&\!\!\!\!\!\!&\!\!\!\!\!\!&\!\!\!1\!\!\!&\!\!\!\!\!\!&\!\!\!1\!\!\!&\!\!\!\!\!\!&\!\!\!3\!\!\!&\!\!\!\!\!\!&\!\!\!3\!\!\!&\!\!\!\!\!\!&\!\!\!3\!\!\!&\!\!\!\!\!\!&\!\!\!1\!\!\!&\!\!\!\!\!\!&\!\!\!\\
\!\!\!&\!\!\!$\frac{7}{2}$\!\!\!&\!\!\!\!\!\!&\!\!\!\!\!\!&\!\!\!\!\!\!&\!\!\!\!\!\!&\!\!\!\!\!\!&\!\!\!\!\!\!&\!\!\!\!\!\!&\!\!\!\!\!\!&\!\!\!\!\!\!&\!\!\!\!\!\!&\!\!\!\!\!\!&\!\!\!\!\!\!&\!\!\!\!\!\!&\!\!\!\!\!\!&\!\!\!\!\!\!&\!\!\!\!\!\!&\!\!\!\!\!\!&\!\!\!\!\!\!&\!\!\!1\!\!\!&\!\!\!\!\!\!&\!\!\!1\!\!\!&\!\!\!\!\!\!&\!\!\!2\!\!\!&\!\!\!\!\!\!&\!\!\!1\!\!\!&\!\!\!\!\!\!&\!\!\!1\!\!\!&\!\!\!\\
\!\!\!&\!\!\!\!4\!\!&\!\!\!\!\!\!&\!\!\!\!\!\!&\!\!\!\!\!\!&\!\!\!\!\!\!&\!\!\!\!\!\!&\!\!\!\!\!\!&\!\!\!\!\!\!&\!\!\!\!\!\!&\!\!\!\!\!\!&\!\!\!\!\!\!&\!\!\!\!\!\!&\!\!\!\!\!\!&\!\!\!\!\!\!&\!\!\!\!\!\!&\!\!\!\!\!\!&\!\!\!\!\!\!&\!\!\!\!\!\!&\!\!\!\!\!\!&\!\!\!\!\!\!&\!\!\!\!\!\!&\!\!\!\!\!\!&\!\!\!1\!\!\!&\!\!\!\!\!\!&\!\!\!1\!\!\!&\!\!\!\!\!\!&\!\!\!1\!\!\!&\!\!\!\!\!\!&\!\!\!\!\!\!\\
\!\!\!&\!\!\!$\frac{9}{2}$\!\!\!&\!\!\!\!\!\!&\!\!\!\!\!\!&\!\!\!\!\!\!&\!\!\!\!\!\!&\!\!\!\!\!\!&\!\!\!\!\!\!&\!\!\!\!\!\!&\!\!\!\!\!\!&\!\!\!\!\!\!&\!\!\!\!\!\!&\!\!\!\!\!\!&\!\!\!\!\!\!&\!\!\!\!\!\!&\!\!\!\!\!\!&\!\!\!\!\!\!&\!\!\!\!\!\!&\!\!\!\!\!\!&\!\!\!\!\!\!&\!\!\!\!\!\!&\!\!\!\!\!\!&\!\!\!\!\!\!&\!\!\!\!\!\!&\!\!\!\!\!\!&\!\!\!\!\!\!&\!\!\!1\!\!\!&\!\!\!\!\!\!&\!\!\!\!\!\!\!\!&\!\!\!\!\\
\!\!\!&\!\!\!5\!\!\!&\!\!\!\!\!\!&\!\!\!\!\!\!&\!\!\!\!\!\!&\!\!\!\!\!\!&\!\!\!\!\!\!&\!\!\!\!\!\!&\!\!\!\!\!\!&\!\!\!\!\!\!&\!\!\!\!\!\!&\!\!\!\!\!\!&\!\!\!\!\!\!&\!\!\!\!\!\!&\!\!\!\!\!\!&\!\!\!\!\!\!&\!\!\!\!\!\!&\!\!\!\!\!\!&\!\!\!\!\!\!&\!\!\!\!\!\!&\!\!\!\!\!\!&\!\!\!\!\!\!&\!\!\!\!\!\!&\!\!\!\!\!\!&\!\!\!\!\!\!&\!\!\!\!\!\!&\!\!\!\!\!\!&\!\!\!\!\!\!&\!\!\!\!\!\!&\!\!\!1\!\!\!\\
\hline
d\!&\!\!$j_L\slash j_R$\!&\!\!0\!&\!\!\!$\frac{1}{2}$\!\!\!&\!\!\!1\!\!\!&\!\!\!$\frac{3}{2}$\!\!\!&\!\!\!2\!\!\!&\!\!\!$\frac{5}{2}$\!\!\!&\!\!\!3\!\!\!&\!\!\!$\frac{7}{2}$\!\!\!&\!\!\!4\!\!\!&
\!\!\!$\frac{9}{2}$\!\!\!&\!\!\!5\!\!\!&\!\!\!$\frac{11}{2}$\!\!\!&\!\!\!6\!\!\!&\!\!\!$\frac{13}{2}$\!\!\!&\!\!\!7\!\!\!&\!\!\!$\frac{15}{2}$\!\!\!&\!\!\!8\!\!\!&\!\!\!$\frac{17}{2}$\!\!\!
&\!\!\!9\!\!\!&\!\!\!$\frac{19}{2}$\!\!\!&\!\!\!10\!\!\!&\!\!\!$\frac{21}{2}$\!\!\!&\!\!\!11\!\!\!&\!\!\!$\frac{23}{2}$\!\!\!&\!\!\!12\!\!\!&\!\!\!$\frac{25}{2}$\!\!\!&\!\!\!13\!\!\!
&\!\!\!$\frac{27}{2}$\!\!\\
\hline
\end{tabular}}}
\caption{Non-vanishing BPS numbers $N^d_{j_L,j_R}$ of local ${\cal O}(-3)\rightarrow \mathbb{P}^2$ up to $d=7$.}
\label{bpstable}
\end{table}

\subsection{ ${\cal O}(-K_{F_0}) \rightarrow F_0$}

With the two-parameter model given by the polyhedron 2 
\begin{figure}[h!] 
\begin{center} 
\includegraphics[angle=0,width=.2\textwidth]{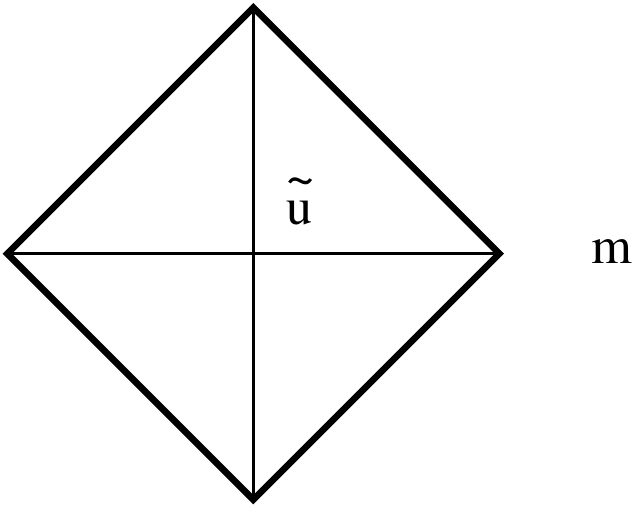}
\begin{quote} 
\caption{The polyhedron 2 with the choice of the mass parameter $m$  
and the modulus $\tilde{u}$. 
\vspace{-1.2cm}} \label{poly10} \end{quote} 
\end{center} 
\end{figure} 
we discuss two perspectives of getting the mirror and perfoming the calculation of the BPS 
numbers. The first starts with the  Mori cone vectors, which correspond to the 
depicted triangulation
\begin{equation} 
 \label{dataf1} 
 \begin{array}{ccrrr|rrl|} 
    \multicolumn{5}{c}{\nu_i }    &l^{(1)}& l^{(2)}&\\ 
    D_u    &&     1&     0&   0&         -2&  -2&       \\ 
    D_1    &&     1&     1&   0&         1&   0&        \\ 
    D_2    &&     1&     0&   1&         0&   1&        \\ 
    D_3    &&     1&    -1&   0&         1&   0&        \\ 
    D_4    &&     1&     0&   -1&        0&   1&       \\  
  \end{array} \ . 
\end{equation} 
Following (\ref{mirrorcurve}) and eliminating coordinates by 
(\ref{modmirrorrelations}) and the $\mathbb{C}^*$-action on 
the $Y_i$ we write the mirror curve in the remaining coordinates 
$x,y$ as 
\begin{equation} \label{curve}
H(x,y)=1+x + \frac{z_1}{x} +y + \frac{z_2}{y}=0\ , 
\end{equation}
where the $z_i$ are defined as in (\ref{batyrevcoordinates}).
The Picard Fuchs equations (\ref{generalPFequations}) become in the case at hand  with $\theta_i:=z_i \frac{d}{d z_i}$   
\begin{equation} 
 \begin{array}{rl} 
{\cal L}^{(1)}=& \theta_1^2-2(\theta_1 +\theta_2-1) (2 \theta_1+2 \theta_2-1)z_1\\ 
{\cal L}^{(2)}=& \theta_2^2-2(\theta_1 +\theta_2-1) (2 \theta_1+2 \theta_2-1)z_2 \ .
\end{array}
\label{PFp1p1} 
\end{equation}
Let us come to the discussion of the mass parameter. To make contact with  the latter  
one finds that at $z_i=0$ one has a constant solution and two 
solutions, which are  linear in $\log(z_i)$ and one solution, 
which is quadratic in $\log(z_i)$. As the two linear logarithmic 
solutions one finds $t_1=\log(z_1)+ \Sigma(z_1,z_2)$ and  $t_2=\log(z_2)+ 
\Sigma(z_1,z_2)$ determining the K\"ahler parameters of the 
$\mathbb{P}^1$'s. Here $\Sigma(z_1,z_2)$ is the same holomorphic 
transcendental function. This suggests to change variables and 
introduce $z=z_1$ and $M=\log(z_1)-\log(z_2)$. The latter is a 
trivial solution. We now consider the differential 
left ideal generated by (\ref{PFp1p1}) up to homogeneous degree 
three in differentations w.r.t. $z$ and $M$. In this ideal one can eliminate all 
differential operators involving derivatives w.r.t. $M$ and 
end up with a  third order differential operator in $z$ 
determining all non-trivial solutions of (\ref{PFp1p1})
\begin{equation} 
\begin{array}{rl} 
{\cal L}=&(60 \left(m-1\right)^2 z^2-18 \left(m+1\right) z+1)\partial_z+
z \left(80 (m-1)^2 z^2-32 (m+1) z+3\right)\partial_z^2 +\\  
         &z^2 \left(16 (m-1)^2 z^2-8 (m+1) z+1\right)\partial_z^3\, . 
\label{PFEp1p1}
\end{array}
\end{equation}
Here we understand $m=e^M$ now as a deformation parameter.
Setting $m=1$ imposes an 
identification of the complexified K\"ahler parameters 
$t_1=t_2$ globally in the quantum moduli space. This 
leads to the diagonal model with $S=\mathbb{P}^1\times \mathbb{P}^1$ 
as base, discussed in section (\ref{diagonalF0}). In particular 
(\ref{PFEp1p1}) restricts for $m=1$ to (\ref{PF2.4}) with the 
appropriate  parameters $c_0=16$ and $a_1=\frac{1}{2}, 
a_2=\frac{1}{2}$. 

After changes of variables we can parametrize the curve 
(\ref{curve}) as
\begin{equation}
y^2+x^2-y-\frac{xy}{\sqrt{z}}-m x^2 y=0
\end{equation}
and bring it into Weierstrass form (\ref{weierstrass}) using 
Nagell's algorithm.

Instead of going over the $l^{(k)}$ vectors the elliptic mirror 
curve is simply associated to the reflexive polyhedron as its 
Newton polynom, i.e. the coordinates of the points determine its 
positive exponents.  In Appendix A we provide 
the Newton polynom of the biggest polyhedra so that 
all polyhedra can be embedded in at least one of them and provide the Weierstrass 
form for them. In this approach it is only necessary to 
specialize the  general Weierstrass forms and to eventually 
rescale the $g_i\rightarrow g_i\lambda^i$ to ensure that the 
closed string period (\ref{nonlogperiod}) has the right 
leading behaviour. Note that according to the $l$ vectors 
the right choice of large complex structure  coordinates we 
get from (\ref{batyrevcoordinates}) is
\begin{equation} 
z_1=\frac{m}{\tilde{u}^2} \quad z_2= \frac{1}{\tilde{u}^2} 
\end{equation}
so that it is immediatly clear that $z_1/z_2=m$ and 
$\tilde u\rightarrow \infty$ is the large radius point. 

In the $\mathbb{P}^1\times \mathbb{P}^1$ case we can use the polytop 
for the cubic or the bi-quartic, the choice does not matter. Let us 
re-define $u=\frac{1}{\tilde u^2}=z_2$. Then we get   
\begin{equation}
\begin{array}{rl}
g_2=& 27 u^4 \left(16 u^4 \left(m^2-m+1\right)-8 u^2 (m+1)+1\right),\\ [2 mm]
g_3=&  -27 u^6 (-1 + 12 (1 + m) u^2 - 24 (2 + m + 2 m^2) u^4 + 32 (2 - 3 m - 3 m^2 + 2 m^3) u^6)\ .
\end{array}
\end{equation}
This yields a $j$-invariant
\begin{equation}
j=\frac{\left(16 \left(m^2-m+1\right) \tilde{u}^2-8 (m+1) \tilde{u}+1\right)^3}{m^2 \tilde{u}^4 \left(16 (m-1)^2 \tilde{u}^2-8 (m+1) \tilde{u}+1\right)}\ .
\label{jfunction}
\end{equation}
At the large radius we require $t(u, {\underline{m}})=\log(u)+{\cal O}(u,{\underline{m}})$ 
and near the single zeros of $\Delta$, $t_c(u,{\underline{m}})= z_c(u,{\underline{m}})+{\cal O}( z_c^2(u,{\underline{m}}))$, 
which fixes the scaling (\ref{giscaling}). We have calculated $F^{(n,g)}$ at the conifold to impose the gap condition. Other 
interesting limits are the Seiberg-Witten limit 
\be 
z_1 \rightarrow  \frac{1}{4} \exp(- 4 \epsilon^2  u),\quad z_2 \rightarrow  \epsilon^4 \Lambda^4 
\label{sw1}
\ee
in which 
\begin{equation}
j= \frac{(3 \Lambda^4 - 4 u^2)^3}{27 \Lambda^8 (\Lambda^4 - u^2)} \ .
\label{jsw}
\end{equation} 
becomes the $j$-function of the massless $SU(2)$ Seiberg-Witten 
curve compare (\ref{SWcurvesnf<3}) and the Chern-Simons limit 
discussed for the refined case in~\cite{CKK}.

We define a single-valued variable near the large radius as 
\be
Q_t=e^{t}=u+ {\cal O}(u^2,{\underline{m}})\ ,
\label{reftf}
\ee
which is easily inverted to $u(Q_f)$. From K\"ahler parameters of 
the two $\mathbb{P}^1$'s we define $Q_i=e^{t_i}$ 
and get the relation $Q_t=Q_2$ and $m=Q_1/Q_2$, which allows us to obtain  
for all expressions defined in section \ref{ellipticdirectintegration}  
the large radius expansion in terms of $Q_i$.  The coefficients in 
(\ref{genus1a},\ref{genus1b}) are  given by  $a=7,b=\frac{7}{2},c=-2$ and $d=-1$.

We have calculated the spin invariants and found the following series   
\begin{equation}
 N^{(1,d)}_{j_L,j_R}=\begin{cases}
1 & {\rm if  }\ \ j_L=0, j_R=\frac{1}{2} + d \\
0 & {\rm otherwise}
\end{cases}
\end{equation}
Up to  $d_1+d_2\le 7$ the refined invariants are reported in Table~\ref{bpstablep1p1}.
\begin{table}[h!]
\centering{{
\begin{tabular}[H]{|c|c|cccccccccccccccccccc|}
\hline
$(d_1,d_2)$\!&\!\!$j_L\backslash j_R$\!&\!\!0\!&\!\!\!$\frac{1}{2}$\!\!\!&\!\!\!1\!\!\!&\!\!\!$\frac{3}{2}$\!\!\!&\!\!\!2\!\!\!&\!\!\!$\frac{5}{2}$\!\!\!&\!\!\!3\!\!\!&\!\!\!$\frac{7}{2}$\!\!\!&\!\!\!4\!\!\!&
\!\!\!$\frac{9}{2}$\!\!\!&\!\!\!5\!\!\!&\!\!\!$\frac{11}{2}$\!\!\!&\!\!\!6\!\!\!&\!\!\!$\frac{13}{2}$\!\!\!&\!\!\!7\!\!\!&\!\!\!$\frac{15}{2}$\!\!\!&\!\!\!8\!\!\!&\!\!\!$\frac{17}{2}$\!\!\!
&\!\!\!9\!\!\!&\!\!\!$\frac{19}{2}$\!\!\! \\
\hline
$(2,2)$\!&\!\!0\!&\!\! \!&\!\!\!\!\!\!&\!\!\!\!\!\!&\!\!\!\!\!\!&\!\!\!\!\!\!&\!\!\!1\!\!\!&\!\!\!\!\!\!&\!\!\!1\!\!\!&\!\!\!\!\!\!&
\!\!\!\!\!\!&\!\!\!\!\!\!&\!\!\!\!\!\!&\!\!\!\!\!\!&\!\!\!\!\!\!&\!\!\!\!\!\!&\!\!\!\!\!\!&\!\!\!\!\!\!&\!\!\!\!\!\!
&\!\!\!\!\!\!&\!\!\!\!\!\! \\
\!&\!\!$\frac{1}{2}$\!&\!\! \!&\!\!\!\!\!\!&\!\!\!\!\!\!&\!\!\!\!\!\!&\!\!\!\!\!\!&\!\!\!\!\!\!&\!\!\!\!\!\!&\!\!\!\!\!\!&\!\!\!1\!\!\!&
\!\!\!\!\!\!&\!\!\!\!\!\!&\!\!\!\!\!\!&\!\!\!\!\!\!&\!\!\!\!\!\!&\!\!\!\!\!\!&\!\!\!\!\!\!&\!\!\!\!\!\!&\!\!\!\!\!\!
&\!\!\!\!\!\!&\!\!\!\!\!\! \\
\hline
\hline
$(2,3)$\!&\!\!0\!&\!\! \!&\!\!\!\!\!\!&\!\!\!\!\!\!&\!\!\!\!\!\!&\!\!\!\!\!\!&\!\!\!1\!\!\!&\!\!\!\!\!\!&\!\!\!1\!\!\!&\!\!\!\!\!\!&
\!\!\!2\!\!\!&\!\!\!\!\!\!&\!\!\!\!\!\!&\!\!\!\!\!\!&\!\!\!\!\!\!&\!\!\!\!\!\!&\!\!\!\!\!\!&\!\!\!\!\!\!&\!\!\!\!\!\!
&\!\!\!\!\!\!&\!\!\!\!\!\! \\
\!&\!\!$\frac{1}{2}$\!&\!\! \!&\!\!\!\!\!\!&\!\!\!\!\!\!&\!\!\!\!\!\!&\!\!\!\!\!\!&\!\!\!\!\!\!&\!\!\!\!\!\!&\!\!\!\!\!\!&\!\!\!1\!\!\!&
\!\!\!\!\!\!&\!\!\!1\!\!\!&\!\!\!\!\!\!&\!\!\!\!\!\!&\!\!\!\!\!\!&\!\!\!\!\!\!&\!\!\!\!\!\!&\!\!\!\!\!\!&\!\!\!\!\!\!
&\!\!\!\!\!\!&\!\!\!\!\!\! \\
\!&\!\!$1$\!&\!\! \!&\!\!\!\!\!\!&\!\!\!\!\!\!&\!\!\!\!\!\!&\!\!\!\!\!\!&\!\!\!\!\!\!&\!\!\!\!\!\!&\!\!\!\!\!\!&\!\!\!\!\!\!&
\!\!\!\!\!\!&\!\!\!\!\!\!&\!\!\!1\!\!\!&\!\!\!\!\!\!&\!\!\!\!\!\!&\!\!\!\!\!\!&\!\!\!\!\!\!&\!\!\!\!\!\!&\!\!\!\!\!\!
&\!\!\!\!\!\!&\!\!\!\!\!\! \\
\hline
\hline
$(2,4)$\!&\!\!0\!&\!\!\! \!&\!\!\!\!\!\!&\!\!\!\!\!\!&\!\!\!\!\!\!&\!\!\!\!\!\!&\!\!\!1\!\!\!&\!\!\!\!\!\!&\!\!\!1\!\!\!&\!\!\!\!\!\!
&\!\!\!2\!\!\!&\!\!\!\!\!\!&\!\!\!2\!\!\!&\!\!\!\!\!\!&\!\!\!\!\!\!&\!\!\!\!\!\!&\!\!\!\!\!\!&\!\!\!\!\!\!&\!\!\!\!\!\!
&\!\!\!\!\!\!&\!\!\!\!\!\! \\                            
\!&\!\!$\frac{1}{2}$\!&\!\! \!&\!\!\!\!\!\!&\!\!\!\!\!\!&\!\!\!\!\!\!&\!\!\!\!\!\!&\!\!\!\!\!\!&\!\!\!\!\!\!&\!\!\!\!\!\!&\!\!\!1\!\!\!&\!\!\!\!\!\!&\!\!\!1\!\!\!&\!\!\!\!\!\!
&\!\!\!2\!\!\!&\!\!\!\!\!\!&\!\!\!\!\!\!&\!\!\!\!\!\!&\!\!\!\!\!\!&\!\!\!\!\!\!
&\!\!\!\!\!\!&\!\!\!\!\!\! \\
\!&\!\!$1$\!&\!\! \!&\!\!\!\!\!\!&\!\!\!\!\!\!&\!\!\!\!\!\!&\!\!\!\!\!\!&\!\!\!\!\!\!&\!\!\!\!\!\!&\!\!\!\!\!\!&\!\!\!\!\!\!&\!\!\!\!\!\!&\!\!\!\!\!\!&\!\!\!1\!\!\!&\!\!\!\!\!\!&\!\!\!1\!\!\!&\!\!\!\!\!\!&\!\!\!\!\!\!&\!\!\!\!\!\!&\!\!\!\!\!\!
&\!\!\!\!\!\!&\!\!\!\!\!\! \\
\!&\!\!$\frac{3}{2}$\!&\!\! \!&\!\!\!\!\!\!&\!\!\!\!\!\!&\!\!\!\!\!\!&\!\!\!\!\!\!&\!\!\!\!\!\!&\!\!\!\!\!\!&\!\!\!\!\!\!&\!\!\!\!\!\!&\!\!\!\!\!\!&\!\!\!\!\!\!&\!\!\!\!\!\!&\!\!\!\!\!\!&\!\!\!\!\!\!&\!\!\!1\!\!\!&\!\!\!\!\!\!&\!\!\!\!\!\!&\!\!\!\!\!\!
&\!\!\!\!\!\!&\!\!\!\!\!\! \\
\hline
\hline
$(2,5)$\!&\!\!0\!&\!\! \!&\!\!\!\!\!\!&\!\!\!\!\!\!&\!\!\!\!\!\!&\!\!\!\!\!\!&\!\!\!1\!\!\!&\!\!\!\!\!\!&\!\!\!1\!\!\!&\!\!\!\!\!\!\!\!&\!\!\!2\!\!\!&\!\!\!\!\!\!&\!\!\!2\!\!\!&\!\!\!\!\!\!&\!\!\!3\!\!\!&\!\!\!\!\!&\!\!\!\!\!\!&\!\!\!\!\!\!
&\!\!\!\!\!\!&\!\!\!\!\!\!&\!\!\!\!\!\! \\
\!&\!\!$\frac{1}{2}$\!&\!\! \!&\!\!\!\!\!\!&\!\!\!\!\!\!&\!\!\!\!\!\!&\!\!\!\!\!\!&\!\!\!\!\!\!&\!\!\!\!\!\!&\!\!\!\!\!\!&\!\!\!1\!\!\!&\!\!\!\!\!\!&\!\!\!1\!\!\!&\!\!\!\!\!\!&\!\!\!2\!\!\!&\!\!\!\!\!\!&\!\!\!2\!\!\!&\!\!\!\!\!\!&\!\!\!\!\!\!&\!\!\!\!\!\!
&\!\!\!\!\!\!&\!\!\!\!\!\! \\
\!&\!\!$1$\!&\!\! \!&\!\!\!\!\!\!&\!\!\!\!\!\!&\!\!\!\!\!\!&\!\!\!\!\!\!&\!\!\!\!\!\!&\!\!\!\!\!\!&\!\!\!\!\!\!&\!\!\!\!\!\!&\!\!\!\!\!\!&\!\!\!\!\!\!&\!\!\!1\!\!\!&\!\!\!\!\!\!&\!\!\!1\!\!\!&\!\!\!\!\!\!&\!\!\!2\!\!\!&\!\!\!\!\!\!&\!\!\!\!\!\!
&\!\!\!\!\!\!&\!\!\!\!\!\! \\
\!&\!\!$\frac{3}{2}$\!&\!\! \!&\!\!\!\!\!\!&\!\!\!\!\!\!&\!\!\!\!\!\!&\!\!\!\!\!\!&\!\!\!\!\!\!&\!\!\!\!\!\!&\!\!\!\!\!\!&\!\!\!\!\!\!&\!\!\!\!\!\!&\!\!\!\!\!\!&\!\!\!\!\!\!&\!\!\!\!\!\!&\!\!\!\!\!\!&\!\!\!1\!\!\!&\!\!\!\!\!\!&\!\!\!1\!\!\!&\!\!\!\!\!\!
&\!\!\!\!\!\!&\!\!\!\!\!\! \\
\!&\!\!$2$\!&\!\! \!&\!\!\!\!\!\!&\!\!\!\!\!\!&\!\!\!\!\!\!&\!\!\!\!\!\!&\!\!\!\!\!\!&\!\!\!\!\!\!&\!\!\!\!\!\!&\!\!\!\!\!\!&\!\!\!\!\!\!&\!\!\!\!\!\!&\!\!\!\!\!\!&\!\!\!\!\!\!&\!\!\!\!\!\!&\!\!\!\!\!\!&\!\!\!\!\!\!&\!\!\!\!\!\!&\!\!\!1\!\!\!
&\!\!\!\!\!\!&\!\!\!\!\!\! \\
\hline
\hline
$(3,3)$\!&\!\!0\!&\!\! \!&\!\!\!\!\!\!&\!\!\!\!\!\!&\!\!\!1\!\!\!&\!\!\!\!\!\!&\!\!\!1\!\!\!&\!\!\!\!\!\!&\!\!\!3\!\!\!&\!\!\!\!\!\!&\!\!\!3\!\!\!&\!\!\!\!\!\!&\!\!\!4\!\!\!&\!\!\!\!\!\!&\!\!\!\!\!\!&\!\!\!\!\!\!&\!\!\!\!\!\!&\!\!\!\!\!\!&\!\!\!\!\!\!
&\!\!\!\!\!\!&\!\!\!\!\!\! \\
\!&\!\!$\frac{1}{2}$\!&\!\! \!&\!\!\!\!\!\!&\!\!\!\!\!\!&\!\!\!\!\!\!&\!\!\!\!\!\!&\!\!\!\!\!\!&\!\!\!1\!\!\!&\!\!\!\!\!\!&\!\!\!2\!\!\!&
\!\!\!\!\!\!&\!\!\!3\!\!\!&\!\!\!\!\!\!&\!\!\!3\!\!\!&\!\!\!\!\!\!&\!\!\!\!1\!\!&\!\!\!\!\!\!&\!\!\!\!\!\!&\!\!\!\!\!\!
&\!\!\!\!\!\!&\!\!\!\!\!\! \\
\!&\!\!$1$\!&\!\! \!&\!\!\!\!\!\!&\!\!\!\!\!\!&\!\!\!\!\!\!&\!\!\!\!\!\!&\!\!\!\!\!\!&\!\!\!\!\!\!&\!\!\!\!\!\!&\!\!\!\!\!\!&
\!\!\!1\!\!\!&\!\!\!\!\!\!&\!\!\!2\!\!\!&\!\!\!\!\!\!&\!\!\!3\!\!\!&\!\!\!\!\!\!&\!\!\!\!\!\!&\!\!\!\!\!\!&\!\!\!\!\!\!
&\!\!\!\!\!\!&\!\!\!\!\!\! \\
\!&\!\!$\frac{3}{2}$\!&\!\! \!&\!\!\!\!\!\!&\!\!\!\!\!\!&\!\!\!\!\!\!&\!\!\!\!\!\!&\!\!\!\!\!\!&\!\!\!\!\!\!&\!\!\!\!\!\!&\!\!\!\!\!\!&
\!\!\!\!\!\!&\!\!\!\!\!\!&\!\!\!\!\!\!&\!\!\!1\!\!\!&\!\!\!\!\!\!&\!\!\!1\!\!\!&\!\!\!\!\!\!&\!\!\!\!\!\!&\!\!\!\!\!\!
&\!\!\!\!\!\!&\!\!\!\!\!\! \\
\!&\!\!$2$\!&\!\! \!&\!\!\!\!\!\!&\!\!\!\!\!\!&\!\!\!\!\!\!&\!\!\!\!\!\!&\!\!\!\!\!\!&\!\!\!\!\!\!&\!\!\!\!\!\!&\!\!\!\!\!\!&
\!\!\!\!\!\!&\!\!\!\!\!\!&\!\!\!\!\!\!&\!\!\!\!\!\!&\!\!\!\!\!\!&\!\!\!\!\!\!&\!\!\!1\!\!\!&\!\!\!\!\!\!&\!\!\!\!\!\!
&\!\!\!\!\!\!&\!\!\!\!\!\! \\
\hline
\hline
$(3,4)$\!&\!\!0\!&\!\! \!&\!\!\!1\!\!\!&\!\!\!\!\!\!&\!\!\!1\!\!\!&\!\!\!\!\!\!&\!\!\!3\!\!\!&\!\!\!\!\!\!&\!\!\!4\!\!\!&\!\!\!\!\!\!&
\!\!\!7\!\!\!&\!\!\!\!\!\!&\!\!\!6\!\!\!&\!\!\!\!\!\!&\!\!\!7\!\!\!&\!\!\!\!\!\!&\!\!\!1\!\!\!&\!\!\!\!\!\!&\!\!\!1\!\!\!
&\!\!\!\!\!\!&\!\!\!\!\!\! \\
\!&\!\!$\frac{1}{2}$\!&\!\! \!&\!\!\!\!\!\!&\!\!\!\!\!\!&\!\!\!\!\!\!&\!\!\!\!\!\!&\!\!\!\!\!\!&\!\!\!1\!\!\!&\!\!\!\!\!\!&\!\!\!2\!\!\!&
\!\!\!\!\!\!&\!\!\!4\!\!\!&\!\!\!\!\!\!&\!\!\!6\!\!\!&\!\!\!\!\!\!&\!\!\!\!8\!\!&\!\!\!\!\!\!&\!\!\!2\!\!\!&\!\!\!\!\!\!
&\!\!\!\!\!\!&\!\!\!\!\!\! \\
\!&\!\!$1$\!&\!\! \!&\!\!\!\!\!\!&\!\!\!\!\!\!&\!\!\!\!\!\!&\!\!\!\!\!\!&\!\!\!\!\!\!&\!\!\!\!\!\!&\!\!\!1\!\!\!&\!\!\!\!\!\!&
\!\!\!2\!\!\!&\!\!\!\!\!\!&\!\!\!5\!\!\!&\!\!\!\!\!\!&\!\!\!6\!\!\!&\!\!\!\!\!\!&\!\!\!7\!\!\!&\!\!\!\!\!\!&\!\!\!1\!\!\!
&\!\!\!\!\!\!&\!\!\!\!\!\! \\
\!&\!\!$\frac{3}{2}$\!&\!\! \!&\!\!\!\!\!\!&\!\!\!\!\!\!&\!\!\!\!\!\!&\!\!\!\!\!\!&\!\!\!\!\!\!&\!\!\!\!\!\!&\!\!\!\!\!\!&\!\!\!\!\!\!&
\!\!\!\!\!\!&\!\!\!\!\!\!&\!\!\!\!\!\!&\!\!\!1\!\!\!&\!\!\!\!\!\!&\!\!\!2\!\!\!&\!\!\!\!\!\!&\!\!\!4\!\!\!&\!\!\!\!\!\!
&\!\!\!1\!\!\!&\!\!\!\!\!\! \\
\!&\!\!$2$\!&\!\! \!&\!\!\!\!\!\!&\!\!\!\!\!\!&\!\!\!\!\!\!&\!\!\!\!\!\!&\!\!\!\!\!\!&\!\!\!\!\!\!&\!\!\!\!\!\!&\!\!\!\!\!\!&
\!\!\!\!\!\!&\!\!\!\!\!\!&\!\!\!\!\!\!&\!\!\!\!\!\!&\!\!\!1\!\!\!&\!\!\!\!\!\!&\!\!\!2\!\!\!&\!\!\!\!\!\!&\!\!\!3\!\!\!
&\!\!\!\!\!\!&\!\!\!\!\!\! \\
\!&\!\!$\frac{5}{2}$\!&\!\! \!&\!\!\!\!\!\!&\!\!\!\!\!\!&\!\!\!\!\!\!&\!\!\!\!\!\!&\!\!\!\!\!\!&\!\!\!\!\!\!&\!\!\!\!\!\!&\!\!\!\!\!\!&
\!\!\!\!\!\!&\!\!\!\!\!\!&\!\!\!\!\!\!&\!\!\!\!\!\!&\!\!\!\!\!\!&\!\!\!\!\!\!&\!\!\!\!\!\!&\!\!\!1\!\!\!&\!\!\!\!\!\!
&\!\!\!1\!\!\!&\!\!\!\!\!\! \\
\!&\!\!$3$\!&\!\! \!&\!\!\!\!\!\!&\!\!\!\!\!\!&\!\!\!\!\!\!&\!\!\!\!\!\!&\!\!\!\!\!\!&\!\!\!\!\!\!&\!\!\!\!\!\!&\!\!\!\!\!\!&
\!\!\!\!\!\!&\!\!\!\!\!\!&\!\!\!\!\!\!&\!\!\!\!\!\!&\!\!\!\!\!\!&\!\!\!\!\!\!&\!\!\!\!\!\!&\!\!\!\!\!\!&\!\!\!\!\!\!
&\!\!\!\!\!\!&\!\!\!1\!\!\! \\
\hline
\end{tabular}}}
\caption{Non-vanishing BPS numbers $N^{(d_1,d_2)}_{j_L,j_R}$ of local ${\cal O}(-2,-2)\rightarrow \mathbb{P}^1\times \mathbb{P}^1$.}
\label{bpstablep1p1}
\end{table}

\subsection{${\cal O}(-K_{\mathcal{B}_1}) \rightarrow \mathcal{B}_1$}
\begin{figure}[h!] 
\begin{center} 
\includegraphics[angle=0,width=.2\textwidth]{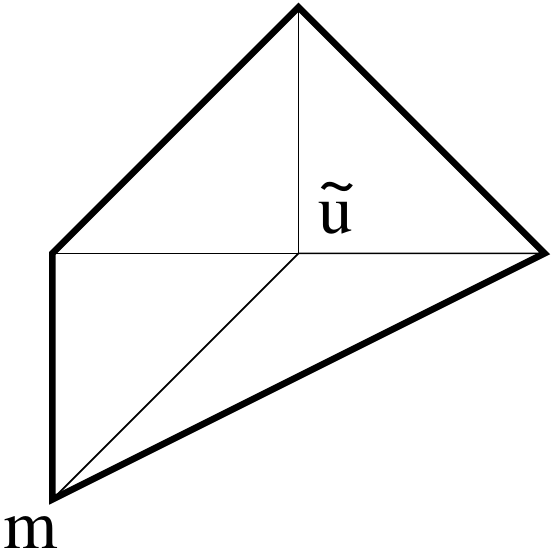}
\begin{quote} 
\caption{The polyhedron 3 with the choice of the mass parameter $m_1$  
and the modulus $\tilde{u}$.
\vspace{-1.2cm}} \label{poly3} \end{quote} 
\end{center} 
\end{figure} 
The Mori cone is given by
\begin{equation} 
 \label{datap2} 
 \begin{array}{cc|crr|rrl|} 
    \multicolumn{5}{c}{\nu_i }    &l^{(1)}=l^{(f)} & l^{(2)}=l^{(b)}&\\ 
    D_u    &&     1&     0&   0&           -2& -1&     \\ 
    D_1    &&     1&     1&   0&                1& 0& \\ 
    D_2    &&     1&     0&   1&                 0& 1& \\ 
    D_3    &&     1&    -1&   0&                  1&  -1& \\ 
    D_4    &&     1&    -1&   -1&                  0& 1& \\ 
  \end{array} \ . 
\end{equation} 

The invariant coordinates are
\be
z_1 = \frac{m}{\tilde{u}^2}, \quad z_2 = \frac{1}{\tilde{u} m} .
\ee
In our choice for the mass parameter we obtain from the embedding 
\be
X Y^2 + Y Z^2 + X^2 Z + \tilde{u} X Y Z + m X Z^2 = 0 
\ee
and its Weierstrass normal form is given by
\ban
y^2 &=& x^3+  \frac{1}{12} \Big(-24 \tilde{u} - \tilde{u}^4 + 8 \tilde{u}^2 m - 16 m^2  
  \Big) x \nn \\ && + \frac{1}{216} \Big(-216 - 36 \tilde{u}^3 - \tilde{u}^6 + 144 \tilde{u} m + 12 \tilde{u}^4 m  - 48 \tilde{u}^2 m^2 + 
   64 m^3 \Big)\, .
   \ean
As in the $\mathbb{P}^1\times \mathbb{P}^1$ case, there  is a trivial 
solution to the Picard Fuchs equation that reads
\be
\varphi_1 = \log (z_1) - 2 \log(z_2) = 3 \log(m) .
\ee
The third order differential operator reads in the case at hand 
\ban
{\cal L}&=& (-12 m^2 + 9 \tilde{u} - 18 m \tilde{u}^2 + 8 m^2 \tilde{u}^3)\partial_{\tilde{u}}+ (-108 m - 128 m^4 + 144 m^2 \tilde{u} + 27 \tilde{u}^2  \nn \\ 
&& - 64 m^3 \tilde{u}^2  - 52 m \tilde{u}^3 + 24 m^2 \tilde{u}^4)\partial_{\tilde{u}}^2 +  
          (-9 + 8 m \tilde{u}) (-27 + 16 m^3 + 36 m \tilde{u}  \nn \\ && 
 - 8 m^2 \tilde{u}^2 - \tilde{u}^3 + m \tilde{u}^4)\partial_{\tilde{u}}^3 \, .
\label{PFEf1}
\ean

We followed the same logic as in the previous section to get the 
large radius expansion and obtain the spin  invariants.  Again there is a series of known numbers    
\begin{equation}
 N^{(d,1)}_{j_L,j_R}=\begin{cases}
1 & {\rm if  }\ \ j_L=0, j_R=\frac 2 d \\
0 & {\rm otherwise}
\end{cases}\, .
\end{equation}
Note that this  equation reduces to the one for the $\mathbb{P}^2$ base in the 
blow-down limit $m=0$. We note that the discriminant reads  
\be 
\Delta = 1 - \tilde{u} - 8 m_1 \tilde{u}^2 + 36 m_1 \tilde{u}^3 - m_1 (27 - 16 m_1) \tilde{u}^4\, .
\ee
The prepotential is given as
\be
\begin{array}{rl}
F= &\text{class} + {\rm L}_{0, 1} - 2 {\rm L}_{1, 0} + 3 {\rm L}_{1, 1} + 5 {\rm L}_{2, 1} - 6 {\rm L}_{2, 2} + 
 7 {\rm L}_{3, 1} - 32 {\rm L}_{3, 2} + 27 {\rm L}_{3, 3} + 9 {\rm L}_{4, 1} \\ & - 110 {\rm L}_{4, 2} + 
 286 {\rm L}_{4, 3} - 192 {\rm L}_{4, 4} + 11 {\rm L}_{5, 1} - 288 {\rm L}_{5, 2} + 
 1651 {\rm L}_{5, 3} - 3038 {\rm L}_{5, 4} + 1695 {\rm L}_{5, 5} \\ & + 13 {\rm L}_{6, 1} - 
 644 {\rm L}_{6, 2} + 6885 {\rm L}_{6, 3} - 25216 {\rm L}_{6, 4} + 35870 {\rm L}_{6, 5} - 
 17064 {\rm L}_{6, 6} \, .
 \end{array}
 \ee
 Again we have denoted $L_\beta = \text{Li}_3(Q^\beta)$.
Generally $N^{d_1,d_2}_{j_L,j_R}=0$ for $d_1<d_2$ and again there is an infinite series  of spin invariants 
that can be given in a closed form 
\begin{equation}
 N^{(d,1)}_{j_L,j_R}=\begin{cases}
1 & {\rm if  }\ \ j_L=0, j_R= 2 d \\
0 & {\rm otherwise}
\end{cases} \, .
\end{equation}

Up to  $d_1+d_2\le 7$ the refined invariants are reported in Table~\ref{bpstable2}.
\begin{table}[h!]
\centering{{
\begin{tabular}[H]{|c|c|cccccccccccccccccccc|}
\hline
$(d_1,d_2)$\!&\!\!$j_L\backslash j_R$\!&\!\!0\!&\!\!\!$\frac{1}{2}$\!\!\!&\!\!\!1\!\!\!&\!\!\!$\frac{3}{2}$\!\!\!&\!\!\!2\!\!\!&\!\!\!$\frac{5}{2}$\!\!\!&\!\!\!3\!\!\!&\!\!\!$\frac{7}{2}$\!\!\!&\!\!\!4\!\!\!&
\!\!\!$\frac{9}{2}$\!\!\!&\!\!\!5\!\!\!&\!\!\!$\frac{11}{2}$\!\!\!&\!\!\!6\!\!\!&\!\!\!$\frac{13}{2}$\!\!\!&\!\!\!7\!\!\!&\!\!\!$\frac{15}{2}$\!\!\!&\!\!\!8\!\!\!&\!\!\!$\frac{17}{2}$\!\!\!
&\!\!\!9\!\!\!&\!\!\!$\frac{19}{2}$\!\!\! \\
\hline
$(2,2)$\!&\!\!0\!&\!\! \!&\!\!\!\!\!\!&\!\!\!\!\!\!&\!\!\!\!\!\!&\!\!\!\!\!\!&\!\!\!1\!\!\!&\!\!\!\!\!\!&\!\!\!\!\!\!&\!\!\!\!\!\!&
\!\!\!\!\!\!&\!\!\!\!\!\!&\!\!\!\!\!\!&\!\!\!\!\!\!&\!\!\!\!\!\!&\!\!\!\!\!\!&\!\!\!\!\!\!&\!\!\!\!\!\!&\!\!\!\!\!\!
&\!\!\!\!\!\!&\!\!\!\!\!\! \\
\hline
\hline
$(3,2)$\!&\!\!0\!&\!\! \!&\!\!\!\!\!\!&\!\!\!\!\!\!&\!\!\!\!\!\!&\!\!\!\!\!\!&\!\!\!1\!\!\!&\!\!\!\!\!\!&\!\!\!1\!\!\!&\!\!\!\!\!\!&\!\!\!\!\!\!&\!\!\!\!\!\!&\!\!\!\!\!\!&\!\!\!\!\!\!&\!\!\!\!\!\!&\!\!\!\!\!\!&\!\!\!\!\!\!&\!\!\!\!\!\!&\!\!\!\!\!\!
&\!\!\!\!\!\!&\!\!\!\!\!\! \\
\!&\!\!$\frac{1}{2}$\!&\!\! \!&\!\!\!\!\!\!&\!\!\!\!\!\!&\!\!\!\!\!\!&\!\!\!\!\!\!&\!\!\!\!\!\!&\!\!\!\!\!\!&\!\!\!\!\!\!&\!\!\!1\!\!\!&
\!\!\!\!\!\!&\!\!\! \!\!\!&\!\!\!\!\!\!&\!\!\!\!\!\!&\!\!\!\!\!\!&\!\!\!\!\!\!&\!\!\!\!\!\!&\!\!\!\!\!\!&\!\!\!\!\!\!
&\!\!\!\!\!\!&\!\!\!\!\!\! \\
\hline
\hline
$(4,2)$\!&\!\!0\!&\!\!\! \!&\!\!\!\!\!\!&\!\!\!\!\!\!&\!\!\!\!\!\!&\!\!\!\!\!\!&\!\!\!1\!\!\!&\!\!\!\!\!\!&\!\!\!1\!\!\!&\!\!\!\!\!\!&\!\!\!2\!\!\!&\!\!\!\!\!\!&\!\!\!\!\!\!&\!\!\!\!\!\!&\!\!\!\!\!\!&\!\!\!\!\!\!&\!\!\!\!\!\!&\!\!\!\!\!\!&\!\!\!\!\!\!
&\!\!\!\!\!\!&\!\!\!\!\!\! \\                            
\!&\!\!$\frac{1}{2}$\!&\!\! \!&\!\!\!\!\!\!&\!\!\!\!\!\!&\!\!\!\!\!\!&\!\!\!\!\!\!&\!\!\!\!\!\!&\!\!\!\!\!\!&\!\!\!\!\!\!&\!\!\!1\!\!\!&\!\!\!\!\!\!&\!\!\!1\!\!\!&\!\!\!\!\!\!&\!\!\!1\!\!\!&\!\!\!\!\!\!&\!\!\!\!\!\!&\!\!\!\!\!\!&\!\!\!\!\!\!&\!\!\!\!\!\!
&\!\!\!\!\!\!&\!\!\!\!\!\! \\
\!&\!\!$1$\!&\!\! \!&\!\!\!\!\!\!&\!\!\!\!\!\!&\!\!\!\!\!\!&\!\!\!\!\!\!&\!\!\!\!\!\!&\!\!\!\!\!\!&\!\!\!\!\!\!&\!\!\!\!\!\!&\!\!\!\!\!\!&\!\!\!\!\!\!&\!\!\!\!\!\!&\!\!\!\!\!\!&\!\!\!1\!\!\!&\!\!\!\!\!\!&\!\!\!\!\!\!&\!\!\!\!\!\!&\!\!\!\!\!\!
&\!\!\!\!\!\!&\!\!\!\!\!\! \\
\hline
\hline
$(5,2)$\!&\!\!0\!&\!\! \!&\!\!\!\!\!\!&\!\!\!\!\!\!&\!\!\!\!\!\!&\!\!\!\!\!\!&\!\!\!1\!\!\!&\!\!\!\!\!\!&\!\!\!1\!\!\!&\!\!\!\!\!\!\!\!&\!\!\!2\!\!\!&\!\!\!\!\!\!&\!\!\!2\!\!\!&\!\!\!\!\!\!&\!\!\!\!\!\!&\!\!\!\!\!&\!\!\!\!\!\!&\!\!\!\!\!\!
&\!\!\!\!\!\!&\!\!\!\!\!\!&\!\!\!\!\!\! \\
\!&\!\!$\frac{1}{2}$\!&\!\! \!&\!\!\!\!\!\!&\!\!\!\!\!\!&\!\!\!\!\!\!&\!\!\!\!\!\!&\!\!\!\!\!\!&\!\!\!\!\!\!&\!\!\!\!\!\!&\!\!\!1\!\!\!&\!\!\!\!\!\!&\!\!\!1\!\!\!&\!\!\!\!\!\!&\!\!\!2\!\!\!&\!\!\!\!\!\!&\!\!\!\!\!\!&\!\!\!\!\!\!&\!\!\!\!\!\!&\!\!\!\!\!\!
&\!\!\!\!\!\!&\!\!\!\!\!\! \\
\!&\!\!$1$\!&\!\! \!&\!\!\!\!\!\!&\!\!\!\!\!\!&\!\!\!\!\!\!&\!\!\!\!\!\!&\!\!\!\!\!\!&\!\!\!\!\!\!&\!\!\!\!\!\!&\!\!\!\!\!\!&\!\!\!\!\!\!&\!\!\!\!\!\!&\!\!\!1\!\!\!&\!\!\!\!\!\!&\!\!\!1\!\!\!&\!\!\!\!\!\!&\!\!\!\!\!\!&\!\!\!\!\!\!&\!\!\!\!\!\!
&\!\!\!\!\!\!&\!\!\!\!\!\! \\
\!&\!\!$\frac{3}{2}$\!&\!\! \!&\!\!\!\!\!\!&\!\!\!\!\!\!&\!\!\!\!\!\!&\!\!\!\!\!\!&\!\!\!\!\!\!&\!\!\!\!\!\!&\!\!\!\!\!\!&\!\!\!\!\!\!&\!\!\!\!\!\!&\!\!\!\!\!\!&\!\!\!\!\!\!&\!\!\!\!\!\!&\!\!\!\!\!\!&\!\!\!1\!\!\!&\!\!\!\!\!\!&\!\!\!\!\!\!&\!\!\!\!\!\!
&\!\!\!\!\!\!&\!\!\!\!\!\! \\
\hline
\hline
$(4,3)$\!&\!\!0\!&\!\! \!&\!\!\!\!\!\!&\!\!\!\!\!\!&\!\!\!\!\!\!&\!\!\!1\!\!\!&\!\!\!\!\!\!&\!\!\!\!1\!\!&\!\!\!\!\!\!&\!\!\!2\!\!\!&\!\!\!\!\!\!&\!\!\!1\!\!\!&\!\!\!\!\!\!&\!\!\!1\!\!\!&\!\!\!\!\!\!&\!\!\!\!\!\!&\!\!\!\!\!\!&\!\!\!\!\!\!&\!\!\!\!\!\!
&\!\!\!\!\!\!&\!\!\!\!\!\! \\                              
\!&\!\!$\frac{1}{2}$\!&\!\! \!&\!\!\!\!\!\!&\!\!\!\!\!\!&\!\!\!\!\!\!&\!\!\!\!\!\!&\!\!\!\!\!\!&\!\!\!\!\!\!&\!\!\!1\!\!\!&\!\!\!\!\!\!&
\!\!\!2\!\!\!&\!\!\!\!\!\!&\!\!\!\!2\!\!&\!\!\!\!\!\!&\!\!\!\!\!\!&\!\!\!\!1\!\!&\!\!\!\!\!\!&\!\!\!\!\!\!&\!\!\!\!\!\!
&\!\!\!\!\!\!&\!\!\!\!\!\! \\
\!&\!\!$1$\!&\!\! \!&\!\!\!\!\!\!&\!\!\!\!\!\!&\!\!\!\!\!\!&\!\!\!\!\!\!&\!\!\!\!\!\!&\!\!\!\!\!\!&\!\!\!\!\!\!&\!\!\!\!\!\!&\!\!\!\!\!\!&\!\!\!1\!\!&\!\!\!\!\!\!&\!\!\!\!1\!\!&\!\!\!\!\!\!&\!\!\!\!\!\!&\!\!\!\!\!\!&\!\!\!\!\!\!&\!\!\!\!\!\!
&\!\!\!\!\!\!&\!\!\!\!\!\! \\                           
\!&\!\!$\frac{3}{2}$\!&\!\! \!&\!\!\!\!\!\!&\!\!\!\!\!\!&\!\!\!\!\!\!&\!\!\!\!\!\!&\!\!\!\!\!\!&\!\!\!\!\!\!&\!\!\!\!\!\!&\!\!\!\!\!\!&\!\!\!\!\!\!&\!\!\!\!\!\!&\!\!\!\!\!\!&\!\!\! \!\!\!&\!\!\!1\!\!\!&\!\!\!1\!\!\!&\!\!\!\!\!\!&\!\!\!\!\!\!&\!\!\!\!\!\!
&\!\!\!\!\!\!&\!\!\!\!\!\! \\
\hline
\hline
$(5,3)$\!&\!\!0\!&\!\! \!&\!\!\!1\!\!\!&\!\!\!\!\!\!&\!\!\!1\!\!\!&\!\!\!\!\!\!&\!\!\!\!\!\!&\!\!\!3\!\!\!&\!\!\!\!\!\!&\!\!\!\!3\!\!&\!\!\!\!\!\!&\!\!\!5\!\!\!&\!\!\!\!\!\!&\!\!\!\!3\!\!&\!\!\!\!\!\!&\!\!\!2\!\!\!&\!\!\!\!\!\!&\!\!\!\!\!\!&\!\!\!\!\!\!
&\!\!\!\!\!\!&\!\!\!\!\!\! \\
\!&\!\!$\frac{1}{2}$\!&\!\! \!&\!\!\!\!\!\!&\!\!\!\!\!\!&\!\!\!\!\!\!&\!\!\!\!\!\!&\!\!\!1\!\!\!&\!\!\!\!\!\!&\!\!\!2\!\!\!&\!\!\!\!\!\!&
\!\!\!4\!\!\!&\!\!\!\!\!\!&\!\!\!5\!\!\!&\!\!\!\!\!\!&\!\!\!\!5\!\!&\!\!\!\!\!\!&\!\!\!1\!\!\!&\!\!\!\!\!\!&\!\!\!\!\!\!
&\!\!\!\!\!\!&\!\!\!\!\!\! \\                
\!&\!\!$1$\!&\!\! \!&\!\!\!\!\!\!&\!\!\!\!\!\!&\!\!\!\!\!\!&\!\!\!\!\!\!&\!\!\!\!\!\!&\!\!\!\!\!\!&\!\!\!\!\!\!&\!\!1\!\!\!\!&
\!\!\!\!\!\!&\!\!\!2\!\!\!&\!\!\!\!\!\!&\!\!\!4\!\!\!&\!\!\!\!\!\!&\!\!\!3\!\!\!&\!\!\!\!\!\!&\!\!\!1\!\!\!&\!\!\!\!\!\!
&\!\!\!\!\!\!&\!\!\!\!\!\! \\    
\!&\!\!$\frac{3}{2}$\!&\!\! \!&\!\!\!\!\!\!&\!\!\!\!\!\!&\!\!\!\!\!\!&\!\!\!\!\!\!&\!\!\!\!\!\!&\!\!\!\!\!\!&\!\!\!\!\!\!&\!\!\!\!\!\!&\!\!\!\!\!\!&\!\!\!\!\!\!&\!\!\!1\!\!\!&\!\!\!\!\!\!&\!\!\!2\!\!\!&\!\!\!\!\!\!&\!\!\!3\!\!\!&\!\!\!\!\!\!&\!\!\!1\!\!\!
&\!\!\!\!\!\!&\!\!\!\!\!\! \\
\!&\!\!$2$\!&\!\! \!&\!\!\!\!\!\!&\!\!\!\!\!\!&\!\!\!\!\!\!&\!\!\!\!\!\!&\!\!\!\!\!\!&\!\!\!\!\!\!&\!\!\!\!\!\!&\!\!\!\!\!\!&\!\!\!\!\!\!&\!\!\!\!\!\!&\!\!\!\!\!\!&\!\!\!\!\!\!&\!\!\!\!\!\!&\!\!\!\!1\!\!&\!\!\!\!\!\!&\!\!\!1\!\!\!&\!\!\!\!\!\!
&\!\!\!\!\!\!&\!\!\!\!\!\! \\
\!&\!\!$\frac{5}{2}$\!&\!\! \!&\!\!\!\!\!\!&\!\!\!\!\!\!&\!\!\!\!\!\!&\!\!\!\!\!\!&\!\!\!\!\!\!&\!\!\!\!\!\!&\!\!\!\!\!\!&\!\!\!\!\!\!&
\!\!\!\!\!\!&\!\!\!\!\!\!&\!\!\!\!\!\!&\!\!\!\!\!\!&\!\!\!\!\!\!&\!\!\!\!\!\!&\!\!\!\!\!\!&\!\!\!\!\!\!&\!\!\!\!\!\!
&\!\!\!1\!\!\!&\!\!\!\!\!\! \\
\hline
\end{tabular}}}
\caption{Non vanishing BPS numbers $N^{(d_1,d_2)}_{j_L,j_R}$ of local ${\cal O}(-2,-1)\rightarrow \mathbb{F}_1$.}
\label{bpstable2}
\end{table}

The Seiberg-Witten limit for $\mathbb{F}_1$ is 
\be 
z_1 \rightarrow  \frac{1}{4} \exp(- 2 \sqrt{2}  \epsilon^2  \tilde{u}),\quad z_2 \rightarrow  \epsilon^4 \Lambda^4\ .
\label{sw2}
\ee

\subsection{${\cal O}(-K_{\mathbb{F}_2}) \rightarrow \mathbb{F}_2$}

We consider the two-parameter model given by the polyhedron 2 
\begin{figure}[h!] 
\begin{center} 
\includegraphics[angle=0,width=.2\textwidth]{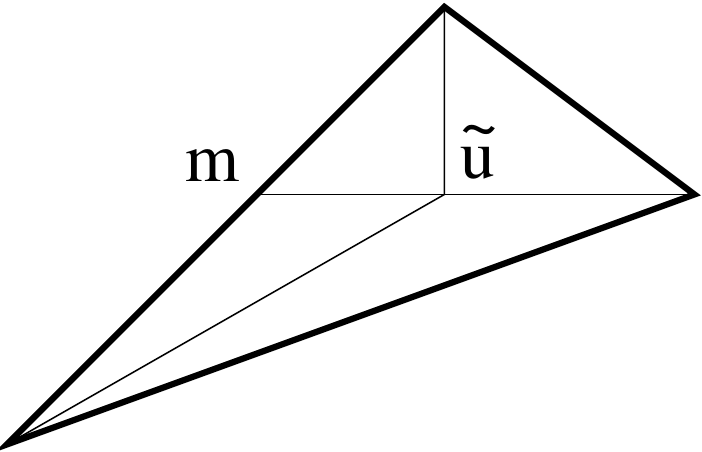}
\begin{quote} 
\caption{The polyhedron 4 with the choice of the mass parameter $m$  
and the modulus $\tilde{u}$.} 
\vspace{-1.2cm} \label{poly10} \end{quote} 
\end{center} 
\end{figure} 
with the  Mori cone vectors, which correspond to the 
depicted triangulation
\begin{equation} 
 \label{dataf1} 
 \begin{array}{ccrrr|rrl|} 
    \multicolumn{5}{c}{\nu_i }    &l^{(1)}=l^{(f)}& l^{(2)}=l^{(b)}&\\ 
    D_u    &&     1&     0&   0&         -2&  0&       \\ 
    D_1    &&     1&     1&   0&         1&   0&        \\ 
    D_2    &&     1&     0&   1&         0&   1&        \\ 
    D_3    &&     1&    -1&   0&         1&   -2&        \\ 
    D_4    &&     1&    -2&   -1&        0&   1&       \\  
  \end{array} \ . 
\end{equation} 

Here we observe a new phenomenon namely a point on the edge, 
which corresponds to an almost del Pezzo surface.    
The large structure coordinates are 
\begin{equation}
z_1=\frac{m}{\tilde{u}^2},\qquad    z_2=\frac{1}{m^2} \ . 
\label{zcoordinatesF2}
\end{equation}
We cannot take simply a ratio between the two coordinates to  get 
the non-dynamical parameter $m$. Let define as before 
$u=\frac{1}{\tilde u^2}$, then we find by specialization of (\ref{generalquartic}) for the appropriate rescaled $g_i$ 
\begin{equation}
\begin{array}{rl}
g_2=& 27 u^4 \left((1-4 m u)^2-48 u^2\right),\\ [2 mm]
g_3=&-27 u^6 \left(64 m^3 u^3-48 m^2 u^2-288 m u^3+12 m u+72 u^2-1\right)  \ ,
\end{array}
\end{equation}
which defines $t_f$. Let us denote the K\"ahler parameter of the 
base $t_2$ and the  one  of the fibre by $t_1$. Then we find 
\be 
t_f=Q_1^\frac{1}{2} Q_2, \quad m=\frac{1 + Q_2}{Q_2^\frac{1}{2}} \ .
\label{mirrormapf2}
\ee 
So typically for the almost del Pezzo surfaces we find one 
transcental mirror map $u(t_f)$ involving an elliptic 
integral and rational mirror maps for the mass 
parameters on the edges. The latter fact is simply 
due to the fact that the geometry on the edges is  a 
rational  geometry involving only Hirzebruch sphere 
trees of resolved $ADE$ singularities. In fact in the 
toric case just $A_n$-singularities. Now remarkably the  
spin invariants  $N^\beta_{j_l,j_r}$ are the same 
however with a shift in the classes so that 
$N^{d_f,d_b}_{j_l,j_r}(\mathbb{F}_2)=N^{d_f-d_b,d_b}_{j_l,j_r}(\mathbb{F}_0)$ 
and $N^{d_f,d_b}_{j_l,j_r}(\mathbb{F}_2)=0$ for $d_f<d_b$.    

The Seiberg-Witten limit for $\mathbb{F}_2$ is 
\be 
z_1 \rightarrow  \frac{1}{4} \exp(- 2 \epsilon^2  \tilde{u}), \quad  z_2 \rightarrow  \epsilon^4 \Lambda^4\ .
\label{sw3}
\ee

\subsection{${\cal O}(-K_{\mathcal{B}_2}) \rightarrow \mathcal{B}_2$}
\begin{figure}[h!] 
\begin{center} 
\includegraphics[angle=0,width=.2\textwidth]{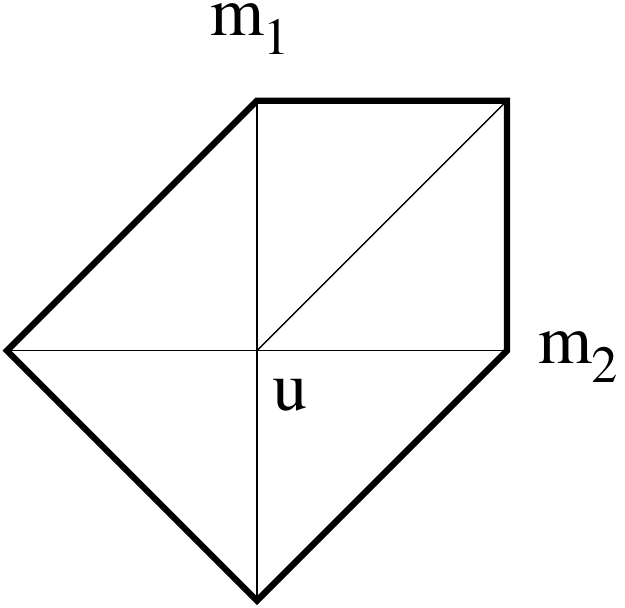}
\begin{quote} 
\caption{The polyhedron 5 with the choice of the mass parameter $m_1, m_2$.
and the modulus $\tilde{u}$.
\vspace{-1.2cm}} \label{poly5} \end{quote} 
\end{center} 
\end{figure} 
The Mori cone is given by
\begin{equation} 
 \label{datadp2} 
 \begin{array}{cc|crr|rrrl|} 
    \multicolumn{5}{c}{\nu_i }    &l^{(1)} & l^{(2)} &l^{(3)}&\\ 
    D_u    &&     1&     0&   0&           -1& -1& -1&     \\ 
    D_1    &&     1&     1&   0&           -1&  1&  0&\\ 
    D_2    &&     1&     1&   1&            1& -1&  1&\\ 
    D_3    &&     1&     0&   1&            0&  1& -1&\\ 
    D_4    &&     1&    -1&   0&            0&  0&  1&\\ 
    D_5    &&     1&     0&  -1&            1&  0&  0&
  \end{array} \ . 
\end{equation} 
The invariant coordinates are given by
\be
z_1 = \frac{m_1 m_2}{\tilde{u}}, \quad z_2 = \frac{1}{\tilde{u} m_2}, \quad z_3 = \frac{m_2}{\tilde{u}^2}.
\ee
The mirror curve reads
\be
X Y^2 + Y Z^2 + X^2 Z + \tilde{u} X Y Z + m_1 X Z^2 + m_2 X^2 Y= 0 
\ee
and the Weierstrass normal form is given by
\ban
y^2 &=& x^3+ \frac{1}{12} \Big(-24 \tilde{u} - \tilde{u}^4 + 8 \tilde{u}^2 m_1 - 16 m_1^2 + 8 \tilde{u}^2 m_2 + 16 m_1 m_2 - 
  16 m_2^2\Big) x \nn \\ && + \frac{1}{216} \Big(-216 - 36 \tilde{u}^3 - \tilde{u}^6 + 144 \tilde{u} m_1 + 12 \tilde{u}^4 m_1  - 48 \tilde{u}^2 m_1^2 + 
   64 m_1^3 + 144 \tilde{u} m_2 \nn \\ && + 12 \tilde{u}^4 m_2  - 24 \tilde{u}^2 m_1 m_2 - 96 m_1^2 m_2 - 
   48 \tilde{u}^2 m_2^2 - 96 m_1 m_2^2 + 64 m_2^3\Big).
\ean
Also in this case a third order differential operator can be constructed. Note however, that in order to derive it one needs two take into account five $l$ vectors out of which only three are linearly independent in order to make the ideal of differential operators close. This is due to the fact that linear dependent relations can give rise to further linear independent differential operators. The full differential operator may be found in the appendix.

Denoting as usual $L_\beta = \text{Li}_3(Q^\beta)$, the prepotential is given as
\be
\begin{array}{rl}
F= &\text{class} + {\rm L}_{0, 0, 1} + {\rm L}_{0, 1, 0} - 2 {\rm L}_{0, 1, 1}  + 3 {\rm L}_{1, 1, 1} - 4 {\rm L}_{1, 2, 1} + 5 {\rm L}_{1, 2, 2}  - 
 6 {\rm L}_{1, 3, 2} \\ &+ 7 {\rm L}_{1, 3, 3} - 8 {\rm L}_{1, 4, 3} + 9 {\rm L}_{1, 4, 4}  - 6 {\rm L}_{2, 2, 2}  + 35 {\rm L}_{2, 3, 2}  - 
 32 {\rm L}_{2, 3, 3}  - 32 {\rm L}_{2, 4, 2} \\ & + 135 {\rm L}_{2, 4, 3} - 
 110 {\rm L}_{2, 4, 4}  + 
 27 {\rm L}_{3, 3, 3}   - 
 400 {\rm L}_{3, 4, 3} + 286 {\rm L}_{3, 4, 4} - 192 {\rm L}_{4, 4, 4} \, .
 \end{array}
 \ee
 Note that there is a symmetry between the first and the third entry, so that we have omitted redundant terms.

 \subsection{${\cal O}(-K_{\mathcal{B}_3}) \rightarrow \mathcal{B}_3$}
 
 This is the maximal still generic toric blow-up of $\mathbb{P}^2$ and is 
 represented by the polyhedron $7$.  
\begin{figure}[h!] 
\begin{center} 
\includegraphics[angle=0,width=.2\textwidth]{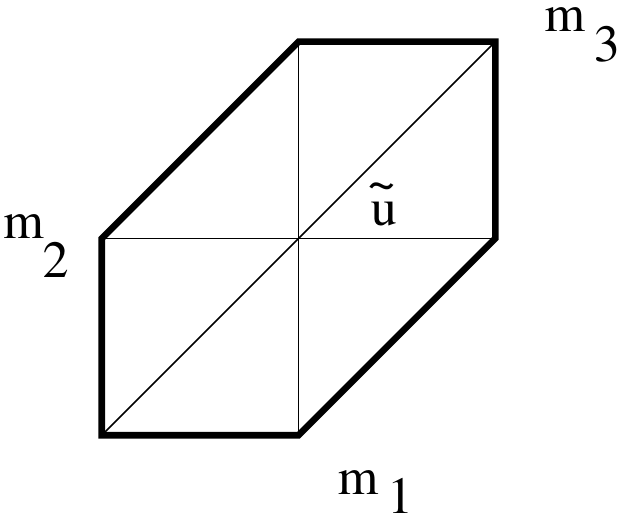}
\begin{quote} 
\caption{The polyhedron 7 with the choice of the mass parameters $m_1,m_2,m_3$  
and the modulus $\tilde{u}$.}
\vspace{-1.2cm} \label{poly7} \end{quote} 
\end{center} 
\end{figure} 
The Mori cone vectors, which correspond to the depicted triangulation are given below 
\begin{equation} 
 \label{datadp3} 
 \begin{array}{ccrrr|rrrrrrl|} 
    \multicolumn{5}{c}{\nu_i }    &l^{(1)}& l^{(2)}& l^{(3)} & l^{(4)}&  l^{(5)}& l^{(6)} &\\ 
    D_u    &&     1&     0&   0&         -1&   -1& -1&  -1& -1&  -1&       \\ 
    D_1    &&     1&     1&   0&         -1&    1& 0&   0&  0&  1&       \\ 
    D_2    &&     1&     1&   1&          1&   -1& 1&   0&  0&  0&     \\ 
    D_3    &&     1&     0&   1&          0&    1& -1&  1&  0&  0&    \\ 
    D_4    &&     1&    -1&   0&          0&    0& 1&   -1& 1&  0&     \\ 
    D_5    &&     1&    -1&  -1&          0&    0& 0&   1&  -1& 1 &   \\ 
    D_6    &&     1&     0&  -1&          1&    0& 0&   0&   1& -1&    \\ 
  \end{array} \ . 
\end{equation} 
One finds the mirror curve
\be
X Y^2 + Y Z^2 + X^2 Z + \tilde{u} X Y Z + m_1 X Z^2 + m_2 X^2 Y + m_3 Y^2 Z= 0 
\ee
and  the Weierstrass normal form
\ban
y^2 &=& 4x^3 + \frac{1}{12} \Big(-16 m_1^2 + 16 m_1 m_2 - 16 m_2^2 + 16 m_1 m_3 + 16 m_2 m_3 - 
   16 m_3^2 - 24 \tilde{u} \nn \\ && - 24 m_1 m_2 m_3 \tilde{u} + 8 m_1 \tilde{u}^2 + 8 m_2 \tilde{u}^2 + 
   8 m_3 \tilde{u}^2 - \tilde{u}^4\Big) \nn \\ && +
   \frac{1}{216} \Big(-216 + 64 m_1^3 - 96 m_1^2 m_2 - 96 m_1 m_2^2 + 64 m_2^3 - 
   96 m_1^2 m_3 - 48 m_1 m_2 m_3 \nn \\ && - 96 m_2^2 m_3 - 96 m_1 m_3^2 - 
   96 m_2 m_3^2 - 216 m_1^2 m_2^2 m_3^2 + 64 m_3^3 + 144 m_1 \tilde{u} + 
   144 m_2 \tilde{u} \nn \\ && + 144 m_3 \tilde{u} + 144 m_1^2 m_2 m_3 \tilde{u} \nn  + 
   144 m_1 m_2^2 m_3 \tilde{u} + 144 m_1 m_2 m_3^2 \tilde{u} - 48 m_1^2 \tilde{u}^2 \nn \\ && - 
   24 m_1 m_2 \tilde{u}^2 - 48 m_2^2 \tilde{u}^2 - 24 m_1 m_3 \tilde{u}^2 - 24 m_2 m_3 \tilde{u}^2 - 
   48 m_3^2 \tilde{u}^2 - 36 \tilde{u}^3 \nn \\ && - 36 m_1 m_2 m_3 \tilde{u}^3 + 12 m_1 \tilde{u}^4 + 
   12 m_2 \tilde{u}^4 + 12 m_3 \tilde{u}^4 - \tilde{u}^6\Big)\, .
\ean
In this case the Mori cone is not simplicial, but we will find a choice of these 
vectors, which truncates in the correct way to all possibilities of embedding 
all lower blow up cases into this model. This is only possible by using one non-integer  
combination of the Mori vectors 
${\tilde l}^{(1)}=\frac{1}{3} \sum_{i=0}^2 (l^{(2 i+1)}-l^{(2 i+2)})$ as well as  $\tilde l^{(2)}=l^{(2)}$, 
${\tilde l}^{(3)}=l^{(4)}$ and $\tilde l^{(4)}=l^{(6)}$. The  corresponding large complex 
structure variables are
\begin{equation} 
 z_1=m_1 m_2 m_3,\  z_2=\frac{1}{m_1 \tilde{u}},\ z_3=\frac{1}{m_2 \tilde{u}},\ z_4=\frac{1}{m_3 \tilde{u}}\ .
 \label{lcp7}
\end{equation}
We can also calculate the ring of intersection numbers for the choice of basis of curves 
defined by ${\tilde l}^{(i)}$ and the dual divisors $J_i$ as
\begin{equation} 
R=J_1^2 + J_1 J_2 + J_1 J_3 + J_2 J_3 + J_1 J_4 + J_2 J_4 + J_3 J_4.
\end{equation} 
With this informations the instantons can be calculated following \cite{Chiang:1999tz}.  
Alternatively we can specialize either polyhedron 15 or 16, redefine $\tilde{u}\rightarrow u = 1/\tilde{u}$ and  
rescale $g_i\rightarrow \lambda^i g_i$ with $\lambda=  18 u^4 $. Then we obtain the 
mirror map (\ref{nonlogperiod}) as
\begin{equation}
u=Q_t + (1 + m_1) Q_t^3 + 2 m_3 Q_t^4 + (1 - m_1 + m_1^2 - 3 m_2) Q_t^5 + {\cal O}(Q_t^6)\, .
\end{equation}
Here we have defined $Q_t=e^t=(Q_1 Q_2 Q_3 Q_4)^\frac{1}{3}$. It follows from (\ref{lcp7}) that 
\begin{equation}
m_1=\frac{(Q_1 Q_3 Q_4)^\frac{1}{3}}{Q_2^\frac{2}{3}},\ \  m_2=\frac{(Q_1 Q_3 Q_4)^\frac{1}{3}}{Q_3^\frac{2}{3}},\ \  m_3=\frac{(Q_1 Q_2 Q_3)^\frac{1}{3}}{Q_4^\frac{2}{3}}. \ \  
\end{equation}
This defines the large radius variables and allows to extract the BPS-numbers directly from 
the curve.  For example we list here the prepotential up to multi-degree 16 in the instantons.
With the notation $L_\beta={\rm Li}_3(Q^\beta)$ we get      
\begin{equation}
\begin{array}{rl}
F=& \text{class}+ {\rm L}_{  0, 0, 0, 1} + {\rm L}_{  1, 0, 0, 1} - 2 {\rm L}_{  1, 0, 1, 1} + 
 3 {\rm L}_{  1, 1, 1, 1} + 3 {\rm L}_{  2, 1, 1, 1} - 4 {\rm L}_{  2, 1, 1, 2} + 5 {\rm L}_{  2, 1, 2, 2}  \\ & -
 6 {\rm L}_{  2, 2, 2, 2} + 5 {\rm L}_{  3, 1, 2, 2} - 6 {\rm L}_{  3, 1, 2, 3} + 7 {\rm L}_{  3, 1, 3, 3} - 
 36 {\rm L}_{  3, 2, 2, 2} + 35 {\rm L}_{  3, 2, 2, 3} - 32 {\rm L}_{  3, 2, 3, 3} \\ & + 27 {\rm L}_{  3, 3, 3, 3} +  
 7 {\rm L}_{  4, 1, 3, 3} - 8 {\rm L}_{  4, 1, 3, 4} + 9 {\rm L}_{  4, 1, 4, 4} - 
 6 {\rm L}_{  4, 2, 2, 2} + 35 {\rm L}_{  4, 2, 2, 3} - 32 {\rm L}_{  4, 2, 2, 4} \\ & - 
 160 {\rm L}_{  4, 2, 3, 3} +  135 {\rm L}_{  4, 2, 3, 4} - 110 {\rm L}_{  4, 2, 4, 4} +
 531 {\rm L}_{  4, 3, 3, 3} - 400 {\rm L}_{  4, 3, 3, 4} + 286 {\rm L}_{  4, 3, 4, 4} \\ &- 
 192 {\rm L}_{  4, 4, 4, 4}\ .
\end{array}
\end{equation}
Note that there is a symmetry in the last three entries, so that we 
present only the $\beta$, which are ordered w.r.t these entries. 

\subsection{Del Pezzos related to the bi-quadratic}

Here we want to discuss the remaining polyhedra that are 
embeddable in the square polyhedron 16 of the bi-quadratic. They 
are physically interesting e.g. as they correspond to the 
five-dimensional $SU(2)$ Seiberg-Witten theories with $N_f=0,1,2,3,4$ 
matter multiplets in the fundamental representation. The Seiberg  
Witten curves with $N_f<3$ are given by, see  
e.g. \cite{Argyres:1995wt} (also for the $N_f=4$ case) 
\be
y^2=(x^2-u^{sw})^2-\Lambda^{4-N_f} \prod_{i=0} ^{N_f} (x+m^{sw}_i)\ . 
\label{SWcurvesnf<3}
\ee       
The geometry ${\cal O}_{\mathbb{F}_2}\rightarrow \mathbb{F}_2$ corresponds to one of 
five-dimensional realization of the $SU(2)$ Seiberg-Witten theory with $N_f=0$. 
The limit in the moduli space, which corresponds to  
$R=\frac{1}{\epsilon}\rightarrow \infty$, i.e. in which the four-dimensional 
$SU(2)$ Seiberg-Witten theory emerges was already given in (\ref{sw3}).    

Polyhedron 6 can be viewed as the blow-up of $\mathbb{F}_2$ and each 
blow-up adds one matter in the fundamental representation.

\begin{figure}[h!] 
\begin{center} 
\includegraphics[angle=0,width=.2\textwidth]{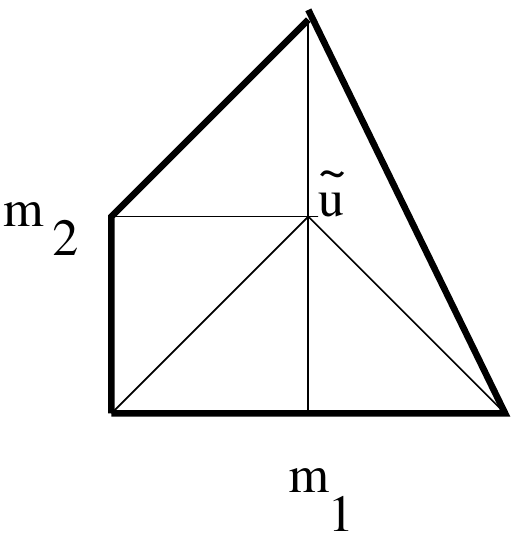}
\begin{quote} 
\caption{The polyhedron 6 with the choice of the mass parameters $m_1,m_2$  
and the modulus $\tilde{u}$.}
\vspace{-1.2cm} \label{poly7} \end{quote} 
\end{center} 
\end{figure} 

The Mori cone vectors for this model are  
\begin{equation} 
 \label{dataf1} 
 \begin{array}{ccrrr|rrrl|} 
    \multicolumn{5}{c}{\nu_i }    &l^{(1)}& l^{(2)}& l^{(3)} &\\ 
    D_u    &&     1&     0&   0&         -1&   -1& 0&        \\ 
    D_1    &&     1&     1&   -1&        0&   0 & 1&          \\ 
    D_2    &&     1&     0&   1&         1&   0& 1&          \\ 
    D_3    &&     1&    -1&   0&        -1&  1&  0&         \\ 
    D_4    &&     1&    -1&   -1&        1&   -1&1&          \\ 
    D_5    &&     1&     0&  -1&         0&   1& -2&         \\  
  \end{array} \ . 
\end{equation} 
From this we get the large volume variables 
\begin{equation} 
z_1=\frac{1}{\tilde{u} m_2} ,\quad z_2=\frac{m_1 m_2}{\tilde{u}}, \quad z_3=\frac{1}{m_1^2} \, .
\label{zcoordinatesp6}
\end{equation}
In this case we define $Q_t= Q_1^{\frac{1}{2}} Q_2 Q_3^{\frac{1}{4}} $, 
so that the transcendental mirror map is 
\be u=Q_t- m_1 Q_t^3 + 2 m_2 Q_t^4 + 
(-3 + m_1^2) Q_t^5+{\cal O}(Q_t^7)\, .
\ee
The rational mirror maps are          
\begin{equation}
\frac{z_1}{z_2} =\frac{Q_1}{(1 + Q_3) Q_2},\quad z_3 = \frac{Q_3}{(1 + Q_3)^2} \ . 
\label{mirrormapp6}
\end{equation}
The Seiberg-Witten limit is given by 
\be 
z_1=\left(\exp{2^{\frac{2}{3}}} m_1^{sw} \epsilon\right), \quad  z_2=\frac{1}{2} \exp\left(-2^{\frac{2}{3}} \epsilon(2^{\frac{2}{3}} \epsilon  u^{sw}+ m^{sw}_1)\right), 
\quad  z_3= \Lambda^3 \epsilon^3\ .
\label{sw4}
\ee

The first rational Gromov-Witten invariants follow then 
from a suitable specialization of the biquartic curve   
(\ref{thebi-quadratic}) as  

\begin{equation}
 \begin{array}{rl}
F=& \text{class}+L_{1, 0, 0} + L_{0, 1, 0} - 2 L_{1, 1, 0} + L_{0, 1, 1} - 
 2 L_{1, 1, 1} + 3 L_{1, 2, 1} - 4 L_{2, 2, 1} + 5 L_{2, 3, 1} \\ &  - 
 6 L_{3, 3, 1} + 7 L_{3, 4, 1} -  8 L_{4, 4, 1} + 9 L_{4, 5, 1} - 
 10 L_{5, 5, 1} + 11 L_{5, 6, 1} - 12 L_{6, 6, 1} + 
 13 L_{6, 7, 1} \\ & + 5 L_{2, 3, 2} - 6 L_{3, 3, 2} - 6 L_{2, 4, 2} + 
 35 L_{3, 4, 2}  - 32 L_{4, 4, 2} - 32 L_{3, 5, 2} + 
 135 L_{4, 5, 2} \\ & - 110 L_{5, 5, 2} - 110 L_{4, 6, 2} +  
 385 L_{5, 6, 2} - 288 L_{6, 6, 2} - 288 L_{5, 7, 2} + 
 7 L_{3, 4, 3} - 8 L_{4, 4, 3} \\ & - 32 L_{3, 5, 3} + 
 135 L_{4, 5, 3} - 110 L_{5, 5, 3} + 27 L_{3, 6, 3} - 
 400 L_{4, 6, 3} + 1100 L_{5, 6, 3} + 286 L_{4, 7, 3} \\ &+ 
 9 L_{4, 5, 4} - 10 L_{5, 5, 4} - 110 L_{4, 6, 4} .
\end{array}
\end{equation}
 
Next we blow it up once more
\begin{figure}[h!] 
\begin{center} 
\includegraphics[angle=0,width=.2\textwidth]{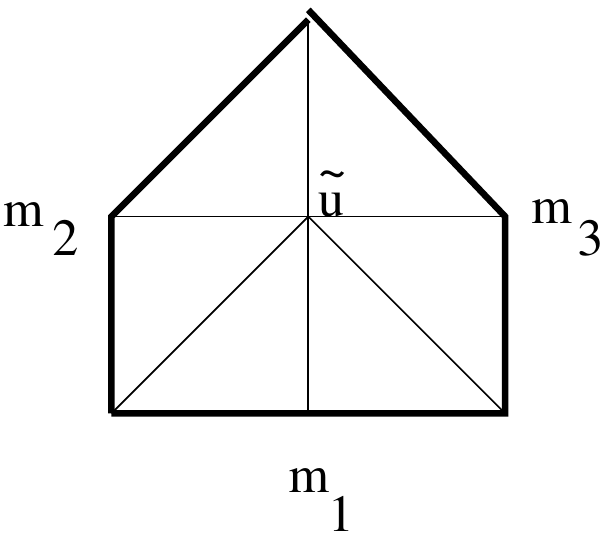}
\begin{quote} 
\caption{The polyhedron 9 with the choice of the mass parameters $m_1,m_2,m_3$  
and the modulus $\tilde{u}$.}
\vspace{-1.2cm} \label{poly8} \end{quote} 
\end{center} 
\end{figure} 
to get a model with three masses. The new feature we want to discuss here  
is a non-simplicial Mori cone in a model with rational mirror maps  
\begin{equation} 
 \label{dataf1} 
 \begin{array}{ccrrr|rrrrrl|} 
    \multicolumn{5}{c}{\nu_i }    &l^{(1)}& l^{(2)}& l^{(3)} &l^{(4)}& l^{(5)} &   \\ 
    D_u    &&     1&     0&   0&         -1&   -1& 0& -1 &-1 & \\ 
    D_1    &&     1&     1&   -1&        0&   0 & 1&  -1&    1 &   \\ 
    D_2    &&     1&     0&   1&         1&   0& 1&   0 &   1 &  \\ 
    D_3    &&     1&    -1&   0&        -1&  1&  0&     0&  0&          \\ 
    D_4    &&     1&    -1&   -1&        1&   -1&1&    0 &   0 &   \\ 
    D_5    &&     1&     0&  -1&         0&   1& -2&   1&    0 &  \\  
    D_6    &&     1&     1&   0&         0&   0&  0&   1&   -1&  \\
  \end{array} \ . 
\end{equation} 
The four large volume coordinates are redundantly given by 
\be 
z_1=\frac{1}{{\tilde u} m_2},\ \ z_2= \frac{m_1 m_2}{\tilde u}, \ \ z_3=\frac{1}{\tilde u}, 
\ \ z_4=\frac{m_1 m_3}{\tilde u},\ \ z_5=\frac{1}{\tilde u m_3}\ .        
\label{variablespoly9}
\ee 
These coordinates fulfill the following non-trivial mirror maps 
$Q_t=(Q_1 Q_2 Q_3 Q_4 Q_5)^{\frac{1}{4}}$ and  
\be
z_3= \frac{Q_3}{(1+Q_3)^2},\ \ 
\frac{z_1}{z_2}=\frac{Q_1}{Q_2 (1+Q_3)},\ \  
\frac{z_5}{z_1}=\frac{Q_5}{Q_4 (1+Q_3)},\ \ Q_1 Q_2=z_1 z_2=z_4 z_5 =Q_4 Q_5\ .  
\label{rationalpoly9}     
\ee 
To extract the Gromov-Witten invariants from the specialized curve 
(\ref{thebi-quadratic}) we can solve the masses $m_1,m_2,m_3$ as well as  
$Q_t$  either for $Q_1,Q_2,Q_3,Q_4$ or $Q_2,Q_3,Q_4,Q_5$, which corresponds 
to two chambers of the non-simplicial K\"ahler cone, which are 
symmetric under the exchange of $Q_1 Q_2\leftrightarrow Q_4 Q_5$ and 
moreover specialize for $Q_4= Q_5=0$ to the previously discussed model. 
In view of the symmetry we list only the invriants for  $Q_1,Q_2,Q_3,Q_4$
\begin{equation}
 \begin{array}{rl}
F=&\text{class}+L_{0, 0, 0, 1} + L_{0, 0, 1, 1} + L_{0, 1, 0, 0} + L_{0, 1, 1, 0} - 
2 L_{0, 1, 1, 1} +  L_{1, 0, 0, 0} - 2 L_{1, 1, 0, 0}\\&  - 2 L_{1, 1, 1, 0} + 
 3 L_{1, 1, 1, 1} + 3 L_{1, 2, 1, 0} - 4 L_{1, 2, 1, 1} - 
 4 L_{1, 2, 2, 1} + 5 L_{1, 2, 2, 2} + 5 L_{1, 3, 2, 1} \\& - 
 6 L_{1, 3, 2, 2} - 4 L_{2, 2, 1, 0} + 5 L_{2, 2, 1, 1} + 
 5 L_{2, 2, 2, 1} - 6 L_{2, 2, 2, 2} + 5 L_{2, 3, 1, 0} -
 6 L_{2, 3, 1, 1}  \\& + 5 L_{2, 3, 2, 0} - 36 L_{2, 3, 2, 1} - 
6 L_{2, 4, 2, 0} - 6 L_{3, 3, 1, 0} + 7 L_{3, 3, 1, 1} - 6 L_{3, 3, 2, 0} +7 L_{3, 4, 1, 0}
 \end{array} \ . 
\end{equation}         
Again it is quite interesting to know the 
Seiberg-Witten limit. We define $z_f=z_1 z_2$ 
and obtain   
\be 
z_1 = \frac{1}{2} \exp( -2\epsilon  m_1^{sw} ),\ \  
z_f = \frac{1}{4} \exp(-   4 \epsilon^2 u^{sw}),\ \  
z_3  = \Lambda^2 \epsilon^2, \ \   
z_4 = \frac{1}{2} \exp(2 \epsilon   m^{sw}_2)\ .
\label{sw5}
\ee

We finally discuss a model with a simplicial Mori cone, which 
can be symmetrized like in the last case to the full $D5$ del 
Pezzo.  We performed the calculation, but leave the details to 
the reader.       
\begin{figure}[h!] 
\begin{center} 
\includegraphics[angle=0,width=.2\textwidth]{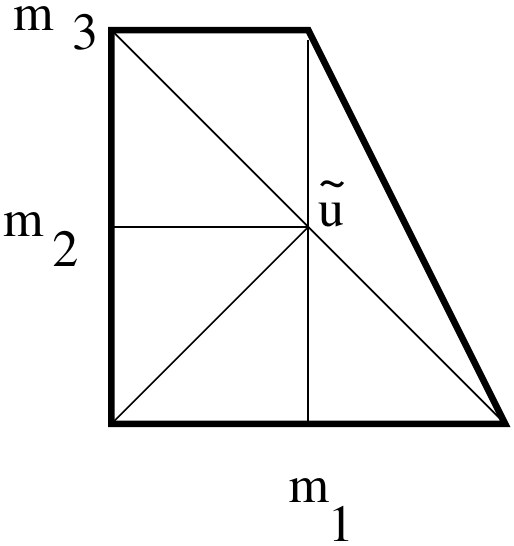}
\begin{quote} 
\caption{The polyhedron 8 with the choice of the mass parameters $m_1,m_2,m_3$  
and the modulus $\tilde{u}$.}
\vspace{-1.2cm} \label{poly8} \end{quote} 
\end{center} 
\end{figure} 

The Mori cone vectors  determine the large volume variables 
\begin{equation} 
 \label{dataf1} 
 \begin{array}{ccrrr|rrrrl|} 
    \multicolumn{5}{c}{\nu_i }    &l^{(1)}& l^{(2)}& l^{(3)} & l^{(4)} &\\ 
    D_u    &&     1&     0&   0&         0&   -1& 0&  -1&       \\ 
    D_1    &&     1&     1&   -1&        1&   0& 0&  0&         \\ 
    D_2    &&     1&     0&   1&         0&   0& 0&  1&         \\ 
    D_3    &&     1&    -1&   1&         0&   0& 1&  -1&        \\ 
    D_4    &&     1&    -1&   0&         0&   1&-2&  1&         \\ 
    D_5    &&     1&    -1&  -1&         1&  -1& 1&  0&       \\ 
    D_6    &&     1&     0&  -1&        -2&   1& 0&  0&        \\ 
  \end{array} \ . 
\end{equation} 
These read in terms of the $m_i$  and $\tilde{u}$
\begin{equation} 
z_1=\frac{1}{m_1^2},\quad z_2=\frac{m_1 m_2}{\tilde{u}}, \quad z_3=\frac{m_3}{m_2^2},  \quad z_4=\frac{m_2}{\tilde{u} m_3} \ . 
\label{zcoordinatesp8}
\end{equation}
With $Q_t= Q_1^{\frac{1}{4}} (Q_2 Q_3 Q_4)^{\frac{1}{2}}$ we see as  before that the variables $z_i$ are not 
independent transcendental functions of the K\"ahler parameters rather one has the following relations
\begin{equation}
z_1 = \frac{Q_1}{(1 + Q_1)^2},\quad  z_3 = \frac{Q_3}{(1 + Q_3)^2},\quad  z_2 =\frac{z_4 (1 + Q_1) Q_2}{Q_4} \, .
\label{mirrormapp8}
\end{equation}

\begin{equation}
 \begin{array}{rl}
F=&\text{class}+ L_{0, 0, 0, 1} + L_{0, 0, 1, 1} + L_{0, 1, 0, 0} +  L_{0, 1, 1, 0} - 2 L_{0, 1, 1, 1} 
+ L_{1, 1, 0, 0} +  L_{1, 1, 1, 0} \\ & - 2 L_{1, 1, 1, 1} - 2 L_{1, 2, 1, 0} +
 3 L_{1, 2, 1, 1} + 3 L_{1, 2, 2, 1} - 4 L_{1, 2, 2, 2} - 
 4 L_{1, 3, 2, 1} + 5 L_{1, 3, 2, 2} \\ & + 5 L_{1, 3, 3, 2} - 
 6 L_{1, 3, 3, 3} - 6 L_{1, 4, 3, 2} + 7 L_{1, 4, 3, 3} + 
 7 L_{1, 4, 4, 3} - 4 L_{2, 3, 2, 1} + 5 L_{2, 3, 2, 2} \\ & + 
 5 L_{2, 3, 3, 2}  - 6 L_{2, 3, 3, 3} + 5 L_{2, 4, 2, 1} - 
 6 L_{2, 4, 2, 2} + 5 L_{2, 4, 3, 1} - 36 L_{2, 4, 3, 2} + 
 35 L_{2, 4, 3, 3} \\ & - 6 L_{2, 4, 4, 2} - 6 L_{2, 5, 3, 1} + 
 35 L_{2, 5, 3, 2} - 6 L_{3, 4, 3, 2} - 6 L_{3, 5, 3, 1} \, .
 \end{array}
\end{equation}

\subsection{A mass deformation of the local $E_8$ del Pezzo surface}

Let us consider the polyhedron $10$.
\begin{figure}[h!] 
\begin{center} 
\includegraphics[angle=0,width=.2\textwidth]{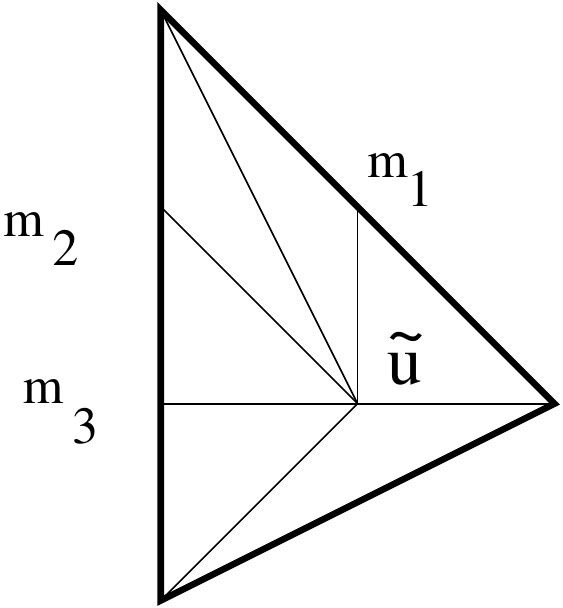}
\begin{quote} 
\caption{The polyhedron 10 with the choice of the mass parameters $m_1,m_2,m_3$  
and the modulus $\tilde{u}$.}
\vspace{-1.2cm} \label{poly10} \end{quote} 
\end{center} 
\end{figure} 
The Mori cone vectors, which correspond to the depicted triangulation are given below 
\begin{equation} 
 \label{dataf1} 
 \begin{array}{ccrrr|rrrrl|} 
    \multicolumn{5}{c}{\nu_i }    &l^{(1)}& l^{(2)}& l^{(3)} & l^{(4)} &\\ 
    D_u    &&     1&     0&   0&         0&   1& 0&  0&       \\ 
    D_1    &&     1&     1&   0&         1&   0& 0&  0&         \\ 
    D_2    &&     1&     0&   1&         -2&  1& 0&  0&         \\ 
    D_3    &&     1&    -1&   2&         1&  -1& 1&  0&        \\ 
    D_4    &&     1&    -1&   1&         0&   1&-2&  1&         \\ 
    D_5    &&     1&    -1&   0&         0&   0& 1& -2&       \\ 
    D_6    &&     1&    -1&  -1&         0&   0& 0&  1&        \\ 
  \end{array} \ . 
\end{equation} 
With the indicated mass parameters of the three non-renormalizable modes  and the para-meter $\tilde{u}$ 
the Mori vectors determine the large volume $B$-model coordinates 
\begin{equation} 
 z_1=\frac{1}{m_1^2},\ \quad z_2=\frac{m_1 m_2}{\tilde{u}},\ z_3=\frac{m_3}{m_2^2},\ z_4=\frac{m_2}{m_3^2}\ .
\label{zcoordinatesp10}
\end{equation}
The anti-canonical class of the $E_8$ del Pezzo corresponds 
to an elliptic curve, which in turn has the following Mori vector  
\begin{equation} 
l_e= 3 l^{(1)} + 6 l^{(2)} + 4 l^{(3)} + 2 l^{(4)}=\sum_i a_i l^{(i)} \ .   
\label{ellipticMori}
\end{equation} 
This equation implies that $z_e=\frac{1}{u^6} = z_1^3 z_2^6 z_3^4 z_4^2$ is the correct large volume 
modulus for this curve independently of the masses. By specializing the expression in Appendix A.8 
as $m_1=0, m_2=1, m_4 =1, m_6=1, m_3= m_2, m_5= 0, a_1= 0, a_2 = m_1, a_3= m_3, 
\tilde u =\frac{1}{u}$ and scaling $g_i\rightarrow \lambda^i g_i$ with $\lambda=  18 u^4 $  
we get
\begin{equation} 
\begin{array}{rl}
g_2=& 27 u^4 (24 m_1 u^3-48 m_2 u^4+16 m_3^2 u^4-8 m_3 u^2+1)\, ,\\
g_3=&27 u^6 (216 m_1^2 u^6+12 m_3 u^2 (-12 m_1 u^3+24 m_2 u^4-1)+\\ 
& 36 m_1 u^3-72 m_2 u^4-64 m_3^3 u^6+48 m_3^2 u^4-864 u^6+1)\, .
\end{array}
\end{equation}

The scaling is chosen so that $\frac{dt}{du} = \frac{1}{u} + 2 m_3 u + O(u^2)$ and $t(u,m)$ 
becomes the logarithmic solution $t(u,m)=\log(u)+ {\cal O}(u)$  at infinity $z_e=0$, 
which corresponds to $\frac{1}{j}\sim q \sim u^6$. Hence we get as the transcedental 
mirror map $u= Q_t- m_3 Q_t^3+ {\cal O}(Q_t^4)$, with $Q_t=(Q_e)^{1\over 6}=\sqrt{Q_1} Q_2 Q_3^{2\over 3} Q_4^{1\over 3}$.  
The non-transcendental rational mirror maps are 
\begin{equation} 
z_1=\frac{ Q_1}{(1 + Q_2)^2}, \ \ z_3=Q_3 \frac{1 + Q_4 + Q_3 Q_4}{(1 + Q_3 + Q_3 Q_4)^2},\ \ z_4=Q_4\frac{1 + Q_3 + Q_3 Q_4}{(1 + Q_4 + Q_3 Q_4)^2}\ . 
\label{mirrormapp10} 
\end{equation}

The existence of these rational solutions for the mirror maps can be 
proven from the system of differential equations that corresponds to the Mori 
vectors listed above. With the knowledge  of these rational solutions the 
system of differential equations can be reduced to one third order differential 
equation in $u$ parametrized by the $m_i$. 

Such rational solutions exist for the differential operators associated 
to Mori vectors describing  the linear relations of points on an 
(outer) edge of a toric diagram. One can understand their existence 
from the fact that this subsystem describes effectively a non-compact 
two-dimensional CY geometry, whose compact part is a Hirzebruch sphere 
tree, which has no non-trivial mirror maps.               

This defines the K\"ahler parameters of the $A$-model geometry and 
relates them to the $u,m_i$. This allows to extract the BPS invariants 
for this mass deformation of the $E_8$ curve.

\section{Local non-rigid geometries}
In contrast to the del Pezzo cases we now discuss local 
geometries which are movable inside the Calabi-Yau  space. We start with 
${\cal L}_1\oplus {\cal L}_2 \rightarrow \Sigma_g$ for $g>0$ and discuss 
special aspects of the $M$-string geometry.

\subsection{Local elliptic curves and quasi-modular forms}   

Recently, Aganagic et al. \cite{Aganagic:2012} have proposed the partition function for the refined topological string on a class of local Calabi-Yau manifolds which are given by two line bundles over a Riemann surface $\mathcal{L}_1\oplus \mathcal{L}_2 \rightarrow \Sigma_g$, where $g$ is the genus of the Riemann surface.
 The Calabi-Yau condition requires the sum of the degrees of the line bundles to be $2g-2$.
 In the case of genus one, i.e. the Riemann surface $\Sigma_g$ is an elliptic curve, we expect the partition function to have some modular properties, which were studied for the unrefined case in e.g. \cite{Dijkgraaf, Kaneko}. 

In the following we consider the Calabi-Yau manifold $\mathcal{O}(1)\oplus \mathcal{O}(-1) \rightarrow \Sigma_{g=1}$. The refined partition function in \cite{Aganagic:2012} simplifies to  
\begin{eqnarray} \label{ellipticref5.1}
Z (\epsilon_1,\epsilon_2,q)= \sum_{R} \exp[\frac{1}{2} (\epsilon_1||R||^2+\epsilon_2 ||R^T||^2)] q^{|R|}, 
\end{eqnarray}
where the sum is over all two-dimensional partitions $R=\{ R_1\geq R_2\geq \cdots\} $.
 Here $R^T$ is the transpose of the partition and the notations are $|R|=\sum_i R_i $, $||R||^2=\sum_i (R_i)^2 $.
 $q=e^{-t}$ is the exponential of the K\"ahler modulus and the small expansion parameter in the large volume limit. 

In the unrefined case $\epsilon\equiv \epsilon_1=-\epsilon_2$, the partition function (\ref{ellipticref5.1}) can be written as an infinite product 
\begin{eqnarray}
Z (\epsilon, -\epsilon, q) =  \textrm{Res}\, \frac{1}{2\pi i z}  \prod_{n=1}^{\infty}  (1-e^{\frac{(2n-1)^2\epsilon }{8}}q^{n-\frac{1}{2}} z)  (1-e^{-\frac{(2n-1)^2\epsilon }{8}} q^{n-\frac{1}{2}} z^{-1}) .
\end{eqnarray}
We can expand around the small parameter $\epsilon$ and compute the free energy for $g\geq 1$
\begin{eqnarray} \label{unrefelliptic5.3}
\log(Z (\epsilon, -\epsilon, q) ) = \sum_{g=1}^\infty \epsilon^{2(g-1)} F^{(g)}(q) . 
\end{eqnarray}  
It is easy to see that the genus one amplitude is given by $F^{(1)} = -\frac{1}{24} \log( \frac{\eta(q)^{24} }{q } )$. Here the refined partition function (\ref{ellipticref5.1}) only includes the world-sheet instanton contributions, and the  $\frac{1}{24} \log( q )$ piece is cancelled by the genus one perturbative contribution so that the total amplitude is modular.
 For higher genus $g>1$, it was proven in \cite{Kaneko} that the amplitudes  $F^{(g)}$ are quasi-modular forms of $SL(2,\mathbb{Z})$ of weight $6(g-1)$. 

The refined partition function (\ref{ellipticref5.1}) is symmetric under the exchange $\epsilon_1\leftrightarrow \epsilon_2$.
 However, unlike the cases of del Pezzo and half K3 Calabi-Yau manifolds studied in the previous sections, it is not symmetric under $\epsilon_{1,2}\rightarrow -\epsilon_{1,2}$.
 As a consequence, the refined Gopakumar-Vafa invariants do not fit in the full $SU(2)_L\times SU(2)_R$ representations of the five-dimensional little group, and they could be negative integers.
 We can still expand the logarithm $\log(Z (\epsilon_1, \epsilon_2, q) )$ as power series of $(\epsilon_1+\epsilon_2)^n(\epsilon_1\epsilon_2) ^{g-1}$.
 We find that the coefficients are generally not quasi-modular forms, except for the unrefined amplitudes $F^{(g)}$ in (\ref{unrefelliptic5.3}) and the even power terms in the Nekrasov-Shatashvili limit where one of $\epsilon_{1,2}$ parameters vanishes. 

More precisely, we can set $\epsilon_2=0$ and $\epsilon\equiv \epsilon_1$, and expand the free energy as 
\begin{eqnarray}
\log(Z (\epsilon, 0 , q) ) = \sum_{n=0}^\infty \epsilon^{n} G^{(n)}(q) . 
\end{eqnarray}  
Here the first term appears also in the unrefined case $G^{(0)}= F^{(1)}= -\frac{1}{24} \log( \frac{\eta(q)^{24} }{q } )$.
 We find that the higher order even terms $G^{(2n)}(q)$ for $n\geq 1$ are quasi-modular forms of weight $6n$. Some formulae at low order read as follows 
\begin{eqnarray}
G^{(2)} &=& \frac{E_2 E_4-E_6}{5760}, \nonumber \\
G^{(4)} &=& \frac{-2 E_2^3 E_6 +6 E_2^2 E_4^2-6 E_2 E_4 E_6 +E_4^3+E_6^2}{1990656},  \nonumber \\
G^{(6)} &=& \frac{1}{5733089280} [99 E_2^5 E_4^2  -495 E_2^4 E_4 E_6 +110 E_2^3 (5 E_4^3+4 E_6^2)-990 E_2^2 E_4^2 E_6 \nonumber \\ && 
+15 E_2 (13 E_4^4+20 E_4 E_6^2) -79 E_4^3 E_6-20 E_6^3] .
\end{eqnarray}

It turns out that the proof of  the quasi-modularity of the refined amplitudes $G^{(2n)}(q)$ in the Nekrasov-Shatashvili limit is much simpler than that  of \cite{Kaneko} for the unrefined case, and we can also find explicit formulae for them.
 First we note that there is also an infinite product formula for (\ref{ellipticref5.1}) in the Nekrasov-Shatashvili limit 
\begin{eqnarray} 
Z (\epsilon, 0 , q)  = \prod_{m=1}^\infty (1-e^{\frac{\epsilon^2}{2} m^2} q^m )^{-1} . 
\end{eqnarray} 
We can then compute the logarithm 
\begin{eqnarray} \label{log5.7}
\log(Z (\epsilon, 0 , q)) = \sum_{m=1}^\infty   \sum_{k=1}^\infty  \frac{e^{\frac{\epsilon}{2} km^2}q^{km}}{k} 
= \sum _ {l=0}^{\infty} \frac{\epsilon^l}{2^l l!} \sum_{m=1}^\infty \sum_{k=1}^\infty k^{l-1} m^{2l} q^{km } . 
\end{eqnarray}
On the other hand, we have the well-known Eisenstein series expansion
\begin{eqnarray}
E_{2n}(q)  = 1-\frac{4n}{B_{2n}} \sum_{m=1}^{\infty}  \sum_{k=1}^{\infty}  m^{2n-1} q^{km}. 
\end{eqnarray} 
It is now easy to check the explicit formulae for the even power terms in (\ref{log5.7}) 
\begin{eqnarray}
G^{(2n)} =-\frac{B_{2n+2}}{2^{2n+2} (n+1) (2n)!} (q \frac{d}{d q})^{2n-1}E_{2n+2}(q) \, .
\label{NSlocalelliptic}
\end{eqnarray} 
Since the derivative $q\frac{d}{d q}$ preserves the quasi-modularity and increases the modular weight by 2, we see that $G^{(2n)}$ is a quasi-modular form of weight $6n$.

\subsection{Modularity in M-Strings and refined constant map contributions}

Recently, the elliptic genus of ``M-Strings", constructed by M2-branes suspended between adjacent parallel M5-branes, has been studied in \cite{Haghighat:2013gba}.  This is related to the partition function of five-dimensional $N=2^*$ $SU(N)$ gauge theory, which is geometrically engineered by topological strings on certain elliptic Calabi-Yau manifolds, given by a deformation of an $A_{N-1}$-fibration over $T^2$.  The topological partition function can be computed by the refined topological vertex formalism \cite{IKV}. There are two choices of the preferred direction of the vertices, along the vertical edge or the horizotal edge in the toric diagrams of the Calabi-Yau manifolds. The calculations based on the different choices of the preferred direction should give the same topological partition function, and the equivalence often provides non-trivial identities involving Macdonald polynomials \cite{Iqbal:2008ra}. 

\begin{figure}
\centering
\includegraphics[angle=0,width=0.4\textwidth]{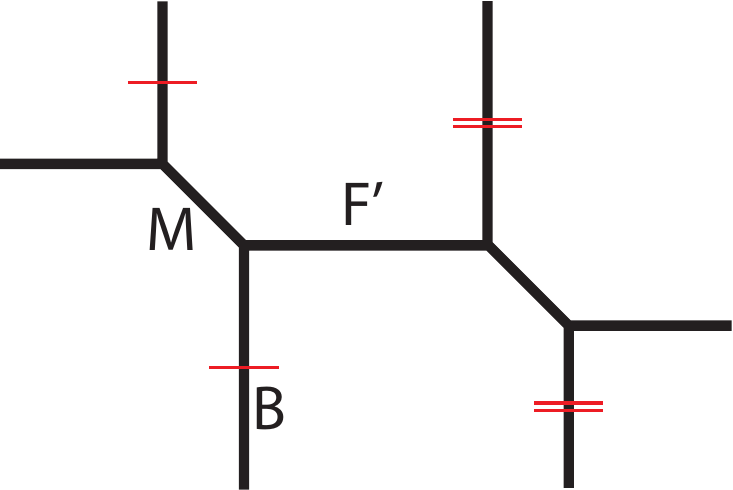} 
\begin{quote} 
\caption{The toric diagram for the five-dimensional ${N}=2^*$ $SU(2)$ gauge theory.} \label{SU2} 
\end{quote} 
\end{figure} 
\vspace{0cm}
Here we consider the case of $SU(2)$ gauge theory. The toric diagram is provided in \cite{Haghighat:2013gba} and displayed here in Figure \ref{SU2}.  We see that there are three curve classes $M, B, F^\prime$ in the geometry. It is convenient to change basis 
\begin{eqnarray}
F = F^\prime + M, ~~~~~ E = B + M, 
\end{eqnarray}  
where $F$ and $E$ denote respectively the fiber and the elliptic curve. In the new basis, the coefficient $r$ in the integral homology class $kF+lE+rM$ can be a negative integer, and the refined BPS invariants may be non-vanishing only  in the range $ |r| \leq k+l$.   

We are interested in the modularity property of the elliptic curve class $E$. We will see that the modularity condition in the class $E$ provides an approach to fix the refinement of the constant map contributions of Gromov-Witten theory. In \cite{Haghighat:2013gba} it has been found that the modularity is manifest if one chooses the preferred direction of the refined vertex to be the horizontal class $F$. The refined BPS invariants for the curve $kF+lE+rM$ with $k>0$ are computed there. 

For the case of zero class $F$, i.e. $k=0$, we can choose the vertical edge as preferred direction and set $Q_f=0$ in the $SU(2)$ partition function in  \cite{Haghighat:2013gba}, which becomes 
\begin{eqnarray}
Z_{inst} =
\big{[}  Z^{(1)} \big{]}^2. 
\end{eqnarray}
Here the partition function factorizes as square of $Z^{(1)}$, which is the partition function of a single M5-brane wrapped on a circle with mass deformation $m$ 
\begin{eqnarray}
Z^{(1)} =\sum_{\nu }Q_{\tau}^{|\nu|} \prod_{(i,j)\in\nu } \frac{(1-Q_{m}\,q^{\nu_{j}^{t}-i+\frac{1}{2}}t^{\nu_{i}-j+\frac{1}{2}})(1-Q_{m}^{-1}\,q^{\nu_{j}^{t}-i+\frac{1}{2}}t^{\nu_{i}-j+\frac{1}{2}})}{(1-\,q^{\nu_{j}^{t}-i}t^{\nu_{i}-j+1})(1-\,t^{\nu_{i}-j}q^{\nu_{j}^{t}-i+1})}, 
\end{eqnarray}
where $Q_\tau$ and $Q_m$ are the corresponding exponentials of the K\"ahler parameters of the classes $E$ and $M$. We are summing over all two-dimensional partitions $\nu$ and the refined parameters are defined by $q=e^{\epsilon_1}, t=e^{-\epsilon_2}$. A different form of the topological partition function, more convenient for extracting the BPS invariants, was given in \cite{Iqbal:2008ra}. 

 We extract the refined BPS invariants from the free energy $F_{inst} = \rm{Log}(Z_{inst})$ 
\begin{eqnarray}
F_{inst} &=& -\sum_{j_L,j_R}\sum_{l,r}\sum_{m=1}^{\infty} 
\frac{(-1)^{2j_L+2j_R}n^{lE+rM }_{j_L,j_R}}{m (q^{\frac{m}{2}} - q^{-\frac{m}{2}})(t^{\frac{m}{2}} - t^{-\frac{m}{2}})} (Q_\tau^l Q_m^r)^m  \nonumber \\ && \cdot
\frac{(tq)^{-mj_L} - (tq)^{mj_L+m} }{1-(tq)^m} \frac{(\frac{t}{q})^{-mj_R} - (\frac{t}{q})^{mj_R+m} }{1-(\frac{t}{q})^m}, 
\end{eqnarray}
and find that the non-vanishing invariants are the following ones 
\begin{eqnarray} \label{refineBPS.5.72}
n^{lE+M}_{(0,0)}=n^{lE-M}_{(0,0)}=n^{lE}_{(\frac{1}{2},0)}=1, ~~~l=1,2,3,\cdots
\end{eqnarray}
We check the refined BPS invariants to some finite degree $l$, but it may be possible to prove the formulas for all degrees from some ingenious identities of Young diagrams. 

We further restrict to the zero mass deformation, i.e. $Q_m=1$, and study the modularity in only the $E$-direction. First we consider the unrefined case $q=t$. The BPS invariants in the integer basis $I_r=([\frac{1}{2}]+2[0])^r$ are 
\begin{eqnarray} \label{unrefineBPS.5.73}
\tilde{n}^{lE}_0=0 , ~~~ \tilde{n}^{lE}_1=1 , ~~~ l=1,2,\cdots
\end{eqnarray}
The topological free energy is expanded as $F_{inst} =  \sum_{g=0}^\infty \epsilon^{2g-2} F_{inst}^{(g)}$. For simplicity we consider the modularity of higher genus amplitudes $F^{(g)}$ with $g\geq 2$, which would have positive modular weights. It is straightforward to compute the higher genus amplitudes from the BPS invariants (\ref{unrefineBPS.5.73}) as 
\begin{eqnarray}
F_{inst}^{(g)}= \tilde{n}^{lE}_0  \frac{B_{2g}}{2g(2g-2)!} \sum_{d=1}^\infty \textrm{Li}_{3-2g} (Q_\tau^d), ~~~~ g\geq 2.
\end{eqnarray}
Here only the invariants $\tilde{n}^{lE}_0$ contribute to the higher genus amplitudes, while $\tilde{n}^{lE}_1=1$ only contribute to the genus one amplitude. The appearance of the polylogarithm function is due to the multi-covering contributions of the BPS invariants, and the Bernoulli numbers $B_{2g}$ come from the familiar expansion 
\begin{eqnarray}
\frac{1}{(2\sinh(\frac{\epsilon}{2}))^2} = \frac{1}{\epsilon^2} - \sum_{g=1}^{\infty} \frac{B_{2g}}{2g(2g-2)!}\epsilon^{2g-2}.
\end{eqnarray}

The well known constant map contribution in Gromov-Witten theory is 
\begin{eqnarray} \label{constantmap.5.76}
C_g = \chi \frac{B_{2g} {B_{2g-2}} }{8g(g-1) (2g-2)!} ,
\end{eqnarray} 
where $\chi$ is the regularized Euler number of the non-compact A-model Calabi-Yau manifold. We see that if we use a regularized Euler number $\chi=\tilde{n}^{lE}_0$, the classical constant map contribution and the instanton contribution can nicely combine into Eisenstein series 
\begin{eqnarray}
C_g + F_{inst}^{(g)} = \chi \frac{B_{2g} {B_{2g-2}} }{8g(g-1) (2g-2)!} E_{2g-2}(Q_\tau) . 
\end{eqnarray}
Here the case $\chi=0$ is somewhat special, but for more general elliptic Calabi-Yau manifolds with non-vanishing Euler character, this is the mechanism for the quasi-modularity of the topological string amplitudes of zero base class.

We should expect that the nice modularity property generalizes to the refined theory. It turns out to provide a proposal for the refinement of the constant map contribution (\ref{constantmap.5.76}), which is different from the one in \cite{IKV}. First we note the following expansion 
\begin{eqnarray}
\frac{1}{(q^{\frac{1}{2}} - q^{-\frac{1}{2}})(t^{\frac{1}{2}} - t^{-\frac{1}{2}})} = 
-\frac{1}{\epsilon_1\epsilon_2} +\frac{\epsilon_1^2+ \epsilon_2^2}{24\epsilon_1\epsilon_2} - \sum_{g_1+g_2\geq 2}^\infty   b_{g_1} b_{g_2} \epsilon_1^{2g_1-1} \epsilon_2^{2g_2-1},
\end{eqnarray} 
where the coefficients are given as $b_g =  \frac{2^{2g-1}-1}{2^{2g-1}}\frac{B_{2g}}{(2g)!}$. Then the proposal in \cite{IKV} for the generating function for the higher genus refined constant map contributions is 
\begin{eqnarray}
C = \chi  \sum_{g_1+g_2\geq 2}^\infty  \frac{B_{2g_1+2g_2-2} }{4(g_1+g_2-1)} b_{g_1} b_{g_2} \epsilon_1^{2g_1-1} \epsilon_2^{2g_2-1}. 
\end{eqnarray}
Suppose the only non-vanishing refined BPS invariants in the $E$ direction were $n^{lE}_{0,0} = -\chi$, then we could easily see that  the classical and instanton contributions for the higher genus refined amplitudes would exactly combine into Eisenstein series, similarly as in the unrefined case. 

However this is not the case here, as we see in (\ref{refineBPS.5.72}) that the non-vanishing refined BPS invariants are actually $n^{lE}_{0,0} =2$ and $n^{lE}_{\frac{1}{2},0} =1$ in the zero mass limit. We can expand the single cover contribution of the BPS states 
\begin{eqnarray}
\frac{(tq)^{-\frac{1}{2}}+ (tq)^{\frac{1}{2}} -2 }{(q^{\frac{1}{2}} - q^{-\frac{1}{2}})(t^{\frac{1}{2}} - t^{-\frac{1}{2}})} &=& 
- \frac{(\epsilon_1-\epsilon_2)^2}{4\epsilon_1\epsilon_2} + \frac{(\epsilon_1^2-\epsilon_2^2)^2 } { 192 \epsilon_1 \epsilon_2} - \frac{1}{7680 \epsilon_1 \epsilon_2} (\epsilon_1^2-\epsilon_2^2)^2 (\epsilon_1^2+ \epsilon_2^2)
 \nonumber \\ && 
 +\frac{1}{15482880 \epsilon_1 \epsilon_2} (\epsilon_1^2-\epsilon_2^2)^2(51\epsilon_1^4+58 \epsilon_1^2\epsilon_2^2 + 51\epsilon_2^4)  + \mathcal{O}(\epsilon^8)  . \nonumber 
\end{eqnarray}
We propose the following generating function for the higher genus ($g_1+g_2\geq 2$) refined constant map contributions
\begin{eqnarray}
C &=&  -\frac{1}{2} [ \frac{B_2}{2}   \frac{(\epsilon_1^2-\epsilon_2^2)^2 } { 192 \epsilon_1 \epsilon_2}  - \frac{B_4}{4}   \frac{1}{7680 \epsilon_1 \epsilon_2} (\epsilon_1^2-\epsilon_2^2)^2 (\epsilon_1^2+ \epsilon_2^2)   \nonumber \\ &&
 +\frac{B_6}{6}  \frac{1}{15482880 \epsilon_1 \epsilon_2} (\epsilon_1^2-\epsilon_2^2)^2(51\epsilon_1^4+58 \epsilon_1^2\epsilon_2^2 + 51\epsilon_2^4)  + \mathcal{O}(\epsilon^8) ],
\end{eqnarray}
where we add an overall factor of $-\frac{1}{2}$ and the factor $\frac{B_{2g-2}}{2g-2}$ for the $\epsilon^{2g-2}$ power term. Our proposal for the refined constant map contributions can combine with instanton contributions into Eisenstein series in the present case. Surprisingly, the refined constant map contribution is not proportional to the Euler character, which is effectively zero here and would imply vanishing constant map contributions. As a result, the formula presented here is not universal for general Calabi-Yau manifolds and has to be determined case by case.

\section{The half K3 surface: massless theory}

We consider the local non-compact Calabi-Yau threefold, constructed by the canonical line bundle over the half K3 surface, similarly as for the del Pezzo surfaces.
The half K3 surface can be embedded in an elliptic fibration over a Hirzebruch surface. In our previous paper \cite{HK2010}, we studied the refined BPS invariants for the homology classes $p+df$ in $H_2(\mathcal{B}_9,\mathbb{Z})$ , where we wrap the 2-brane once around the base $p$ and $d$ times around the fiber $f$.
Here we will further provide the formulae for refined topological string amplitudes in terms of modular forms, and give a refinement of the modular anomaly equation in \cite{HST}. The refined modular anomaly equation enables us to compute the refined GV invariants for the more general two-parameter classes $n_bp+df$.

The topological string amplitudes on the half K3 Calabi-Yau threefold are equivalent to the partition function of the six-dimensional non-critical E-string compactified on a circle.
The winding and momentum  numbers of the E-string on the compactified circle correspond to the wrapping numbers $n_b$ and $d$ on the base and fiber in the homology classes $n_bp+df$ in the half K3 surface.  

Recalling the results about intersection form of the homology lattice of the $\frac{1}{2}K3$ from the discussion in \ref{delpezzohalfk3}, one finds that the base class $p$ and the fiber class $f$
\begin{eqnarray} \label{basefiber}
p=[e_9],~~~~~ f= 3[l]-\sum_{i=1}^9 [e_i]
\end{eqnarray}
intersect as
\begin{eqnarray}
p\cdot p=-1, ~~~p\cdot f=1, ~~~f\cdot f=0.
\end{eqnarray} 

In addition to the  classes $p$ and $f$, we can turn on eight additional mass parameters corresponding to the other homology classes in $H_2(\mathcal{B}_9,\mathbb{Z})$. The Seiberg-Witten curve description of the prepotential was studied in \cite{Minahan:1997a,Eguchi:2002}.
 The higher genus  topological string amplitudes without refinement have been studied e.g. in \cite{Hosono:2002, Sakai:2011}.  The amplitudes have an $E_8$ symmetry with respect to the eight mass parameters and can be constructed from the theta functions of the $E_8$ lattice, which reduce to modular forms of $SL(2,\mathbb{Z})$ in the massless limit.

In this section we first review the  refined G\"{o}ttsche formula in subsection \ref{sec.refinedGottsche}, which is an essential ingredient for the refined topological string amplitudes with wrapping number on the base $n_b=1$.
 In the subsections \ref{sec.nb=1} and \ref{sec.nb>1} we study the massless theory for cases of wrapping number $n_b=1$ and $n_b>1$, where the (refined) topological string amplitudes can be written in terms of quasi-modular forms of $SL(2,\mathbb{Z})$.  In the next section 
we discuss the refined amplitudes with non-vanishing $E_8$ mass parameters.

\subsection{The refined G\"{o}ttsche formula} \label{sec.refinedGottsche}
The G\"{o}ttsche formula \cite{Gottsche} is the generating function for the Betti numbers of the Hilbert scheme of $d$ points on a complex surface $S$. The refinement of the formula appears in  \cite{HST}. We consider the case of the surface $S$ with Betti numbers  $b_0(S)=1, b_1(S)=0$, which is the case for both, the K3 and the half K3 surfaces.
 The inverse of the refined formula is an infinite product
\begin{eqnarray} \label{Gottsche3.3}
\frac{1}{G^{S} (q, y_L,y_R) } &=& \prod_{n=1}^{\infty} (1- y_L y_Rq^n) (1- y_L y_R^{-1}q^n) (1- y_L^{-1} y_Rq^n) 
  (1- y_L^{-1} y_R^{-1} q^n) (1-q^n) ^{b_2(S)-2}
  \nonumber \\ &=& 
  \prod_{n=1}^{\infty} (1- e^{\epsilon_1}q^n) (1- e^{-\epsilon_1}q^n) (1- e^{\epsilon_2}q^n)  (1- e^{-\epsilon_2} q^n) (1-q^n)  ^{b_2(S)-2} \, ,
\end{eqnarray}
where we use the notation $y_{R,L}=\exp(\frac{\epsilon_1\pm \epsilon_2}{2})$.

The refined product $G^{S} (q, y_L,y_R)$ is the generating function for the refinement of the Betti numbers of the Hilbert scheme in the representations of  two $SU(2)$ Lefschetz actions 
\begin{eqnarray} \label{Gottsche3.4}
G^{S} (q, y_L,y_R) =  \sum_{d=0}^{\infty}  \sum_{j_L,j_R}  (-1)^{2j_L+2j_R} n^{d}_{j_L,j_R} 
(\sum_{k=-j_L}^{j_L} y_L^{2k} ) (\sum_{k=-j_R}^{j_R} y_R^{2k} ) q^d. 
\end{eqnarray}
Here  $j_L, j_R$ are non-negative half integers labeling the $SU(2)_L\times SU(2)_R$ representations and  we denote  $n^d_{j_L,j_R}$ as the refined Betti numbers of the Hilbert scheme of $d$ points on $S$. 

We can compute the refined Betti numbers from (\ref{Gottsche3.3}), (\ref{Gottsche3.4}) by plugging in the number for $b_2(S)$. We list the refined numbers for the K3 surface ($b_2=22$) and the half K3 surface $\mathcal{B}_9$ ($b_2=10$) in tables \ref{tableBettiK3} and \ref{tableBettihalfK3}.
 The cases without refinement have been studied in \cite{KKV} for the K3 surface and in \cite{HST, Hosono:2002} for the half K3 surface. The numbers in tables \ref{tableBettiK3} and \ref{tableBettihalfK3} reduce to the previous results in the unrefined limit. 

More generally in the topological string models, we may multiply the G\"{o}ttsche product by a modular form of a subgroup of $SL(2,\mathbb{Z})$ and extract the corresponding BPS numbers $n^d_{j_L,j_R}$.
 We mentioned that since the refined BPS invariants count the multiplicity of BPS states without sign, they should be non-negative integers.
 However we find that this constraint is not very strong, and it is satisfied for a large class of models, e.g. when $b_2\geq 2$ and the modular form is a power of the $E_4$ Eisenstein series.
 On the other hand, some BPS numbers become negative when we include the $E_6$ series in the modular form.  This is because the coefficients in the $E_4$ series are positive while they are negative in the $E_6$ series.
 We will see that the topological string amplitudes on the local K3 Calabi-Yau model correspond to the choice of the $E_4$ modular form as the pre-factor in the G\"{o}ttsche product.
 We speculate that the other choices which give non-negative integers for the refined BPS 
numbers $n^d_{j_
L,j_R}$ are possible candidates for consistent refined topological string models.

\subsection{Wrapping number on the base $n_b=1$}  \label{sec.nb=1}
We now consider refined topological string theory on the local half K3 Calabi-Yau, with the wrapping number for the base class $p$ fixed as one.
 In this case there are no multiple cover contributions within the classes $p+df$.
 Therefore the genus zero Gromov-Witten invariants are the same as the Gopakumar-Vafa invariants $n_0^{p+df}$.
 The generating function \cite{Klemm:1996hh}  is known to be
\begin{eqnarray} \label{genuszero3.1}
\sum_{d=0}^{+\infty} n_0^{p+df} q^d = \frac{q^{\frac{1}{2}}E_4(q)}{\eta^{12}(q)}.
\end{eqnarray}

The higher genus refined GV invariants can be obtained from a generating function $\mathcal{G}(\epsilon_1,\epsilon_2,q)$ \cite{HK2010}, defined by the product of the genus zero generating function (\ref{genuszero3.1}) and the refined G\"{o}ttsche formula in (\ref{Gottsche3.3}) as follows
\begin{eqnarray}  \label{generate3.2}
\mathcal{G}(\epsilon_1,\epsilon_2,q) = E_4(q) G^{\mathcal{B}_9} (q, y_L,y_R) .
\end{eqnarray}

We see that the formula for the generating function $\mathcal{G}(\epsilon_1,\epsilon_2,q)$ is an A-model expression.
  The exponents of $q$ in $\mathcal{G}(\epsilon_1,\epsilon_2,q)$ count the self-intersection numbers of the second cohomology classes.
 In our case the class $p+df$ corresponds to the term $q^d$.
 On the other hand, similar to the formula for the refined Betti numbers in (\ref{Gottsche3.4}), the generating function $\mathcal{G}(\epsilon_1,\epsilon_2,q)$ can be also expressed in terms of the refined Gopakumar-Vafa invariants 

\begin{eqnarray} \label{GV3.4}
\mathcal{G}(\epsilon_1,\epsilon_2,q) =  \sum_{d=0}^{\infty}  \sum_{j_L,j_R}  (-1)^{2j_L+2j_R} n^{p+df}_{j_L,j_R} 
(\sum_{k=-j_L}^{j_L} y_L^{2k} ) (\sum_{k=-j_R}^{j_R} y_R^{2k} ) q^d, 
\end{eqnarray}
where $j_L,j_R$ take values in the non-negative half integers.
 
One can extract the refined GV invariants from the equations (\ref{Gottsche3.3}, \ref{generate3.2}, \ref{GV3.4}) with the Betti number for half K3 surface $b_2(\mathcal{B}_9)=10$. This was done in our previous paper \cite{HK2010}. Here we also list the integers in table \ref{tableB9} for completeness.
 We note that the refined GV invariants for the class $p+f$ are the same as the refined GV invariants for the del Pezzo $E_8$ for degree $d=1$ displayed in the table in section \ref{e8delpezzo}.
 This is expected since the  $E_8$ is a sub-family in the half K3, and the class $p+f$ in the half K3 corresponds to the class of degree $d=1$ in the del Pezzo $E_8$.

The (refined) topological string amplitudes can be expressed in terms of the (refined) GV invariants as explained in (\ref{productrefined}).
Due to multiple cover contributions, in general there is no simple relation to compute the topological string amplitude generating function $F(\epsilon_1,\epsilon_2,q)$ in (\ref{topo2.1}) directly from the G\"{o}ttsche generating function $\mathcal{G}(\epsilon_1,\epsilon_2,q)$, and one has to first compute the (refined) GV invariants as an intermediate step.
 However, in our case since we only consider the topological string amplitudes for the classes $p+df$ and there is no multiple cover contribution, we can write down a simple relation between the two generating functions 
\begin{eqnarray} \label{rela3.5}
 F(\epsilon_1,\epsilon_2,q) = \frac{\mathcal{G}(\epsilon_1,\epsilon_2,q) }{4\sinh(\frac{\epsilon_1}{2})\sinh(\frac{\epsilon_2}{2})} .
\end{eqnarray}
  
It turns out that the higher genus refined topological string amplitudes can be expressed as quasi-modular forms of $SL(2,\mathbb{Z})$.
 To show this, we first provide an useful identity for the refined G\"{o}ttsche  formula which is the refinement of an identity mentioned in \cite{HST}, 
\begin{eqnarray} \label{identity3.6}
&& \frac{\epsilon_1\epsilon_2}{4\sinh(\frac{\epsilon_1}{2})\sinh(\frac{\epsilon_2}{2})} 
\prod_{n=1}^{\infty} \frac{(1-q^n) ^4}{ (1- e^{\epsilon_1}q^n) (1- e^{-\epsilon_1}q^n) (1- e^{\epsilon_2}q^n)  (1- e^{-\epsilon_2} q^n) } \nonumber \\
& = &  \exp [ -\sum_{k=1}^{\infty} \frac{B_{2k}}{2k(2k)!} (\epsilon_1^{2k} + \epsilon_2^{2k}) E_{2k} (q) ]. 
\end{eqnarray}
It is a straightforward exercise to demonstrate this identity by taking the logarithm on both sides and using the well-known formula for the Eisenstein series $E_{2k}(q) = 1-\frac{4k}{B_{2k}}\sum_{n=1}^{\infty} \frac{n^{2k-1}q^n}{1-q^n}$.  

Utilizing the identity (\ref{identity3.6}) and the relation (\ref{rela3.5}), we can write the topological string generating function for the class $p+df$ as 
\begin{eqnarray}  \label{above3.7}
F(\epsilon_1,\epsilon_2,q) =  \frac{q^{\frac{1}{2}}E_4(q)}{\eta^{12}(q)} \frac{1}{\epsilon_1\epsilon_2} \exp [ -\sum_{k=1}^{\infty} \frac{B_{2k}}{2k(2k)!} (\epsilon_1^{2k} + \epsilon_2^{2k}) E_{2k} (q) ]. 
\end{eqnarray}
The Eisenstein series $E_{2k}$ for $k\geq 2$ are modular forms of $SL(2,\mathbb{Z})$ and can be written as polynomials of $E_4$ and $E_6$.
   $E_{2}$ is quasi-modular, and we can easily see the modular anomaly equation for $F(\epsilon_1,\epsilon_2,q)$ with respect to $E_{2}$ from the above formula,
\begin{eqnarray} \label{moduano3.8}
\partial_{E_2} \log(F(\epsilon_1,\epsilon_2,q)) = -\frac{\epsilon_1^{2} + \epsilon_2^{2}}{24},
\end{eqnarray}
where we have used the Bernoulli number $B_2=\frac{1}{6}$.
 Our modular anomaly equation provides a refinement for the modular anomaly equation in \cite{HST} for the class $p+df$.\footnote{Our convention has a factor of three difference comparing to that of \cite{HST}.} 

We can decompose the topological string generating function as in (\ref{topo2.1}), 
\begin{eqnarray}  
F(\epsilon_1,\epsilon_2,q)= \sum_{n,g=0}^{+\infty} (\epsilon_1+\epsilon_2)^{2n}(\epsilon_1\epsilon_2)^{g-1}
 F^{(n,g)}(q),
\end{eqnarray}
where $F^{(0,g)}$ are the conventional unrefined amplitudes. It is easy to see from (\ref{above3.7}) that the genus zero amplitude is 
\begin{eqnarray}
F^{(0,0)}(q)=  \frac{q^{\frac{1}{2}}E_4(q)}{\eta^{12}(q)}
\end{eqnarray}
and the higher genus amplitudes divided by the genus zero amplitude $\frac{F^{(n,g)}(q)}{F^{(0,0)}(q) }$ are quasi-modular forms of weight $2(n+g)$. We can write down some explicit formulae at low genus 
\begin{eqnarray} \label{lowgenus}
&& F^{(0,1)} =\frac{E_2}{12} F^{(0,0)}, ~~~  F^{(1,0)} =- \frac{E_2}{24} F^{(0,0)},  
~~~  F^{(0,2)} =\frac{5E_2^2+E_4}{1440} F^{(0,0)},  \nonumber \\&&  
F^{(1,1)} =- \frac{5E_2^2+2E_4}{1440} F^{(0,0)},~~~
F^{(2,0)} =\frac{5E_2^2+2E_4}{57600} F^{(0,0)}  . 
\end{eqnarray} 
The modular anomaly equation (\ref{moduano3.8}) can be written more explicitly for the higher genus amplitudes as
\begin{eqnarray} \label{holoone3.11}
\partial_{E_2} F^{(n,g)} = \frac{1}{12}F^{(n,g-1)} -\frac{1}{24}F^{(n-1,g)} ,
\end{eqnarray}
where the term $F^{(n,g-1)}$ is defined to be zero if $g=0$, and similarly for $F^{(n-1,g)}$ if $n=0$. We note that this modular anomaly equation seems quite different from the one we used before (\ref{holo2.11}). It would be interesting to elucidate the connection.

\subsection{Wrapping number on the base $n_b>1$}   \label{sec.nb>1}
We propose a refinement of the HST modular anomaly equation in \cite{HST} as follows 
\begin{eqnarray} \label{refinedmodular}
\partial_{E_2} F^{(n,g,n_b)} &=& \frac{1}{24} \sum_{n_1=0}^n\sum_{g_1=0}^g \sum_{s=1}^{n_b-1} s(n_b-s) 
F^{(n_1,g_1,s)} F^{(n-n_1,g-g_1,n_b-s)} \nonumber \\
&& +\frac{n_b(n_b+1)}{24} F^{(n,g-1,n_b)} - \frac{n_b}{24} F^{(n-1,g,n_b)} .
\end{eqnarray}
Here $F^{(n,g,n_b)}$ are the refined topological string amplitudes with wrapping number $n_b$ on the base, as appearing in the generating function 
\begin{eqnarray} \label{generate}
F=\sum_{n,g,n_b=0}^{\infty} (\epsilon_1+\epsilon_2)^{2n} (\epsilon_1\epsilon_2)^{g-1} e^{2\pi i t_b n_b} F^{(n,g,n_b)}(q),
\end{eqnarray}
where $t_b$ is the K\"ahler modulus of the base $\mathbb{P}^1$, and $q=e^{2\pi i \tau}$ is the modulus of the fiber. It is conjectured that $(\frac{\eta(q)^{12}}{\sqrt{q}})^{n_b}F^{(n,g,n_b)}(q)$ are quasi-modular forms of $SL(2,\mathbb{Z})$ of weight $6n_b+2(g+n)-2$, i.
e.  polynomials of the Eisenstein series $E_2(q), E_4(q), E_6(q)$, so that the partial derivative with respect to $E_2$ is well defined. 

Our proposal (\ref{refinedmodular}) reduces to the HST modular anomaly equation \cite{HST} in the unrefined case of  $n=0$, which can be further reduced by setting $g=0$ to the genus zero equation studied earlier in \cite{Minahan:1997}. The generalizations to more elliptic Calabi-Yau manifolds have been studied recently in \cite{Alim:2012, Klemm:2012}. The genus zero prepotential with base wrapping number $n_b$ is also equivalent to the partition function of topological ${N}=4$ $U(n_b)$ super Yang-Mills theory \cite{Minahan:1998}.
 Another special case is the case of wrapping number $n_b=1$ studied in the previous subsection \ref{sec.nb=1}, where our proposal becomes the equation (\ref{holoone3.11}) derived there. We have guessed the factor $ \frac{n_b}{24} $ in the last term in (\ref{refinedmodular}) by trials and tested by higher genus calculations, which we now discuss.

We can solve the higher genus refined amplitudes recursively in $n,g,n_b$ by integrating the refined holomorphic anomaly equation.
 The integration constant is a polynomial of $E_4(q)$ and $E_6(q)$, and needs to be fixed by some boundary conditions. Here we will not use the gap conditions at the conifold locus, because it would require a careful analysis of the moduli space, which is quite complicated in multi-parameter models.
 Instead, we will utilize the boundary conditions due to vanishing GV invariants. We have mentioned before that if a GV integer vanishes 
$\tilde{n}^{\beta}_{g_L,g_R}=0$, then the higher genus neighbors also vanish $\tilde{n}^{\beta}_{g_L+1,g_R}=\tilde{n}^{\beta}_{g_L,g_R+1}=0$.
  In addition, at genus zero it is also known \cite{Klemm:1996hh} that the GV invariants vanish  $\tilde{n}^{n_bp+df}_{0,0}=0$ if $n_b>d$, except for the case of $n_b=1,d=0$ where we have $\tilde{n}^{p}_{0,0}=1$. The property was used in \cite{Minahan:1997} to fix the integration constants for the genus zero modular anomaly equation. 

In the special case of $n_b=1$ studied in the previous subsection  \ref{sec.nb=1}, the integration of the modular anomaly equation by fixing the integration constants by vanishing GV invariants leads to a nice A-model type formula (\ref{rela3.5}), related to the refined G\"{o}ttsche formula (\ref{Gottsche3.3}).
 The A-model type formulae are exact to all genera, and are rather convenient for extracting the complete GV invariants $n^\beta_{j_L,j_R}$  for a fixed homology class $\beta$.
 Here in the general case with wrapping number $n_b>1$, the A-model type formula is currently not available and therefore B-model calculations need to be done in order  to extract the BPS invariants recursively. 

We are able to compute the refined topological string amplitudes to some high genus $n+g$ and some high wrapping number $n_b$, and we can extract the complete refined GV invariants to some high degree for the base $n_b$ and fiber.
 We provide some refined amplitudes $F^{(n,g,n_b)}$ for low genus $n+g$ and wrapping number $n_b\geq 2$. The genus zero amplitudes \cite{Minahan:1997} for $n_b=2,3$ are 
\begin{eqnarray}
&& F^{(0,0,2)}  = \frac{q  }{\eta(q)^{24}}  \frac{ E_4 }{24}  (E_2 E_4+2 E_6),  \label{masslessformulae1} \\
&& F^{(0,0,3)} =  \frac{q^{\frac{3}{2}} }{ \eta(q)^{36}}  \frac{ E_4 }{15552 }    (54 E_2^2 E_4^2+216 E_2 E_4 E_6+109 E_4^3+197 E_6^2).  \nonumber 
\end{eqnarray}
The genus one amplitudes for $n_b=2,3$ are 
\begin{eqnarray}
&& F^{(0,1,2)}  = \frac{q  }{\eta(q)^{24}}  \frac{1}{1152} (10 E_2^2 E_4^2+9 E_4^3+24 E_2 E_4 E_6+5 E_6^2), 
 \label{masslessformulae2}  \\ 
&& F^{(1,0,2)}  = - \frac{q  }{\eta(q)^{24}}  \frac{1}{1152} (4 E_2^2 E_4^2+7 E_4^3+8 E_2 E_4 E_6+ 5 E_6^2), 
\nonumber \\ 
&& F^{(0,1,3)} =  \frac{q^{\frac{3}{2}} }{ \eta(q)^{36}}  \frac{ 78 E_2^3 E_4^3+299 E_2 E_4^4+360 E_2^2 E_4^2 E_6+472 E_4^3 E_6+439 E_2 E_4 E_6^2+80 E_6^3 }{62208}    ,   \nonumber \\ 
&& F^{(1,0,3)} =  - \frac{q^{\frac{3}{2}} }{ \eta(q)^{36}}  \frac{ 54 E_2^3 E_4^3+ 235 E_2 E_4^4 +216 E_2^2 E_4^2 E_6+776 E_4^3 E_6+287 E_2 E_4 E_6^2+160 E_6^3 }{124416}  .   \nonumber  
\end{eqnarray}
The genus two amplitudes for $n_b=2$ are 
\begin{eqnarray}
&& F^{(0,2,2)}  = \frac{q  }{\eta(q)^{24}}  \frac{190 E_2^3 E_4^2+417 E_2 E_4^3+540 E_2^2 E_4 E_6+356 E_4^2 E_6+225 E_2 E_6^2}{207360 }, 
\nonumber  \\
&& F^{(1,1,2)}  = - \frac{q  }{\eta(q)^{24}}  \frac{25 E_2^3 E_4^2+79 E_2 E_4^3+60 E_2^2 E_4 E_6+122 E_4^2 E_6+50 E_2 E_6^2}{34560}, 
\nonumber \\ 
&& F^{(2,0,2)}  =  \frac{q  }{\eta(q)^{24}}  \frac{10 E_2^3 E_4^2+37 E_2 E_4^3+20 E_2^2 E_4 E_6+76 E_4^2 E_6+25 E_2 E_6^2}{69120}.  \label{masslessformulae3}
\end{eqnarray}

We list some results for refined BPS invariants for $n_b>1$ in the tables \ref{tableB9n_b=2} and \ref{tableB9n_b=3,4}. The diagonal classes $n_b(p+f)$ correspond to the homology classes with degrees $d=n_b$ in the one-parameter $E_8$ del Pezzo model, and we check the matching of the corresponding BPS integers for the diagonal classes $n_b(p+f)$ with those in section \ref{e8delpezzo}. 
Our results demonstrate the compatibility of the refined version of the HST modular anomaly equation we proposed in (\ref{refinedmodular}) with the refined version of the BCOV holomorphic anomaly equation in  (\ref{holo2.11}).

In two special cases, namely the genus zero case $n+g=0$ and the case of wrapping number $n_b=1$, it is clear that one can in principle solve the refined topological strings to all 
orders in the other directions, i.e. to all orders in $n_b$ in the case of  $n+g=0$ and all orders in $n,g$ in the case of $n_b=1$.  It is tempting to wonder whether the vanishing conditions 
for GV invariants are sufficient in principle to fix the topological string amplitudes to all orders in both $n+g$ and $n_b$. Unfortunately this is not the case, even for the unrefined case 
$n=0$. The integration constant for the modular anomaly equation for $F^{(n,g,n_b)}$ is a polynomial of $E_4$ and $E_6$ of weighted degree $2(n+g)+6n_b-2$, so the number of unknown constant 
we need to fix is given by $[\frac{2(n+g)+6n_b-2}{12}]$ for odd $n+g+n_b$, or by $[\frac{2(n+g)+6n_b-8}{12}]$ for even $n+g+n_b$. On the other hand, we can also estimate the number of 
vanishing GV invariants $\tilde{n}_{g_L,g_R}^{n_bp+df}$ for given $n_b,g_L,g_R$. 
It is clear that the GV invariants vanish for $d<n_b$ since the genus zero invariants already vanish $\tilde{n}_{0,0}^{n_bp+df}=0$ if $d<n_b$. Based on arguments from algebraic geometry, 
e.g. in \cite{KKV,HKQ}, we know that the left top genus for the diagonal class $n_bp+n_bf$ goes like quadratic power $n_b^2$ for large $n_b$, and this is also confirmed by our explicit 
results. The GV invariant $\tilde{n}_{g_L,g_R}^{n_b(p+f)}$ would not vanish if the total genus $g_L+g_R$ scaled as quadratic power  $n_b^2$ at large $n_b$ but was smaller than the left 
top genus. In this case we would have just $n_b$ vanishing GV invariants $\tilde{n}_{0,0}^{n_bp+df}$ from $d=0,1,\cdots,n_b-1$, but the number of unknown constants could be of the order 
of the quadratic power $n_b^2$ at large $n_b$. So eventually the boundary conditions from vanishing GV invariants would not be enough to fix the (refined) topological string amplitudes at large genus. 

In fact, we can be a little more precise and show that the situation regarding fixing the ambiguity for the diagonal classes $n_b(p+f)$ is not better or worse than that of the $E_8$ model.
 This is rather surprising because in the case of the $E_8$ model we have utilized the conifold gap condition which is very powerful in fixing the holomorphic ambiguity, while in the half K3 model we do not use the conifold gap condition.
 In the $E_8$ model, we have $[\frac{2(n+g)-2}{6}]$ unknown constants in the holomorphic ambiguity in $F^{(n,g)}$ due to the singularity at the orbifold point.
 Suppose for the genus $g_L+g_R=n+g$ we can fix the (refined) GV invariants $\tilde{n}^\beta_{g_L,g_R}$ up to degree $d-1$, then in this case the number of conditions from GV invariants minus the number of  unknown constants is $d-1- [\frac{2(n+g)-2}{6}]$.
 On the other hand, for the refined amplitude $F^{(n,g,n_b)}$ with $n_b=d$ in the half K3 model, based on the discussions in the previous paragraph, we can easily check that the number of 
conditions from GV invariants minus the number of ambiguous constants is up to $\pm 1$ basically $\frac{1}{2}(d-1-[\frac{2(n+g)-2}{6}])$, i.e. one half of that of the $E_8$ model. So we see that the boundary conditions from GV invariants become insufficient  to fix the ambiguity at about the same time for the $E_8$ model and for the diagonal classes in the half K3 model. 

These arguments indicate that the vanishing of BPS invariants in the two-parameter half K3 model should imply the conifold gap conditions in the one-parameter $E_8$ model. It would interesting to understand this point more carefully.

\section{The half K3 surface: massive theory}  \label{sec.turning}
Based on the experience from  the studies of higher genus terms in the Nekrasov function for $SU(2)$ Seiberg-Witten theory with $N_f=4$ fundamental flavors or with one adjoint hypermultiplet in \cite{HKK}, we may hope that the mass deformation provides more boundary conditions for fixing the ambiguity comparing to the massless theory.
  There is one crucial difference with the studies in \cite{HKK}, where the mass parameters are not moduli parameters, and there is no enumerative geometric interpretation of the refined amplitudes in terms of the BPS invariants.
 In the present case, the mass parameters represent K\"ahler moduli of the $E_8$ part of the half  K3 surface, and are part of the geometry of the moduli space.
 The refined BPS invariants have extra labels in terms of the corresponding wrapping degrees which are Weyl orbits of the $E_8$ lattice.
 In the mirror geometry, the mass parameters appear as polynomials in the Seiberg-Witten curve in \cite{HKK}, and in the present case they appear in exponential 
or  
trigonometric functions.  This point was explained in \cite{Minahan:1997a}. 

We will need to use the theta function of the $E_8$ lattice $\Gamma_8$ 
\begin{eqnarray} \label{thetaE8}
\Theta(\vec{m},\tau) =\sum_{\vec{w}\in \Gamma_8}\exp(\pi i \tau \vec{w}^2+2\pi i \vec{m}\cdot \vec{w}) =
\frac{1}{2} \sum_{k=1}^4 \prod_{j=1}^8 \theta_k(m_j,\tau)\, ,
\end{eqnarray} 
where $\vec{m}=(m_1,m_2,\cdots,m_8) $ are the $E_8$ mass parameters. $\theta_k(m,\tau)$ is the ordinary Jacobi theta function and we provide the conventions in appendix \ref{appendixJacobi}.  The theta function $\Theta(\vec{m},\tau)$ of the $E_8$ lattice is a power series in $q=e^{2\pi i\tau}$, and from the well-known transformation properties of Jacobi theta functions, one can check that it has the following transformation behavior under a $SL(2,\mathbb{Z})$ transformation 
\begin{eqnarray}
\Theta(\frac{\vec{m}}{c\tau+d},\frac{a\tau+b}{c\tau+d}) =(c\tau+d)^4 \exp(\frac{i\sum_i m_i^2 }{c\tau+d}) \Theta(\vec{m},\tau). 
\end{eqnarray}
We see that except for the modular phase, 
it has modular weight four under the $SL(2,\mathbb{Z})$ transformation. 
Furthermore, in the massless limit $\vec{m}=0$, the theta function $\Theta(\vec{m},\tau)$ is simply the $E_4$ Eisenstein series.

One can easily see that the $E_8$ theta function (\ref{thetaE8}) is invariant under a Weyl group action\footnote{The Weyl group of $E_8$ is explicitly described in \ref{delpezzohalfk3}.} on the 
mass vector $\vec{m}$ since the lattice $\Gamma_8$ and the norm of a lattice vector are invariant under the action. 

The Weyl orbit of a lattice vector consists of the lattice vectors generated by acting all the elements of the Weyl group $W(E_8)$ on the lattice vector. We can classify the $E_8$ lattice points into classes of Weyl orbits.
 It is clear that the lattice vectors in the same Weyl orbit have the same norm. We denote by $\mathcal{O}_{p,k}$ the Weyl orbit whose element $\vec{w}$ has norm square $\vec{w}\cdot \vec{w}=2p$ and which has order $|\mathcal{O}_{p,k}|=k$. 
Since the Weyl orbits of the same norm usually have different numbers of elements, the notation should not cause any confusion. It is known that some low Weyl orbits  are given as follows
\begin{eqnarray} \label{Weylorbit}
&& \mathcal{O}_{0,1}, ~~\mathcal{O}_{1,240},~~ \mathcal{O}_{2,2160},~~ \mathcal{O}_{3,6720}, ~~
 \mathcal{O}_{4,240}, \mathcal{O}_{4,17280}, ~~ \mathcal{O}_{5,30240} \nonumber \\
&& \mathcal{O}_{6,60480}, ~~   \mathcal{O}_{7,13440},   \mathcal{O}_{7,69120},  ~~   \mathcal{O}_{8,2160},   \mathcal{O}_{8,138240},  ~~  \mathcal{O}_{9,240},   \mathcal{O}_{9,181440},    ~~  \nonumber \\
&&   \mathcal{O}_{10,30240},  \mathcal{O}_{10,241920},   ~~   \mathcal{O}_{11,138240},    \mathcal{O}_{11,181440}, 
~~  \mathcal{O}_{12,6720},   \mathcal{O}_{12,483840},  \nonumber \\
&& \mathcal{O}_{13,13440},   \mathcal{O}_{13,30240}, \mathcal{O}_{13,483840},  
~~ \cdots  .
\end{eqnarray}
We see that the lattice vectors with norm square $0,2,4,6,10,12$ consist of a single Weyl orbit respectively, while the lattice vectors with norm square $8,14,16,18,20, \cdots$ fall into multiple orbits. Any Weyl orbit $\mathcal{O}_{p,k}$ can be multiplied with a positive integer $n$ and generates another Weyl orbit with the same order $\mathcal{O}_{n^2p,k}= \{n\vec{w} | ~\vec{w}\in \mathcal{O}_{p,k} \}$. 

The $E_8$ theta function (\ref{thetaE8}) can be written as sums over Weyl orbits 
\begin{eqnarray} \label{E8weyl}
\Theta(\vec{m},q) =\sum_{\mathcal{O}_{p,k}}  q^p  \sum_{\vec{w}\in \mathcal{O}_{p,k}} \exp(2\pi i \vec{m}\cdot\vec{w}), 
\end{eqnarray} 
where $q=e^{2\pi i\tau}$. We will see that the (refined) topological string amplitudes in the massive half K3 model are constructed from  the $E_8$ theta function, and can be also written as sums over the Weyl orbits. The refined BPS invariants are labelled by the class $n_bp+df$ and the Weyl orbit $\mathcal{O}_{p,k}$, and denoted as $n^{n_bp+df, \mathcal{O}_{p,k}}_{j_L,j_R}$. We can compute the refined amplitudes in terms of refined BPS invariants with the formula (\ref{productrefined}) and by summing over the homology classes $\beta=( n_bp+df, \mathcal{O}_{p,k})$. As in the massless theory we denote the refined amplitudes by  $F^{(n,g,n_b)}(q,\vec{m})$ as appearing in the generating function in (\ref{generate}).
 The arguments $q$ and $\vec{m}$ represent the K\"ahler moduli of the fiber class $f$ and Weyl orbits. 

In the massless limit, the sum over  a Weyl orbit $\mathcal{O}_{p,k}$, is simply the order $|\mathcal{O}_{p,k}|=k$. So the refined BPS invariants in the massless limit can be computed from the more general massive invariants by 
\begin{eqnarray} \label{masslessweyl}
(n^{n_bp+df}_{j_L,j_R} )_{massless} =\sum_{\mathcal{O}_{p,k}} k  \cdot n^{n_bp+df, \mathcal{O}_{p,k}}_{j_L,j_R} \, .
\end{eqnarray}

First we consider the case of wrapping number $n_b=1$. In this case the refined G\"{o}ttsche formula is still available as in the massless theory.
 We can simply replace the Eisenstein series $E_4$ with the theta function $\Theta(\vec{m},\tau)$ for the generating function in (\ref{generate3.2}) in the massless theory
\begin{eqnarray}  
\mathcal{G}(\epsilon_1,\epsilon_2,q) =\Theta(\vec{m},q) G^{\mathcal{B}_9} (q, y_L,y_R). 
\end{eqnarray}
Using the formula for the $E_8$ theta function (\ref{E8weyl}), we find 
\begin{eqnarray} \label{Gottsche1}
\mathcal{G}(\epsilon_1,\epsilon_2,q)  &=&  \sum_{d=0}^{\infty}  \sum_{ \mathcal{O}_{p,k}}   \sum_{j_L,j_R}  (-1)^{2j_L+2j_R} (n^{\mathcal{B}_9})^{d}_{j_L,j_R}  (\sum_{k=-j_L}^{j_L} y_L^{2k} ) (\sum_{k=-j_R}^{j_R} y_R^{2k} )  
 \nonumber \\ && \times  
q^{d+p}  \sum_{\vec{w}\in \mathcal{O}_{p,k}}  \exp(2\pi i \vec{m}\cdot\vec{w})  .
\end{eqnarray}

We can then extract the refined Gopakumar-Vafa BPS invariants  similarly as in  (\ref{GV3.4}) for the massless theory, with the additional sum over Weyl orbits 
\begin{eqnarray}  \label{Gottscheorbit}
\mathcal{G}(\epsilon_1,\epsilon_2,q)  &=&  \sum_{d=0}^{\infty}  \sum_{ \mathcal{O}_{p,k}}    \sum_{j_L,j_R}  (-1)^{2j_L+2j_R} n^{p+df, \mathcal{O}_{p,k}}_{j_L,j_R}   (\sum_{k=-j_L}^{j_L} y_L^{2k} ) (\sum_{k=-j_R}^{j_R} y_R^{2k} )  \nonumber \\ && \times 
q^d \sum_{\vec{w}\in \mathcal{O}_{p,k}}  \exp(2\pi i \vec{m}\cdot\vec{w})  , 
\end{eqnarray}
where the sums over $j_L,j_R$ are over non-negative half integers.
 To extract the BPS integers, we first use the formula 
(\ref{Gottsche3.4}) to write the G\"{o}ttsche product $G^{\mathcal{B}_9} (q, y_L,y_R)$ in terms of the refined Betti numbers of the Hilbert schemes which have been computed in table \ref{tableBettihalfK3}, and which we denote here by $(n^{\mathcal{B}_9})^{d}_{j_L,j_R}$ with the superscript $\mathcal{B}_9$ to avoid confusion with other BPS invariants. 

 Comparing (\ref{Gottsche1}) and  (\ref{Gottscheorbit}),  we find the BPS invariants in terms of the refined Betti numbers 
 \begin{eqnarray} \label{nb=1weyl}
n^{p+df, \mathcal{O}_{p,k}}_{j_L,j_R}  =   (n^{\mathcal{B}_9})^{d-p}_{j_L,j_R}, ~~~ \textrm{for}~p\leq d
\end{eqnarray}
and it is understood that $n^{p+df, \mathcal{O}_{p,k}}_{j_L,j_R} =0$ in the case of $p>d$. 

We see from the formula (\ref{nb=1weyl}) that the refined BPS invariants $n^{p+df,\mathcal{O}_{p,k}}_{j_L,j_R}$ are identical for the homology classes with the same $d-p$.  
Furthermore for a class $p+df$, the refined BPS invariants vanish if the  square length for the Weyl orbits are sufficiently large. The Weyl orbit with maximal 
length and non-vanishing BPS invariants will be called the top Weyl orbit(s). Here the top Weyl orbit for the class $p+df$ is the orbit $\mathcal{O}_{d,k}$.

As a check of the formalism, we can compute the refined BPS invariants for the half K3 model in the massless limit using the formula (\ref{masslessweyl}) and the Weyl orbits (\ref{Weylorbit}), in terms of the refined Betti numbers in table \ref{tableBettihalfK3}. The results agree with those from direct calculations in table \ref{tableB9}.  This is simply due to the fact that the $E_8$ theta function $\Theta(\vec{m},q)$ is the Eisenstein series $E_4$ in the massless limit.

Similarly as in the massless theory, we can write down the genus zero amplitude for $n_b=1$ by setting $y_L=y_R=1$ in the G\"{o}ttsche product 
\begin{eqnarray}
F^{(0,0,1)}=  \frac{q^{\frac{1}{2}}  \Theta(\vec{m},q) }{\eta(q)^{12}}. 
\end{eqnarray}
The formulae for the refined higher genus amplitudes and the modular anomaly equation for $n_b=1$ are the same as in the massless theory in (\ref{lowgenus}), (\ref{holoone3.11}). 

Now we consider the case of wrapping number $n_b>1$.
 The higher genus refined amplitude with an $\eta$ function factor $(\frac{\eta(q)^{12 }}{\sqrt{q}})^{n_b} F^{(n,g,n_b)}$ has modular weight $2(n+g)+6n_b-2$ as in the massless theory.
  However, the modular ambiguity is not simply a modular form of $SL(2,\mathbb{Z})$.
  Instead, the modular ambiguity can be written as a linear combination of level $n_b$  $E_8$ characters, and the coefficients are mass-independent modular forms of $\Gamma_1(n_b)$ \cite{Minahan:1998}.   

There is a convenient way to write the ansatz for the refined amplitudes.
 It is known that there are nine Weyl invariant Jacobi modular forms of the $E_8$ lattice, which can be constructed from the $E_8$ theta function (\ref{thetaE8}), see e.g. \cite{Sakai:2011}.
 The nine Jacobi forms are denoted by $A_1,  A_2,  A_3, A_4, A_5$ and $B_2, B_3, B_4, B_6$.  Here $A_1=\Theta(\vec{m}, \tau)$ is simply the $E_8$ theta function and we provide the detailed formulae for the other forms in appendix \ref{appendixJacobi}.  All the characters of the fundamental representation of the higher level $E_8$ algebra can then be written as polynomials in $A_n$ and $B_n$, where the generators $A_n$ or $B_n$ contribute an $E_8$ level number $n$.
 For example, at level one there is only one polynomial $A_1$. There are three polynomials $A_1^2, A_2, B_2$  at level two, five polynomials $A_1^3,  A_1A_2,  A_1B_2,  A_3, B_3$ at  level three, and ten polynomials at level four.

The Jacobi form $A_n$ has modular weight four and $B_n$ has modular weight six, and they are simply the Eisenstein series $E_4$ and $E_6$ in the massless limit.
 Together with the quasi-modular forms $E_2, E_4, E_6$ of $SL(2,\mathbb{Z})$ which are independent of the mass parameters and have $E_8$ level number zero, we have all the ingredients for constructing  the refined amplitudes. 

It is natural to guess that similarly as in the massless theory, the scaled refined amplitude with $\eta$ function factor $(\frac{\eta(q)^{12 }}{\sqrt{q}})^{n_b} F^{(n,g,n_b)}$ can be written as polynomials of the nine $E_8$ Jacobi forms $A_n$, $B_n$ and the $SL(2,\mathbb{Z})$  quasi-modular forms $E_2$, $E_4$, $E_6$, and it has $E_8$ level number $n_b$ and modular weight $2(n+g)+6n_b-2$.
 We find that this is true for level $n_b\leq 4$.
 However, for level $n_b\geq 5$, the scaled amplitude $(\frac{\eta(q)^{12 }}{\sqrt{q}})^{n_b} F^{(n,g,n_b)}$ is not exactly a modular form, but a rational function of modular forms with the powers in $E_4$ as the denominator.
 For example for the case of $n_b=5$, we check that a denominator of $E_4$ is sufficient and  the scaled amplitude $E_4 (\frac{\eta(q)^{12 }}{\sqrt{q}})^{n_b} F^{(n,g,n_b)}$ is again always a modular form, i.e. can be written as a polynomial of $A_n$, $B_n$ and $E_n$.

In any case there is only a finite number of unknown constants in the ansatz for the amplitude.
 There is no further algebraic relation among the generators for generic mass parameters, so that the expression for a refined amplitude is unique.
  The $E_2$-dependent part of the  amplitudes is determined by the refined modular anomaly equation we proposed in (\ref{refinedmodular}).
 We can further fix the modular ambiguity by vanishing conditions of the refined BPS invariants.
 Here the vanishing conditions  are $n^{n_bp+df,\mathcal{O}_{p,k}}_{0,0}=0$ for $d<n_b$, except for the case $n^{p,\mathcal{O}_{0,1}}_{0,0}=1$.

The generators $A_n$ and $B_n$ are sums over Weyl orbits, similarly as the theta function of the $E_8$ lattice in  (\ref{E8weyl}), as can be seen using their formulae in appendix \ref{appendixJacobi}.
 On the other hand, the refined amplitudes are polynomials of the generators, and in order to extract the refined BPS invariants from the amplitudes, we must write the refined amplitudes as sums over Weyl orbits as in the general formula (\ref{productrefined}). So we need to decompose the product of sums over Weyl orbits into a sum of the sums over Weyl orbits as 
\begin{eqnarray}
(\sum_{\vec{w} \in \mathcal{O}_1} e^{2\pi i \vec{m}\cdot\vec{w}} ) (\sum_{\vec{w} \in \mathcal{O}_2} e^{2\pi i \vec{m}\cdot\vec{w}} ) = \sum_i m_i  \sum_{\vec{w} \in \mathcal{O}^\prime_i} e^{2\pi i \vec{m}\cdot\vec{w}},
\end{eqnarray}
where $m_i$ are non-negative integers for the multiplicity in the decomposition.
 It is straightforward to compute the decompositions of the $E_8$ Weyl orbits.
 The product with the zero-length orbit is trivial $\mathcal{O}_{p,k} \otimes \mathcal{O}_{0,1} = \mathcal{O}_{p,k}$.
 Here we provide some decompositions for the low orbits 
\begin{eqnarray}  \label{decomexamples}
\mathcal{O}_{1,240} \otimes \mathcal{O}_{1,240} &=& 240\cdot\mathcal{O}_{0,1}\oplus56\cdot\mathcal{O}_{1,240}\oplus14\cdot\mathcal{O}_{2,2160}\oplus2\cdot\mathcal{O}_{3,6720}\oplus\mathcal{O}_{4,240}, \nonumber \\ \mathcal{O}_{2,2160} \otimes \mathcal{O}_{1,240} &=& 126\cdot\mathcal{O}_{1,240}\oplus64\cdot\mathcal{O}_{2,2160}\oplus27\cdot\mathcal{O}_{3,6720}\oplus8\cdot\mathcal{O}_{4,17280}\oplus\mathcal{O}_{5,30240}, \nonumber \\ \mathcal{O}_{2,2160} \otimes \mathcal{O}_{2,2160} &=& 2160\cdot\mathcal{O}_{0,1}\oplus576\cdot\mathcal{O}_{1,240}\oplus280\cdot\mathcal{O}_{2,2160}\oplus144\cdot\mathcal{O}_{3,6720}  \nonumber \\ && \oplus126\cdot\mathcal{O}_{4,240}  \oplus70\cdot\mathcal{O}_{4,17280}\oplus32\cdot\mathcal{O}_{5,30240}\oplus10\cdot\mathcal{O}_{6,60480} \nonumber \\ &&   \oplus2\cdot\mathcal{O}_{7,69120}\oplus\mathcal{O}_{8,2160}, \nonumber \\ \mathcal{O}_{3,6720} \otimes \mathcal{O}_{1,240} &=& 56\cdot\mathcal{O}_{1,240}\oplus84\cdot\mathcal{O}_{2,2160}\oplus54\cdot\mathcal{O}_{3,6720}\oplus56\cdot\mathcal{O}_{4,240}\oplus28\cdot\mathcal{O}_{4,17280}  \nonumber \\ &&   \oplus12\cdot\mathcal{O}_{5,30240}\oplus3\cdot\mathcal{O}_{6,60480}\oplus\mathcal{O}_{7,13440}, \nonumber \\ \mathcal{O}_{3,6720} \otimes \mathcal{O}_{2,2160} &=& 756\cdot\mathcal{O}_{1,240}\oplus448\cdot\mathcal{O}_{2,2160}\oplus270\cdot\mathcal{O}_{3,6720}\oplus168\cdot\mathcal{O}_{4,17280}  \nonumber \\ &&  \oplus92\cdot\mathcal{O}_{5,30240}\oplus48\cdot\mathcal{O}_{6,60480}\oplus27\cdot\mathcal{O}_{7,13440}\oplus21\cdot\mathcal{O}_{7,69120}
\nonumber \\ && 
\oplus7\cdot\mathcal{O}_{8,138240}\oplus\mathcal{O}_{9,181440}.
\end{eqnarray}

We mention an identity for the order $|\mathcal{O}_{p_1,k_1}|=k_1$ and related multiplicities. Denote by $m^{p_3,k_3}_{p_1,k_1;p_2,k_2}$  the multiplicity of the Weyl orbit $\mathcal{O}_{p_3,k_3}$ in the decomposition of the product of the orbits $\mathcal{O}_{p_1,k_1}$ and $\mathcal{O}_{p_2,k_2}$.
 We can fix an element in $\mathcal{O}_{p_3,k_3}$, subtract from it all the elements in $\mathcal{O}_{p_1,k_1}$, and check which orbits the subtracted vectors belong to. We find an one-to-one correspondence of the elements in  $\mathcal{O}_{p_1,k_1}$ with all multiplicities of the orbit  $\mathcal{O}_{p_3,k_3}$ in the decomposition of $\mathcal{O}_{p_1,k_1}$ with another orbit 
\begin{eqnarray}
|\mathcal{O}_{p_1,k_1}| = \sum_{\mathcal{O}_{p_2,k_2}} m^{p_3,k_3}_{p_1,k_1;p_2,k_2},
\end{eqnarray}
which is valid for any two orbits $\mathcal{O}_{p_1,k_1}$ and $\mathcal{O}_{p_3,k_3}$.

Different Weyl orbits with the same norm can be distinguished by their multiplicities in the decomposition of the product of two orbits.
 For example, we can see in (\ref{decomexamples}) in the decomposition of  $\mathcal{O}_{3,6720} \otimes \mathcal{O}_{2,2160}$, that the multiple orbits $\mathcal{O}_{p,k}$ in the cases of $p=4,7,8,9$ appear with different multiplicities.

We find that the vanishing conditions of the refined GV invariants over-determine the modular ambiguity at low genus.
 The redundancy provides non-trivial tests of the consistency of the refined modular anomaly equation (\ref{refinedmodular}) and the refined amplitudes with generic mass parameters. As an example of an explicit check, we find that if we change the factor of $\frac{n_b}{24}$ in the last term in (\ref{refinedmodular}), there would be no solution at genus two to the modular ambiguity that satisfies the vanishing conditions.

We provide some low order formulae for the refined amplitudes.
 The genus zero results have been written down in \cite{Minahan:1998}, we also include them here for completeness.
 The formulae in terms of the Jacobi forms $A_n$ and $B_n$ are simpler than those in terms of the $E_8$ characters originally presented for genus zero case in \cite{Minahan:1998}.
 The genus zero formulae are
\begin{eqnarray}
F^{(0,0,2)} &=&  \frac{ q}{96 \cdot\eta^{24}} [4 E_2 A_1^2 + 3 E_6 A_2 + 5 E_4 B_2],  \nonumber \\
F^{(0,0,3)} &=&   \frac{q^{\frac{3}{2}}}{15552 \cdot\eta^{36} } [54 A_1^3 E_2^2 - 54 A_1^3 E_4 + 135 A_1 B_2 E_2 E_4 + 135 A_1 A_2 E_4^2  \nonumber \\ &&  +  28 A_3 E_4^3 + 225 A_1 B_2 E_6 + 81 A_1 A_2 E_2 E_6 - 28 A_3 E_6^2]. 
\end{eqnarray}
Some higher genus formulae are 
\begin{eqnarray}
F^{(1,0,2)} &=&  - \frac{ q}{1152 \cdot\eta^{24}} [4 A_1^2 E_2^2 + 4 A_1^2 E_4 + 5 B_2 E_2 E_4 + 3 A_2 E_4^2 
+ 5 B_2 E_6 +  3 A_2 E_2 E_6],  \nonumber \\
F^{(0,1,2)} &=&  \frac{ q}{1152 \cdot\eta^{24}} [10 A_1^2 E_2^2 + 6 A_1^2 E_4 + 15 B_2 E_2 E_4 + 3 A_2 E_4^2 + 5 B_2 E_6 + 
 9 A_2 E_2 E_6],  \nonumber \\ 
F^{(1,0,3)} &=&  - \frac{ q^{\frac{3}{2}}}{124416 \cdot\eta^{36}} [54 A_1^3 E_2^3 + 18 A_1^3 E_2 E_4 
+ 135 A_1 B_2 E_2^2 E_4 + 630 A_1 B_2 E_4^2 \nonumber \\  && 
+ 189 A_1 A_2 E_2 E_4^2 - 160 B_3 E_4^3 + 28 A_3 E_2 E_4^3 - 72 A_1^3 E_6 + 
 315 A_1 B_2 E_2 E_6  \nonumber \\  &&  + 81 A_1 A_2 E_2^2 E_6 + 378 A_1 A_2 E_4 E_6 
 + 160 B_3 E_6^2 - 28 A_3 E_2 E_6^2],  \nonumber \\
F^{(0,1,3)} &=& \frac{ q^{\frac{3}{2}}}{62208 \cdot\eta^{36}} [78 A_1^3 E_2^3 - 54 A_1^3 E_2 E_4 
+ 225 A_1 B_2 E_2^2 E_4 + 360 A_1 B_2 E_4^2  \nonumber \\  && 
+  297 A_1 A_2 E_2 E_4^2 - 80 B_3 E_4^3 + 56 A_3 E_2 E_4^3 - 24 A_1^3 E_6 + 
 495 A_1 B_2 E_2 E_6  \nonumber \\  &&  + 135 A_1 A_2 E_2^2 E_6 + 216 A_1 A_2 E_4 E_6 + 80 B_3 E_6^2 - 
 56 A_3 E_2 E_6^2], 
\end{eqnarray}
\begin{eqnarray}
F^{(2,0,2)} &=&   \frac{ q}{138240 \cdot\eta^{24}} [20 A_1^2 E_2^3 + 44 A_1^2 E_2 E_4 
+ 25 B_2 E_2^2 E_4 + 65 B_2 E_4^2 +  30 A_2 E_2 E_4^2   \nonumber \\  &&  + 48 A_1^2 E_6 \
+ 50 B_2 E_2 E_6 + 15 A_2 E_2^2 E_6 + 39 A_2 E_4 E_6],  \nonumber \\
F^{(1,1,2)} &=&  - \frac{ q}{69120  \cdot\eta^{24}} [50 A_1^2 E_2^3 + 98 A_1^2 E_2 E_4 
+ 75 B_2 E_2^2 E_4 + 105 B_2 E_4^2 +  60 A_2 E_2 E_4^2   \nonumber \\  &&
+ 76 A_1^2 E_6 + 100 B_2 E_2 E_6 + 45 A_2 E_2^2 E_6 + 
 63 A_2 E_4 E_6],  \nonumber \\
 F^{(0,2,2)} &=&   \frac{ q}{414720 \cdot\eta^{24}} [380 A_1^2 E_2^3 + 564 A_1^2 E_2 E_4 + 675 B_2 E_2^2 E_4 + 315 B_2 E_4^2 +  270 A_2 E_2 E_4^2  \nonumber \\  &&
 + 208 A_1^2 E_6 + 450 B_2 E_2 E_6 + 405 A_2 E_2^2 E_6 + 
 189 A_2 E_4 E_6]. 
\end{eqnarray}
In the massless limit the $E_8$ Jacobi forms simplify as $A_n=E_4, B_n=E_6$, and these formulae reduce to the ones previously obtained in (\ref{masslessformulae1}),   (\ref{masslessformulae2}) and  (\ref{masslessformulae3}).

We list the refined BPS invariants for the case $n_b=2$ in the tables \ref{tablehalfK3massive1} and \ref{tablehalfK3massive2}.
 We see that the top Weyl orbit $\mathcal{O}_{p,k}$ for the class $2p+df$ is $p=2d-2$.
 This is easy to understand from the formula (\ref{E8weyl}), which shows the Weyl orbits with maximal norm in the coefficient of $q^d$ are $\mathcal{O}_{2d,k}$ in the level two $E_8$ characters  $\Theta(2\vec{m}, 2\tau)$, $\Theta(\vec{m}, \frac{\tau}{2})$ and  $\Theta(\vec{m}, \frac{\tau+1}{2})$, which appear in the formulae for the Jacobi forms $A_2$ and $B_2$.
 Our explicit results further show that the contributions of the Weyl orbits $\mathcal{O}_{2d,k}$ and $\mathcal{O}_{2d-1,k}$ vanish, so that the top Weyl orbit turns out to be $\mathcal{O}_{2d-2,k}$.
 Another interesting feature is  that the refined BPS invariants are similar for the classes $\beta=(2p+df,\mathcal{O}_{p,k})$ with the same value of $2d-2-p$.
 In particular the refined BPS invariants have the smallest top genus, and are 
exactly identical for the top Weyl orbits $p=2d-2$.
 The exact identifications gradually disappear for classes with higher values of $2d-2-p$.

Some refined BPS invariants for the cases $n_b=3,4$ are listed in the tables \ref{massivenb=3} and \ref{massivenb=4}.
 We find that the general empirical formula of the top Weyl orbit $\mathcal{O}_{p,k}$ for the class $n_bp+df$ reads
\begin{eqnarray} \label{topweylmassive}
p=n_bd-\frac{n_b(n_b+1)}{2}+1.
\end{eqnarray}
Below the top Weyl orbit, the refined BPS invariants $n^{n_bp+df, \mathcal{O}_{p,k}}_{j_L,j_R}$ may still completely vanish for some homology classes with small non-negative values of $n_bd-\frac{n_b(n_b+1)}{2}+1-p$.
 However, there is at least one Weyl orbit $\mathcal{O}_{p,k}$ at each integer $p\leq n_bd-\frac{n_b(n_b+1)}{2}+1$, where the refined BPS invariants do not completely vanish. 

The non-vanishing refined BPS invariants at the  top Weyl orbit have the top genus  pair $(g_L,g_R)^{top}=(0,n_b-1)$.
 Furthermore, the top genus pair for the classes with non-vanishing BPS invariants  increases exactly at the same place as we lower the Weyl orbit.
  So the top genus pair for the class $(n_bp+df, \mathcal{O}_{p,k})$ with non-vanishing BPS invariants is 
\begin{eqnarray}   \label{topgenusmassive}
(g_L,g_R)^{top} = (n_bd-\frac{n_b(n_b+1)}{2}+1-p)(1,1) + (0,n_b-1). 
\end{eqnarray}
These formulae should come from the algebraic geometric properties of holomorphic curves in the half K3 surface. 

Again as in the $n_b=1$ case, we can reproduce the results for the massless theory of the half K3 model in tables \ref{tableB9n_b=2}  and \ref{tableB9n_b=3,4} using the formula (\ref{masslessweyl}), the Weyl orbits (\ref{Weylorbit}) and the more general refined BPS invariants in tables \ref{tablehalfK3massive1}, \ref{tablehalfK3massive2},  \ref{massivenb=3} and \ref{massivenb=4} of the massive theory. 

It is clear that in the massive theory it is easier to fix the modular ambiguity than in the massless theory.
 For example, for the case of genus $n+g=1$ and $n_b=1$, there is one modular ambiguity in the massless theory which is proportional to $E_6$, but there is no modular ambiguity in the massive theory since there is no $E_8$ level one holomorphic modular form of weight 6.

We discuss whether the vanishing conditions are sufficient for fixing the modular ambiguity in more detail.
 We have used the vanishing conditions $n^{n_bp+df,\mathcal{O}_{p,k}}_{0,0}=0$ for $d<n_b$ (with the exception  $n^{p,\mathcal{O}_{0,1}}_{0,0}=1$).
 Surprisingly it turns out that there are also vanishing conditions available even for arbitrarily large fiber degree $d$, due to the the empirical formula for the top Weyl orbits (\ref{topweylmassive}).
 For a level $n_b$ Jacobi modular form which is a polynomial of the generators $A_n$ and $B_n$, one can check that the highest Weyl orbit appearing in the coefficient of $q^d$ has the half norm square $p=n_bd$.
 The sum of the form $q^d \sum_{\vec{w}\in \mathcal{O}_{p,k}}e^{2\pi i \vec{w}\cdot \vec{m}}$ with $p=n_bd$ is non-vanishing in the Jacobi form for infinitely many fiber degrees $d$.
  For $n_b>1$, this is higher than the top Weyl orbit with non-vanishing BPS invariants according to our formula (\ref{topweylmassive}). So the vanishing of the BPS 
invariants  $n^{n_bp+df,\mathcal{O}_{p,k}}_{j_L,j_R}=0$ for $p>n_bd-\frac{n_b(n_b+1)}{2}+1$ should impose constraints on the ansatz for the level $n_b$ refined topological amplitudes for these fiber degrees.

However, one can construct a certain ansatz for the modular ambiguity at level $n_b\geq 2$, such that the sums of the form $q^d \sum_{\vec{w}\in \mathcal{O}_{p,k}}e^{2\pi i \vec{w}\cdot \vec{m}}$ do not appear for all degrees $d$ and Weyl orbits  with $p>n_bd-\frac{n_b(n_b+1)}{2}+1$.
 For example, we can consider the case of $n_b=2$ and genus $n+g=3$.
 The modular ambiguity is a Jacobi form of level $n_b=2$ and weight $16$, multiplying by the factor of $\frac{q}{\eta(q)^{24}}$.
 The level 2 forms $A_2$, $B_2$ and $A_1^2$ can be written in terms of sums of the form  $q^d \sum_{\vec{w}\in \mathcal{O}_{p,k}}e^{2\pi i \vec{w}\cdot \vec{m}}$, where the Weyl orbit $p\leq 2d$.
 We can look at an example of a modular ambiguity $\frac{q(E_6^2-E_4^3)}{\eta^{24}}A_2\sim qA_2$.
 Due to the extra factor of $q$, now the modular ambiguity $qA_2$ is written as a sum of the   form  $q^d \sum_{\vec{w}\in \mathcal{O}_{p,k}}e^{2\pi i \vec{w}\cdot \vec{m}}$, where $p\leq 2d-2$.
 So this ambiguity cannot be fixed by the vanishing 
conditions of BPS 
invariants due to the top Weyl orbit formula (\ref{topweylmassive}).

\subsection{Flow to the del Pezzo models} 
One can take some limits of the mass parameters and flow from the $E_8$ model to $E_n$ ($n=5,6,7$) models  \cite{Minahan:1997a}.
  We consider the diagonal class  $\beta=(d(p+f),\mathcal{O}_{p,k})$ where the base number $n_b$ equals $d$, and denote the parameter $q$ as the combined K\"ahler parameter of the base and fiber classes.
 We have discussed that the $E_8$ model is simply the massless limit of the diagonal classes in half K3 model.
 For the $E_n$ ($n=5,6,7$) model,  one performs the following scalings of $q$ and the mass parameters  
\begin{eqnarray} \label{scaling} 
&& q\rightarrow e^{2\pi (8-n)\Lambda} q, ~~~  m_j\rightarrow i\Lambda +\mu, ~~(j=n, \cdots, 7),  \nonumber 
\\ &&  m_8 \rightarrow -i (8-n) \Lambda +\mu \, .
\end{eqnarray}

The refined amplitudes of the half K3 model consist of sums over the Weyl orbits of the $E_8$ lattice of the form $\sum_{\vec{w}\in \mathcal{O}_{p,k}} q^d e^{2\pi i \vec{w}\cdot \vec{m}}$.
 To flow to the $E_n$ model, we keep only the terms which are independent of the scaling parameter $\Lambda$ under the scaling (\ref{scaling}).
 We denote by $\mathcal{O}^{E_n,d}_{p,k}$ the subset of the $E_8$ Weyl orbit $\mathcal{O}_{p,k}$, whose elements satisfy the condition 
\begin{eqnarray} \label{subsetorbit}
\mathcal{O}^{E_n,d}_{p,k} := \{\vec{w}\in \mathcal{O}_{p,k} |~  \sum_{j=n}^7 w_j - (8-n)w_8 =(8-n)d ~\}. 
\end{eqnarray} 
This condition is also compatible with multiple cover contributions in the refined topological amplitudes, namely if a lattice vector $\vec{w}\in \mathcal{O}^{E_n,d}_{p,k} $, then it is also true that $r \vec{w}\in \mathcal{O}^{E_n,rd}_{r^2p,k} $ for any multi-covering positive integer $r$.

To compare with the results in section \ref{sec.delpezzo}, we further take the massless limit for the remaining mass parameters $\mu\rightarrow 0$ and $m_j\rightarrow 0$ ($j=1,\cdots, n-1$).
 The sum over the subset $\mathcal{O}^{E_n,d}_{p,k}$ of the $E_8$ Weyl orbit is then simply the order, i.e. the number of elements of the subset.
 Similar to the formula  (\ref{masslessweyl}), we can compute the refined BPS invariants for the $E_n$ models by the formula 
\begin{eqnarray} \label{masslessEn} 
(n^d_{j_L,j_R})_{E_n} = \sum_{\mathcal{O}_{p,k}} |\mathcal{O}^{E_n,d}_{p,k}|\cdot  n^{d(p+f),\mathcal{O}_{p,k}}_{j_L,j_R}  .
\end{eqnarray}

It is straightforward to check the elements in the $E_8$ Weyl orbits (\ref{Weylorbit}) and to compute the subset $\mathcal{O}^{E_n,d}_{p,k}$ for various degrees $d$ and  $E_n$ models.
 We list the data for the orders of the subset for some low orbits and degrees for the $D_5, E_6, E_7$ models in the table \ref{Enorbit}.

We can then compute the refined BPS invariants for the $E_n$ models using the  formula  (\ref{masslessEn}), the orders of the Weyl orbits in table \ref{Enorbit},  and the refined BPS invariants for the diagonal classes $n_b=d$ for the $E_8$ model in tables  \ref{tablehalfK3massive1}, \ref{tablehalfK3massive2}, \ref{massivenb=3} and \ref{massivenb=4}.
 We reproduce the results in the corresponding tables for $D_5$, $E_6$ and $E_7$ up to $d\leq 5$ in section \ref{sec.delpezzo}.

\begin{table}
\begin{center} {\footnotesize 
\begin{tabular} {|c|c|c|c|c|c|c|c|c|c|c|c|c|} \hline d $\backslash$ 
orbits  & $\mathcal{O}_{0,1}$  & $\mathcal{O}_{1,240}$  & 
$\mathcal{O}_{2,k}$  & $\mathcal{O}_{3,k}$  & $\mathcal{O}_{4,k_1}$  
& $\mathcal{O}_{4,k_2}$  & $\mathcal{O}_{5,k}$  & $\mathcal{O}_{6,k}$ 
 & $\mathcal{O}_{7,k_1}$  & $\mathcal{O}_{7,k_2}$  & 
$\mathcal{O}_{8,k_1}$  & $\mathcal{O}_{8,k_2}$  \\  \hline  1 & 0 & 
16 & 176 & 640 & 0 & 1296 & 2416 & 4336 & 976 & 4960 & 0 & 8272 \\  
\hline2 & 0 & 0 & 10 & 140 & 16 & 576 & 1052 & 2710 & 508 & 2704 & 
176 & 6336 \\  \hline3 & 0 & 0 & 0 & 0 & 0 & 16 & 176 & 640 & 208 & 
1088 & 0 & 2416 \\  \hline4 & 0 & 0 & 0 & 0 & 0 & 0 & 0 & 1 & 6 & 40 
& 10 & 320 \\  \hline5 & 0 & 0 & 0 & 0 & 0 & 0 & 0 & 0 & 0 & 0 & 0 & 
0 \\  \hline \end{tabular} \vskip 8pt The $D_5$ model \vskip 15pt 
\begin{tabular} {|c|c|c|c|c|c|c|c|c|c|c|c|c|} \hline d $\backslash$ 
orbits  & $\mathcal{O}_{0,1}$  & $\mathcal{O}_{1,240}$  & 
$\mathcal{O}_{2,k}$  & $\mathcal{O}_{3,k}$  & $\mathcal{O}_{4,k_1}$  
& $\mathcal{O}_{4,k_2}$  & $\mathcal{O}_{5,k}$  & $\mathcal{O}_{6,k}$ 
 & $\mathcal{O}_{7,k_1}$  & $\mathcal{O}_{7,k_2}$  & 
$\mathcal{O}_{8,k_1}$  & $\mathcal{O}_{8,k_2}$  \\  \hline  1 & 0 & 
27 & 270 & 891 & 54 & 1944 & 3564 & 5724 & 1350 & 7560 & 432 & 12096 
\\  \hline2 & 0 & 0 & 27 & 270 & 27 & 864 & 1998 & 3564 & 972 & 4752 
& 270 & 8640 \\  \hline3 & 0 & 0 & 0 & 1 & 2 & 72 & 414 & 1260 & 434 
& 1944 & 144 & 4032 \\  \hline4 & 0 & 0 & 0 & 0 & 0 & 0 & 0 & 27 & 54 
& 216 & 27 & 864 \\  \hline5 & 0 & 0 & 0 & 0 & 0 & 0 & 0 & 0 & 0 & 0 
& 0 & 0 \\  \hline \end{tabular} \vskip 8pt The $E_6$ model \vskip 
15pt \begin{tabular} {|c|c|c|c|c|c|c|c|c|c|c|c|c|} \hline d 
$\backslash$ orbits  & $\mathcal{O}_{0,1}$  & $\mathcal{O}_{1,240}$  
& $\mathcal{O}_{2,k}$  & $\mathcal{O}_{3,k}$  & $\mathcal{O}_{4,k_1}$ 
 & $\mathcal{O}_{4,k_2}$  & $\mathcal{O}_{5,k}$  & 
$\mathcal{O}_{6,k}$  & $\mathcal{O}_{7,k_1}$  & $\mathcal{O}_{7,k_2}$ 
 & $\mathcal{O}_{8,k_1}$  & $\mathcal{O}_{8,k_2}$  \\  \hline  1 & 0 
& 56 & 576 & 1512 & 0 & 4032 & 5544 & 12096 & 1568 & 12096 & 0 & 
24192 \\  \hline2 & 0 & 1 & 126 & 756 & 56 & 2016 & 4158 & 7560 & 
1512 & 10080 & 576 & 16128 \\  \hline3 & 0 & 0 & 0 & 56 & 0 & 576 & 
1512 & 4032 & 1512 & 4032 & 0 & 12096 \\  \hline4 & 0 & 0 & 0 & 0 & 1 
& 0 & 126 & 756 & 56 & 2016 & 126 & 4032 \\  \hline5 & 0 & 0 & 0 & 0 
& 0 & 0 & 0 & 0 & 56 & 0 & 0 & 576 \\  \hline \end{tabular} \vskip 
8pt The $E_7$ model \vskip 15pt
 \begin{tabular} {|c|c|c|c|c|c|c|c|c|c|c|c|c|} \hline d $\backslash$ 
orbits  & $\mathcal{O}_{0,1}$  & $\mathcal{O}_{1,240}$  & 
$\mathcal{O}_{2,k}$  & $\mathcal{O}_{3,k}$  & $\mathcal{O}_{4,k_1}$  
& $\mathcal{O}_{4,k_2}$  & $\mathcal{O}_{5,k}$  & $\mathcal{O}_{6,k}$ 
 & $\mathcal{O}_{7,k_1}$  & $\mathcal{O}_{7,k_2}$  & 
$\mathcal{O}_{8,k_1}$  & $\mathcal{O}_{8,k_2}$  \\  \hline  2 & 0 & 0 
& 2 & 84 & 0 & 422 & 784 & 2184 & 420 & 2380 & 0 & 5266 \\  \hline4 & 
0 & 0 & 0 & 0 & 0 & 0 & 0 & 0 & 0 & 1 & 2 & 42 \\  \hline6 & 0 & 0 & 
0 & 0 & 0 & 0 & 0 & 0 & 0 & 0 & 0 & 0 \\  \hline \end{tabular} \vskip 
8pt The $\mathbb{P}^1	\times \mathbb{P}^1$ model \vskip 15pt
 \begin{tabular} {|c|c|c|c|c|c|c|c|c|c|c|c|c|} \hline d $\backslash$ 
orbits  & $\mathcal{O}_{0,1}$  & $\mathcal{O}_{1,240}$  & 
$\mathcal{O}_{2,k}$  & $\mathcal{O}_{3,k}$  & $\mathcal{O}_{4,k_1}$  
& $\mathcal{O}_{4,k_2}$  & $\mathcal{O}_{5,k}$  & $\mathcal{O}_{6,k}$ 
 & $\mathcal{O}_{7,k_1}$  & $\mathcal{O}_{7,k_2}$  & 
$\mathcal{O}_{8,k_1}$  & $\mathcal{O}_{8,k_2}$  \\  \hline  1 & 0 & 0 
& 0 & 0 & 0 & 1 & 56 & 420 & 168 & 728 & 70 & 2296 \\  \hline2 & 0 & 
0 & 0 & 0 & 0 & 0 & 0 & 0 & 0 & 0 & 0 & 0 \\  \hline \end{tabular} 
\vskip 8pt The $\mathbb{P}^2$ model \vskip 15pt
}
\caption{The orders of subsets $|\mathcal{O}^{X,d}_{p,k}|$ for  $X=D_5, E_6, E_7, \mathbb{P}^1\times \mathbb{P}^1,  \mathbb{P}^2$ models.
 Here $k_i=|\mathcal{O}_{p,k_i}|$ are the orders of the $E_8$ Weyl orbits available in (\ref{Weylorbit}), and we sort them by increasing order in the case of multiple orbits with the same norm.}
\label{Enorbit}
 \end{center}
\end{table}

We note the following inequality for the element $\vec{w}$ in $\mathcal{O}^{E_n,d}_{p,k} $ 
\begin{eqnarray} \label{inequal}
(8-n)^2d^2 &=&   ( \sum_{j=n}^7 w_j - (8-n)w_8 )^2
~\leq~  (  \sum_{j=n}^71   + (8-n)^2  )( \sum_{j=n}^8 w_j^2 )  \nonumber \\
&\leq & 2 (8-n)(9-n) p
\end{eqnarray}
Therefore the orbit must satisfy $p\geq \frac{(8-n) d^2 }{2(9-n)} $, otherwise the subset $\mathcal{O}^{E_n,d}_{p,k} $ would be empty.
 This is also confirmed explicitly by the data in table \ref{Enorbit}. Furthermore, the top Weyl orbit with non-vanishing BPS invariants is $p\leq \frac{d(d-1)}{2}+1$ for the diagonal classes $n_b=d$ according to  (\ref{topweylmassive}).
 So the argument in the sum in the formula for the $E_n$ model (\ref{masslessEn}) is only non-vanishing for the $E_8$ Weyl orbits  $\mathcal{O}_{p,k}$ in the range $ \frac{(8-n) d^2 }{2(9-n)} \leq p\leq \frac{d(d-1)}{2}+1$ for the half square length $p$ of the orbit. 

We can also flow to the $\mathbb{P}^1\times \mathbb{P}^1$ and $\mathbb{P}^2$ models.
 For the  $\mathbb{P}^1\times \mathbb{P}^1$ model, we set $n=1$ in the subset  (\ref{subsetorbit}) of the $E_8$ Weyl orbits   
\begin{eqnarray}
\mathcal{O}^{\mathbb{P}^1\times \mathbb{P}^1 ,d}_{p,k} := \{\vec{w}\in \mathcal{O}_{p,k} |~  \sum_{j=1}^7 w_j -  7 w_8 =7 d ~\}. 
\end{eqnarray} 
As the point $\vec{w}$ lies in the $E_8$ lattice, the sum $\sum_{j=1}^8 w_j$ is an even integer and $w_j$ is   
an integer or half integer, we see that $\mathcal{O}^{\mathbb{P}^1\times \mathbb{P}^1 ,d}_{p,k}$ is an empty set for odd degree $d$.
 The formula for the refined BPS invariants (\ref{masslessEn}) is modified to depend on only even degree invariants from the half K3 model
\begin{eqnarray} \label{masslessP1P1}
 (n^d_{j_L,j_R})_{\mathbb{P}^1\times \mathbb{P}^1 } = \sum_{\mathcal{O}_{p,k}} |\mathcal{O}^{\mathbb{P}^1\times \mathbb{P}^1 ,2d}_{p,k}|\cdot  n^{2d(p+f),\mathcal{O}_{p,k}}_{j_L,j_R}  .
\end{eqnarray}
We also list the data for $ |\mathcal{O}^{\mathbb{P}^1\times \mathbb{P}^1 ,d}_{p,k}|$ for even $d$ in table \ref{Enorbit}, and reproduce the results for $d=1,2$ in table \ref{tableP1P1} in section \ref{diagonalF0}. Similarly as for  the $E_n$ models, we find that the argument in the sum (\ref{masslessP1P1}) is non-vanishing in the range $\frac{7}{4}d^2 \leq p \leq 2d^2-d+1$. 

For the $\mathbb{P}^2$ model, the subset of Weyl orbits is defined as 
\begin{eqnarray}
\mathcal{O}^{\mathbb{P}^2 ,d}_{p,k} := \{\vec{w}\in \mathcal{O}_{p,k} |~  \sum_{j=1}^7 w_j -   w_8 = 8d ~\}.
\end{eqnarray} 
We also list the data for $|\mathcal{O}^{\mathbb{P}^2 ,d}_{p,k}|$ in table \ref{Enorbit}.
 The refined invariants from the half K3 model only contribute when the degree $d$ is divisible by 3
\begin{eqnarray} \label{masslessP2}
 (n^d_{j_L,j_R})_{\mathbb{P}^2 } = \sum_{\mathcal{O}_{p,k}} |\mathcal{O}^{\mathbb{P}^2 ,d}_{p,k}|\cdot  n^{3d(p+f),\mathcal{O}_{p,k}}_{j_L,j_R}  .
\end{eqnarray}
Similarly as in  the other models, the argument in the sum (\ref{masslessP2}) is non-vanishing in the range $4 d^2 \leq p \leq \frac{3d(3d-1)}{2}+1$.
   Our results for the massive half K3 model are available for checking only the $d=1$ result in table \ref{bpstable} in section \ref{toricandmass}.

We can check the top genus pairs for the refined BPS invariants $n^d_{j_L,j_R}$ using the formulae (\ref{masslessEn}), (\ref{masslessP1P1}) and (\ref{masslessP2}) for the del Pezzo models.
 Using the general top genus formula  (\ref{topgenusmassive}) for the massive half K3 model and specializing to the diagonal classes $n_b=d$, we find that the top genus pairs for $n^d_{j_L,j_R}$ are realized at the smallest integer $p$ for which the orbit $\mathcal{O}^{X, d}_{p,k}$ is non-empty, where $X=D_5, E_6, E_7, E_8,\mathbb{P}^1\times \mathbb{P}^1, \mathbb{P}^2$ represents the del Pezzo models.
 We have discussed all the models except $E_8$, for which there is no constraint for Weyl orbits and the  lower bound is simply $p\geq 0$.
 From our discussions we find the top genus pairs for various models
\begin{eqnarray}
(g_L,g_R)^{top} =  \left\{
\begin{array}{cl}
([\frac{d^2}{2(9-n)}-\frac{d}{2}])(1,1)+(1,d),    ~~~&  \textrm{for the}~ D_5, E_6, E_7, E_8 ~\textrm{models,} \\
 ([\frac{d^2}{4} ] -d)(1,1)+(1,2d),    ~~~&  \textrm{for the}~  \mathbb{P}^1\times \mathbb{P}^1 ~\textrm{model,}  \\
(\frac{(d-1)(d-2)}{2},\frac{d(d+3)}{2}),    ~~~&  \textrm{for the}~ \mathbb{P}^2 ~\textrm{model.}  
\end{array} 
\right.
\end{eqnarray}
The formula agrees with all results in the corresponding tables that can be found in section \ref{sec.delpezzo} for the groups $D_5$, $E_6$, $E_7$ and $E_8$, in section \ref{toricandmass} for the group $\mathbb{P}^2$ and in section \ref{diagonalF0} for the diagonal $\mathbb{P}^1 \times \mathbb{P}^1$-model.


Furthermore, we can now explain certain patterns for the refined BPS invariants at the top genus for the del Pezzo models from the general formulae (\ref{masslessEn}), (\ref{masslessP1P1}) and (\ref{masslessP2}).
One can observe in the just cited tables 
that the top genus refined invariants follow a periodicity of $9-n$ for the $E_n$ ($n=5,6,7$) models and  a periodicity of two for the $\mathbb{P}^1\times \mathbb{P}^1$ model, while  the top genus numbers are always 1 for the $E_8$ and $\mathbb{P}^2$ models.
 The top genus numbers in the $D_5$, $E_6$, $E_7$ models are exactly the dimensions of the smallest irreducible representations of the groups $SO(10)$, $E_6$, $E_7$. 

We observe from the tables \ref{tablehalfK3massive1}, \ref{tablehalfK3massive2}, \ref{massivenb=3}, \ref{massivenb=4} that for the massive half K3 model, the top genus invariants are always 1.
 So the patterns for the del Pezzo models should come from the orders of orbits $|\mathcal{O}^{X,d}_{p,k}|$ in the formulae (\ref{masslessEn}) and (\ref{masslessP2}), or $|\mathcal{O}^{X,2d}_{p,k}|$ in (\ref{masslessP1P1}) for the $X=\mathbb{P}^1\times\mathbb{P}^1$ case.
 According to (\ref{topgenusmassive}) we should consider the lowest Weyl orbits which have the largest top genus.
 In the following we will assume that the top genus numbers from the massive half K3 models for the lowest non-empty orbits are always 1, so the top genus number for the del Pezzo models is simply  the order of the lowest non-empty orbit, or their sum if there are multiple non-empty lowest orbits with the same length.
 With this assumption for the massive half K3 model, we explain the patterns and compute the top genus numbers case by case. 

For the $E_8$ model there is no constraint therefore the lowest orbit is simply $\mathcal{O}_{0,1}$, and the top genus number is always $|\mathcal{O}_{0,1}|=1$. 

For the $E_7$ model, we have discussed that according to the inequality (\ref{inequal}) the lowest orbit $\mathcal{O}^{E_7,d}_{p,k}$ for degree $d$ is achieved for the smallest integer $p\geq \frac{d^2}{4}$.
 We discuss the situation according to the divisibility of $d$ by 2. 
\begin{enumerate}
\item $d$ is an even integer.
 In this case the norm square of the lowest orbit is $L^2=2p=\frac{d^2}{2}$.
 There is a unique lattice vector $\vec{w} \in \mathcal{O}^{E_7,d}_{p,k}$, which is $\vec{w}=(0,0,0,0,0,0, \frac{d}{2},-\frac{d}{2})$, and the order is  $\sum_k |\mathcal{O}^{E_7,d}_{p,k}|=1$.
 So  the top genus number is 1.  
\item $d$ is an odd integer.
 In this case the norm square of the lowest orbit is $L^2=2p=\frac{d^2+3}{2}$.
 There are two classes of the lattice vectors $\vec{w} \in \mathcal{O}^{E_7,d}_{p,k}$.
 Firstly, we can use $(w_7,w_8)=(\frac{d}{2},-\frac{d}{2})\pm (\frac{1}{2},\frac{1}{2})$, and all   $w_i$ ($i=1,2,\cdots, 6$) are 0 except one of them is $1$ or $-1$.
  There are $2\cdot6\cdot 2=24$ elements in this class.
 Secondly, we can use  $(w_7,w_8)=(\frac{d}{2},-\frac{d}{2})$, and the $w_i$ ($i=1,2,\cdots, 6$) are either $\frac{1}{2}$ or $-\frac{1}{2}$ with an odd number of them being positive to satisfy the $E_8$ lattice condition.
 This class contributes 32 elements.
 In total we find  $\sum_k |\mathcal{O}^{E_7,d}_{p,k}|=24+32=56$, which is exactly the top genus number for odd degrees observed in the tables 
for $E_7$ in subsection \ref{e7delpezzo}.
\end{enumerate}

For the $E_6$ model, the lowest orbit $\mathcal{O}^{E_6,d}_{p,k}$ of degree $d$ lies at the smallest integer $p\geq \frac{d^2}{3}$.
 We discuss the situation according to the remainder of $d$ divided by 3.
 We successfully derive the top genus numbers in tables for $E_6$ in subsection \ref{e6delpezzo} for all cases. 
\begin{enumerate}
\item $d\equiv 0$ mod 3.
 In this case the norm square of the lowest orbit is $L^2=2p=\frac{2d^2}{3}$. 
There is a unique lattice vector $\vec{w} \in \mathcal{O}^{E_6,d}_{p,k}$, which is $\vec{w}=(0,0,0,0,0,\frac{d}{3}, \frac{d}{3},-\frac{2d}{3})$, and the order is  $\sum_k |\mathcal{O}^{E_6,d}_{p,k}|=1$.
 So  the top genus number is 1.  

\item $d\equiv 1$ mod 3.
 In this case the norm square of the lowest orbit is $L^2=2p=\frac{2d^2+4}{3}$.
 There are several classes of the lattice vectors $\vec{w} \in \mathcal{O}^{E_6,d}_{p,k}$.
 Firstly, we can set $(w_6,w_7,w_8)=(\frac{d}{3}, \frac{d}{3},-\frac{2d}{3})-\frac{1}{3}(1,1,1)$, and the other $w_i$ ($i=1,2,\cdots, 5$) being 0 except one of them is $1$ or $-1$.
 There are 10 such vectors.
 Secondly, we look at $(w_6,w_7,w_8)=(\frac{d}{3}, \frac{d}{3},-\frac{2d}{3})+\frac{1}{6}(1,1,1)$, the  other $w_i$ ($i=1,2,\cdots, 5$) are $\pm \frac{1}{2}$ with even number of them positive.
 There are 16 such vectors.
 Finally there is the vector with $(w_6,w_7,w_8)=(\frac{d}{3}, \frac{d}{3},-\frac{2d}{3})+\frac{2}{3}(1,1,1)$ and all other  $w_i=0$ ($i=1,2,\cdots, 5$).
  In total we find  $\sum_k |\mathcal{O}^{E_6,d}_{p,k}|=10+16+1=27$. 

\item $d\equiv 2$ mod 3.
 The norm square of the lowest orbit is also  $L^2=2p=\frac{2d^2+4}{3}$.
 This case is similar to the case of $d\equiv 1$ mod 3.  By completely similar reasoning we also find $\sum_k |\mathcal{O}^{E_6,d}_{p,k}|=27$.
 
\end{enumerate}

For the $D_5$ model, the lowest orbit $\mathcal{O}^{D_5,d}_{p,k}$ of degree $d$ lies at the smallest integer $p\geq \frac{3d^2}{8}$.
 We discuss the situation according to the remainder of $d$ divided by 4, and find complete agreement with the pattern in table \ref{tableE5}. 
\begin{enumerate}
\item $d\equiv 0$ mod 4.
 In this case the norm square of the lowest orbit is $L^2=2p=\frac{3d^2}{4}$. 
There is a unique lattice vector $\vec{w} \in \mathcal{O}^{D_5,d}_{p,k}$, which is $\vec{w}=(0,0,0,0,\frac{d}{4},\frac{d}{4}, \frac{d}{4},-\frac{3d}{4})$, and the order is  $\sum_k |\mathcal{O}^{D_5,d}_{p,k}|=1$.
 So  the top genus number is 1.  

\item $d\equiv 1$ mod 4.
 In this case the norm square of the lowest orbit is $L^2=2p=\frac{3d^2+5}{4}$.
 There are two classes of the lattice vectors $\vec{w} \in \mathcal{O}^{D_5,d}_{p,k}$.
 Firstly, we can set $(w_5,w_6,w_7,w_8)=(\frac{d}{4},\frac{d}{4}, \frac{d}{4},-\frac{3d}{4})-\frac{1}{4}(1,1,1,1)$, and the other $w_i$ ($i=1,2,3, 4$) are 0 except one of them is 1 or $-1$.
 There are 8 such vectors.
 Secondly, we look at $(w_5,w_6,w_7,w_8)=(\frac{d}{4},\frac{d}{4}, \frac{d}{4},-\frac{3d}{4})+\frac{1}{4}(1,1,1,1)$, the  other $w_i$ ($i=1,2,3,4$) are $\pm \frac{1}{2}$ with odd number of them positive to satisfy the $E_8$ lattice condition.
 There are 8 such vectors. In total we find  $\sum_k |\mathcal{O}^{D_5,d}_{p,k}|=8+8=16$. 

\item $d\equiv 2$ mod 4.
 The norm square of the lowest orbit is also  $L^2=2p=\frac{3d^2+4}{4}$.
 Firstly, we find 2 vectors in the orbit with $(w_1,w_2,w_3,w_4)=(0,0,0,0)$ and $(w_5,w_6,w_7,w_8)=(\frac{d}{4},\frac{d}{4}, \frac{d}{4},-\frac{3d}{4})\pm \frac{1}{2}(1,1,1,1)$.
 Secondly, we consider $(w_5,w_6,w_7,w_8)=(\frac{d}{4},\frac{d}{4}, \frac{d}{4},-\frac{3d}{4})$, the  other $w_i$ ($i=1,2,3,4$) are $\pm \frac{1}{2}$ with even number of them positive to satisfy the $E_8$ lattice condition.
 There are 8 such vectors.
 In total we find $\sum_k |\mathcal{O}^{D_5,d}_{p,k}|=2+8=10$ in this case. 

\item $d\equiv 3$ mod 4.
  The norm square of the lowest orbit is also  $L^2=2p=\frac{3d^2+5}{4}$. This case is similar to the case of $d\equiv 1$ mod 4.
  By completely similar reasoning we also find $\sum_k |\mathcal{O}^{D_5,d}_{p,k}|=16$. 

\end{enumerate}

For the $\mathbb{P}^1\times \mathbb{P}^1$ model, the lowest orbit $\mathcal{O}^{\mathbb{P}^1\times \mathbb{P}^1,2d}_{p,k}$ of degree $d$ lies at the smallest integer $p\geq \frac{7d^2}{4}$.
 If $d$ is even, the norm square is $L^2=2p=\frac{7d^2}{2}$.
 There is a unique lattice vector in the orbit $\vec{w} =\frac{d}{4}(1,1,1,1,1,1,1,-7) $.
 If $d$ is odd, the norm square is $L^2=2p=\frac{7d^2+1}{2}$.
  There are 2 lattice vectors $\vec{w} =\frac{d}{4}(1,1,1,1,1,1,1,-7)\pm \frac{1}{4}(1,1,1,1,1,1,1,1)$.
 This agrees with the top genus numbers, which are 2 for odd degrees and 1 for even degrees in table \ref{tableP1P1}.

Finally for the $\mathbb{P}^2$ model,   the lowest orbit $\mathcal{O}^{\mathbb{P}^2,d}_{p,k}$ of degree $d$ lies at the smallest integer $p\geq 4d^2$.
 The norm square of the lattice vectors is $L^2=2p=8d^2$.
 There is a unique vector $\vec{w}=(d,d,d,d,d,d,d,-d)$ in the orbit, implying that the top genus numbers are always 1.

\section{Conclusion}

We have calculated in a systematic way the refined BPS invariants 
for the local del Pezzo geometries in which the compact part is 
rigid. The physical spin decomposition of the BPS numbers reflects  
the beautiful geometry governed by group theory and the modularity 
that is encoded in the elliptic curve of the mirror (B-model) 
geometry. In particular for the tensionless E-string it 
naturally provides a partition function which encodes the group 
theoretical as well as the spin content of the BPS spectrum. 

One very useful aspect of our formalism is the high degree 
of universality with which it produces geometrical invariants 
from a Riemann surface and a meromorphic differential $\lambda$. 
In fact the choice of the latter is hidden in our formalism in the 
correct scaling (\ref{giscaling}). I.e. the $J$-function, 
despite its name is not the invariant that determines 
the model. 

We give the limit from the five-dimensional theories to the four-dimensional Seiberg-Witten 
theories in many concrete cases. These limits are limits in the 
moduli space parameters only. As a consequence  they apply 
directly to refined amplitudes, which due to the  
analytic nature of our results are well defined and 
immediately expandable everywhere in the moduli space.  
This is a quite different quality then the limits often 
performed in the literature where the coordinates of 
the curves and the moduli are rescaled and it is merely 
checked that one gets the correct limiting curve.         

We then discussed important aspects of local Calabi-Yau manifolds   
in which the compact part is movable and which are modular due to 
the occurance of an elliptic curve in the A-model geometry, such as 
${\cal L}_1\oplus {\cal L}_2 \rightarrow \Sigma_g$ for $g=1$, where 
we obtained closed expressions in the Nekrasov-Shatashvili limit 
(\ref{NSlocalelliptic}), and  the M-string geometry. In particular we give evidence 
that a refined recursion and the modular structure extend to these 
cases.   

The $\frac{1}{2} K3$, which we analyzed throughoutly for the massless 
and the massive  case, has with the affine $\hat E_8$ the largest symmetry 
group and combines both aspects. It is the master geometry encoding all 
local del Pezzo cases, which are sucessively reached by flop transitions 
and blow-downs. The idea to calculate newly defined refined stable pair invariants for the 
elliptic singularities is mathematically very natural and could be 
generalized to the  extended elliptic singularities~\cite{Saito}. 

One more speculative but potentially very far reaching physical 
application is that these invariants seems not only to count the 
massless vector bosons related to  the $[p,q]$-string but also 
generate the (exceptional) gauge symmetry in F-theory, but also 
infinite towers of massive BPS states that are present in the 
geometry of the mutually non-local 7-branes. This holds similarly 
in models with non-vanishing Wilson lines, i.e. when the gauge group is 
partially broken and matter is generated. After some early 
results for example for the $F^4$-coupling for F-theory on 
K3~\cite{Lerche:1998nx, Lerche:1998gz}, which are in fact 
also closely related to elliptic curves with special monodromy,  
there are not too many facts known for F-theory as a string theory. 

It seems that the refined stable pair invariants provide a window 
into the microscopic working of F-theory.  Moreover the improved 
understanding of the mirror geometry, their GKZ systems 
and the role of non-renormalizable deformations can be be useful to  
study open string amplitudes and Wilson line amplitudes for the 
geometry of the tensionless string, the $[p,q]$-strings or the 
geometries more formally associated to more general Chern-Simons quiver gauge 
theories in the spirit of the remodeling conjecture~\cite{Marino:2006hs, Bouchard:2007ys}
 and the ABJM/topological string 
connection~\cite{Kapustin:2009kz, Drukker:2010nc, Klemm:2012ii}.

Another obvious application of the result for  $\frac{1}{2} K3$ is to use 
T-duality on the elliptic fibre~\cite{Minahan:1997,Minahan:1997a} to obtain 
a refinement of the partition of super Yang-Mills theory on four-manifolds 
with $b_{2}^+(S)=1$~\cite{Vafa:1994tf} for higher rank gauge theory\footnote{We 
are grateful to Jan Manschot for checking already that the classes with $n_b=2,3$ 
lead to the results obtained by direct calculation~ \cite{Alim:2010cf, Klemm:2012}.}.

Finally we have used mirror symmetry and the construction of Looijenga 
as well as  Friedman, Morgan and Witten to see in some detail that the 
phenomena of reduction of the structure group for Yang-Mills bundles 
on the two-torus and the stable pairs invariants connected to the 
$E_n$ gauge symmetry enhancements are effects governed by the same moduli. 
It well known~\cite{Dijkgraaf, Kaneko} that the partition function 
of two-dimensional Yang-Mills theory is given in terms of quasi-modular forms akin
to the expressions we obtain for the $E_n$ groups related to del Pezzo surfaces. 
It would be interesting to see if this is more then a formal coincidence 
or if the partition function of two-dimensional Yang-Mills theory with exceptional groups can be refined 
and has a direct relation to the results we obtain for the del Pezzo 
surfaces.

\vspace{0.20in} {\leftline {\bf Acknowledgements}}

We thank Murad Alim, Ga\"etan Borot, Jinwon Choi, J\"urgen Fuchs,  Ori Gannor, Babak Haghighat, Hans Jockers,  
Denis Klevers,  Marcos Mari\~no,  Cumrun Vafa and S.T.Yau   
for discussions. We especially thank Sheldon Katz for comments on geometric interpretations of refined BPS 
states. Parts of MH's work were conducted during his membership 
at Kavli IPMU. He thanks the Young Thousand People grant for support and 
the BCTP and HCM  in Bonn 
for hospitability. AK and MH like to thank AIMS for the opportunity to 
finalize this work. The work of MP is supported by a full scholarship of the Deutsche Telekom 
Stiftung, by a partial scholarship of the graduate school BCGS and by an ideational 
scholarship of the German National Academic Foundation.

\appendix

\section{The general Weierstrass forms for the cubic, the quartic and the bi-quadratic}   
\label{section:weierstrass}  
In this section we discuss the Weierstrass normal forms corresponding to the curves that are 
associated to the polyhedra 13, 15 and 16. This is turn corresponds to determing the 
Weierstrass normal form of a general quartic, bi-quadratic and cubic curve respectively. 
The corresponding algorithms are well-known in the literature. We briefly present 
them here for convenience and completing the discussion. The respective coefficients are translated 
as follows 
 
 \begin{center}
 \begin{tabular}{|c|c|c|c|c|c|c|c|c|c|c|c|c|}
 \hline
 Curve & $ \tilde u$ & $a_1$ &$ a_2$ & $ a_3$ & $m_1$ & $m_2 $& $m_3$ &$ m_4$ & $m_5 $&$ m_6 $& $m_7$ & $m_8$\\ \hline
 Cubic   &    $s_6 $& $s_5 $& $s_3 $ & $s_9$ &$ s_8 $& $s_2 $& $s_7$ & $s_{10}$ & $s_1 $& $s_4$ & - & -\\ \hline 
 Quartic   & $l_7$& $l_9$ & $l_6$ &$ l_4$ & $l_8$ & - & $l_3$ &$ l_5$ & - &$ l_2 $& $ l_1$ & - \\ \hline 
 Biquadratic   & $a_{11}$& $a_{22}$ & $a_{01}$ &$ a_{10}$ & $a_{21}$ & $a_{12}$ & $a_{00}$ &$ a_{20}$ & - &-& - &$ a_{02}$ \\ \hline 
\end{tabular} \, .
\end{center}

\subsection{The Weierstrass normal form of cubic curves}
We consider a cubic curve in projective space $\mathbb{P}^2 = \{[x:y:z]\}$ that is given by
\ba
0 &=& s_1 x^3 + s_2 x^2 y + s_3 x y^2 + s_4 y^3 + s_5 x^2 z + s_6 x y z + s_7 y^2 z + s_8 x z^2 + s_9 y z^2 + s_{10} z^3 \nn \\
&=& F_1(x,y) + \mathbb{F}_2(x,y) z + F_3 (x,y) z^2 + s_{10} z^3 \, .
\label{generalcubic}
\ea

The algorithm that brings this curve into Weierstrass normal form is called Nagell's algorithm. We review only the most important steps here and refer to the literature \cite{Connell} for an extensive discussion.
By a coordinate transformation $x\rightarrow x +p$, we can achieve that $s_{10}=0$. Without loss of generality we assume that $s_9 \neq 0$. We define
\be
e_i = F_i(s_9, -s_8) .
\ee
Next, the coordinate transformation
\be
(x,y) \longmapsto \begin{cases} \left(x- s_9 \frac{e_2}{e_3}\, y, y+ s_8 \frac{e_2}{e_3}\,  x \right) & \text{if $e_3 \neq 0$} \\\left(x-s_9 \, y, y+ s_8  x \right) & \text{if $e_3 = 0$} \end{cases}
\ee
is applied in order to map the rational point to the origin. Having brought the curve into this form, it is re-written as
\be
x^2 f_3'(1,t) + x' f_2'(1,t) +f_1' (1,t) =0 , \quad t= y/x ,
\ee
where the $f_i$ denote the homogeneous parts of degree $i$. The solution to this quadratic equation is given by
\be
x = \frac{-f_2'(1,t) \pm \sqrt{\delta}}{2 f_3'(1,t)}, \quad \delta = f_2'(1,t)^2 - 4 f_3'(1,t) f_1'(1,t) \, .
\ee
One zero of $\delta$ is given by $t_0 = - s_8/s_9$. By introducing 
\be
t = t_0 + \frac{1}{\tau} ,
\ee
one obtains a cubic polynomial
\be
\rho = \tau^4 \delta ,
\ee
which is easily brought into Weierstrass normal form
\be
Y^2 = 4X^3 - g_2 X  - g_3 \, .
\ee
Its explicit form in terms of the coefficients $s_i$ reads
\ban 
Y^2 &=& 4 X^3 - \frac{1}{12} \biggl(   s_6^4 - 8 s_5 s_6^2 s_7 + 16 s_5^2 s_7^2 + 24 s_4 s_5 s_6 s_8 - 8 s_3 s_6^2 s_8 - 
 16 s_3 s_5 s_7 s_8 + 24 s_2 s_6 s_7 s_8 \nn \\ &&- 48 s_1 s_7^2 s_8 + 16 s_3^2 s_8^2 - 
 48 s_2 s_4 s_8^2 - 48 s_4 s_5^2 s_9  + 24 s_3 s_5 s_6 s_9 - 8 s_2 s_6^2 s_9 - 
 16 s_2 s_5 s_7 s_9\nn \\ && + 24 s_1 s_6 s_7 s_9 - 16 s_2 s_3 s_8 s_9 + 144 s_1 s_4 s_8 s_9 + 
 16 s_2^2 s_9^2 - 48 s_1 s_3 s_9^2 - 48 s_3^2 s_5 s_{10} \nn \\ &&  + 144 s_2 s_4 s_5 s_{10} + 
 24 s_2 s_3 s_6 s_{10}  - 216 s_1 s_4 s_6 s_{10} - 48 s_2^2 s_7 s_{10} + 
 144 s_1 s_3 s_7 s_{10} \biggr) X \nn \\ && - \frac{1}{216} \biggl( s_6^6 - 12 s_5 s_6^4 s_7 + 48 s_5^2 s_6^2 s_7^2 - 64 s_5^3 s_7^3 + 
 36 s_4 s_5 s_6^3 s_8 - 12 s_3 s_6^4 s_8 - 144 s_4 s_5^2 s_6 s_7 s_8 \nn \\ && + 
 24 s_3 s_5 s_6^2 s_7 s_8 + 36 s_2 s_6^3 s_7 s_8 + 96 s_3 s_5^2 s_7^2 s_8 - 
 144 s_2 s_5 s_6 s_7^2 s_8 - 72 s_1 s_6^2 s_7^2 s_8 \nn \\ && + 288 s_1 s_5 s_7^3 s_8 + 
 216 s_4^2 s_5^2 s_8^2 - 144 s_3 s_4 s_5 s_6 s_8^2 + 48 s_3^2 s_6^2 s_8^2 - 
 72 s_2 s_4 s_6^2 s_8^2 + 96 s_3^2 s_5 s_7 s_8^2  \nn \\ &&- 144 s_2 s_4 s_5 s_7 s_8^2 - 
 144 s_2 s_3 s_6 s_7 s_8^2 + 864 s_1 s_4 s_6 s_7 s_8^2 + 216 s_2^2 s_7^2 s_8^2 - 
 576 s_1 s_3 s_7^2 s_8^2 \nn \\ && - 64 s_3^3 s_8^3 + 288 s_2 s_3 s_4 s_8^3 - 
 864 s_1 s_4^2 s_8^3 - 72 s_4 s_5^2 s_6^2 s_9 + 36 s_3 s_5 s_6^3 s_9 - 
 12 s_2 s_6^4 s_9 \nn \\ && + 288 s_4 s_5^3 s_7 s_9 - 144 s_3 s_5^2 s_6 s_7 s_9 + 
 24 s_2 s_5 s_6^2 s_7 s_9 + 36 s_1 s_6^3 s_7 s_9 + 96 s_2 s_5^2 s_7^2 s_9 \nn \\ && - 
 144 s_1 s_5 s_6 s_7^2 s_9 - 144 s_3 s_4 s_5^2 s_8 s_9 - 144 s_3^2 s_5 s_6 s_8 s_9 + 
 720 s_2 s_4 s_5 s_6 s_8 s_9 + 24 s_2 s_3 s_6^2 s_8 s_9 \nn \\ && - 648 s_1 s_4 s_6^2 s_8 s_9 + 
 48 s_2 s_3 s_5 s_7 s_8 s_9 - 1296 s_1 s_4 s_5 s_7 s_8 s_9 - 
 144 s_2^2 s_6 s_7 s_8 s_9 \nn \\ &&+ 720 s_1 s_3 s_6 s_7 s_8 s_9 - 
 144 s_1 s_2 s_7^2 s_8 s_9 + 96 s_2 s_3^2 s_8^2 s_9 - 576 s_2^2 s_4 s_8^2 s_9 + 
 864 s_1 s_3 s_4 s_8^2 s_9 \nn \\ && + 216 s_3^2 s_5^2 s_9^2 - 576 s_2 s_4 s_5^2 s_9^2 - 
 144 s_2 s_3 s_5 s_6 s_9^2 + 864 s_1 s_4 s_5 s_6 s_9^2 + 48 s_2^2 s_6^2 s_9^2\nn \\ && - 
 72 s_1 s_3 s_6^2 s_9^2 + 96 s_2^2 s_5 s_7 s_9^2 - 144 s_1 s_3 s_5 s_7 s_9^2 - 
 144 s_1 s_2 s_6 s_7 s_9^2 + 216 s_1^2 s_7^2 s_9^2 \nn \\ &&+ 96 s_2^2 s_3 s_8 s_9^2 - 
 576 s_1 s_3^2 s_8 s_9^2 + 864 s_1 s_2 s_4 s_8 s_9^2 - 64 s_2^3 s_9^3 + 
 288 s_1 s_2 s_3 s_9^3 - 864 s_1^2 s_4 s_9^3 \nn \\ && - 864 s_4^2 s_5^3 s_{10} + 
 864 s_3 s_4 s_5^2 s_6 s_{10} - 72 s_3^2 s_5 s_6^2 s_{10} - 
 648 s_2 s_4 s_5 s_6^2 s_{10} + 36 s_2 s_3 s_6^3 s_{10} \nn \\ && + 540 s_1 s_4 s_6^3 s_{10} - 
 576 s_3^2 s_5^2 s_7 s_{10} + 864 s_2 s_4 s_5^2 s_7 s_{10} + 
 720 s_2 s_3 s_5 s_6 s_7 s_{10} \nn \\ && - 1296 s_1 s_4 s_5 s_6 s_7 s_{10} - 
 72 s_2^2 s_6^2 s_7 s_{10} - 648 s_1 s_3 s_6^2 s_7 s_{10} - 576 s_2^2 s_5 s_7^2 s_{10} \nn \\ && + 
 864 s_1 s_3 s_5 s_7^2 s_{10} + 864 s_1 s_2 s_6 s_7^2 s_{10} - 864 s_1^2 s_7^3 s_{10} + 
 288 s_3^3 s_5 s_8 s_{10} \nn \\ && - 1296 s_2 s_3 s_4 s_5 s_8 s_{10} + 
 3888 s_1 s_4^2 s_5 s_8 s_{10} - 144 s_2 s_3^2 s_6 s_8 s_{10} + 
 864 s_2^2 s_4 s_6 s_8 s_{10} \nn \\ &&- 1296 s_1 s_3 s_4 s_6 s_8 s_{10} - 
 144 s_2^2 s_3 s_7 s_8 s_{10} + 864 s_1 s_3^2 s_7 s_8 s_{10} - 
 1296 s_1 s_2 s_4 s_7 s_8 s_{10}\nn \\ && - 144 s_2 s_3^2 s_5 s_9 s_{10} + 
 864 s_2^2 s_4 s_5 s_9 s_{10} - 1296 s_1 s_3 s_4 s_5 s_9 s_{10} - 
 144 s_2^2 s_3 s_6 s_9 s_{10} \nn \\ &&+ 864 s_1 s_3^2 s_6 s_9 s_{10} - 
 1296 s_1 s_2 s_4 s_6 s_9 s_{10} + 288 s_2^3 s_7 s_9 s_{10} - 
 1296 s_1 s_2 s_3 s_7 s_9 s_{10} \nn \\ &&+ 3888 s_1^2 s_4 s_7 s_9 s_{10} + 
 216 s_2^2 s_3^2 s_{10}^2 - 864 s_1 s_3^3 s_{10}^2 - 864 s_2^3 s_4 s_{10}^2 + 
 3888 s_1 s_2 s_3 s_4 s_{10}^2\nn \\ &&  - 5832 s_1^2 s_4^2 s_{10}^2 \biggr) \, .
 \label{thecubic} 
 \ean
 
 \subsection{The Weierstrass normal form of quartic curves}
 We start by considering homogeneous quartic curves \cite{Connell} in $\mathbb{P}^{(1,1,2)}$ that have the form
 \be
0= l_1 p^4 + l_2 p^3 q + l_3 p^2 q^2 + l_4 p q^3 + l_5 q^4 + l_6 p^2 r + l_7 p q r + l_8 q^2 r + l_9 r^2 \, .
\label{generalquartic}
\ee 
By setting $q=1$ we obtain the affine part, which reads after a change of coordinates
\be \label{intermediatequartic}
v^2 = \tilde l_1 p^4 + \tilde l_2 p^3 + \tilde l_3 p^2 + \tilde l_4 p + \tilde l_5 .  
\ee
The constant term can be eliminated by shifting $p \longmapsto p+ a$,  $a$ being a root of \eqref{intermediatequartic}.  After applying the final coordinate transformation
\be
v \longmapsto \frac{v}{p^2}, \quad p \longmapsto \frac{1}{p}
\ee
the curve is takes the form of a cubic
\be
y^2 + a_1 x y + a_3 y = x^3 + a_2 x^2 + a_4 x + a_6,
\ee
where the $a_i$ can be expressed in terms of $\tilde l_i$ as 
\be
a_1 = \frac{\tilde l_4}{\sqrt{\tilde l_5}}, \quad a_2 = c- \frac{\tilde l_4^2}{4 \tilde l_5}, \quad a_3 = 2 \sqrt{\tilde l_5} \tilde l_2, \quad a_4 = -4 \tilde l_5 \tilde l_1, \quad a_6 = a_2 a_4 \, .
\ee
Nagell's algorithm can be applied to this form and one finds the Weierstrass normal form
\ba
Y^2 &=& 4 X^3 - \frac{1}{12}\biggl(l_7^4 - 8 l_6 l_7^2 l_8 + 16 l_6^2 l_8^2 + 48 l_5 l_6^2 l_9 - 
  24 l_4 l_6 l_7 l_9 + 8 l_3 l_7^2 l_9 + 16 l_3 l_6 l_8 l_9 - 24 l_2 l_7 l_8 l_9 \nn \\ &&  + 
  48 l_1 l_8^2 l_9 + 16 l_3^2 l_9^2 - 48 l_2 l_4 l_9^2 + 192 l_1 l_5 l_9^2\biggr) X - \frac{1}{216}\biggl(l_7^6 - 12 l_6 l_7^4 l_8 + 48 l_6^2 l_7^2 l_8^2 \nn \\ && - 64 l_6^3 l_8^3 + 
  72 l_5 l_6^2 l_7^2 l_9 - 36 l_4 l_6 l_7^3 l_9 + 12 l_3 l_7^4 l_9 - 
  288 l_5 l_6^3 l_8 l_9 + 144 l_4 l_6^2 l_7 l_8 l_9 - 24 l_3 l_6 l_7^2 l_8 l_9 \nn \\ && - 
  36 l_2 l_7^3 l_8 l_9 - 96 l_3 l_6^2 l_8^2 l_9 + 144 l_2 l_6 l_7 l_8^2 l_9 + 
  72 l_1 l_7^2 l_8^2 l_9 - 288 l_1 l_6 l_8^3 l_9 + 216 l_4^2 l_6^2 l_9^2 \nn \\ && - 
  576 l_3 l_5 l_6^2 l_9^2 - 144 l_3 l_4 l_6 l_7 l_9^2 + 864 l_2 l_5 l_6 l_7 l_9^2 + 
  48 l_3^2 l_7^2 l_9^2 - 72 l_2 l_4 l_7^2 l_9^2 - 576 l_1 l_5 l_7^2 l_9^2 \nn \\ && + 
  96 l_3^2 l_6 l_8 l_9^2 - 144 l_2 l_4 l_6 l_8 l_9^2 - 1152 l_1 l_5 l_6 l_8 l_9^2 - 
  144 l_2 l_3 l_7 l_8 l_9^2 + 864 l_1 l_4 l_7 l_8 l_9^2 + 216 l_2^2 l_8^2 l_9^2 \nn \\ &&  - 
  576 l_1 l_3 l_8^2 l_9^2 + 64 l_3^3 l_9^3 - 288 l_2 l_3 l_4 l_9^3 + 
  864 l_1 l_4^2 l_9^3 + 864 l_2^2 l_5 l_9^3 - 2304 l_1 l_3 l_5 l_9^3\biggr)\, .
\label{thequartic}  
\ea

\subsection{The Weierstrass normal form for a bi-quadratic curve}

We follow the discussion in \cite{Duistermaat} and consider a general homogeneous bi-quadratic curve $p$ in $\mathbb{P}^1=\{[s:t]\} \times \mathbb{P}^1=\{[v:w]\} $.
\ban
0&=&a_{00}s^2 w^2+ a_{10}s t w^2 + a_{01}s^2 v w+a_{20} t^2 w^2 + a_{11}s t v w+ a_{02}s^2 v^2 + a_{21} t^2 v w \nn \\ && + a_{12}s t v^2 + a_{22} t^2  v^2.
\label{generalbiquartic}
\ean
The affine part of $p$ reads in the chart $s=1, w=1$
\be
0=a_{00}+ a_{10}t + a_{01}v+a_{20} t^2 + a_{11}t v+ a_{02} v^2 + a_{21} t^2 v + a_{12}t v^2 + a_{22} t^2 v^2 \, .
\ee
We denote by 
\be
\Delta_2(p) = \Big(\sum_{i=0}^2 s^i t^{2-i} a_{i1} \Big)^2 - 4\Big(s^i t^{2-i}a_{i0} \Big) \Big(s^i t^{2-i} a_{i2} \Big)
\ee
the discriminant with respect to the second variable $(v,w)$. The discriminant with respect to the first variable $\Delta_1(p)$ is defined analogously.
To proceed we need to introduce some more notation. Consider a homogeneous quartic in two variables $(x_0, x_1)$ 
\be \label{quarticpol}
f = a_0 x_1^4 + 4 a_1 x_0 + 6 a_2 x_0^2 x_1^2 + 4 a_3 x_0^3 x_1 + a_4 x_0^4 \, .
\ee
Next we introduce the so-called Eisenstein invariants of \eqref{quarticpol} which are projective invariants under the action of $GL(2, \mathbb{C})$ and defined as 
\ban
D & =& a_0 a_4 3 a_2^2 -4 a_1 a_3 \, , \nn \\ 
E &=& a_0 a_3^2 + a_1^2 a_4 - a_0 a_2 a_4 - 2 a_1 a_2 a_3 + a_2^3 \, .
\ean
It can be shown that the coefficients $g_2, g_3$ of the Weierstrass normal form are given as
\be
g_2 = D(\Delta_2(p)), \quad g_3 = -E(\Delta_2(p)) \, .
\ee
The general Weierstrass form of a bi-quadratic curve is finally found to be
\ban
Y^2 &=& 4 X^3 -\frac{1}{12} \biggl( a_{11}^4 - 8 a_{10} a_{11}^2 a_{12} + 
   16 a_{10}^2 a_{12}^2 - 8 a_{02} a_{11}^2 a_{20} - 
   16 a_{02} a_{10} a_{12} a_{20} \nn \\ && + 
   24 a_{01} a_{11} a_{12} a_{20}  
   -48 a_{00} a_{12}^2 a_{20} + 16 a_{02}^2 a_{20}^2 + 
   24 a_{02} a_{10} a_{11} a_{21} - 8 a_{01} a_{11}^2 a_{21} \nn \\ &&- 
   16 a_{01} a_{10} a_{12} a_{21} + 
   24 a_{00} a_{11} a_{12} a_{21} - 
   16 a_{01} a_{02} a_{20} a_{21} + 16 a_{01}^2 a_{21}^2 - 
   48 a_{00} a_{02} a_{21}^2\nn \\ && - 48 a_{02} a_{10}^2 a_{22} + 
   24 a_{01} a_{10} a_{11} a_{22} - 8 a_{00} a_{11}^2 a_{22} - 
   16 a_{00} a_{10} a_{12} a_{22} - 
   48 a_{01}^2 a_{20} a_{22}\nn \\ &&  + 
   224 a_{00} a_{02} a_{20} a_{22} - 
   16 a_{00} a_{01} a_{21} a_{22} + 16 a_{00}^2 a_{22}^2\biggr)X- \frac{1}{216} 
\bigg( a_{11}^6 - 12 a_{10} a_{11}^4 a_{12} \nn \\ && + 
   48 a_{10}^2 a_{11}^2 a_{12}^2 - 64 a_{10}^3 a_{12}^3 - 
   12 a_{02} a_{11}^4 a_{20} + 
   24 a_{02} a_{10} a_{11}^2 a_{12} a_{20} + 
   36 a_{01} a_{11}^3 a_{12} a_{20} \nn \\ &&+ 
   96 a_{02} a_{10}^2 a_{12}^2 a_{20} - 
   144 a_{01} a_{10} a_{11} a_{12}^2 a_{20} - 
   72 a_{00} a_{11}^2 a_{12}^2 a_{20} + 
   288 a_{00} a_{10} a_{12}^3 a_{20} \nn \\ && + 
   48 a_{02}^2 a_{11}^2 a_{20}^2 + 
   96 a_{02}^2 a_{10} a_{12} a_{20}^2 - 
   144 a_{01} a_{02} a_{11} a_{12} a_{20}^2 + 
   216 a_{01}^2 a_{12}^2 a_{20}^2 \nn \\ &&- 
   576 a_{00} a_{02} a_{12}^2 a_{20}^2 - 64 a_{02}^3 a_{20}^3 + 
   36 a_{02} a_{10} a_{11}^3 a_{21} - 
   12 a_{01} a_{11}^4 a_{21} - 
   144 a_{02} a_{10}^2 a_{11} a_{12} a_{21} \nn \\ &&+ 
   24 a_{01} a_{10} a_{11}^2 a_{12} a_{21} + 
   36 a_{00} a_{11}^3 a_{12} a_{21} + 
   96 a_{01} a_{10}^2 a_{12}^2 a_{21} - 
   144 a_{00} a_{10} a_{11} a_{12}^2 a_{21}\nn \\ && - 
   144 a_{02}^2 a_{10} a_{11} a_{20} a_{21} + 
   24 a_{01} a_{02} a_{11}^2 a_{20} a_{21} + 
   48 a_{01} a_{02} a_{10} a_{12} a_{20} a_{21}\nn \\ && - 
   144 a_{01}^2 a_{11} a_{12} a_{20} a_{21} + 
   720 a_{00} a_{02} a_{11} a_{12} a_{20} a_{21} - 
   144 a_{00} a_{01} a_{12}^2 a_{20} a_{21} \nn \\ &&+ 
   96 a_{01} a_{02}^2 a_{20}^2 a_{21} + 
   216 a_{02}^2 a_{10}^2 a_{21}^2 - 
   144 a_{01} a_{02} a_{10} a_{11} a_{21}^2 + 
   48 a_{01}^2 a_{11}^2 a_{21}^2\nn \\ && - 
   72 a_{00} a_{02} a_{11}^2 a_{21}^2 + 
   96 a_{01}^2 a_{10} a_{12} a_{21}^2 - 
   144 a_{00} a_{02} a_{10} a_{12} a_{21}^2  - 
   144 a_{00} a_{01} a_{11} a_{12} a_{21}^2\nn \\ && + 
   216 a_{00}^2 a_{12}^2 a_{21}^2 + 
   96 a_{01}^2 a_{02} a_{20} a_{21}^2 - 
   576 a_{00} a_{02}^2 a_{20} a_{21}^2 - 64 a_{01}^3 a_{21}^3 + 
   288 a_{00} a_{01} a_{02} a_{21}^3 \nn \\ &&- 
   72 a_{02} a_{10}^2 a_{11}^2 a_{22} + 
   36 a_{01} a_{10} a_{11}^3 a_{22} - 
   12 a_{00} a_{11}^4 a_{22} + 
   288 a_{02} a_{10}^3 a_{12} a_{22} 
\ean 
 
 \ban
   && - 
   144 a_{01} a_{10}^2 a_{11} a_{12} a_{22} + 
   24 a_{00} a_{10} a_{11}^2 a_{12} a_{22} + 
   96 a_{00} a_{10}^2 a_{12}^2 a_{22} - 
   576 a_{02}^2 a_{10}^2 a_{20} a_{22} \nn \\ &&+ 
   720 a_{01} a_{02} a_{10} a_{11} a_{20} a_{22} - 
   72 a_{01}^2 a_{11}^2 a_{20} a_{22} - 
   480 a_{00} a_{02} a_{11}^2 a_{20} a_{22} - 
   144 a_{01}^2 a_{10} a_{12} a_{20} a_{22} \nn \\ &&- 
   960 a_{00} a_{02} a_{10} a_{12} a_{20} a_{22} + 
   720 a_{00} a_{01} a_{11} a_{12} a_{20} a_{22} - 
   576 a_{00}^2 a_{12}^2 a_{20} a_{22} \nn \\ &&- 
   576 a_{01}^2 a_{02} a_{20}^2 a_{22} + 
   2112 a_{00} a_{02}^2 a_{20}^2 a_{22} - 
   144 a_{01} a_{02} a_{10}^2 a_{21} a_{22}- 
   144 a_{01}^2 a_{10} a_{11} a_{21} a_{22} \nn \\ &&+ 
   720 a_{00} a_{02} a_{10} a_{11} a_{21} a_{22} + 
   24 a_{00} a_{01} a_{11}^2 a_{21} a_{22} + 
   48 a_{00} a_{01} a_{10} a_{12} a_{21} a_{22} \nn \\ &&- 
   144 a_{00}^2 a_{11} a_{12} a_{21} a_{22} + 
   288 a_{01}^3 a_{20} a_{21} a_{22} - 
   960 a_{00} a_{01} a_{02} a_{20} a_{21} a_{22} + 
   96 a_{00} a_{01}^2 a_{21}^2 a_{22} \nn \\ &&- 
   576 a_{00}^2 a_{02} a_{21}^2 a_{22} + 
   216 a_{01}^2 a_{10}^2 a_{22}^2 - 
   576 a_{00} a_{02} a_{10}^2 a_{22}^2 - 
   144 a_{00} a_{01} a_{10} a_{11} a_{22}^2 \nn \\ &&+ 
   48 a_{00}^2 a_{11}^2 a_{22}^2 + 
   96 a_{00}^2 a_{10} a_{12} a_{22}^2 - 
   576 a_{00} a_{01}^2 a_{20} a_{22}^2 + 
   2112 a_{00}^2 a_{02} a_{20} a_{22}^2 \nn \\ && + 
   96 a_{00}^2 a_{01} a_{21} a_{22}^2 - 64 a_{00}^3 a_{22}^3\bigg).
   \label{thebi-quadratic}
\ean

\section{Some more details on del Pezzo surfaces}
In this appendix we discuss some further interesting aspects of del Pezzo surfaces. In particular, we concentrate on the relation to the $E_n$-curves that are considered in \cite{Eguchi:2002nx} starting with the $E_8$-curve.
\subsection{$E_n$-curves as Cubic curves}
The $E_8$-curve has been given in~\cite{Eguchi:2002nx}
{\footnotesize{
\begin{eqnarray}
y^2 &=& 4 x^3 + \Big(-\frac{1}{12}\tilde{u}^4 + \big(\frac{2 }{3}{\chi_1^{E_8}}-\frac{50
   }{3}{\chi_8^{E_8}}+1550\big)\tilde{u}^2 + \big(-70 {\chi_1^{E_8}}+2 {\chi_2^{E_8}}-12
   {\chi_7^{E_8}} \nn \\ && +1840 {\chi_8^{E_8}}-115010\big)\tilde u -\frac{4} {3}{\chi_1^{E_8}}{\chi_1^{E_8}}+\frac{8 }{3}{\chi_1^{E_8}}
   {\chi_8^{E_8}}+1824 {\chi_1^{E_8}}-112
   {\chi_2^{E_8}}+4 {\chi_3^{E_8}}-4 {\chi_6^{E_8}}\nn \\ &&+680
   {\chi_7^{E_8}}-\frac{28
   }{3}{\chi_8^{E_8}}{\chi_8^{E_8}}-50744 {\chi_8^{E_8}}+2399276  \Big)x \nn \\ && -\frac{1}{216}\tilde{u}^6 + 4\tilde{u}^5 + \big(\frac{1}{18}\chi_1^{E_8}+\frac{47
   }{18}\chi_8^{E_8}-\frac{5177}{6} \big)\tilde{u}^4 + \big(-\frac{107
   }{6}\chi_1^{E_8}+\frac{1}{6}\chi_2^{E_8}+3
   \chi_7^{E_8} \nn \\ &&-\frac{1580
   }{3}\chi_8^{E_8}+\frac{504215}{6} \big)\tilde{u}^3 + \big(-\frac{2 }{9}\chi_1^{E_8}\chi_1^{E_8}-\frac{20
   }{9}\chi_1^{E_8} \chi_8^{E_8}+\frac{5866
   }{3}\chi_1^{E_8}-\frac{112
   }{3}\chi_2^{E_8}+\frac{1}{3} \chi_3^{E_8} \nn \\ &&+\frac
   {11 }{3}\chi_6^{E_8}-\frac{1450
   }{3}\chi_7^{E_8}+\frac{196
   }{9}\chi_8^{E_8}\chi_8^{E_8}+39296
   \chi_8^{E_8}-\frac{12673792}{3} \big)\tilde{u}^2 + \big(\frac{94}{3}\chi_1^{E_8}\chi_1^{E_8}\nn \\ &&-\frac{2 }{3}\chi_1^{E_8}
   \chi_2^{E_8}+\frac{718}{3}\chi_1^{E_8}
   \chi_8^{E_8}-\frac{270736
   }{3}\chi_1^{E_8}-\frac{10 }{3}\chi_2^{E_8}
   \chi_8^{E_8}+2630 \chi_2^{E_8}-52
   \chi_3^{E_8}+4 \chi_5^{E_8} \nn \\ && -416 \chi_6^{E_8}+16
   \chi_7^{E_8} \chi_8^{E_8}+25880
   \chi_7^{E_8}-\frac{7328
   }{3}\chi_8^{E_8} \chi_8^{E_8} -\frac{3841382
   }{3}\chi_8^{E_8}+107263286 \big)\tilde{u} \nn \\ && \frac{8}{27}\chi_1^{E_8}\chi_1^{E_8}\chi_1^{E_8}+\frac{28
   }{9}\chi_1^{E_8}\chi_1^{E_8} \chi_8^{E_8}-1065
   \chi_1^{E_8}\chi_1^{E_8}+\frac{118 }{3}\chi_1^{E_8}
   \chi_2^{E_8}-\frac{4 }{3}\chi_1^{E_8}
   \chi_3^{E_8}+\frac{4 }{3}\chi_1^{E_8} \nn \\ && 
   \chi_6^{E_8}-\frac{8 }{3}\chi_1^{E_8}
   \chi_7^{E_8}-\frac{40}{9}\chi_1^{E_8}
   \chi_8^{E_8}\chi_8^{E_8}-\frac{19264 }{3}\chi_1^{E_8}
   \chi_8^{E_8}+\frac{4521802
   }{3}\chi_1^{E_8}-\chi_2^{E_8}\chi_2^{E_8} \nn \\ &&  +\frac{572
   }{3}\chi_2^{E_8} \chi_8^{E_8}-59482
   \chi_2^{E_8}-\frac{20 }{3}\chi_3^{E_8}
   \chi_8^{E_8}+1880 \chi_3^{E_8}+4
   \chi_4^{E_8}-232 \chi_5^{E_8}+\frac{8
   }{3}\chi_6^{E_8} \chi_8^{E_8}\nn \\ &&  +11808
   \chi_6^{E_8}-\frac{2740 }{3}\chi_7^{E_8}
   \chi_8^{E_8}-460388
   \chi_7^{E_8}+\frac{136
   }{27}\chi_8^{E_8}\chi_8^{E_8}\chi_8^{E_8}+\frac{205492
   }{3}\chi_8^{E_8}\chi_8^{E_8}\nn \\ && +\frac{45856940
   }{3}\chi_8^{E_8}-1091057493 \label{sakaie8}
\end{eqnarray}}}
We have denoted the characters of $E_n$ by
\be \label{character}
\chi_i^{E_n} = \sum_{\vec \nu \in \mathcal{R}_i}e^{i \vec {\tilde m} \vec \nu} .
\ee
$\mathcal{R}_i$ denotes the representation with highest weight being the $i$th fundamental one and $\vec m$ takes values in the $\mathbb{C}$-extended root space and have an 
interpretation as Wilson line parameters. In particular if all the masses are zero the $\chi_i$ become just the dimensions of the corresponding 
weight modules  given in (\ref{e8dimensions}).
Starting from the $E_8$-curve, one obtains successively $E_n$-curve whose maximal singularity is an $E_n$-singularity. Fur this purpose, one decomposes the characters of $E_n$ into representations of
\be
E_n \longrightarrow E_{n-1} \times U(1) ,
\ee
 factors out the $U(1)$  part $L$ and finally takes the leading part in $L$. For the present case of $E_8$ the scaling relations are explicitly given as
\ban
&\big( \chi_1^{E_8},  \chi_2^{E_8},  \chi_3^{E_8}, \chi_4^{E_8},  \chi_5^{E_8},  \chi_6^{E_8}, \chi_7^{E_8}, \chi_8^{E_8}  \big)  \nn \\& \longmapsto \big(L^2 \chi_1^{E_7}, L^3 \chi_2^{E_7}, L^4 \chi_3^{E_7}, L^6 \chi_4^{E_7}, L^5 \chi_5^{E_7}, L^4 \chi_6^{E_7}, L^3\chi_7^{E_7}, L^2 \big),   \nn \\
& (\tilde u,x,y)  \longmapsto  (L \tilde u, L^2 x, L^3 y), \quad L\rightarrow \infty \, .
\ean
Accordingly one finds the $E_7$-curve
\ban
y^2 &=& 4x^3 +\Big(-\frac{1}{12}\tilde{u}^4 + \big(\frac{2 }{3}\chi_1^{E_7}-\frac{50}{3} \big)\tilde{u}^2 + \big(2 \chi_2^{E_7}-12 \chi_7^{E_7} \big)\tilde{u} -\frac{4 }{3}\chi_1^{E_7}\chi_1^{E_7}+\frac{8
   }{3}\chi_1^{E_7}\nn \\ &&+4 \chi_3^{E_7}-4
   \chi_6^{E_7}-\frac{28}{3} \Big) x \nn \\ && -\frac{1}{216}\tilde{u}^6 + \big(\frac{1}{18}\chi_1^{E_7}+\frac{47}{18} \big)\tilde{u}^4 + \big(\frac{1}{6}\chi_2^{E_7}+3 \chi_7^{E_7} \big)\tilde{u}^3 + \big(-\frac{2 }{9}\chi_1^{E_7}\chi_1^{E_7}-\frac{20
   }{9}\chi_1^{E_7}+\frac{1}{3}\chi_3^{E_7} \nn \\ && +\frac
   {11 }{3}\chi_6^{E_7}+\frac{196}{9} \big)\tilde{u}^2 + \big( -\frac{2}{3} \chi_1^{E_7} \chi_2^{E_7}-\frac{10
   }{3}\chi_2^{E_7}+4 \chi_5^{E_7}+16 \chi_7^{E_7} \big) \tilde{u} + \frac{8 }{27}\chi_1^{E_7}\chi_1^{E_7}\chi_1^{E_7} \nn \\ && +\frac{28
   }{9}\chi_1^{E_7}\chi_1^{E_7}-\frac{4 }{3}\chi_1^{E_7}
   \chi_3^{E_7}+\frac{4 }{3}\chi_1^{E_7}
   \chi_6^{E_7}-\frac{40
   }{9}\chi_1^{E_7}-\chi_2^{E_7}\chi_2^{E_7}-\frac{20
   }{3}\chi_3^{E_7}+4 \chi_4^{E_7}\nn \\ &&+\frac{8
   }{3}\chi_6^{E_7}+\frac{136}{27}\, ,
\ean
as well as the $E_6$-curve
\ban
y^2 &=& 4x^3 + \Big( - \frac{1}{12} \tilde u^4 + \frac{2}{3}  \chi_1^{E_6} \tilde u^2 +  \big(-12 + 2 \chi_2^{E_6}\big) \tilde u  + 4 \chi_3^{E_6} - 4 \chi_6^{E_6} - \frac{4}{3} \chi_1^{E_6} \chi_1^{E_6} \Big) x  \nn \\  && - \frac{1}{216}\tilde u^6 + \frac{1}{18} \chi_1^{E_6} \tilde u^4  + \big(3 + \frac{1}{6} \chi_2^{E_6}  \big)\tilde u^3  + \big(-\frac{2}{9}\chi_1^{E_6}\chi_1^{E_6} + \frac{1}{3} \chi_3^{E_6} + \frac{11}{3} \chi_6^{E_6} \big)\tilde u^2 \nn \\ && + \big(4 \chi_5^{E_6} - \frac{2}{3}  \chi_1^{E_6} \chi_2^{E_6} \big) \tilde u + \frac{8}{27} \chi_1^{E_6} \chi_1^{E_6} \chi_1^{E_6}   - \chi_2^{E_6} \chi_2^{E_6}   - \frac{4}{3} \chi_1^{E_6} \chi_3^{E_6} + 4 \chi_4^{E_6} + \frac{4}{3} \chi_1^{E_6} \chi_6^{E_6} \, . \nn \\ &&
\ean
The $E_6$-curve can be mapped onto the general Weierstrass normal form of the cubic \eqref{thecubic} provided one chooses a different gauge. 
Instead of setting three of the inner points of the one-dimensional faces to 1, which is convenient for the description of local $\mathbb{P}^2$ and its blow-ups, 
one has to set the coefficients of the monomials $x^3, y^3, z^3$ to $-1$. The characters $\chi_i$ may then be written in terms of $a_1, a_2, a_3, m_1, m_2, m_3$ as follows
\ban
\chi_1 &=& m_3 a_1 + m_1 a_2 + m_2 a_3, \nn \\
\chi_2 &=& -3 - m_1 m_2 m_3 + m_1 a_1 + m_2 a_2 + m_3 a_3 - a_1 a_2 a_3, \nn \\
 \chi_3 &=& -m_1^2 m_2 - m_2^2 m_3 - m_1 m_3^2 - 2 m_2 a_1 - 
  2 m_3 a_2 + m_1 m_3 a_1 a_2 - a_1 a_2^2 - 2 m_1 a_3 \nn \\ && + m_2 m_3 a_1 a_3 - a_1^2 a_3 + 
  m_1 m_2 a_2 a_3 - a_2 a_3^2, \nn \\
  \chi_4 &=&
 9 - m_1^3 - m_2^3 - m_3^3 - 6 m_1 a_1 - m_1^2 m_2 m_3 a_1 - m_2^2 m_3^2 a_1 - 
  2 m_2 m_3 a_1^2 - a_1^3 \nn \\ && - 6 m_2 a_2 - m_1 m_2^2 m_3 a_2 - m_1^2 m_3^2 a_2 + 
  m_1 m_2 a_1 a_2 - 2 m_3^2 a_1 a_2 - 2 m_1 m_3 a_2^2 \nn \\ &&- m_3 a_1^2 a_2^2 - a_2^3 - 
  m_1^2 m_2^2 a_3 - 6 m_3 a_3 - m_1 m_2 m_3^2 a_3 - 2 m_2^2 a_1 a_3 + 
  m_1 m_3 a_1 a_3 \nn \\ &&- 2 m_1^2 a_2 a_3 + m_2 m_3 a_2 a_3 + m_1 m_2 m_3 a_1 a_2 a_3 - 
  m_1 a_1^2 a_2 a_3 - m_2 a_1 a_2^2 a_3\nn \\ && - 2 m_1 m_2 a_3^2 - m_2 a_1^2 a_3^2 - 
  m_3 a_1 a_2 a_3^2 - m_1 a_2^2 a_3^2 - a_3^3, \nn \\
\chi_5 &=& -m_1 m_2^2 - m_1^2 m_3 - 
  m_2 m_3^2 - 2 m_3 a_1 - 2 m_1 a_2 + m_2 m_3 a_1 a_2 - a_1^2 a_2 - 2 m_2 a_3 \nn \\ &&+ 
  m_1 m_2 a_1 a_3 + m_1 m_3 a_2 a_3 - a_2^2 a_3 - a_1 a_3^2, \nn \\
\chi_6 &=& m_2 a_1 + m_3 a_2 + m_1 a_3.
\ean
Repeating the above procedure one finds the $D_5$-curve
\ban \label{SakaiD5curve}
y^2 &=& 4x^3 + \Big( - \frac{1}{12} \tilde u^4 + \frac{2}{3} \chi_1^{D_5}  \tilde u^2 + 2 \chi_2^{D_5} \tilde u   - \frac{4}{3} \chi_1^{D_5} \chi_1^{D_5} + 4 \chi_3^{D_5} -4 \Big) x  \nn \\  && - \frac{1}{216}\tilde u^6 + \frac{1}{18} \chi_1^{D_5}  \tilde u^4 + \frac{1}{6} \chi_2^{D_5} \tilde u^3   + \big( \frac{11}{3} - \frac{2}{9} \chi_1^{D_5} \chi_1^{D_5} + \frac{1}{3} \chi_3^{D_5} \big)\tilde u^2 +\big(4  \chi_5^{D_5} - \frac{2}{3} \chi_1^{D_5} \chi_2^{D_5} \big) \tilde u \nn \\ && + \frac{4}{3} \chi_1^{D_5}   + 
 \frac{8}{27} \chi_1^{D_5} \chi_1^{D_5} \chi_1^{D_5}    - \chi_2^{D_5} \chi_2^{D_5}  - 
 \frac{4}{3} \chi_1^{D_5} \chi_3^{D_5} + 4 \chi_4^{D_5} ,
 \ean
the $E_4$-curve
\ban
y^2 &=& 4x^3 + \Big(-\frac{1}{12}\tilde u^4 + \frac{2}{3} \chi_1^{E_4} \tilde u^2 + 2 \chi_2^{E_4} \tilde u  + 4 \chi_3^{E_4} - \frac{4}{3} \chi_1^{E_4} \chi_1^{E_4}\Big)x \nn \\ && -\frac{1}{216}\tilde u^6 + \frac{1}{18} \chi_1^{E_4} \tilde u^4 + \frac{1}{6} \chi_2^{E_4} \tilde u^3 + \big(\frac{1}{3} \chi_2^{E_3} -\frac{2}{9} \chi_1^{E_4} \chi_1^{E_4}   \big) \tilde u^2 + \big(4 - \frac{2}{3} \chi_1^{E_4} \chi_2^{E_4} \big) \tilde u \nn \\ && + \frac{8}{27} \chi_1^{E_4} \chi_1^{E_4} \chi_1^{E_4} - \chi_2^{E_4} \chi_2^{E_4} - \frac{4}{3} \chi_1^{E_4} \chi_3^{E_4} + 4 \chi_4^{E_4} ,
\ean
and finally the $E_3$-curve
\ban \label{E3curve}
y^2 &=& 4x^3 + \Big(-\frac{1}{12}\tilde u^4 + \frac{2}{3} \chi_1^{E_3} \tilde u^2 + 2 \chi_2^{E_3} \tilde u  + 4 \chi_3^{E_3} - \frac{4}{3} \chi_1^{E_3} \chi_1^{E_3}\Big)x \nn \\ && -\frac{1}{216}\tilde u^6 + \frac{1}{18} \chi_1^{E_3} \tilde u^4 + \frac{1}{6} \chi_2^{E_3} \tilde u^3 + \big(\frac{1}{3}\chi_3^{E_3} -\frac{2}{9} \chi_1^{E_3} \chi_1^{E_3} \big)\tilde u^2 -\frac{2}{3} \chi_1^{E_3} \chi_2^{E_3} \tilde u \nn \\ && -\frac{4}{3} \chi_1^{E_3} \chi_3^{E_3} + \frac{8}{27} \chi_1^{E_3} \chi_1^{E_3} \chi_1^{E_3} - \chi_2^{E_3} \chi_2^{E_3} + 4.
\ean

The case of the $E_3$ is distinguished in the following sense.
 It is the last curve that is toric, but it is the first curve for which the identification of the orthogonal complement to the canonical class inside the homology lattice can be identified with the root lattice of $E_3$.
 In the following we illustrate this correspondence explicitly. 
As a first step we recall the toric data of $\mathcal{B}_3$, i.e. the generators of the Mori cone \eqref{datadp3} and make the identification of points in the toric diagram with divisors explicit. Note that we omit the non-compact direction, i.e. we are not considering the CY geometry here, but just its base for simplicity.
 Therefore curves and divisors are the same in this case.

\begin{center}
$\begin{array}{|c|c|c|c|c|c|c|} \hline 
l^{(1)} & l^{(2)} & l^{(3)} & l^{(4)} & l^{(5)} & l^{(6)} & \text{Divisor Class} \\ \hline\hline
-1 & -1 & -1 & -1 & -1& -1 & \\
-1 & 1 & 0 & 0 & 0& 1 & e_1 \\
1 & -1 & 1 & 0 & 0 & 0 & h - e_1 - e_2 \\
0 & 1 &-1 &1 & 0 & 0 & e_2 \\
0 & 0& 1 & -1 & 1 & 0& h-e_2-e_3 \\
0& 0& 0 & 1 & -1 & 1 & e_3 \\
1 & 0 & 0& 0 & 1 & -1 & h -e_1 -e_3 \\\hline \hline
e_1 & h -e_1-e_2 & e_2 & h-e_2-e_3 & e_3 & h-e_1-e_3 & \\ \hline
 \end{array}$
 \end{center}
 here we have denoted by $h$ the hyperplane class in $\mathbb{P}^2$ and by $e_1, e_2, e_3$ the classes of the three blow-up divisors.

As it was already explained in the discussion in the main text, the Mori as well as the K\"ahler cone are non-simplicial.
 It is sufficient for our purposes to only search for a dual basis of the first four generators $l^{(1)},..., l^{(4)}$. The dual generators read in terms of $h, e_1, e_2, e_3$
\begin{center}
$\begin{array}{|c|c|} 
\text{Generator} & \text{Dual generator} \\
e_1 & h - e_1 -e_3 \\
h-e_1 - e_2 & h-e_3 \\
e_2 & h-e_2 \\
h-e_2-e_3 & e_3\\
\end{array}$ \, .
\end{center}
The K\"ahler form enjoys accordingly an expansion
\be
J = v_1 (h - e_1 -e_3) + v_2(h-e_3) + v_3(h-e_2) + v_4 e_3 \, .
\ee
The moduli are given by the relations corresponding to the $l^{(i)}$
\be
\log(v_1) = \frac{a_1 a_3}{\tilde{u} m_1}, \quad \log(v_2)= \frac{m_1 m_2}{\tilde{u} a_1}, \quad \log(v_3) = \frac{a_1 a_2}{\tilde{u} m_2}, \quad \log(v_4) = \frac{m_2 m_3}{\tilde{u} a_2},
\ee
where we have used the leading mirror map at large radius.
Using this, one easily computes the volumes of a divisor $D$ as
\be
\text{vol}(D) = \int_D J = J \cdot D. 
\ee
and obtains explicitly
\be
\text{vol}(h) = \log(v_1 v_2 v_3), \quad \text{vol}(e_1) = \log(v_1), \quad \text{vol}(e_2) = \log(v_3), \quad \text{vol}(e_3) = \log \Big(\frac{v_1 v_2}{v_4}\Big) \, .
\ee

For convenience we recall explicitly the homology of the $\mathcal{B}_3$ surface. 
The orthogonal complement of the canonical class of $\mathcal{B}_3$ reads
\be
K= - 3h + e_1 +e_2 + e_3 \, .
\ee
The simple roots of $\mathcal{B}_3$ are given in terms of divisors as
\be
\alpha_1 = e_1 - e_2, \quad \alpha_2 = e_2 - e_3,\quad \alpha_3 =  h - e_1-e_2-e_3.
\ee
Note that the roots intersect precisely as the Cartan matrix of $A_2 \times A_1$ 
\be
\alpha_i \cdot \alpha_j = \begin{pmatrix} -2 & 1 &0\\1 & -2 &0\\ 0& 0 & -2 \end{pmatrix} \, .
\ee
From these one can determine the corresponding fundamental weights $\mathcal{V}_i$ which are defined as
\be
2 \frac{\alpha_i \cdot \mathcal{V}_j}{\alpha_i \cdot \alpha_i} = \delta_{ij} \quad \Leftrightarrow \quad \alpha_i \cdot \mathcal{V}_j = - \delta_{ij} \, .
\ee
One obtains
\be
\mathcal{V}_1 = \frac{1}{3}\big(2e_1 -e_2-e_3 \big), \quad \mathcal{V}_2 = \frac{1}{3} \big(e_1 + e_2 -2e_3 \big), \quad \mathcal{V}_3 = - \frac{1}{2}\big(-h +e_1 +e_2 +e_3 \big)\, .
\ee
Acting with the roots on the highest weights, one can work out the representations. E.g. taking $\mathcal{V}_1$  as the highest weight, one obtains the fundamental representation of $A_2$ which consists out of the following weights
\be 
\nu_1 = \mathcal{V}_1, \quad \nu_2=\mathcal{V}_1 - \alpha_1, \quad \nu_3 = \mathcal{V}_1 -\alpha_1 - \alpha_2 \, .
\ee
By pairing the roots (that are still given in terms of divisors) with  the K\"ahler form and exponentiating the result, one can work out the characters of the representation and compare with \eqref{character}. For $\mathcal{V}_1$ one obtains
\ban
\chi_1 &=& \Big(\frac{{a}_1 {a}_3 {m}_2 {m}_3}{{a}_2^2 {m}_1^2}\Big)^{\frac{1}{3}} + \Big(\frac{{a}_1 {a}_2 {m}_1 {m}_3}{{a}_3^2 {m}_2^2}\Big)^{\frac{1}{3}} + \Big(\frac{{a}_2 {a}_3 {m}_1 {m}_2}{{a}_1^2 {m}_3^2}\Big)^{\frac{1}{3}}, \nn \\
\chi_2 &=& \Big(\frac{{a}_1^2 {m}_3^2}{{a}_2 a_3 {m}_1 m_2}\Big)^{\frac{1}{3}} + \Big(\frac{{a}_3^2  {m}_2^2}{{a}_1 a_2 {m}_1 m_3}\Big)^{\frac{1}{3}} +\Big(\frac{{a}_2^2 {m}_1^2}{{a}_1 {a}_3 {m}_2 {m}_3}\Big)^{\frac{1}{3}}, \nn \\
\chi_3 &=& \Big(\frac{a_1 a_2 a_3}{m_1 m_2 m_3})^{-\frac{1}{2}} + \Big(\frac{a_1 a_2 a_3}{m_1 m_2 m_3})^{\frac{1}{2}} \, .
\ean
We obtain a matching of the curve \eqref{E3curve} with \eqref{thecubic}, with $m_4,m_5,m_6$ vanishing,  provided we make the following identification
\be
{a}_1 \mapsto \frac{1}{{m}_1}, \quad {a}_2 \mapsto \frac{1}{{m}_2}, \quad {a}_3 \mapsto \frac{1}{{m}_3}. 
\ee
In this case the characters read explicitly
\ban
\chi_1 &=& \frac{m_1}{m_3} + \frac{m_2}{m_1} + \frac{m_3}{m_2}\, , \nn \\
\chi_2 &=& \frac{m_1}{m_2} + \frac{m_2}{m_3} + \frac{m_3}{m_1}\, , \nn \\
\chi_3 &=& \frac{1}{m_1 m_2 m_3} + {m_1 m_2 m_3}\, .
\ean
Note that this is just one possible identification, e.g. the identification
\be
{a}_1 \mapsto \frac{1}{{m}_2}, \quad {a}_2 \mapsto \frac{1}{{m}_3}, \quad {a}_3 \mapsto \frac{1}{{m}_1}. 
\ee
leads to the characters of the complex conjugate representation.
We conclude by noting that the identification on the level of Wilson line parameters $\tilde m_1, \tilde m_2, \tilde m_3$ is given as
\be
M_1 = \frac{m_3}{m_1}, \quad M_2 = \frac{m_2}{m_1}, \quad M_3=\frac{1}{m_1 m_2 m_3}, \quad M_i = e^{\tilde m_i} .
\ee
\subsection{The third order differential operator for $\mathcal{B}_2$}
The third order differential operator for $\mathcal{B}_2$ is given by
\ban
{\cal L}&=& (-12 m_1^2 + 12 m_1 m_2 + 16 m_1^4 m_2 - 12 m_2^2 - 12 m_1^3 m_2^2 - 
  12 m_1^2 m_2^3 + 16 m_1^5 m_2^3 + 16 m_1 m_2^4  \nn \\ &&  - 32 m_1^4 m_2^4 + 
  16 m_1^3 m_2^5 + 9 \tilde u + 24 m_1^2 m_2 \tilde u + 24 m_1 m_2^2 \tilde u - 32 m_1^4 m_2^2 \tilde u + 
  56 m_1^3 m_2^3 \tilde u  \nn \\ &&  - 32 m_1^2 m_2^4 \tilde u - 18 m_1 \tilde u^2 - 18 m_2 \tilde u^2 - 
  4 m_1^3 m_2 \tilde u^2 - 68 m_1^2 m_2^2 \tilde u^2 - 4 m_1 m_2^3 \tilde u^2 - 
  4 m_1^4 m_2^3 \tilde u^2  \nn \\ &&  - 4 m_1^3 m_2^4 \tilde u^2  + 8 m_1^2 \tilde u^3 + 36 m_1 m_2 \tilde u^3 + 
  8 m_2^2 \tilde u^3 + 28 m_1^3 m_2^2 \tilde u^3 + 28 m_1^2 m_2^3 \tilde u^3 + 
  8 m_1^4 m_2^4 \tilde u^3  \nn \\ &&  - 16 m_1^2 m_2 \tilde u^4 - 16 m_1 m_2^2 \tilde u^4 - 
  16 m_1^3 m_2^3 \tilde u^4 + 7 m_1^2 m_2^2 \tilde u^5)\partial_{\tilde u}+ 
(-108 m_1 - 128 m_1^4 \nn \\ && - 108 m_2   + 64 m_1^3 m_2 - 264 m_1^2 m_2^2 - 
  192 m_1^5 m_2^2 + 64 m_1 m_2^3 + 128 m_1^4 m_2^3 - 128 m_2^4 \nn \\ && + 
  128 m_1^3 m_2^4  - 128 m_1^6 m_2^4 - 192 m_1^2 m_2^5 + 256 m_1^5 m_2^5 - 
  128 m_1^4 m_2^6 + 144 m_1^2 \tilde u\tilde u \nn \\ && + 450 m_1 m_2  + 288 m_1^4 m_2 \tilde u    + 
  144 m_2^2 \tilde u + 240 m_1^3 m_2^2 \tilde u + 240 m_1^2 m_2^3 \tilde u + 288 m_1^5 m_2^3 \tilde u \tilde u \nn \\ && + 
  288 m_1 m_2^4 \tilde u - 320 m_1^4 m_2^4 \tilde u   + 288 m_1^3 m_2^5 \tilde u + 27 \tilde u^2 - 
  64 m_1^3 \tilde u^2 - 384 m_1^2 m_2 \tilde u^2 \tilde u \nn \\ && - 384 m_1 m_2^2 \tilde u^2 - 
  336 m_1^4 m_2^2 \tilde u^2 - 64 m_2^3 \tilde u^2    - 240 m_1^3 m_2^3 \tilde u^2 - 
  336 m_1^2 m_2^4 \tilde u^2 - 64 m_1^5 m_2^4 \tilde u^2 \tilde u \nn \\ && - 64 m_1^4 m_2^5 \tilde u^2 - 
  52 m_1 \tilde u^3 - 52 m_2 \tilde u^3    + 112 m_1^3 m_2 \tilde u^3 + 172 m_1^2 m_2^2 \tilde u^3 + 
  112 m_1 m_2^3 \tilde u^3 \tilde u \nn \\ && + 112 m_1^4 m_2^3 \tilde u^3 + 112 m_1^3 m_2^4 \tilde u^3 + 
  24 m_1^2 \tilde u^4    + 100 m_1 m_2 \tilde u^4 + 24 m_2^2 \tilde u^4 + 12 m_1^3 m_2^2 \tilde u^4 \tilde u \nn \\ && + 
  12 m_1^2 m_2^3 \tilde u^4 + 24 m_1^4 m_2^4 \tilde u^4 - 46 m_1^2 m_2 \tilde u^5  
  46 m_1 m_2^2 \tilde u^5 - 46 m_1^3 m_2^3 \tilde u^5 + 21 m_1^2 m_2^2 \tilde u^6)\partial_{\tilde u}^2  \nn \\ &&
          +(-9 - 4 m_1^2 m_2 - 4 m_1 m_2^2 + 8 m_1 \tilde u + 8 m_2 \tilde u  + 8 m_1^2 m_2^2 \tilde u - 
     7 m_1 m_2 \tilde u^2) (-27 + 16 m_1^3 \tilde u \nn \\ && - 24 m_1^2 m_2 - 24 m_1 m_2^2 + 
    16 m_1^4 m_2^2 + 16 m_2^3 - 32 m_1^3 m_2^3   + 16 m_1^2 m_2^4 + 36 m_1 \tilde u + 
    36 m_2 \tilde u \tilde u \nn \\ && - 16 m_1^3 m_2 \tilde u + 64 m_1^2 m_2^2 \tilde u - 16 m_1 m_2^3 \tilde u - 
    8 m_1^2 \tilde u^2 - 46 m_1 m_2 \tilde u^2    - 8 m_2^2 \tilde u^2 - 8 m_1^3 m_2^2 \tilde u^2 \tilde u \nn \\ && - 
    8 m_1^2 m_2^3 \tilde u^2 - \tilde u^3 + 8 m_1^2 m_2 \tilde u^3 + 8 m_1 m_2^2 \tilde u^3 + m_1 \tilde u^4 + 
    m_2 \tilde u^4  + m_1^2 m_2^2 \tilde u^4 - m_1 m_2 \tilde u^5)\partial_{\tilde u}^3 \, . \nn \\ &&
\label{PFEf2}
\ean
We note the limits $\mathcal{B}_1$ and $\mathbb{P}^2$ in the case of one respectively two vanishing mass parameters.

\subsubsection{The vertical $E_6$  del Pezzo and its transition} 
Here we recall shortly the transition from the $(5,101)$ 
elliptic Calabi-Yau  with a zero section and two rational sections, 
by first contracting all sections in the elliptic surface and then 
shrinking the $E_6$ del Pezzo to an elliptic singularity that 
deforms to the  $(4,112)$ model with the same fiber 
properties~\cite{Klemm:1996hh}.

{\footnotesize 
\begin{equation} 
 \label{dataratfib} 
 \begin{array}{ccrrrr|rrrrrl|} 
    \multicolumn{6}{c}{\nu_i }    &l^{(T)}& l^{(S)}& l^{(U)} & l^{(\mathbb{F}_1^F)}&  l^{(\mathbb{F}^B_1)} &\\ 
    D_0    &&     0&     0&   0&   0&      -1&    -1& -1&   0&   0&    \\ 
    D_1    &&     0&     0&   1&   1&      -1&    1&  1&  -1& -2&        \\ 
    D_2    &&    -1&     0&   1&   1&       0&    0&  0&   -1&  1&       \\ 
    D_3    &&    -1&    -1&   1&   1&       0&    0&  0&   1&  0&      \\ 
    D_4    &&     0&     1&   1&   1&       0&    0&  0 &  1&  0&     \\ 
    D_5    &&     1&     0&   1&   1&       0&    0&  0&    0&  1&      \\ 
    D_6    &&     0&     0&   0&   1&       1&   -1&  -1&   0&  0&   \\ 
    D_7    &&     0&     0&   1&   0&       1&    0&  0&    0&  0&    \\  
    D_8    &&     0&     0&   0&   -1&      0&    1&  0&    0&  0&      \\ 
    D_9    &&     0&     0&  -1&   0&       0&    0&  1&    0&  0&   \\ 
  \end{array} \ . 
\end{equation} }

The most interesting property is the splitting of the $E_8$ representations 
of the BPS-invariants into  $U(1)_1\times E_7$  and further into 
$U(1)_1\times U(1)_2\times E_6$ representations. We quote the pattern 
from~\cite{Klemm:1996hh}     
$$
\vbox{\offinterlineskip\tabskip=0pt
\halign{\strut\vrule#
&\hfil~$#$
&\vrule#&~
\hfil ~$#$~
&\hfil ~$#$~
&\hfil $#$~
&\hfil $#$~
&\hfil $#$~
&\hfil $#$~
&\hfil $#$~
&\hfil $#$~
&\vrule#
&\hfil ~$#$~
&\vrule#\cr
\noalign{\hrule}
& && U(1)_1\times U(1)_2\times E_6  \quad d_{W_1} 
       & 0    &  1  &  2   &  3    &     4 &  5    &   6     &&U(1)_1\times E_7 \quad  \sum &\cr
\noalign{\hrule}
&d_{W_2}
    &&  &     &      &      &       &       &       &         && &\cr
& 0 &&  &    1&      &      &       &       &       &         &&1&\cr
& 1 &&  &    1&    27&   27 &      1&       &       &         &&56&\cr
& 2 &&  &     &      &   27 &     84&     27&       &         &&138&\cr
& 3 &&  &     &      &      &      1&     27& 27    &     1   &&56&\cr
& 4 &&  &     &      &      &       &       &       &     1   &&1 &\cr
\noalign{\hrule}
&   &&  &     &      &      &   E_8 &       &       &        &&\sum=252 &\cr
\noalign{\hrule}}
\hrule}$$
\vskip-7pt
\noindent
Invariants of $X_{(\Delta^F(5)\times_F^2\Delta^B(3))}$  for rational curves of degree $d_{\mathbb{F}^B_1}=0$, 
$d_{\mathbb{F}^F_1}=1$ and $d_T=1$.

The classical  intersection ring is   
\begin{equation}
 \begin{array}{rl}
{\cal R}=&2 B^2 S+3 B^2 T+2 B^2 U+2 B F S+3 B F T+2 B F U+6 B S^2+9 B S T+9 B S U \\ &  + 9 B T^2+
 9 B T U+6 B U^2+4 F S^2+6 F S T+  6 F S U+6 F T^2+6 F T U+4 F U^2 \\ & +16 S^3+24 S^2 T+24 S^2 U+ 
 24 S T^2+24 S T U+24 S U^2+24 T^3+24 T^2 U+24 T U^2\\ & +16 U^3
\end{array}
\end{equation}
supplemented by intersection with the cycle defined by the second chern class  
\be 
F c_2=24,\quad B c_2=36,\quad T c_2=84, \quad S c_2=88, \quad U c_2=88\ .
\ee
The Wilson line classes correspond to $W_1=T-S$ and $W_2=U$. We see that 
$X_{(\Delta^F(5)\times_F^2\Delta^B(3))}$ is a K3-fibration over the base of  $\mathbb{F}_1$,
an elliptic fibration over $\mathbb{F}_1$ and in particular an elliptic 
surface $S$ over the base of $\mathbb{F}_1$. The flopped
phase is defined by the classes   
\be
l^{(T)},l^{(S)}+l^{(\mathbb{F}_1^F)},l^{(U)}+l^{(\mathbb{F}_1^F)}, -l^{(\mathbb{F}_1^F)}, l^{(\mathbb{F}_1^B)} +l^{(\mathbb{F}_1^F)} 
\ee 
and flops all sections out of $S$, which becomes an elliptic pencil 
with three basepoints, i.e. an $E_6$ del Pezzo which can be shrunken and 
deformed to $X_{(\Delta^F(5)\times_F^2\Delta^B(1))}$.

\section{Weyl invariant Jacobi modular forms for $E_8$ lattice} \label{appendixJacobi}
Our convention for the theta functions are 
\begin{eqnarray}
&& \theta_1(m,\tau)= -i \sum_{n\in \mathbb{Z}} (-1)^n \exp[\pi i m(2n+1) ] \exp[\pi i \tau (n+\frac{1}{2})^2], \nonumber \\
&& \theta_2(m,\tau)= \sum_{n\in \mathbb{Z}} \exp[\pi i m(2n+1) ] \exp[\pi i \tau (n+\frac{1}{2})^2],  \nonumber \\
&& \theta_3(m,\tau)= \sum_{n\in \mathbb{Z}} \exp(2\pi i mn ) \exp (\pi i \tau n^2),  \nonumber \\
&& \theta_4(m,\tau)= \sum_{n\in \mathbb{Z}} (-1)^n\exp(2\pi i mn ) \exp (\pi i \tau n^2) . 
\end{eqnarray} 
We use the notation  $\theta_i(\tau) \equiv \theta_i(0,\tau)$ for the massless theta functions.  We define a modular form $h(\tau)$ as 
\begin{eqnarray}
h(\tau) =\theta_2(2\tau)  \theta_2(6\tau)  + \theta_3(2\tau)  \theta_3(6\tau) = 1+6q+ 6q^3 +6 q^4 + 12 q^7 + \mathcal{O}(q^9) ,
\end{eqnarray} 
where the parameter is $q=e^{2\pi i\tau}$.

The nine Weyl invariant Jacobi forms can be written in terms of the theta function of the $E_8$ lattice 
\begin{eqnarray} 
\Theta(\vec{m},\tau) =\sum_{\vec{w}\in \Gamma_8}\exp(\pi i \tau \vec{w}^2+2\pi i \vec{m}\cdot \vec{w}) . 
\end{eqnarray} 
The formulae read as follows 
\begin{eqnarray}
 A_1 &=&  \Theta(\vec{m},\tau) , ~~~~~~ A_4 ~~ =~~  \Theta(2\vec{m}, \tau),  \nonumber \\
A_n &=&  \frac{n^3}{n^3 + 1}[(\Theta(n\vec{m},n\tau)+ \frac{1}{n^4}
\sum_{k=0}^{n-1} \Theta(\vec{m},\frac{\tau+k}{n})],   ~~~~ n= 2, 3,5,    \nonumber \\
 B_2  &=&  \frac{8}{15} [ (\theta_3(\tau)^4 + \theta_4(\tau)^4)\Theta(2\vec{m}, 2\tau)  
- \frac{1}{2^4}(\theta_2(\tau)^4 + \theta_3(\tau)^4)\Theta(\vec{m}, \frac{\tau}{2}) 
 \nonumber \\ && 
+\frac{1}{2^4}(\theta_2(\tau)^4 - \theta_4(\tau)^4)\Theta(\vec{m}, \frac{\tau+1}{2}) ],  \nonumber \\
 B_3 &=&  \frac{81}{80}[ h(\tau)^2\Theta(3\vec{m},3\tau) - \frac{1}{3^5}
  \sum_{k=0}^2 h(\frac{\tau+k}{3})^2  \Theta (\vec{m}, \frac{\tau+k}{3})],  \nonumber \\
B_4 &=&    \frac{16}{15}  [  \theta_4(2\tau)^4 \Theta(4\vec{m}, 4\tau) 
- \frac{1}{2^4}\theta_4(2\tau)^4 \Theta(2\vec{m}, \tau+\frac{1}{2}) 
 - \frac{1}{4^5 } \sum_{k=0}^3 \theta_2(\frac{\tau+k}{2})^4 \Theta(\vec{m},  \frac{\tau+k}{4})],  \nonumber \\
  B_6 &=&  \frac{9}{10} [ h(\tau)^2 \Theta(6\vec{m},6\tau) + \frac{ h(\tau)^2  }{2^4}
 \sum_{k=0}^1 \Theta (3\vec{m},\frac{3\tau+3k}{2}) 
 - \frac{1}{3^5} \sum_{k=0}^2 h(\frac{\tau+k}{3})^2  \Theta (2\vec{m}, \frac{2(\tau+k)}{3})
 \nonumber \\ &&  -  \frac{1}{3\cdot 6^4}  \sum_{k=0}^5 h(\frac{\tau+k}{3})^2  \Theta (\vec{m}, \frac{\tau+k}{6}).
   \end{eqnarray}

\section{The BPS invariants for the half  K3 and the diagonal classes of $\mathbb{P} \times  \mathbb{P}^1$}
 
Here we list the BPS invariants for the half K3  and the  diagonal $\mathbb{P}\times \mathbb{P}^1$ 

\subsection{The diagonal $\mathbb{P}\times \mathbb{P}^1$ model} 
\label{diagonalF0} 
\begin{table}[H]
\begin{center}
{\footnotesize 
\begin{tabular} {|c|c|c|} \hline $2j_L \backslash 2j_R$  & 0 & 1 \\  \hline  0 &  & 2 \\  \hline 
\noalign{\vskip 2mm} \multispan{3} $d$=1 \\ \end{tabular}} \hspace{0.5cm}
\footnotesize{
\begin{tabular} {|c|c|c|c|c|} \hline $2j_L \backslash 2j_R$  & 0 & 1 & 2 & 3 \\  
\hline  0 &  &  &  & 1 \\  \hline 
\noalign{\vskip 2mm} \multispan{5} $d$=2 \\ \end{tabular}} \hspace{0.5cm}
\footnotesize{
\begin{tabular} {|c|c|c|c|c|c|c|} \hline $2j_L \backslash 2j_R$  & 0 & 1 & 2 & 3 & 4 & 5 \\  \hline  0 &  &  &  &  &  & 2 \\  \hline \noalign{\vskip 2mm} \multispan{7} $d$=3 \\ \end{tabular}} 
\end{center}
\end{table}
\begin{table}[H]
\begin{center}
{\footnotesize
\begin{tabular} {|c|c|c|c|c|c|c|c|c|c|} \hline $2j_L \backslash 2j_R$  & 0 & 1 & 2 & 3 & 4 & 5 & 6 & 7 & 8 \\  \hline  0 &  &  &  &  &  & 1 &  & 3 &  \\  \hline1 &  &  &  &  &  &  &  &  & 1 \\  \hline \end{tabular} \vskip 3pt  $d=4$ \vskip 15pt   \begin{tabular} {|c|c|c|c|c|c|c|c|c|c|c|c|c|} \hline $2j_L \backslash 2j_R$  & 0 & 1 & 2 & 3 & 4 & 5 & 6 & 7 & 8 & 9 & 10 & 11 \\  \hline  0 &  &  &  &  &  & 2 &  & 2 &  & 6 &  &  \\  \hline1 &  &  &  &  &  &  &  &  & 2 &  & 2 &  \\  \hline2 &  &  &  &  &  &  &  &  &  &  &  & 2 \\  \hline 
\end{tabular} \vskip 3pt  $d=5$ \vskip 15pt   \begin{tabular} {|c|c|c|c|c|c|c|c|c|c|c|c|c|c|c|c|c|} \hline $2j_L \backslash 2j_R$  & 0 & 1 & 2 & 3 & 4 & 5 & 6 & 7 & 8 & 9 & 10 & 11 & 12 & 13 & 14 & 15 \\  \hline  0 &  &  &  & 1 &  & 3 &  & 5 &  & 7 &  & 10 &  &  &  
&  \\  \hline1 &  &  &  &  &  &  & 1 &  & 4 &  & 5 &  & 7 &  & 1 &  \\  \hline2 &  &  &  &  &  &  &  &  &  & 1 &  & 4 &  & 5 &  &  \\  
\hline3 &  &  &  &  &  &  &  &  &  &  &  &  & 1 &  & 3 &  \\  \hline4 &  &  &  &  &  &  &  &  &  &  &  &  &  &  &  & 1 \\  \hline \end{tabular} \vskip 3pt  $d=6$ \vskip 15pt   \begin{tabular} {|c|c|c|c|c|c|c|c|c|c|c|c|c|c|c|c|c|c|c|c|c|} \hline $2j_L \backslash \
2j_R$  & 0 & 1 & 2 & 3 & 4 & 5 & 6 & 7 & 8 & 9 & 10 & 11 & 12 & 13 & 14 & 15 & 16 & 17 & 18 & 19 \\  \hline  0 &  & 2 &  & 2 &  & 8 &  & 10 &  & 18 &  & 16 &  & 22 &  & 2 &  & 2 &  &  \\  \hline1 &  &  &  &  & 2 &  & 4 &  & 10 &  & 14 &  & 20 &  & 18 &  & 4 &  &  &  \\  \hline2 &  &  &  &  &  &  &  & 2 &  & 4 &  & 12 &  & 14 &  & 18 &  & 2 &  &  \\  \hline3 &  &  &  &  &  &  &  &  &  &  & 2 &  & 4 &  & 10 &  & 10 &  & 2 &  \\  \hline4 &  &  &  &  &  &  &  &  &  &  &  &  &  & 2 &  & 4 &  & 8 &  &  \\  \hline5 &  &  &  &  &  &  &  &  &  &  & &  &  &  &  &  & 2 &  & 2 &  \\  \hline6 &  &  &  &  &  &  &  &  &  &  &  &  &  &  &  &  &  &  &  & 2 \\  \hline \end{tabular} \vskip 3pt  $d=7$ \vskip 10pt    
\caption{The GV invariants $n^{d}_{j_L,j_R}$ for $d=1,2,\cdots,7$ for the local  $\mathbb{P}^1\times\mathbb{P}^1$ model}}
\end{center}
\label{tableP1P1}
\end{table}

\subsection{The massless half K3} 
\begin{table}[H]
\begin{center} {\footnotesize 
\begin{tabular} {|c|c|} \hline $2j_L \backslash 2j_R$  & 0 \\  \hline  0 & 1 \\  \hline \noalign{\vskip 2mm} \multispan{2} $d$=0 \\ \end{tabular}} \hspace{0.5cm}
\footnotesize{
\begin{tabular} {|c|c|c|} \hline $2j_L \backslash 2j_R$  & 0 & 1 \\  
\hline  0 & 20 &  \\  \hline1 &  & 1 \\  \hline \noalign{\vskip 2mm} \multispan{3} $d$=1 \\ \end{tabular}} \hspace{0.5cm}
\footnotesize{\begin{tabular} {|c|c|c|c|} \hline $2j_L 
\backslash 2j_R$  & 0 & 1 & 2 \\  \hline  0 & 231 &  &  \\  \hline1 & 
 & 21 &  \\  \hline2 &  &  & 1 \\  \hline 
\noalign{\vskip 2mm} \multispan{4} $d$=2 \\ \end{tabular}} \hspace{0.0cm}
\end{center}
\end{table}
\begin{table}[H]
\begin{center}
\footnotesize{
 \begin{tabular} {|c|c|c|c|c|} \hline $2j_L 
\backslash 2j_R$  & 0 & 1 & 2 & 3 \\  \hline  0 & 1981 &  & 1 &  \\  
\hline1 &  & 252 &  &  \\  \hline2 & 1 &  & 21 &  \\  \hline3 &  &  & 
 & 1 \\  \hline \noalign{\vskip 2mm} \multispan{5} $d$=3 \\ \end{tabular}} \hspace{0.5cm}
\footnotesize{
\begin{tabular} {|c|c|c|c|c|c|} \hline $2j_L \backslash 2j_R$  & 0 & 
1 & 2 & 3 & 4 \\  \hline  0 & 13938 &  & 21 &  &  \\  \hline1 &  & 
2233 &  & 1 &  \\  \hline2 & 21 &  & 253 &  &  \\  \hline3 &  & 1 &  
& 21 &  \\  \hline4 &  &  &  &  & 1 \\  \hline \noalign{\vskip 2mm} \multispan{6} $d$=4 \\ \end{tabular}}
\end{center}
\end{table}

\begin{table}[H]
\begin{center} 
\footnotesize{
\begin{tabular} {|c|c|c|c|c|c|c|} \hline 
$2j_L \backslash 2j_R$  & 0 & 1 & 2 & 3 & 4 & 5 \\  \hline  0 & 84777 
&  & 253 &  &  &  \\  \hline1 &  & 16171 &  & 22 &  &  \\  \hline2 & 
253 &  & 2254 &  & 1 &  \\  \hline3 &  & 22 &  & 253 &  &  \\  
\hline4 &  &  & 1 &  & 21 &  \\  \hline5 &  &  &  &  &  & 1 \\  
\hline \end{tabular} \vskip 3pt  $d=5$ \vskip 15pt   \begin{tabular} 
{|c|c|c|c|c|c|c|c|} \hline $2j_L \backslash 2j_R$  & 0 & 1 & 2 & 3 & 
4 & 5 & 6 \\  \hline  0 & 460272 &  & 2254 &  & 1 &  &  \\  \hline1 & 
 & 100949 &  & 274 &  &  &  \\  \hline2 & 2254 &  & 16424 &  & 22 &  
&  \\  \hline3 &  & 274 &  & 2255 &  & 1 &  \\  \hline4 & 1 &  & 22 & 
 & 253 &  &  \\  \hline5 &  &  &  & 1 &  & 21 &  \\  \hline6 &  &  &  
&  &  &  & 1 \\  \hline \end{tabular} \vskip 3pt  $d=6$ \vskip 10pt  
}\caption{The refined Betti numbers $n^{d}_{j_L,j_R}$ for $d=0,1,\cdots,6$ for the K3 surface}
\label{tableBettiK3}
 \end{center}
\end{table}

\begin{table}[H]
\begin{center} {\footnotesize 
\begin{tabular} {|c|c|} \hline $2j_L \backslash 2j_R$  & 0 \\  \hline  0 & 1 \\  \hline \noalign{\vskip 2mm} \multispan{2} $d$=0 \\ \end{tabular}} \hspace{0.5cm}
{\footnotesize
 \begin{tabular} {|c|c|c|} \hline $2j_L \backslash 2j_R$  & 0 & 1 \\  
\hline  0 & 8 &  \\  \hline1 &  & 1 \\  \hline \noalign{\vskip 2mm} \multispan{3} $d$=1 \\ \end{tabular}} \hspace{0.5cm}
{\footnotesize 
\begin{tabular} {|c|c|c|c|} \hline $2j_L 
\backslash 2j_R$  & 0 & 1 & 2 \\  \hline  0 & 45 &  &  \\  \hline1 &  
& 9 &  \\  \hline2 &  &  & 1 \\  \hline 
\noalign{\vskip 2mm} \multispan{4} $d$=2 \\ \end{tabular}} 
 \end{center}
\end{table}

\begin{table}[H]
\begin{center} {\footnotesize 
\begin{tabular} {|c|c|c|c|c|} \hline $2j_L 
\backslash 2j_R$  & 0 & 1 & 2 & 3 \\  \hline  0 & 201 &  & 1 &  \\  
\hline1 &  & 54 &  &  \\  \hline2 & 1 &  & 9 &  \\  \hline3 &  &  &  
& 1 \\   \hline \noalign{\vskip 2mm} \multispan{5} $d$=3 \\ \end{tabular}} \hspace{0.5cm}
{\footnotesize    
\begin{tabular} {|c|c|c|c|c|c|} \hline $2j_L \backslash 2j_R$  & 0 & 
1 & 2 & 3 & 4 \\  \hline  0 & 781 &  & 9 &  &  \\  \hline1 &  & 255 & 
 & 1 &  \\  \hline2 & 9 &  & 55 &  &  \\  \hline3 &  & 1 &  & 9 &  \\ 
 \hline4 &  &  &  &  & 1 \\  \hline \noalign{\vskip 2mm} \multispan{5} $d$=4 \\ \end{tabular}} 
\end{center}
\end{table}

\begin{table}[H]
\begin{center}
{\footnotesize 
\begin{tabular} {|c|c|c|c|c|c|c|} \hline $2j_L 
\backslash 2j_R$  & 0 & 1 & 2 & 3 & 4 & 5 \\  \hline  0 & 2727 &  & 
55 &  &  &  \\  \hline1 &  & 1036 &  & 10 &  &  \\  \hline2 & 55 &  & 
264 &  & 1 &  \\  \hline3 &  & 10 &  & 55 &  &  \\  \hline4 &  &  & 1 
&  & 9 &  \\  \hline5 &  &  &  &  &  & 1 \\  \hline
\noalign{\vskip 2mm} \multispan{7} $d$=5 \\ \end{tabular}} \hspace{0.5cm}
{\footnotesize
\begin{tabular} {|c|c|c|c|c|c|c|c|} 
\hline $2j_L \backslash 2j_R$  & 0 & 1 & 2 & 3 & 4 & 5 & 6 \\  \hline 
 0 & 8785 &  & 264 &  & 1 &  &  \\  \hline1 &  & 3764 &  & 64 &  &  & 
 \\  \hline2 & 264 &  & 1091 &  & 10 &  &  \\  \hline3 &  & 64 &  & 
265 &  & 1 &  \\  \hline4 & 1 &  & 10 &  & 55 &  &  \\  \hline5 &  &  
&  & 1 &  & 9 &  \\  \hline6 &  &  &  &  &  &  & 1 \\  \hline
\noalign{\vskip 2mm} \multispan{8} $d$=6 \\ \end{tabular}} \vskip 10pt  
\caption{The refined Betti numbers $n^{d}_{j_L,j_R}$ for $d=0,1,\cdots,6$ for the half K3 surface}
\label{tableBettihalfK3}
\end{center}
\end{table}

\begin{table}[H]
\begin{center} {\footnotesize 
\begin{tabular} {|c|c|} \hline $2j_L \backslash 2j_R$  & 0 \\  \hline  0 & 1 \\  \hline \noalign{\vskip 2mm} \multispan{2} $d$=0 \\ \end{tabular}} \hspace{0.5cm}
{\footnotesize     \begin{tabular} {|c|c|c|} \hline $2j_L \backslash 2j_R$  & 0 & 1 \\  \hline  0 & 248 &  \\  \hline1 &  & 1 \\  \hline 
    \noalign{\vskip 2mm} \multispan{3} $d$=1 \\ \end{tabular}} \hspace{0.5cm}
{\footnotesize 
\begin{tabular} {|c|c|c|c|} \hline $2j_L \backslash 2j_R$  & 0 & 1 & 2 \\  \hline  0 & 4125 &  &  \\  \hline1 &  & 249 &  \\  \hline2 &  &  & 1 \\  \hline \noalign{\vskip 2mm} \multispan{4} $d$=2 \\ \end{tabular}} \hspace{0.5cm}
\end{center}
\end{table}

\begin{table}[H]
\begin{center}
{\footnotesize     \begin{tabular} {|c|c|c|c|c|} \hline $2j_L \backslash 2j_R$  & 0 & 1 & 2 & 3 \\  \hline  0 & 35001 &  & 1 &  \\  \hline1 &  & 4374 &  &  \\  \hline2 & 1 &  & 249 &  \\  \hline3 &  &  &  & 1 \\  \hline \noalign{\vskip 2mm} \multispan{5} $d$=3 \\ \end{tabular}} \hspace{0.5cm}
{\footnotesize      \begin{tabular} {|c|c|c|c|c|c|} \hline $2j_L \backslash 2j_R$  & 0 & 1 & 2 & 3 & 4 \\  \hline  0 & 217501 &  & 249 &  &  \\  \hline1 &  & 
39375 &  & 1 &  \\  \hline2 & 249 &  & 4375 &  &  \\  \hline3 &  & 1 &  & 249 &  \\  \hline4 &  &  &  &  & 1 \\  \hline \noalign{\vskip 2mm} \multispan{6} $d$=4 \\ \end{tabular}} 
\end{center}
\end{table}

\begin{table}[H]
\begin{center}
{\footnotesize      \begin{tabular} {|c|c|c|c|c|c|c|} \hline $2j_L \backslash 2j_R$  & 0 & 1 & 2 & 3 & 4 & 5 \\  \hline  0 & 1097127 &  & 4375 &  &  &  \\  \hline1 &  & 256876 &  & 250 &  &  \\  \hline2 & 4375 &  & 39624 &  & 1 &  \\  \hline3 &  & 250 &  & 4375 &  &  \\  \hline4 &  &  & 1 &  & 249 &  \\  \hline5 &  &  &  &  &  & 1 \\  \hline \end{tabular} \vskip 3pt  $d=5$ \vskip 15pt}
\end{center}
\end{table}

\begin{table}[H]
\begin{center}
{\footnotesize
\begin{tabular} {|c|c|c|c|c|c|c|c|} \hline $2j_L \backslash 2j_R$  & 0 & 1 & 2 & 3 & 4 & 5 & 6 \\  \hline  0 & 4791745 &  & 39624 &  & 1 &  &  \\  \hline1 &  & 1354004 &  & 4624 &  &  &  \\  \hline2 & 39624 &  & 261251 &  & 250 &  &  \\  \hline3 &  & 4624 &  & 39625 &  & 1 &  \\  \hline4 & 1 &  & 250 &  & 4375 &  &  \\  \hline5 &  &  &  & 1 &  & 249 &  \\  \hline6 &  &  &  &  &  &  & 1 \\  \hline \end{tabular} \vskip 3pt  $d=6$ \vskip 10pt   }
\caption{The GV invariants $n^{p+df}_{j_L,j_R}$ for $d=0,1,\cdots,6$ for the local half K3 model}
\label{tableB9}
 \end{center}
\end{table}

\begin{table}[H]
\begin{center} {\footnotesize 
 \begin{tabular} {|c|c|c|c|c|} \hline $2j_L \backslash 2j_R$  & 0 & 1 
& 2 & 3 \\  \hline  0 &  & 3876 &  &  \\  \hline1 &  &  & 248 &  \\  
\hline2 &  &  &  & 1 \\  \hline \noalign{\vskip 2mm} \multispan{5} $n_b$=2, 
$d$=2 \\ \end{tabular}} \hspace{0.5cm}
{\footnotesize     \begin{tabular} {|c|c|c|c|c|c|c|} \hline $2j_L 
\backslash 2j_R$  & 0 & 1 & 2 & 3 & 4 & 5 \\  \hline  0 &  & 186126 & 
 & 249 &  &  \\  \hline1 & 4124 &  & 38877 &  & 1 &  \\  \hline2 &  & 
249 &  & 4373 &  &  \\  \hline3 &  &  & 1 &  & 249 &  \\  \hline4 &  
&  &  &  &  & 1 \\   
\hline \noalign{\vskip 2mm} \multispan{5} $n_b$=2, $d$=3 \\ \end{tabular}}
\end{center}
\end{table}

\begin{table}[H]
\begin{center}
{\footnotesize
 \begin{tabular} {|c|c|c|c|c|c|c|c|c|} \hline $2j_L 
\backslash 2j_R$  & 0 & 1 & 2 & 3 & 4 & 5 & 6 & 7 \\  \hline  0 &  & 
3884370 &  & 39374 &  & 1 &  &  \\  \hline1 & 225003 &  & 1287378 &  
& 4623 &  &  &  \\  \hline2 &  & 43499 &  & 260503 &  & 250 &  &  \\  
\hline3 & 249 &  & 4623 &  & 39623 &  & 1 &  \\  \hline4 &  & 1 &  & 
250 &  & 4375 &  &  \\  \hline5 &  &  &  &  & 1 &  & 249 &  \\  
\hline6 &  &  &  &  &  &  &  & 1 \\  \hline \end{tabular} \vskip 3pt  
$n_b=2, d=4$ \vskip 15pt   \begin{tabular} {|c|c|c|c|c|c|c|c|c|c|c|} 
\hline $2j_L \backslash 2j_R$  & 0 & 1 & 2 & 3 & 4 & 5 & 6 & 7 & 8 & 
9 \\  \hline  0 &  & 52369748 &  & 1357878 &  & 4375 &  &  &  &  \\  
\hline1 & 5171499 &  & 22839873 &  & 300624 &  & 250 &  &  &  \\  
\hline2 &  & 1587254 &  & 6304873 &  & 44248 &  & 1 &  &  \\  \hline3 
& 39624 &  & 304749 &  & 1397006 &  & 4625 &  &  &  \\  \hline4 &  & 
4624 &  & 44248 &  & 261498 &  & 250 &  &  \\  \hline5 & 1 &  & 250 & 
 & 4625 &  & 39625 &  & 1 &  \\  \hline6 &  &  &  & 1 &  & 250 &  & 
4375 &  &  \\  \hline7 &  &  &  &  &  &  & 1 &  & 249 &  \\  \hline8 
&  &  &  &  &  &  &  &  &  & 1 \\  \hline \end{tabular} \vskip 3pt  
$n_b=2, d=5$ \vskip 10pt
}
\caption{The GV invariants $n^{n_bp+df}_{j_L,j_R}$ for $n_b=2$ and $d=2,3,4,5$ for the local half K3 model}
\label{tableB9n_b=2}
 \end{center}
\end{table}

\begin{table}[H]
\begin{center} {\footnotesize 
 \begin{tabular} {|c|c|c|c|c|c|c|c|} \hline $2j_L \backslash 2j_R$  & 
0 & 1 & 2 & 3 & 4 & 5 & 6 \\  \hline  0 & 30628 &  & 151374 &  & 248 
&  &  \\  \hline1 &  & 4124 &  & 34504 &  & 1 &  \\  \hline2 & 1 &  & 
248 &  & 4124 &  &  \\  \hline3 &  &  &  & 1 &  & 248 &  \\  \hline4 
&  &  &  &  &  &  & 1 \\  \hline \end{tabular} \vskip 3pt  $n_b=3, 
d=3$ \vskip 15pt   \begin{tabular} {|c|c|c|c|c|c|c|c|c|c|c|} \hline 
$2j_L \backslash 2j_R$  & 0 & 1 & 2 & 3 & 4 & 5 & 6 & 7 & 8 & 9 \\  
\hline  0 & 3694119 &  & 11393622 &  & 252004 &  & 249 &  &  &  \\  
\hline1 &  & 1434130 &  & 4880618 &  & 43498 &  & 1 &  &  \\  \hline2 
& 39125 &  & 295005 &  & 1286881 &  & 4623 &  &  &  \\  \hline3 &  & 
4622 &  & 43747 &  & 256377 &  & 250 &  &  \\  \hline4 & 1 &  & 250 & 
 & 4623 &  & 39374 &  & 1 &  \\  \hline5 &  &  &  & 1 &  & 250 &  & 
4374 &  &  \\  \hline6 &  &  &  &  &  &  & 1 &  & 249 &  \\  \hline7 
&  &  &  &  &  &  &  &  &  & 1 \\  \hline \end{tabular} \vskip 3pt  
$n_b=3, d=4$ \vskip 15pt   \begin{tabular} {|c|c|c|c|c|c|c|c|c|c|c|c|} \hline $2j_L \backslash 
2j_R$  & 0 & 1 & 2 & 3 & 4 & 5 & 6 & 7 & 8 & 9 & 10 \\  \hline  0 &  
& 3480992 &  & 7726504 &  & 212879 &  & 248 &  &  &  \\  \hline1 & 
185878 &  & 1209127 &  & 3632614 &  & 38876 &  & 1 &  &  \\  \hline2 
&  & 38876 &  & 251755 &  & 1030753 &  & 4373 &  &  &  \\  \hline3 & 
248 &  & 4373 &  & 39125 &  & 217003 &  & 249 &  &  \\  \hline4 &  & 
1 &  & 249 &  & 4373 &  & 35000 &  & 1 &  \\  \hline5 &  &  &  &  & 1 
&  & 249 &  & 4125 &  &  \\  \hline6 &  &  &  &  &  &  &  & 1 &  & 
248 &  \\  \hline7 &  &  &  &  &  &  &  &  &  &  & 1 \\  \hline 
\end{tabular} \vskip 3pt  $n_b=4, d=4$ \vskip 10pt 
}
\caption{The GV invariants $n^{n_bp+df}_{j_L,j_R}$ for $(n_b,d)=(3,3),(3,4),(4,4)$ for the local half K3 model}
\label{tableB9n_b=3,4}
 \end{center}
\end{table}

\subsection{The massive half K3} 
\begin{table}[H]
\begin{center} {\footnotesize 
\begin{tabular} {|c|c|c|} \hline $2j_L \backslash 2j_R$  & 0 & 1 \\  
\hline  0 &  & 1 \\  \hline \noalign{\vskip 1mm} \multispan{1} $\beta = (2p+2f, 
\mathcal{O}_{2,2160})$, \\ \noalign{\vskip 1mm} \multispan{1} $(2p+3f, 
\mathcal{O}_{4,17280})$,\\ \noalign{\vskip 1mm} \multispan{1} $(2p+4f, 
\mathcal{O}_{6,60480})$, \\ \noalign{\vskip 1mm} \multispan{1} $(2p+5f, 
\mathcal{O}_{8,138240})$ \end{tabular}} \hspace{0.5cm}
{\footnotesize 
\begin{tabular} {|c|c|c|c|} \hline $2j_L \backslash 2j_R$  & 0 & 1 & 
2 \\  \hline  0 &  & 7 &  \\  \hline1 &  &  & 1 \\  \hline 
\noalign{\vskip 1mm} \multispan{4} $\beta =(2p+2f, \mathcal{O}_{1,240})$, \\ \noalign{\vskip 1mm} \multispan{4} $(2p+3f, \mathcal{O}_{3,6720})$,\\ \noalign{\vskip 1mm} \multispan{4} $(2p+4f, \mathcal{O}_{5,30240})$, \\ \noalign{\vskip 1mm} \multispan{4} $(2p+5f, \mathcal{O}_{7,69120}) $, \\ \noalign{\vskip 1mm} \multispan{4} $(2p+5f, \mathcal{O}_{7,13440})$ \end{tabular}} \hspace{0.5cm}
{\footnotesize 
\begin{tabular} {|c|c|c|c|c|} \hline $2j_L \backslash 2j_R$  & 0 & 1 
& 2 & 3 \\  \hline  0 &  & 36 &  &  \\  \hline1 &  &  & 8 &  \\  
\hline2 &  &  &  & 1 \\  \hline
\noalign{\vskip 1mm} \multispan{5} $\beta 
=(2p+2f, \mathcal{O}_{0,1})$, \\
\noalign{\vskip 1mm} \multispan{5} $(2p+4f, \mathcal{O}_{4,240} )$ \end{tabular}}
\end{center}
\end{table}



\begin{table}[H]
\begin{center}{\footnotesize
\begin{tabular} {|c|c|c|c|c|} \hline $2j_L \backslash 2j_R$  & 0 & 1 
& 2 & 3 \\  \hline  0 &  & 38 &  &  \\  \hline1 & 1 &  & 9 &  \\  
\hline2 &  &  &  & 1 \\  \hline \noalign{\vskip 1mm} \multispan{5} $\beta 
=(2p+3f, \mathcal{O}_{2,2160})$, \\ \noalign{\vskip 1mm} \multispan{5} $(2p+4f, \mathcal{O}_{4,17280})$,\\ \noalign{\vskip 1mm} \multispan{5} $(2p+5f, \mathcal{O}_{6,60480})$ \end{tabular}} \hspace{0.5cm}
{\footnotesize
\begin{tabular} {|c|c|c|c|c|c|} \hline $2j_L \backslash 2j_R$  & 0 & 
1 & 2 & 3 & 4 \\  \hline  0 &  & 163 &  & 1 &  \\  \hline1 & 8 &  & 
52 &  &  \\  \hline2 &  & 1 &  & 9 &  \\  \hline3 &  &  &  &  & 1 \\  
\hline \noalign{\vskip 1mm} \multispan{6} $\beta =(2p+3f, 
\mathcal{O}_{1,240})$, \\ \noalign{\vskip 1mm} \multispan{6} $(2p+4f, 
\mathcal{O}_{3,6720})$,\\ \noalign{\vskip 1mm} \multispan{6} $(2p+5f, 
\mathcal{O}_{5,30240})$ \end{tabular}}
\end{center}
\end{table}

\begin{table}[H]
\begin{center} {\footnotesize 
 \begin{tabular} {|c|c|c|c|c|c|c|} \hline $2j_L \backslash 2j_R$  & 0 & 1 & 2 & 3 & 4 & 5 \\  \hline  0 &  & 606 &  & 9 &  &  \\  \hline1 & 44 &  & 237 &  & 1 &  \\  \hline2 &  & 9 &  & 53 &  &  \\  \hline3 &  &  & 1 &  & 9 &  \\  \hline4 &  &  &  &  &  & 1 \\  \hline \noalign{\vskip 1mm} \multispan{7} $\beta =(2p+3f, \mathcal{O}_{0,1}),~ (2p+5f, \mathcal{O}_{4,240})$ \end{tabular}} \hspace{0.5cm}
{\footnotesize  
 \begin{tabular} {|c|c|c|c|c|c|c|} \hline $2j_L \backslash 2j_R$  & 0 
& 1 & 2 & 3 & 4 & 5 \\  \hline  0 &  & 619 &  & 9 &  &  \\  \hline1 & 
47 &  & 240 &  & 1 &  \\  \hline2 &  & 10 &  & 55 &  &  \\  \hline3 & 
 &  & 1 &  & 9 &  \\  \hline4 &  &  &  &  &  & 1 \\  \hline 
\noalign{\vskip 1mm} \multispan{7}  $\beta =(2p+4f, \mathcal{O}_{2,2160}), ~ (2p+5f, \mathcal{O}_{4,17280})$ \end{tabular}}
\caption{The GV invariants $n^{\beta}_{j_L,j_R}$ for the classes $\beta=(n_bp+df,\mathcal{O}_{p,k}) $ with $n_b=2$ and $d\leq 5$ for the massive local half K3 model}
\label{tablehalfK3massive1}
\end{center}
\end{table}

\begin{table}[H]
\begin{center} {\footnotesize
\begin{tabular} {|c|c|c|c|c|c|c|c|} \hline $2j_L \backslash 2j_R$  & 
0 & 1 & 2 & 3 & 4 & 5 & 6 \\  \hline  0 &  & 2116 &  & 54 &  &  &  \\ 
 \hline1 & 215 &  & 952 &  & 10 &  &  \\  \hline2 &  & 62 &  & 261 &  
& 1 &  \\  \hline3 & 1 &  & 10 &  & 55 &  &  \\  \hline4 &  &  &  & 1 
&  & 9 &  \\  \hline5 &  &  &  &  &  &  & 1 \\  \hline \end{tabular} 
\vskip 8pt  $\beta =(2p+4f, \mathcal{O}_{1,240}), ~ (2p+5f, \mathcal{O}_{3,6720} )$ \vskip 10pt} \end{center}
\end{table}
\begin{table}[H]
\begin{center}
{\footnotesize
 \begin{tabular} {|c|c|c|c|c|c|c|c|c|} \hline $2j_L \backslash 2j_R$  & 0 & 1 & 2 & 3 & 4 & 5 & 6 & 7 \\  
 \hline  0 &  & 6690 &  & 254 &  & 1 &  &  \\  \hline1 & 843 &  & 3378 &  & 63 &  &  &  \\  \hline2 &  & 299 &  & 1063 &  & 10 &  &  \\  \hline3 & 9 &  & 63 &  & 263 &  & 1 &  \\  \hline4 &  & 1 &  & 10 &  & 55 &  &  \\  \hline5 &  &  &  &  & 1 &  & 9 &  \\  \hline6 &  &  &  &  &  &  &  & 1 \\  \hline \end{tabular} \vskip 8pt  $\beta =(2p+4f, \mathcal{O}_{0,1})$ 
} \end{center}
\end{table}
\begin{table}[H]
\begin{center}
{\footnotesize  
  \begin{tabular} {|c|c|c|c|c|c|c|c|c|} \hline $2j_L \backslash 2j_R$  
& 0 & 1 & 2 & 3 & 4 & 5 & 6 & 7 \\  \hline  0 &  & 6717 &  & 256 &  & 
1 &  &  \\  \hline1 & 859 &  & 3395 &  & 64 &  &  &  \\  \hline2 &  & 
304 &  & 1068 &  & 10 &  &  \\  \hline3 & 9 &  & 65 &  & 265 &  & 1 & 
 \\  \hline4 &  & 1 &  & 10 &  & 55 &  &  \\  \hline5 &  &  &  &  & 1 
&  & 9 &  \\  \hline6 &  &  &  &  &  &  &  & 1 \\  \hline 
\end{tabular} \vskip 8pt  $\beta =(2p+5f, \mathcal{O}_{2,2160})$ 
} \end{center}
\end{table}
\begin{table}[H]
\begin{center}
{\footnotesize
\begin{tabular} {|c|c|c|c|c|c|c|c|c|c|} \hline $2j_L \backslash 2j_R$ 
 & 0 & 1 & 2 & 3 & 4 & 5 & 6 & 7 & 8 \\  \hline  0 &  & 19999 &  & 
1043 &  & 9 &  &  &  \\  \hline1 & 3067 &  & 11132 &  & 318 &  & 1 &  
&  \\  \hline2 &  & 1267 &  & 3897 &  & 65 &  &  &  \\  \hline3 & 55 
&  & 326 &  & 1096 &  & 10 &  &  \\  \hline4 &  & 10 &  & 65 &  & 265 
&  & 1 &  \\  \hline5 &  &  & 1 &  & 10 &  & 55 &  &  \\  \hline6 &  
&  &  &  &  & 1 &  & 9 &  \\  \hline7 &  &  &  &  &  &  &  &  & 1 \\  
\hline \end{tabular} \vskip 8pt  $\beta =(2p+5f, 
\mathcal{O}_{1,240})$ 
} \end{center}
\end{table}
\begin{table}[H]
\begin{center}
{\footnotesize
 \begin{tabular} {|c|c|c|c|c|c|c|c|c|c|c|} \hline $2j_L \backslash 
2j_R$  & 0 & 1 & 2 & 3 & 4 & 5 & 6 & 7 & 8 & 9 \\  \hline  0 &  & 
56468 &  & 3798 &  & 55 &  &  &  &  \\  \hline1 & 10059 &  & 34113 &  
& 1344 &  & 10 &  &  &  \\  \hline2 &  & 4694 &  & 13033 &  & 328 &  
& 1 &  &  \\  \hline3 & 264 &  & 1389 &  & 4046 &  & 65 &  &  &  \\  
\hline4 &  & 64 &  & 328 &  & 1098 &  & 10 &  &  \\  \hline5 & 1 &  & 
10 &  & 65 &  & 265 &  & 1 &  \\  \hline6 &  &  &  & 1 &  & 10 &  & 
55 &  &  \\  \hline7 &  &  &  &  &  &  & 1 &  & 9 &  \\  \hline8 &  & 
 &  &  &  &  &  &  &  & 1 \\  \hline \end{tabular} \vskip 8pt  $\beta 
=(2p+5f, \mathcal{O}_{0,1})$ 
}
\caption{The GV invariants $n^{\beta}_{j_L,j_R}$ continued from table \ref{tablehalfK3massive1}}
\label{tablehalfK3massive2}
 \end{center}
\end{table}

\begin{table}[H]
\begin{center}
{\footnotesize \begin{tabular} {|c|c|c|c|} \hline $2j_L \backslash 2j_R$  & 0 & 1 & 
2 \\  \hline  0 &  &  & 1 \\  \hline \noalign{\vskip 1mm} \multispan{4}  $\beta 
=(3p+3f, \mathcal{O}_{4,17280})$ \\ \noalign{\vskip 1mm} \multispan{4}  $(3p+4f, \mathcal{O}_{7,69120})$ \\ \noalign{\vskip 1mm} \multispan{4}  $(3p+5f, \mathcal{O}_{10,241920} )$ \end{tabular}} \hspace{0.5cm} 
{\footnotesize \begin{tabular} {|c|c|c|c|c|} \hline $2j_L \backslash 2j_R$  & 0 & 1 
& 2 & 3 \\  \hline  0 & 1 &  & 6 &  \\  \hline1 &  &  &  & 1 \\  
\hline \noalign{\vskip 1mm} \multispan{5}  $\beta =(3p+3f, \mathcal{O}_{3,6720})$ \\ \noalign{\vskip 1mm} \multispan{5}  $(3p+4f, \mathcal{O}_{6,60480})$ \\ \noalign{\vskip 1mm} \multispan{5}  $(3p+5f, \mathcal{O}_{9,181440})$ \end{tabular}} \hspace{0.5cm}
{\footnotesize
\begin{tabular} {|c|c|c|c|c|c|} \hline $2j_L \backslash 2j_R$  & 0 & 
1 & 2 & 3 & 4 \\  \hline  0 & 7 &  & 30 &  &  \\  \hline1 &  & 1 &  & 
8 &  \\  \hline2 &  &  &  &  & 1 \\  \hline \noalign{\vskip 1mm} \multispan{6}  $\beta =(3p+3f, \mathcal{O}_{2,2160})$ \\ \noalign{\vskip 1mm} \multispan{6}  $(3p+4f, \mathcal{O}_{5,30240})$ \\ \noalign{\vskip 1mm} \multispan{6}  $(3p+5f, \mathcal{O}_{8,2160})$ \\ \noalign{\vskip 1mm} \multispan{6}  $(3p+5f, \mathcal{O}_{8,138240})$ \end{tabular}}
\end{center}
\end{table}

\begin{table}[H]
\begin{center}
{\footnotesize
\begin{tabular} {|c|c|c|c|c|c|c|} \hline $2j_L \backslash 2j_R$  & 0 
& 1 & 2 & 3 & 4 & 5 \\  \hline  0 & 36 &  & 119 &  & 1 &  \\  \hline1 
&  & 8 &  & 43 &  &  \\  \hline2 &  &  & 1 &  & 8 &  \\  \hline3 &  & 
 &  &  &  & 1 \\  \hline \noalign{\vskip 1mm} \multispan{7}  $\beta =(3p+3f, \mathcal{O}_{1,240}),
 ~ (3p+4f, \mathcal{O}_{4,240})$, \\ \noalign{\vskip 1mm} \multispan{7}  $(3p+5f, \mathcal{O}_{7,13440} )$ \end{tabular}} \hspace{0.5cm} 
{\footnotesize
\begin{tabular} {|c|c|c|c|c|c|c|} \hline $2j_L \backslash 2j_R$  & 0 
& 1 & 2 & 3 & 4 & 5 \\  \hline  0 & 37 &  & 129 &  & 1 &  \\  \hline1 
&  & 10 &  & 46 &  &  \\  \hline2 &  &  & 1 &  & 9 &  \\  \hline3 &  
&  &  &  &  & 1 \\  \hline \noalign{\vskip 1mm} \multispan{7} $\beta 
=(3p+4f, \mathcal{O}_{4,17280}), ~ (3p+5f, \mathcal{O}_{7,69120} )$ \end{tabular}}
\end{center}
\end{table}

\begin{table}[H]
\begin{center}
{\footnotesize
\begin{tabular} {|c|c|c|c|c|c|c|c|} \hline $2j_L \backslash 2j_R$  & 
0 & 1 & 2 & 3 & 4 & 5 & 6 \\  \hline  0 & 148 &  & 414 &  & 8 &  &  
\\  \hline1 &  & 44 &  & 184 &  & 1 &  \\  \hline2 & 1 &  & 8 &  & 44 
&  &  \\  \hline3 &  &  &  & 1 &  & 8 &  \\  \hline4 &  &  &  &  &  & 
 & 1 \\  \hline \noalign{\vskip 1mm} \multispan{7} $\beta =(3p+3f, 
\mathcal{O}_{0,1})$ \end{tabular}} \hspace{0.5cm} 
{\footnotesize
\begin{tabular} {|c|c|c|c|c|c|c|c|} \hline $2j_L \backslash 2j_R$  & 
0 & 1 & 2 & 3 & 4 & 5 & 6 \\  \hline  0 & 156 &  & 473 &  & 9 &  &  
\\  \hline1 &  & 58 &  & 205 &  & 1 &  \\  \hline2 & 1 &  & 10 &  & 
52 &  &  \\  \hline3 &  &  &  & 1 &  & 9 &  \\  \hline4 &  &  &  &  & 
 &  & 1 \\  \hline \noalign{\vskip 1mm} \multispan{8} $\beta =(3p+4f, \mathcal{O}_{3,6720}),
 ~(3p+5f, \mathcal{O}_{6,60480}) $ \end{tabular}}
\caption{The GV invariants $n^{\beta}_{j_L,j_R}$ for some classes $\beta=(3p+df,\mathcal{O}_{p,k})$ with $d\leq 5$ for the massive local half K3 model}
\label{massivenb=3}
\end{center}
\end{table}

\begin{table}[H]
\begin{center}
{\footnotesize
\begin{tabular} {|c|c|c|c|c|} \hline $2j_L \backslash 2j_R$  & 0 & 1 
& 2 & 3 \\  \hline  0 &  &  &  & 1 \\  \hline \noalign{\vskip 1mm} \multispan{5}  $\beta =(4p+4f, \mathcal{O}_{7,69120})$, \\ \noalign{\vskip 1mm} \multispan{5}  $(4p+5f, \mathcal{O}_{11,138240})$ \end{tabular}} \hspace{0.4cm} 
{\footnotesize
\begin{tabular} {|c|c|c|c|c|c|} \hline $2j_L \backslash 2j_R$  & 0 & 
1 & 2 & 3 & 4 \\  \hline  0 &  & 1 &  & 5 &  \\  \hline1 &  &  &  &  
& 1 \\  \hline \noalign{\vskip 1mm} \multispan{6}  $\beta =(4p+4f, \mathcal{O}_{6,60480})$, \\ \noalign{\vskip 1mm} \multispan{6}  $(4p+5f, \mathcal{O}_{10,241920})$ \end{tabular}} \hspace{0.4cm} 
{\footnotesize
\begin{tabular} {|c|c|c|c|c|c|c|} \hline $2j_L \backslash 2j_R$  & 0 
& 1 & 2 & 3 & 4 & 5 \\  \hline  0 &  & 7 &  & 23 &  &  \\  \hline1 &  
&  & 1 &  & 7 &  \\  \hline2 &  &  &  &  &  & 1 \\  \hline \noalign{\vskip 1mm} \multispan{6}
$\beta =(4p+4f, \mathcal{O}_{5,30240})$, \\ \noalign{\vskip 1mm} \multispan{6}  $(4p+5f, \mathcal{O}_{9,181440})$ \end{tabular}}
\end{center}
\end{table}

\begin{table}[H]
\begin{center}
{\footnotesize
\begin{tabular} {|c|c|c|c|c|c|c|c|} \hline $2j_L \backslash 2j_R$  & 
0 & 1 & 2 & 3 & 4 & 5 & 6 \\  \hline  0 &  & 35 &  & 84 &  & 1 &  \\  
\hline1 &  &  & 8 &  & 35 &  &  \\  \hline2 &  &  &  & 1 &  & 7 &  \\ 
 \hline3 &  &  &  &  &  &  & 1 \\  \hline \noalign{\vskip 1mm} \multispan{8}  $
\beta =(4p+4f, \mathcal{O}_{4,240})$ \end{tabular}} \hspace{0.5cm} 
{\footnotesize
\begin{tabular} {|c|c|c|c|c|c|c|c|} \hline $2j_L \backslash 2j_R$  & 
0 & 1 & 2 & 3 & 4 & 5 & 6 \\  \hline  0 &  & 36 &  & 92 &  & 1 &  \\  
\hline1 & 1 &  & 9 &  & 37 &  &  \\  \hline2 &  &  &  & 1 &  & 8 &  
\\  \hline3 &  &  &  &  &  &  & 1 \\  \hline \noalign{\vskip 1mm} \multispan{8} $\beta =(4p+4f, \mathcal{O}_{4,17280}),  
 ~(4p+5f, \mathcal{O}_{8,138240})$ \end{tabular}}
\end{center}
\end{table}

\begin{table}[H]
\begin{center}
{\footnotesize
\begin{tabular} {|c|c|c|c|c|c|c|c|} \hline $2j_L \backslash 2j_R$  & 
0 & 1 & 2 & 3 & 4 & 5 & 6 \\  \hline  0 &  & 37 &  & 102 &  &  &  \\  
\hline1 & 1 &  & 9 &  & 38 &  &  \\  \hline2 &  &  &  & 1 &  & 9 &  
\\  \hline3 &  &  &  &  &  &  & 1 \\  \hline  \noalign{\vskip 1mm} \multispan{8}  $\beta =(4p+5f, \mathcal{O}_{8,2160})$ \end{tabular}} \hspace{0.5cm} 
{\footnotesize
\begin{tabular} {|c|c|c|c|c|c|c|c|c|} \hline $2j_L \backslash 2j_R$  
  & 0 & 1 & 2 & 3 & 4 & 5 & 6 & 7 \\  \hline  0 &  & 148 &  & 318 &  
  & 8 &  &  \\  \hline1 & 7 &  & 50 &  & 154 &  & 1 &  \\  \hline2 &  & 1
   &  & 9 &  & 43 &  &  \\  \hline3 &  &  &  &  & 1 &  & 8 &  \\  \hline4 &  
   &  &  &  &  &  &  & 1 \\  \hline \noalign{\vskip 1mm} \multispan{9} $\beta =(4p+4f, \mathcal{O}_{3,6720}), ~ (4p+5f, \mathcal{O}_{7,13440})$ \end{tabular}}
\caption{The GV invariants $n^{\beta}_{j_L,j_R}$ for some classes $\beta=(4p+df,\mathcal{O}_{p,k})$ with $d=4,5$ for the massive local half K3 model}
\label{massivenb=4}
\end{center}
\end{table}

\end{document}